\documentclass{elsarticle}

\usepackage{graphicx}
\usepackage{amssymb}
\usepackage{amsmath}
\usepackage{float}
\usepackage{color}

\def\be{\begin{equation}}
\def\ee{\end{equation}}
\def\bea{\begin{eqnarray}}
\def\eea{\end{eqnarray}}

\def\nn{\nonumber}
\def\nnn{\nonumber  \\}

\def\ma{\mathbf{a}}
\def\map{\mathbf{a'}}
\def\mb{\mathbf{b}}
\def\mbp{\mathbf{b'}}
\def\si{\sum \hspace{-15pt} \int}

\journal{Physics Reports}
\graphicspath{{fig/}}
\begin{document}

\begin{frontmatter}

\title{Linear response of homogeneous nuclear matter with energy density functionals}

\author[1]{A. Pastore}
\address[1]{Institut d'Astronomie et d'Astrophysique, CP 226, Universit\'e Libre de Bruxelles,
B-1050 Bruxelles, Belgium }

\author[2]{D. Davesne}
\address[2]{Institut de Physique Nucl\'eaire de Lyon, CNRS-IN2P3, UMR 5822, Universit\'e Lyon 1, F-69622 Villeurbanne, France}

\author[3]{J. Navarro}
\address[3]{IFIC (CSIC University of Valencia), Apdo. Postal 22085, E-46071 Valencia}

\begin{abstract}
Response functions of infinite nuclear matter with arbitrary isospin asymmetry are studied in the framework of the random phase approximation. The residual interaction is derived from a general nuclear Skyrme energy density functional. Besides the usual central, spin-orbit and tensor terms it could also include other components as new density-dependent terms or three-body terms. Algebraic expressions for the response functions are obtained from the Bethe-Salpeter equation for the particle-hole propagator. 
Applications to symmetric nuclear matter, pure neutron matter and asymmetric nuclear matter are presented and discussed. Spin-isospin strength functions are analyzed for varying conditions of density, momentum transfer, isospin asymmetry, and temperature for some representative Skyrme functionals. 
Particular attention is paid to the discussion of instabilities, either real or unphysical,  which could manifest in finite nuclei.
\end{abstract}


\end{frontmatter}

\tableofcontents


\section{Introduction}\label{introgenerale}

Innovative experiments performed at radioactive ion beam facilities are continuously revealing properties of nuclei closer and closer to the limit of stability imposed by the nuclear strong force. Such experimental data represent a fundamental test for presently existing theories. However there is not yet a single model able to describe the entire nuclear chart. In the low-mass region, {\em ab-initio} few-body methods provide an accurate reproduction and prediction of many
nuclear observables. Recent progress in this domain has allowed to extend the range of validity of the model up to the medium-mass region~\cite{hag10,hol12,wie13,som13}. Large-scale semi-microscopic \emph{shell-model} calculations can also be performed for medium-heavy nuclei in the vicinity of shell closure~\cite{hjo95,bro98,cau05}.
Nevertheless, for the medium and heavy mass region, the tool of choice is the Nuclear Energy Density Functional theory (NEDF) which allows a systematic and quantitative description of the properties of atomic nuclei  from drip-line to drip-line \cite{ben03}. These properties include ground state binding energies, radii and one-body observables, as well as low-energy spectroscopy, fission barriers and decay-probabilities. 

Several families of functionals are nowadays available each one derived according to different hypothesis, either relativistic, non-relativistic or derived from an effective pseudo-potential. Among the non-relativistic ones, the most widely used are those related to the zero-range non-local Skyrme interaction, whose {\em standard} structure was established by Vautherin and Brink \cite{vau72}. It contains central and spin-orbit parts, as well as a density-dependent term, and relies on a limited number of parameters. They are determined by fitting some selected experimental properties of finite nuclei as well as some homogeneous nuclear matter properties deduced from the empirical mass-formula or from microscopic calculations based on bare { two- and three-body interactions~\cite{akm98,zho04,heb10}.} 
In practice, due to the limitations of the symmetries of the wave functions, the self-consistent mean-field calculations based on such effective interactions have been restricted for a long time to ground-state properties of even-even nuclei. Nowadays nearly symmetry-unrestricted calculations are possible~\cite{bal14}, and the  method has been tested with success for instance in studies of rotational bands of heavy nuclei{~\cite{gal94,shi12}} or in systematic calculations of superheavy nuclei {\cite{rut97,typ03,cwi05,erl11}}. { A very interesting result of the NEDF has been discussed in Ref.~\cite{erl12}: despite large differences in various optimization procedures, current Skyrme functionals give a rather consistent picture about the position of the proton and neutron drip-lines. For a more detailed discussion on the predictive power and theoretical uncertainties of NEDF models, we refer to Ref.~\cite{dob14G}. }

A natural extension of this phenomenological approach is the study of excitations, including giant resonances,
by means of the random phase approximation (RPA) \cite{pin66,fet71,lip08,suh07}. 
Its basic ingredients are the particle-hole 
propagator or Green's function, whose poles are the energies of the excited states, and the particle-hole interaction. 
This gives the response function of the 
system to an external field, out of which one can obtain quantities as the transition strengths or the inelastic scattering cross sections for different probes of interest.  
Well-known examples of physical studies that require the knowledge of the response functions are the electron scattering by nuclei or the propagation of neutrinos in nuclear matter.
In the theory of Fermi liquids, the particle-hole interaction $V_{ph}$, or residual interaction between quasiparticles,
 is uniquely defined as the second functional derivative of the energy functional \cite{mig67,bro71,ber75}. 
With the Skyrme-type functionals, $V_{ph}$ turns out to be a contact interaction in coordinate space, 
but it is density-dependent as well as velocity-dependent. By summing the series of bubble diagrams, 
one obtains the RPA linear response to an external field. 
It is worth noting that this RPA response is self-consistent, in the sense that the mean field and the residual interaction $V_{ph}$ are derived from the same energy density functional. Detailed RPA calculations \cite{ber75,giai83} have shown 
that this approach can give a good description of low-lying collective states as well as giant resonances 
in finite nuclei. Furthermore, it enables one to relate the RPA results to the more global quantities 
which can be obtained by sum rule methods.

The study of quantum systems having a large or infinite number of fermions is of interest in many physical situations, 
involving fields as diverse as quantum chemistry, atomic physics, condensed matter or nuclear physics.
 Liquid $^3$ He or the conduction electrons in metals are familiar examples of these systems 
 (see references \cite{pin66,fet71,lip08,dic08} for a non-exhaustive account of them). 
Infinite nuclear matter is just an idealized system, which cannot be experimentally studied in the laboratory. It is nevertheless a useful and broadly used concept, because of its connection with the inner part of atomic nuclei and also with some regions of compact stars. Due to its relative simplicity, it is a very useful testing ground for various theories. It is therefore important to have a description of infinite nuclear matter based on nucleon-nucleon interactions. 
Concerning the response function,  the rather simple analytic form of the Skyrme energy density functional and of its related $V_{ph}$ allows 
one to obtain closed form expressions for the RPA response function, as shown in Ref. \cite{gar92} for the case of symmetric nuclear matter. The formalism on Ref. \cite{gar92} has been extended to finite temperature \cite{bra95,her96}, asymmetric nuclear matter \cite{her97,bra00} including the isospin-flip case \cite{her99}. It has also been used to study 
neutrino propagation in hot neutron matter \cite{nav99} and symmetry energy of nuclear matter \cite{bra05}. 
All the mentioned applications were restricted to the central part of the effective interaction. The effect of the spin-orbit
part on the response function was shown to be rather small \cite{mar06}, vanishing in the limit of zero transferred momentum. 
In the past decade several studies pointed out that the standard Skyrme interaction is not flexible enough to improve the level of accuracy in describing sets of available experimental data \cite{kor13}. 
In particular it was pointed out that the origin of magic numbers in exotic nuclei is connected to the spin-isospin dependent
terms of the interaction \cite{ots01}. 
The original form of the Skyrme interaction~\cite{sky59} contains other terms, including zero-range tensor terms, which are relevant for the spin and spin-isospin channels. In fact, the bare nucleon-nucleon interaction contains an important tensor part, necessary to reproduce not only  the phase shifts of the nucleon-nucleon scattering, but also the quadrupole moment of the deuteron. However, apart from some exploratory studies~\cite{sta77}, these terms have been omitted in the study of finite nuclei up to recently.

Nowadays, there are several zero- and finite-range effective interactions containing tensor terms. Among the finite-range ones, let us mention the M3Y type effective interaction of Nakada \cite{nak03}, which includes finite-range spin-orbit and tensor terms, whose parameters are fitted to the Brueckner's $G-$matrix based on a bare interaction. 
Much closer to the spirit of an effective interaction, in \cite{ots06}, a tensor term was added to a Gogny-type interaction to describe the evolution of nuclear shells for exotic nuclei as well as stable ones. In a similar viewpoint, the authors of \cite{ang11} have also included a tensor term into Gogny interactions. Finally, it is worth mentioning the recent results on a new form of pseudo-potential presented in Ref.~\cite{rai14}. Concerning Skyrme interactions, zero-range tensor terms as originally proposed by Skyrme \cite{sky59} have been either included perturbatively to existing central ones  \cite{bri07,col07,bai09,bai09a}, or with a complete refit of the parameters \cite{ton83,liu91,bro06,les07}. For a general discussion on the impact of tensor terms on ground state properties of finite nuclei see Refs.~\cite{les07,ben09B}. 

The effects of a zero-range tensor interaction on the RPA response function of symmetric nuclear matter have been analyzed in \cite{dav09}. As compared with the central case, the structure of the response function becomes more involved although they look formally the same. The complexity is related to the fact that the tensor interaction couples all channels in a nontrivial way. At least for the T44 interaction used in \cite{dav09}, the effects of the tensor contributions are strong in the vector channels. 
The formalism of linear response theory has also been applied to the case of a Skyrme energy density functional, both for symmetric nuclear matter \cite{pas12} and for pure neutron matter \cite{pas12a}. 
Although several functionals are available \cite{rob10,hup11,car08} the most often used is the one derived from an effective Skyrme interaction \cite{per04}. The interest of using energy density functionals resides in the possibility of including other  interaction or correlation terms guided by physical motivations, which are not easy to consider when dealing with an effective interaction. A significant effort is nowadays devoted to the introduction of new terms in the Skyrme NEDF, as the tensor \cite{les07}, new density dependent couplings \cite{cha09,mar09c}, higher order derivative terms \cite{car08,bec14} or three-body interactions \cite{sad11,sad12,sad13}. Hopefully, 
the present formalism can be easily extended to include  these new terms in the calculation of the response.

The effective interactions are reasonably well controlled around the saturation density $\rho_0$ of symmetric 
nuclear matter, for moderate isospin asymmetries and zero temperature. 
They have also been extrapolated to conditions of density and isospin 
asymmetry which are not experimentally accessible, as for instance nuclear matter with a large neutron 
to proton ratio or pure neutron matter, both at densities up to several times $\rho_0$ and finite temperatures. 
However, such an extrapolation can lead to unphysical instabilities  of nuclear matter.
For instance, most Skyrme parametrizations predict that the isospin asymmetry
energy $\varepsilon_I$ becomes negative when the density is increased~\cite{sam04,cen09,dan09,roc11,rep13,blo13}. Consequently, the symmetric system would be unstable at some density beyond the saturation one, preferring a largely asymmetric system made by an excess of either protons or neutrons. Another type of instability refers to the magnetic properties of neutron matter. Most Skyrme interactions predict that even in the absence of a 
magnetic field a spontaneous magnetization arises in pure neutron  matter at some critical density,  
which can be as low as $\simeq 1.1\rho_0$,~\cite{vid84,kut89,isa04,rio05} depending on the specific parameterization. 
The magnetic instability is related to the Landau parameter in the spin channel.  However, Gogny-type effective interaction calculations \cite{lop06}, relativistic mean-field calculations \cite{nie90,nie91,ber95} as well as Monte Carlo simulations \cite{fan01} and Brueckner Hartree-Fock calculations \cite{vid02,vid02a} exclude such an instability, at least at densities up to $(5 -6)\, \rho_0$, thus indicating that the predicted ferromagnetic instability is unphysical. It is thus important to discern whether or not it is relevant for finite nuclei.

As is well-known~\cite{bay91}, the Landau parameters must satisfy a set of  inequalities to insure that the energy of the system has a minimum stable against small fluctuations. 
A general study of the Skyrme interactions in terms of their Landau parameters was done in Ref.~\cite{mar02} for both symmetric nuclear matter and pure neutron matter. It was shown that the existence of a critical density is inherent to the Skyrme interactions, although it could be possible to construct a Skyrme interaction with a critical density 
 value around $(3.5 - 4)\, \rho_0$ for a reasonable choice of empirical inputs. However the inequalities related to the Landau parameters are not considered in the usual fit of parameters, and the value of the critical density is usually closer to $\rho_0$. 
The tensor interaction produces a deformation of the Fermi surface \cite{dab76}, and the stability of the system results in a new set of inequalities to be fulfilled by the Landau parameters \cite{bac79}. 
The spin and spin-isospin instabilities of  symmetric nuclear matter have been analyzed in Ref.~\cite{cao10} for several Skyrme interactions with tensor components. 

A description of the nuclear properties based on the Landau parameters is reliable in the zero frequency and long wavelength limits, when the quasiparticles interact near the Fermi surface. 
However, other instabilities could occur at nonzero transferred momentum $q$, 
with the appearance of domains with typical size $\lambda \simeq 2 \pi/q$. The first example of such kind
of instability was encountered and examined in detail by
Lesinski \emph{et al.}~\cite{les06}. In a study of the effective mass splitting in the scalar-isoscalar channel of doubly magic nuclei, 
it has been shown that the system converges towards an unphysical
configuration where protons are separated from neutrons. This observation has also been confirmed by RPA calculations in
finite nuclei \cite{ter06}. Another recent example of instability was found in the vector 
channel of several Skyrme functionals \cite{fra12,hel12,sch10b} . They have shown that, for particular values of the time-odd coupling constants, the system can spontaneously polarize.

To improve the existing functionals, it is therefore mandatory to find a tool which is able to detect these instabilities
in all scalar (vector) and isoscalar (isovector) channels. It has already been demonstrated by Lesinski \emph{et al.} \cite{les06} 
that the linear response formalism applied to the Skyrme energy density functional could be used to predict the appearance 
of some finite-size instabilities in nuclei. However, only the central part of the Skyrme interaction was taken into account 
for the building of the linear response. The same formalism for the case of a Skyrme interaction including tensor and 
spin-orbit terms was studied by Davesne \emph{et al.} \cite{dav09}, and afterwards extended to a general Skyrme functional. In particular is was shown that a pole in the response function corresponds to
a zero in the denominator of the inverse energy-weighted sum rule, which is easily written in terms of the coupling constants of the energy density functional~\cite{hel13}. This greatly simplifies the process of pole detection since one just 
has to find the roots of a real function. 

The purpose of the present work is to present and discuss a general formalism to derive the full linear response for 
homogeneous nuclear systems described by means of a general Skyrme energy density functional. It contains of course spin-orbit and tensor terms, but could also include other components such as three- or four-body terms.
The formalism includes temperature in a natural way. Applications to 
symmetric nuclear matter, pure neutron matter and asymmetric nuclear matter are presented and discussed. The calculations are done for varying densities, temperatures and transferred momenta. Among the existing NEDF we have selected a few of them to illustrate the different descriptions encountered to describe the response functions, and in particular the presence of instabilities which could manifest in the calculation of finite nuclei. Some attention will also be paid to the simplified case in which the particle-hole interaction is described in terms of Landau parameters. The effect of currently considered extra terms is also discussed.

In Sec.\ref{Sec:formalism}, we present the basic formalism of the linear response theory and the symbolic method adopted to calculate the strength function of the system. In Sec.\ref{sec:saturated}, we present the results for the case of homogeneous matter saturated in spin and isospin. Adopting some representative Skyrme functionals, we discuss the properties of the strength functions at different values of density, transferred momentum and temperature.
In Sec.\ref{Sec:landau}, we consider the Landau limit of the residual interaction. Taking advantage of the universal character of the Landau parameters, we generalize our results for the case of Landau parameters derived from other effective interactions as Gogny or Nakada and also from realistic microscopic calculations.
In Sec.\ref{sect:asym}, we generalize the results of Sec.\ref{sec:saturated} for the case of arbitrary isospin asymmetry.
In Sec.\ref{sec:extra}, we discuss how to extend our formalism to take into account possible extensions of the Skyrme functionals as for example the inclusion of extra density dependent terms or three- and four-body terms. Finally in Sec.\ref{sec:conclusions}, we present our conclusions and future perspectives.
Some specific technical details are given in the Appendices.


\section{Formalism}
\label{Sec:formalism}

In this  section, we describe a method to obtain the response functions of homogeneous fermion systems, both at zero and at finite temperature. It is based on the use of Green's functions describing the propagation of a particle-hole ($ph$) pair, also called quasiparticle. The response function is simply obtained as a momentum average of the $ph$ propagator. The starting point of this procedure is the calculation of the $ph$ propagator in the mean-field or Hartree-Fock (HF) approximation. Afterwards, the residual interaction between $ph$ pairs is taken into account in the random phase approximation, and the associated $ph$ propagator is obtained as the solution of the so-called Bethe-Salpeter equation. The method consists in writing down 
a set of algebraic equations for various momentum averages of the RPA propagator, whose solution gives the desired response function. 
Since only homogeneous systems are considered here, the proper definitions of the response functions, as well as other related quantities, should be understood as divided by a normalization volume. Along this paper the employed units are such that constants $\hbar$, $c$ and $k_B$ are equal to 1. 
 
\subsection{Linear response to an external spin-isospin perturbation}
The excited states of a system can be studied by measuring its response to an external probe, characterized by an oscillating field, which can exchange momentum $\mathbf{ q}$ and energy $\omega$ with the system~\cite{pin66,fet71,lip08,dic08,rin80,row70}.
Experimental measurements are usually related to the strength function (also called dynamical structure function or dynamical form factor) of the system, defined as
\be
S(\mathbf{q},\omega) = \sum_{n \neq 0} | \langle n | Q | 0 \rangle |^2 \delta(E_n - E_0 - \omega) \, ,
\label{strength1}
\ee
where $|n\rangle$ and $E_n$ are the eigenstates and eigenvalues of the nuclear Hamiltonian, and the sum over $n$ includes both discrete and continuum contributions. $E_{n0}\equiv E_n - E_0$ is the excitation energy and the operator $Q$  is a well-chosen operator that couples directly to states with the desired quantum numbers. 
The interest of having a detailed knowledge of the strength function can be seen as follows (see {\it e.g.} \cite{pin66}). 
Assume that the probe weakly interacts with the system, so that the scattering process may be described within
the Born approximation. For simplicity, assume also that the probe does not act on the spin/isospin degrees of freedom. The Fermi's golden rule stands that the probability per unit time that the probe transfers momentum 
$\mathbf{ q}$ and energy $\omega$ with the system is given by
\be
{\cal P}(\mathbf{ q}, \omega) = 2 \pi |{\cal V}_{\mathbf{q}}|^2 \,  S(\mathbf{q},\omega) \, ,
\ee
where ${\cal V}_{\mathbf{q}}$ is the Fourier transform of the probe-system interaction, and the strength function corresponds to the density-fluctuation operator. In general, cross sections are thus expressed as the product of a kinematic factor, associated to the specific probe, and the strength function, which contains all the relevant physical properties of the system. For instance, the double differential cross section for the inelastic scattering of neutrinos off neutron matter, written  in Eq.~(\ref{cross:nu}), involves the strength functions calculated for three operators, according to the different quantum numbers of the states excited in the process.  

Let us now consider density fluctuations in the spin-isospin channels $(\alpha) = (S,T)$ excited by one-body operators  of the type
\be
Q^{(\alpha)} = \sum_j {\rm e}^{i \mathbf{q} \cdot \mathbf{r}_j} \Theta^{(\alpha)}_j \, , 
\ee
where the index $j$ stands for the particle and
\be
\Theta^{(0,0)}_j = \hat{1} \ , \  
\Theta^{(1,0)}_j = \hat{\text{\boldmath{$\sigma$}}}_j \ , \ 
\Theta^{(0,1)}_j = \hat{\text{\boldmath{$\tau$}}}_j \  \ , \ 
\Theta^{(1,1)}_j = \hat{\text{\boldmath{$\sigma$}}}_j \, \hat{\text{\boldmath{$\tau$}}}_j \, ,
\ee
with $\hat{\text{\boldmath{$\sigma$}}}_j$ and $\hat{\text{\boldmath{$\tau$}}}_j$ being the spin and isospin Pauli matrices.  

If the interaction between the probe and the system is sufficiently weak, the change in the density is proportional to the perturbation induced by the external probe, the factor being the dynamical susceptibility. 
It is calculated at first-order perturbation theory, with the result
\be
\chi^{(\alpha)}(\mathbf{q},\omega) = \sum_{n \neq 0}  
\left\{ \frac{|\langle n | Q^{(\alpha)} | 0 \rangle|^2 }{ \omega - E_{n0} + i \eta} 
- \frac{|\langle n | Q^{(\alpha)} | 0 \rangle|^2 }{ \omega + E_{n0} + i \eta} \right\} \, ,
\label{response1}
\ee
where $\eta$ is an arbitrarily small  positive quantity associated with the adiabatic condition on the external field.
Using the standard relation
\be
\lim_{\eta \to 0^+} \frac{1}{x-a+i \eta} = {\cal P} \frac{1}{x-a} - i \pi \delta(x-a) \, ,
\ee
one can write the following connection between the strength function and the response function
\be
S^{(\alpha)}(\mathbf{q},\omega)  - S^{(\alpha)}(\mathbf{q},- \omega) = - \frac{1}{\pi} {\rm Im} \, \chi^{(\alpha)}(\mathbf{q},\omega) \, .
\label{fd}
\ee
At zero temperature, the system is initially in the ground state, and the sole possible effect of the probe is to excite the system, {\em i.e.} $\omega \ge 0$. In that case $S^{(\alpha)}(\mathbf{q},-\omega)$ is identically zero and the above equation reduces to
\be
S^{(\alpha)}(\mathbf{q},\omega)  = - \frac{1}{\pi} {\rm Im} \, \chi^{(\alpha)}(\mathbf{q},\omega) \, .
\label{strength2}
\ee
However at finite temperature $T$, the ground state of the system at equilibrium corresponds to a statistical mixture characterized by the probability 
\be
p_m = \frac{1}{Z} {\rm e}^{-\beta E_m}\,,
\ee
of finding the system in the state $|m\rangle$. In the previous expression $Z= \sum_m {\rm e}^{-\beta E_m}$ is the partition function and $\beta = 1/ T$. At $T \neq 0$ both the strength function and the response function must be redefined accordingly as
\be
S^{(\alpha)}(\mathbf{q},\omega,T) = \sum_{n, m \neq n} p_m | \langle n | Q^{(\alpha)} | m \rangle |^2 \delta(E_{nm}-  \omega) \, ,
\label{strength3}
\ee
and
\be
\chi^{(\alpha)}(\mathbf{q},\omega,T) = \sum_{n, m \neq n}  p_m \, 
\left\{ \frac{| \langle n | Q^{(\alpha)} | m \rangle|^2}{ \omega - E_{nm} + i \eta} 
- \frac{| \langle n | Q^{(\alpha)} | m \rangle|^2}{ \omega + E_{nm} + i \eta} \right\} \, .
\label{response2}
\ee
At finite temperature it is possible to transfer energy from the system to the probe, so that negative values of $\omega$ are admissible. Assuming time-reversal invariance, the principle of detailed balance establishes the following relationship
\be
S^{(\alpha)}(\mathbf{q},\omega,T)  = {\rm e}^{\beta \omega}  S^{(\alpha)}(\mathbf{q},- \omega,T)\,, 
\label{detailed-balance}
\ee
and the fluctuation-dissipation theorem Eq.~\ref{fd} gives
\be
 - \frac{1}{\pi} {\rm Im} \, \chi^{(\alpha)}(\mathbf{q},\omega,T) = \left( 1 -  {\rm e}^{- \beta \omega} \right) S^{(\alpha)}(\mathbf{q}, \omega,T) \, .
 \label{fluctuat-dissip}
 \ee
 A more detailed discussion on the properties of the strength function $\chi^{(\alpha)}$ for a system of fermions can be found for example in \cite{fet71,lip08,dic08}.

\subsection{Particle-hole propagators and linear response}
\label{ph-propagators}
To sketch the method it suffices to consider a system with only one Fermi surface, as symmetric nuclear matter or 
non-polarized pure neutron matter. 
A basic ingredient is the retarded (or causal) particle-hole Green's function (or $ph$-propagator) which describes the propagation of a $ph$ pair forward in time and which is formally defined as 
\be
G(\mathbf{r}, \mathbf{r'}, t-t') = - i \, \theta(t-t') \, \langle 0 | \left[ \psi^+ (\mathbf{r}, t) \psi (\mathbf{r}, t),
\psi^+ (\mathbf{r'}, t') \psi (\mathbf{r'}, t') \right] | 0 \rangle \, ,
 \label{Gphgen}
 \ee
 where $\psi^+$ and $\psi$ are particle creation and annihilation operators in the Heisenberg representation and $|0\rangle$ stands for the ground state. This is also called the density-density correlation function since the operator product $\psi^+ (\mathbf{r}, t) \psi (\mathbf{r}, t)$ measures the density at point $\mathbf{r}$ and time $t$. As in homogeneous matter the invariance of space holds, it is convenient to work in the momentum-energy space by performing a suitable Fourier transform. The first required element of the method is the retarded propagator $G_{HF}$ of a non-interacting $ph$ pair, which can be obtained by inserting in Eq. (\ref{Gphgen}) complete sets of single-particle states at the appropriate places. It can be expressed as
\be
G_{HF}(\mathbf{k},\mathbf{q},\omega) = \frac{n(\mathbf{k})-n(\mathbf{q}+\mathbf{k})}{ \omega +\varepsilon(\mathbf{k})- \varepsilon(\mathbf{q}+\mathbf{k})+i \eta} \, ,
 \label{GHF}
 \ee
where
\be
n(\mathbf{k}) = \left\{ {\rm e}^{(\varepsilon(\mathbf{k})-\mu)/T} + 1 \right\}^{-1}
\ee
is the Fermi-Dirac occupation number, which reduces to the step function $\theta(k_F-k)$ at zero temperature, and 
 $\varepsilon(\mathbf{k})$ is the single-particle energy
\be
\varepsilon(\mathbf{k}) = \frac{k^2}{2 m^*} + U\,,
\label{sp-energy}
\ee
where $U$ is the mean field, excluding the $k^2$ dependence (such a term contributes to the effective mass $m^*$). Actually, our formalism is directly used only if the effective mass is independent of $k$, because in that case many of the integrations can be done analytically. 

The HF response function is obtained by taking the sum over all states in the Fermi sea 
\be
\chi_{HF}(\mathbf{q},\omega) = n_d \int \frac{d^3 \mathbf{k}}{(2 \pi)^3} \, G_{HF}(\mathbf{k},\mathbf{q},\omega) \, ,
\label{chiHF}
\ee
where $n_d$ is the degeneracy factor (4 for symmetric nuclear matter, 2 for pure neutron matter). $\chi_{HF}(\mathbf{q},\omega)$ is usually called the Lindhardt function~\cite{lin54}, although in a strict way the Lindhardt function refers to the response of a free Fermi gas, that is to the case $m^*=m, U=0$ in (\ref{sp-energy}). 
The absence of index $(\alpha)$ indicates that the response function is independent of the spin-isospin channel.  
In the following, we will often deal with momentum averages similar to the previous one, for which we will adopt the shorthand notation  
\be
 \int \frac{d^3 \mathbf{k}}{(2 \pi)^3} \, f(\mathbf{k}) G_{HF}(\mathbf{k},\mathbf{q},\omega) \equiv \langle f \; G_{HF} \rangle  \, ,
\ee
hiding sometimes the $(\mathbf{q},\omega)$-dependence. 
With this notation, we write $\chi_{HF}(\mathbf{q},\omega) = n_d 
\langle G_{HF} \rangle$. In \ref{app:beta} are given explicit expressions of the required averages used in this report. 
If the effective mass depends on momentum ${\bf k}$ those averages should be numerically calculated from the very beginning.

\begin{figure}[H]
\begin{center}
  \includegraphics[clip,scale=0.12,angle=0]{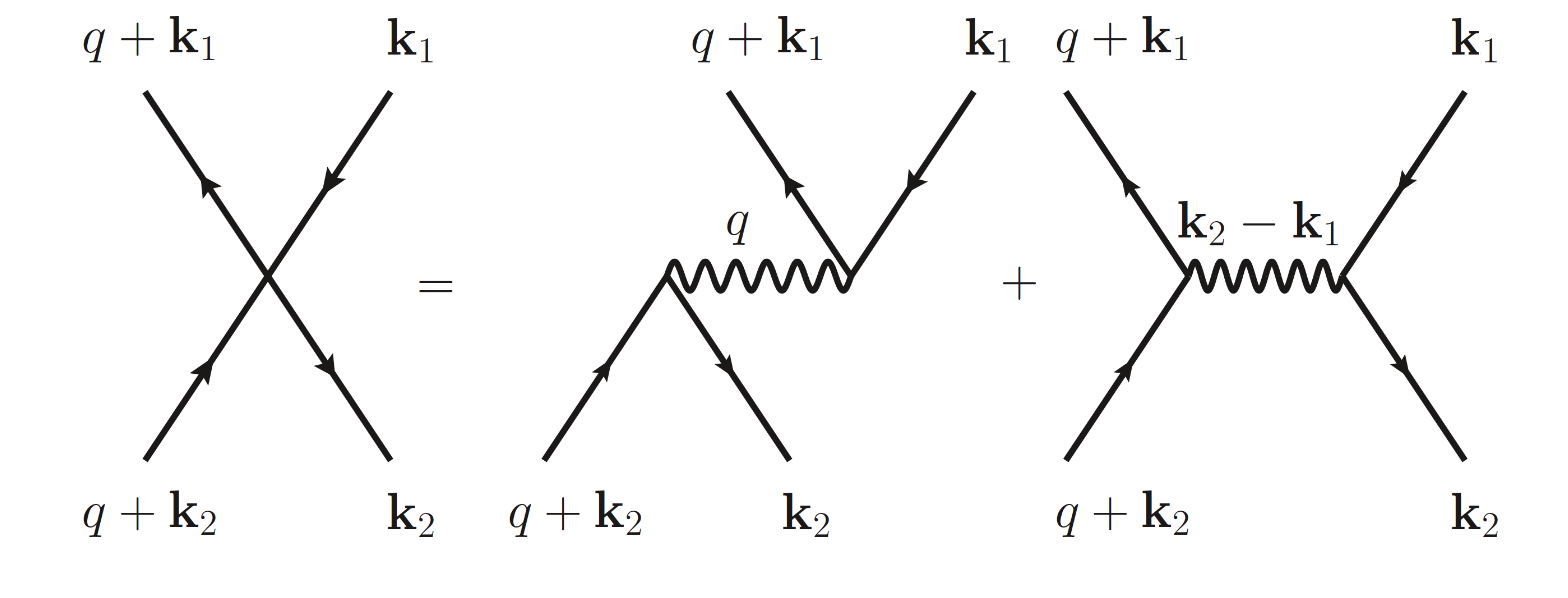}  \\
\caption{Graphical representation of the direct and exchange part of the residual \emph{particle-hole} interaction.}
\label{dir:exch}
\end{center}
\end{figure}

To go beyond the HF description one has to include the residual interaction. The correlated $ph$ propagator is the sum of an infinite number of terms involving the interaction and the HF $ph$ propagator. In the RPA this sum is restricted to the class of so-called ring or bubble diagrams, which allows for a formal summation of the whole series and leads to the Bethe-Salpeter equation for 
the correlated $ph$ propagator, which can be written as  
\begin{eqnarray}
G^{(\alpha)}_{RPA}(\mathbf{k}_{1},\mathbf{q},\omega) &=& G_{HF}(\mathbf{k}_{1},\mathbf{q},\omega) \nnn
&+& 
G_{HF}(\mathbf{k}_{1},\mathbf{q},\omega) \sum_{(\alpha')} \int \frac{d^{3}\mathbf{k}_{2}}{(2 \pi)^3} \,
V_{ph}^{(\alpha,\alpha')}(\mathbf{k}_1, \mathbf{k}_2) G^{(\alpha')}_{RPA}(\mathbf{k}_{2},\mathbf{q},\omega) \, , 
\label{bethe-salpeter}
\end{eqnarray}
where $V_{ph}^{(\alpha,\alpha')}(\mathbf{k}_1, \mathbf{k}_2)$ is the residual interaction matrix element which describes the $ph$ excitations of the system built on a mean-field (Hartree-Fock) ground state. The interaction links two $ph$ pairs with quantum numbers $(\alpha)$ and $(\alpha')$, and hole momenta $\mathbf{k}_1$ and $\mathbf{k}_2$, respectively. It is depicted in Fig.~\ref{dir:exch}. From the solution of this equation one gets the linear response function
\be
\chi^{(\alpha)}_{RPA}(\mathbf{q},\omega)  = n_d  \langle G^{(\alpha)}_{RPA} \rangle \, .
\ee

As illustrated in Fig. \ref{dir:exch}, the $ph$ interaction is the sum of a direct contribution, which only depends on the transferred momentum $\mathbf{q}$, and an exchange one, which depends on the relative $ph$ momentum 
$\mathbf{k}_1-\mathbf{k}_2$. 
The difficulty in solving the Bethe-Salpeter equation thus comes mainly from the exchange term. 
When it is treated in some approximation so that its $\mathbf{k}_1-\mathbf{k}_2$ dependence simplifies, then the resolution of (\ref{bethe-salpeter}) becomes easier. In the case of a central Skyrme pseudo-potential, the $ph$ interaction depends quadratically on the relative hole momentum. This particular dependence allows us to solve the Bethe-Salpeter equation in a rather compact form. It is worth stressing that the usual RPA ignores the exchange terms, and the name of ring approximation can be immediately visualized by iterating the first term in Fig. \ref{dir:exch}. What we call RPA  is sometimes called \emph{extended} or \emph{generalized} RPA.  
The same method can be extended to situations where the $ph$ interaction can be written in a polynomial form in the hole momenta, as it happens with currently used zero-range interactions or energy density functionals. 
A similar approach can also be adopted for finite-range interactions as Gogny~\cite{dec80} or M3Y~\cite{nak03}, but  in such a case the difficulty comes from the range. However, as already shown in ref.~\cite{mar05}, the interaction can be reduced in polynomials by a multipolar expansion and thus treated as in the case of a zero-range interaction.

\subsection{Domain of responses and collective states}

\begin{figure}[H]
\begin{center}
    \includegraphics[clip,scale=0.8,angle=0]{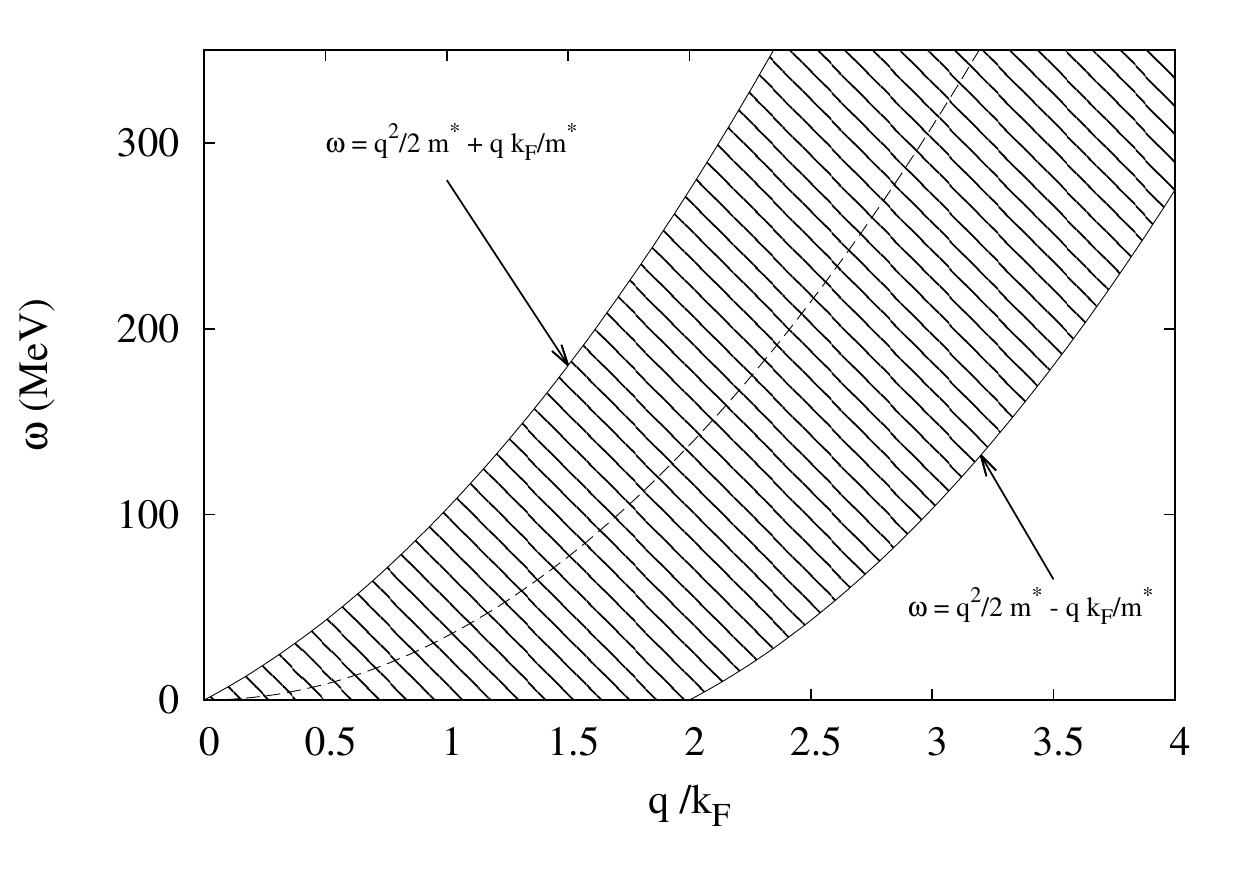}
\caption{Domain of allowed excitation energies  associated to a particle-hole excitation. For illustration, calculations have been done at  $k_F$=1.33~fm$^{-1}$ and $m^*/m$=1.}
\label{excitation}
\end{center}
\end{figure}

\begin{figure}[H]
\begin{center}
     \includegraphics[clip,scale=0.23,angle=-90]{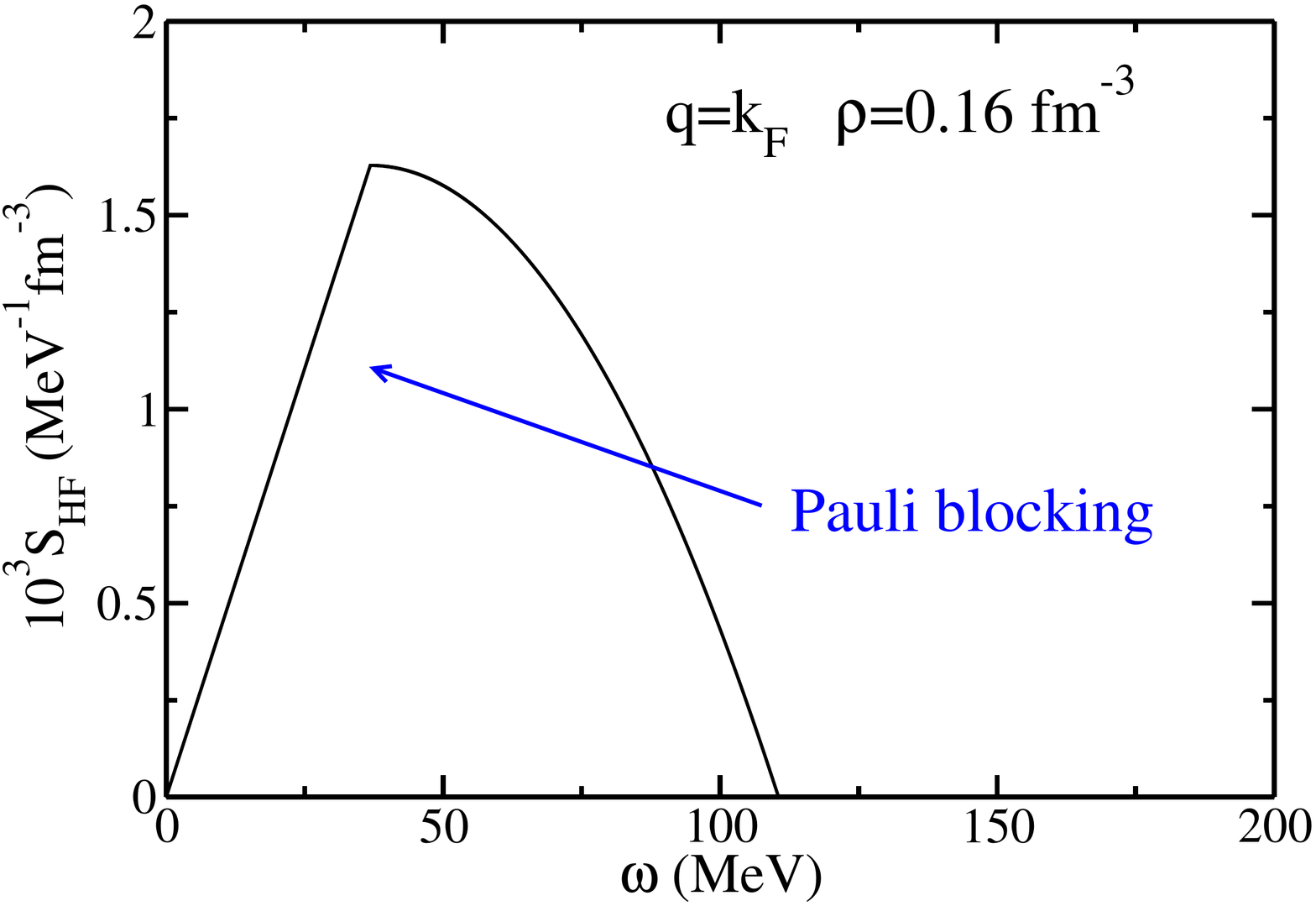}
     \includegraphics[clip,scale=0.23,angle=-90]{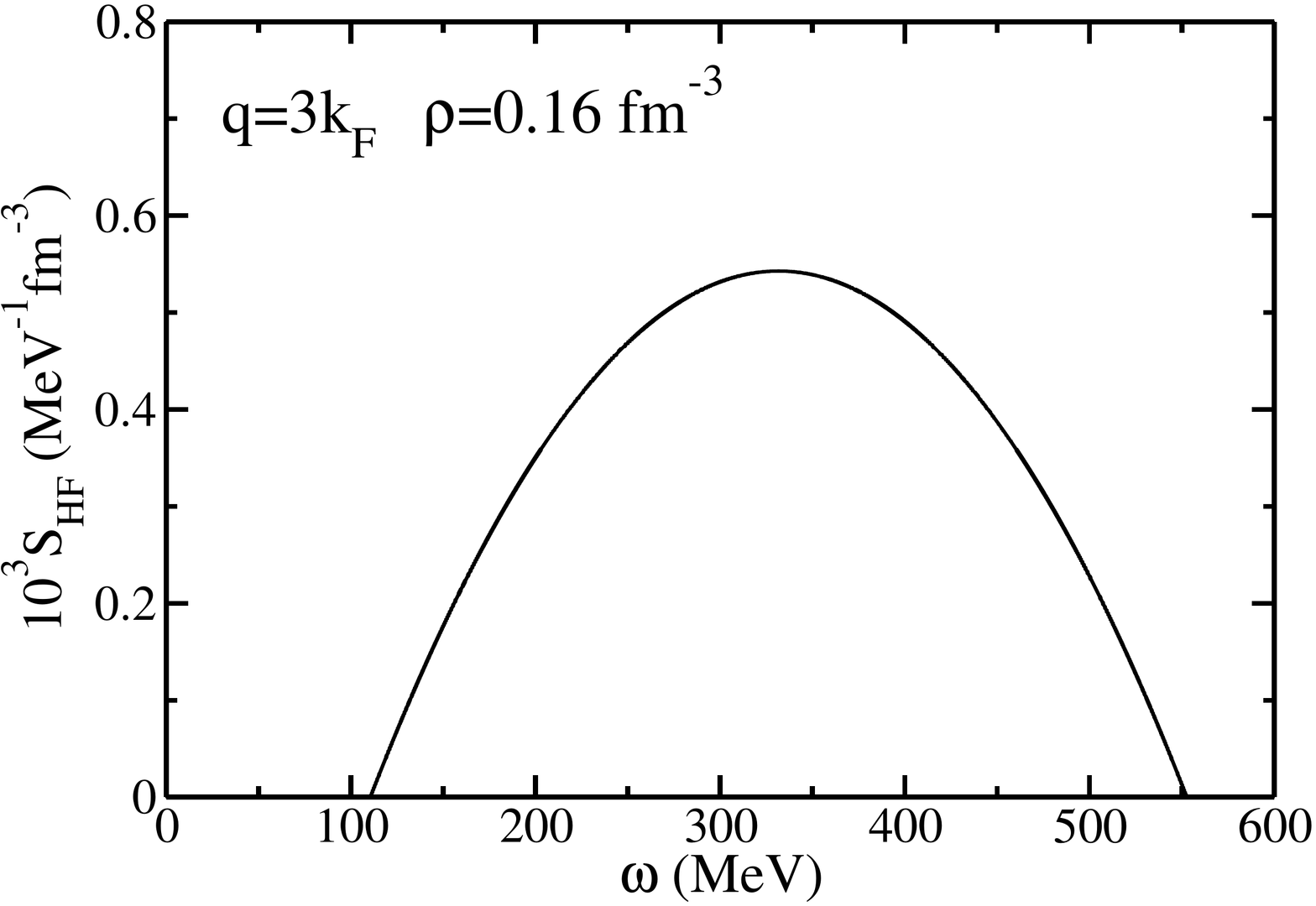}
\caption{Response function of a free Fermi gas at zero temperature in the regions $q<2k_F$ (left panel) and $q>2k_F$ (right panel). In particular the calculations have been done at  $k_F$=1.33~fm$^{-1}$ and $m^*/m$=1.}
\label{reponselibre}
\end{center}
\end{figure}

Let us explore the $(q,\omega)$-domain of possible $ph$ excitations. For simplicity, we assume $T=0$, and discuss thermal effects later on along the paper. Consider a particle with momentum $\mathbf{k}$ in the Fermi sea and energy ${k^2}/{2 m^*} + U$.  If a momentum ${\bf q}$ is transferred to it, its energy changes to $(\mathbf{k}+\mathbf{q})^2/2m^*+U$. Depending on the relative angle between $\mathbf{k}$ and $\mathbf{q}$, the excitation energy $\omega = (\mathbf{k}+\mathbf{q})^2/2m^* - k^2/2m$ is thus inside the shaded region of Fig.~\ref{excitation}. The dotted line inside this region corresponds to the value $q^2/2m^*$. This region defines the 1p1h excitation domain, which correspond to the case we will consider in RPA calculations. 
%
%
To have a physical insight on the response functions, we recall some general features~\cite{fet71}
of the HF response function at $T=0$. It is displayed in Fig.~\ref{reponselibre} as a function of the energy transfer $\omega$ for two momentum values.
%
%
In the region $q<2k_F$, the response first increases linearly with $\omega$. This behavior is due to Pauli blocking, since the momentum of the excited particle can be inside the Fermi sphere. The response reaches its maximum value at a transferred energy $\omega = q k_F/m^* - q^2/2 m^*$, vanishing at and beyond the upper limit of the shade region in Fig.~\ref{excitation}. In the region $q>2k_F$ the response is different from zero only for values of $\omega$ inside the lower and upper limits of the 1p1h excitation domain.

Further physical insight is gained considering the residual interaction in the simplest case when the exchange term is ignored. Assuming $V_{ph}^{(\alpha,\alpha')}(\mathbf{k}_1, \mathbf{k}_2)= \delta(\alpha,\alpha') V_{ph}^{(\alpha)}(q,\omega)$, 
one can immediately solve the Bethe-Salpeter equation (\ref{bethe-salpeter}) and the response function takes the simple form~\cite{ber05} 
\begin{equation}
\chi^{(\alpha)}_{RPA}(q,\omega)=\frac{\chi_{HF}(q,\omega)}{1-V_{ph}^{(\alpha)}(q,\omega) \chi_{HF}(q,\omega)}
\, .
\label{ring}
\end{equation}
This is the so-called ring approximation, which has been largely employed in studies of the response function considering mesonic degrees of freedom \cite{alb80,alb82,alb84}. 
As a matter of terminology, it is worth keeping in mind that in condensed matter textbooks, the term RPA corresponds to neglecting the exchange term, whereas ``extended RPA'' refers to including the full residual interaction. 

Some general properties of the RPA response function can be deduced from this equation. At first, we notice that ${\rm Im}\chi_{RPA}(q,\omega) \propto {\rm Im} \chi_{HF}(q,\omega)$, it follows that the domain of definition of both functions is the same and it corresponds to the  shaded region depicted in Fig.~\ref{excitation}. The residual interaction not only modifies the RPA strength, but can also give rise to excitations outside the energy domain where  ${\rm Im}\chi_{RPA}(q,\omega)\ne0$. Indeed, the response has a pole when the denominator of (\ref{ring}) vanishes:  $1-V_{ph}^{(\alpha)} \chi_{HF} = 0$. This defines a collective excitation~\cite{fet71} (also called zero sound, for historical reasons), which appears outside of the shaded domain of  Fig.~\ref{excitation}. Increasing the value of $q$, the excitation energy eventually crosses the upper $ph$ edge and the excitation is absorbed into the continuum.  

\subsection{Nuclear Energy Density Functional}
\label{EDF}

The nuclear energy density functional  is the tool of choice to perform systematic calculations of binding energies and one-body observables in the region of the nuclear chart that ranges from medium- to heavy-mass atomic nuclei from drip-line to drip-line~\cite{ben03}.  The functional is taken as an expansion of the energy density in powers of local densities (as defined in \ref{loc-den}) and their derivatives up to to a given order. This effective approach relies on a limited number of universal parameters, usually fitted on experimental data of finite nuclei (observables) along with properties of infinite nuclear matter (pseudo-observables)~\cite{mey03} or derived from realistic interactions trough density-matrix expansion (DME)~\cite{car10,sto10,neg75}.

Although several functionals are available~\cite{rob10,hup11,car08,kor10}, the one derived from the Skyrme effective  pseudo-potential~\cite{per04} is the most often used. This functional is expressed in terms of  local densities (defined in  \ref{loc-den}) and coupling constants, that could also depend on density themselves. The most general second order functional given in \cite{per04} contains up to 28 free coupling constants. This number is reduced to 18 when Galilean and gauge invariance~\cite{rai11} are imposed, and further reduced to 12 independent coupling constants if one also requires that such a functional can be derived from an effective Skyrme interaction. Several extensions can be found in the literature: expansion up to 6th order in the momentum (N3LO) has been given in \cite{car08}, showing that is it possible through a DME to convert a finite-range effective interaction into a quasi-local functional \cite{car10,dob10}; inclusion of new density dependence as done in \cite{les07,cha09,mar09c,mar09d}, or inclusion of three-body terms in the effective interaction \cite{sad11}. All these approaches aim at solving specific drawbacks or simply improving the quality and accuracy of the existing functionals.

In the following, we deal with a general second-order NEDF, $i.e.$ including up to a quadratic dependence on local densities and their derivatives, without any reference to a Skyrme interaction, but respecting Galilean and gauge invariance. It can be written as
\be
E = \int d^3 {\mathbf r} \, \left\{ \frac{1}{2m} \tau + \epsilon_{IS} + \epsilon_{IV} \right\} \; ,
\ee
where $\tau$ is the local kinetic density and $ \epsilon_{IS}, \epsilon_{IV}$  are the terms of the functional which depends respectively on isoscalar and isovector densities as discussed in Ref~\cite{per04}. In the general case they are written as
\begin{eqnarray}\label{edf:is}
\epsilon_{IS} &=&  
C^{\rho}_{0}\left[ \rho_{0}\right] 
\, \rho_{0}^{2} + C_{0}^{\Delta \rho} \, \rho_{0} \, \Delta\rho_{0} + 
C^{\tau}_{0} \, \rho_{0} \, \tau_{0} + C^{j}_{0} \ \mathbf{j}_{0}^{2}  + C^{s}_{0}\left[\rho_{0}\right] 
\, \mathbf{s}^{2}_{0} 
+ C^{\nabla s}_{0} \, (\nabla \cdot \mathbf{s}_{0})^{2}
\nnn
&& + C^{\Delta s}_{0} \, \mathbf{s}_{0}\cdot \Delta \mathbf{s}_{0} + C_{0}^{T} \, \mathbf{s}_{0}\cdot \mathbf{T}_{0} + C^{F}_{0} \, \mathbf{s}_{0}\cdot \mathbf{F}_{0} + 
C^{\nabla J}_{0} \, \rho_{0}\nabla \cdot \mathbf{J}_{0} \nnn
&&+ C_{0}^{\nabla j} \, \mathbf{s}_{0}\cdot \left( \nabla \times \mathbf{j}_{0}\right) + C^{J^{(0)}}_{0}(J^{(0)}_{0})^{2}+C^{J^{(1)}}_{0}(\mathbf{J}^{(1)}_{0})^{2}+C^{J^{(2)}}_{0}\sum_{\mu\nu=x}^{z}J_{0\mu\nu}^{(2)}J_{0\mu\nu}^{(2)} \,,
\end{eqnarray}
and
\begin{eqnarray}\label{edf:iv}
\epsilon_{IV} &=&  
C^{\rho}_{1}\left[ \rho_{0}\right] 
\, \vec{\rho}^{\, 2} + C_{1}^{\Delta \rho} \, \vec{\rho} \, \Delta\vec{\rho} + 
C^{\tau}_{1} \, \vec{\rho} \, \vec{\tau} + C^{j}_{1} \, \vec{\mathbf{j}}^{\, 2}  
+ C^{s}_{1}\left[\rho_{0}\right] 
\, \vec{\mathbf{s}}^{\, 2}  + C^{\nabla s}_{1} \, (\nabla\cdot \vec{\mathbf{s}})^{2} \nnn
&& + C^{\Delta s}_{1} \, \vec{\mathbf{s}}\cdot \Delta \vec{\mathbf{s}}  + C_{1}^{T} \, \vec{\mathbf{s}}\cdot \vec{\mathbf{T}} 
+ C^{F}_{1} \, \vec{\mathbf{s}}\cdot \vec{\mathbf{F}} + C^{\nabla J}_{1} \, \vec{\rho} \, \nabla \cdot \vec{\mathbf{J}} 
+ C_{1}^{\nabla j} \, \vec{\mathbf{s}}\cdot \left( \nabla \times \vec{\mathbf{j}}\right)  \nnn
&&+ C^{J^{(0)}}_{1} \, (\vec{J}^{(0)})^{2} + C^{J^{(1)}}_{1} \, (\vec{\mathbf{J}}^{(1)}_{0})^{2}
+ C^{J^{(2)}}_{1} \, \sum_{\mu\nu=x}^{z}\vec{J_{\mu\nu}^{(2)}}\vec{J_{\mu\nu}^{(2)}}  \,.
\end{eqnarray}
Note that the coefficients $C_t^{\rho}[\rho]$ and $C_t^{s}[\rho]$ include also a density dependence, usually taken as
\begin{eqnarray}
C_t^{\rho}[\rho_0] &=& C_t^{\rho,0} + C_t^{\rho,\gamma} \rho_0^{\gamma}\,, \\ 
C_t^{s}[\rho_0] &=& C_t^{s,0} + C_t^{s,\gamma} \rho_0^{\gamma} \, ,
\end{eqnarray}
which is typical of Skyrme interactions~\cite{per04}. In some cases a supplementary $\rho^{\gamma '}$ dependence has been considered, as given in Ref.~\cite{les06}.  The inclusion of such terms in the formulae presented in this paper is immediate: one simply has to duplicate the contribution of the term with power $\gamma$ and replace $\gamma$ by $\gamma'$. 
The functional written in Eqs.\ref{edf:is}-\ref{edf:iv} has been constructed to be consistent with time reversal, space and rotational invariance in agreement with basic symmetries of nuclear interaction~\cite{rin80}.
Imposing also Galilean and gauge invariance, we obtain some relations among coupling constants
\begin{eqnarray}\label{gauge}
C_{t}^{j}&=&-C^{\tau}_{t}\,,\\
C^{J^{(0)}}&=&-\frac{1}{3}C_{t}^{T}-\frac{2}{3}C_{t}^{F}\,,\nonumber\\
C^{J^{(1)}}&=&-\frac{1}{2}C_{t}^{T}+\frac{1}{4}C_{t}^{F}\,,\nonumber\\
C^{J^{(2)}}&=&-C_{t}^{T}-\frac{1}{2}C_{t}^{F}\,,\nonumber\\
C^{\nabla j}_{t}&=&C^{\nabla J}_{t}\,,\nonumber
\end{eqnarray}
\noindent where $t=0,1$ stands for isoscalar and isovector coupling constants. This implies immediately that only some special combinations such as $\rho \tau - \mathbf{j}^2$ can enter the NEDF. 
The Galilean invariance  is a particular case of the gauge invariance~\cite{dob95}. This feature is important for example in Time Dependent Hartree-Fock or RPA calculations, since it ensures that the collective translational mass is correctly equal to the total mass~\cite{eng75,car08}. 
 The original Skyrme interaction~\cite{sky59} respects these symmetries by construction, thus for similarity a functional derived from this interaction should respect them as well. Whether the gauge symmetry should be respected or not for a Skyrme functional is still an open question and we refer to the discussion in ~\cite{car08}. To clarify this point, one considers a general 2-body operator $V(\mathbf{r'}_1,\mathbf{r'}_2, \mathbf{r}_1,\mathbf{r}_2)$ (one drops again spin and isospin indices for simplicity).  Assuming a general Slater determinant wave function, the interaction energy can then be written as  
\begin{equation}
E = \int d^{3}\mathbf{r'}_1 d^{3}\mathbf{r'}_2 d^{3}\mathbf{r}_1 d^{3}\mathbf{r}_2 
V(\mathbf{r'}_1,\mathbf{r'}_2, \mathbf{r}_1,\mathbf{r}_2)
\left[ \rho(\mathbf{r}_1,\mathbf{r'}_1)\rho(\mathbf{r}_2,\mathbf{r'}_2)-\rho(\mathbf{r}_2,\mathbf{r'}_1)\rho(\mathbf{r}_1, \mathbf{r'}_2)\right] \, .
\end{equation}
Let us define a gauge transformation on the $A$-body wave function as
\begin{equation}
|\Psi'\rangle = \exp \left(-i\sum_{i=1}^A \phi(\mathbf{r}_j) \right) \, |\Psi\rangle \, ,
\end{equation}
where $\phi(\mathbf{r}_j)$ is an arbitrary real function. Then the density matrix transforms as 
\begin{equation}
\rho'(\mathbf{r},\mathbf{r'})= \exp(i(\phi(\mathbf{r})-\phi(\mathbf{r'}))) \, \rho(\mathbf{r},\mathbf{r'}) \, .
\end{equation}
When the interaction is local 
\begin{equation}
V(\mathbf{r'}_1,\mathbf{r'}_2, \mathbf{r}_1,\mathbf{r}_2)
=\delta(\mathbf{r'}_1 - \mathbf{r}_1)\delta(\mathbf{r'}_2 - \mathbf{r}_2)  V(\mathbf{r'}_1,\mathbf{r'}_2, \mathbf{r}_1,\mathbf{r}_2)\,,
\end{equation}
\noindent the gauge invariance is automatically satisfied by construction. 
With the help of the transformation law written above for the density matrix, it is easy to deduce that  under a gauge transformation the kinetic energy and current densities transform as $\tau' = \tau + 2 \mathbf{j} \nabla \phi + \rho (\nabla \phi)^2$ and $\mathbf{j'} = \mathbf{j} +\rho \nabla \phi$ . As quoted before, one can immediately see that only the special combination 
$\rho \tau - \mathbf{j}^2$ is invariant. From a more general point of view, it has been shown \cite{rai11} that Galilean invariance and gauge invariance are no longer equivalent at higher orders and that gauge invariance is much more constraining  (by instance at 4th order, the total number of terms can be reduced to 26 with Galilean invariance and to 6 with gauge invariance). The problem of gauge invariance has been addressed in \cite{dav13}. It appears to be linked to the problem of the continuity equation \cite{bla80}. However, at fourth order, this continuity equation contains up to fourth order derivatives so that it is no longer a Schr\"odinger-type equation \cite{bec14}. For practical purposes, the solution becomes non trivial and work on that subject is in progress.

Although we present here the case of a general functional, it is useful to identify the nature of the different coupling constants entering (\ref{edf:is}-\ref{edf:iv}). The central terms of the Skyrme interaction are associated with the coupling constants $C^{\rho}_{t},C^{\Delta \rho}_{t},C^{\tau}_{t},C^{s}_{t}$, the spin orbit terms with $C^{\nabla J}$ and the tensor terms with $C^{\nabla s}_{t},C^{F}_{t}$. The association between the interaction parameters to the different parts of the functional can also be more complicated since, for example, the coupling constants $C^{T}_{t},C^{\Delta s}_{t}$ receive contributions from both the central and tensor part. For completeness, we give in \ref{coupling-SK} the expressions of the above coupling constants in terms of Skyrme parameters. 

Since the importance of the tensor component of the effective interaction has been enlightened by several authors concerning both ground state properties~\cite{ots01,ots06,ang11,bri07,bro06,les07} and excited states~\cite{bai09,dav09,pas12,hel12,fra12,cao11}, the standard second-order NEDF which incorporates the tensor part will be chosen as our reference and thus discussed in details for the different infinite systems treated in the following sections. The possible extensions will be discussed separately and some results presented only for some specific cases.

\subsection{The residual $ph$ interaction}

In the RPA approximation, the excitations of the system result from the residual $ph$ interaction. In the theory of Fermi liquids, this $ph$ interaction can be obtained from the second functional derivative with respect to densities taken at the Hartree-Fock solution \cite{mig67,bro71,ber75}
\begin{equation}\label{secondderivative}
\langle \map , \mbp | \hat{V}_{ph} | \ma , \mb \rangle = \frac{\delta^{2 }E[\rho]}{\delta  \hat{\rho}(\map, \ma)\,
\, \delta \hat{\rho}(\mbp, \mb)} \, .
\end{equation}
The symbol $\ma$ is a shorthand notation for $({\mathbf x_a}, \sigma_a, q_a)$, where $\mathbf{x_a}$ is the spatial coordinate, while $\sigma_a$ and $q_a$ are the spin and isospin variables. It has been shown in \cite{rin80} that such a relation holds in general at HF level. In the general case of a density dependent interaction, it is the only way to obtain the right answer \cite{mig67,bro71,ber75}. However, when there is no density dependence, one can determine more efficiently the matrix elements directly from the pseudo-potential $\hat{V}$ itself with no reference to the EDF by a simple antisymmetrization of the state:
$\langle \map , \mbp | \hat{V}_{ph} | \ma , \mb \rangle \equiv \langle \map , \mbp | \hat{V} A_{\ma  \mb} | \ma , \mb \rangle$ where $A_{\ma  \mb}$ represents here the complete antisymmetrisation operator with respect to state $ | \ma , \mb \rangle$. 

We are now in position to determine explicitly the  residual interaction induced by our general second order functional. In order to simplify the notations, we omit the global implicit factor $\delta(\map,\ma) \delta(\mbp, \mb) \delta(\mathbf{x}_a-\mathbf{x}_b) $. 
Moreover, we use
\begin{equation}\label{m1}
\mathbf{k}=-i \text{\boldmath$\nabla$} \, ,\, \mathbf{k}'=i \text{\boldmath$\nabla$}' \, .
\end{equation}
Following~\cite{gar92}, it is more convenient to express the result in terms of $\mathbf{k}_{1},\mathbf{k}_{2}$ (hole momenta) and $\mathbf{q}$ (transferred momentum)
\begin{eqnarray}\label{changeq}
\begin{cases}
\mathbf{k}_{a}'=\mathbf{k}_{1}+\mathbf{q}\, ,\\
\mathbf{k}_{a}=\mathbf{k}_{1}\nonumber\, ,\\
\mathbf{k}_{b}'=\mathbf{k}_{2}\nonumber\, ,\\
\mathbf{k}_{b}=\mathbf{k}_{2}+\mathbf{q}\, ,
\end{cases}
\end{eqnarray}
which is actually the notation employed in Fig.~\ref{dir:exch}. 
We use the standard components of $\mathbf{k}_{12} \equiv \mathbf{k}_{1}-\mathbf{k}_{2}$ 
\begin{eqnarray}\label{k12def}
(k_{12})_{\mu}^{(1)}&=&\sqrt{\frac{4\pi}{3}}\left[ k_{1}Y_{1,\mu}(\hat{k}_{1})-k_{2}Y_{1,\mu}(\hat{k}_{2})\right] \, ,
\end{eqnarray}
where $(k_{12})_{\mu}^{(1)}$ is a rank-1 tensor and $\mu=-1,0,+1$ is the index of the spherical basis~\cite{var88}.
Finally, after some simple manipulations, we obtain the general expression
\begin{eqnarray}\label{res}
V_{ph}&=&
\frac{1}{4} W_{1}^{(0,0)} + \frac{1}{4} W_{1}^{(0,1)} \hat{\tau}_{a}\circ \hat{\tau}_{b} 
+ \frac{1}{4} W_{1}^{(1,0)} \text{\boldmath$\sigma$}_{a} \cdot\text{\boldmath$\sigma$}_{b} 
+ \frac{1}{4} W_{1}^{(1,1)} \text{\boldmath$\sigma$}_{a} \cdot\text{\boldmath$\sigma$}_{b}\hat{\tau}_{a}\circ \hat{\tau}_{b} \nonumber\\
&+& \frac{1}{4} \left( W_{2}^{(0,0)} + W_{2}^{(0,1)} \hat{\tau}_{a}\circ \hat{\tau}_{b} 
+ W_{2}^{(1,0)} \text{\boldmath$\sigma$}_{a} \cdot\text{\boldmath$\sigma$}_{b}
+ W_{2}^{(1,1)} \text{\boldmath$\sigma$}_{a} \cdot\text{\boldmath$\sigma$}_{b}\hat{\tau}_{a}\circ \hat{\tau}_{b}\right) {\bf k}_{12}^2 \nnn
&+& 2 
C^{\rho,\gamma}_{1} \gamma \rho_{0}^{\gamma-1} 
\vec{\rho} \circ \left( \hat{\tau}_{a}+\hat{\tau}_{b}\right) 
+2 C^{s,\gamma}_{0} \gamma \rho_{0}^{\gamma-1} 
\mathbf{s}_{0}\cdot \left(\text{\boldmath$\sigma$}_{a}+\text{\boldmath$\sigma$}_{b} \right) \nonumber\\
&+& 2 C^{s,\gamma}_{0} \gamma \rho_{0}^{\gamma-1}
\vec{\mathbf{s}}\cdot \left(\text{\boldmath$\sigma$}_{a}\tau_{a}+ \text{\boldmath$\sigma$}_{b}\tau_{b} \right) \nonumber\\
&+&2\left( C_{0}^{\nabla s}+C_{1}^{\nabla s} \hat{\tau}_{a}\circ \hat{\tau}_{b} \right) 
(\mathbf{q} \cdot\text{\boldmath$\sigma$}_{a}) (\mathbf{q}\cdot\text{\boldmath$\sigma$}_{b}) \nonumber\\
&+&\left( C_{0}^{F} +C_{1}^{F}\hat{\tau}_{a}\circ \hat{\tau}_{b}\right) \left\{ 
(\mathbf{k}_{12} \cdot \text{\boldmath$\sigma$}_{a}) 
(\mathbf{k}_{12}\cdot \text{\boldmath$\sigma$}_{b}) 
- \frac{1}{2} (\mathbf{q}\cdot \text{\boldmath$\sigma$}_{a}) 
(\mathbf{q}\cdot \text{\boldmath$\sigma$}_{b}) \right\}\nonumber\\
&-&i\left(C_{0}^{\nabla J}+C_{1}^{\nabla J}\hat{\tau}_{a}\circ \hat{\tau}_{b} \right)\left(\text{\boldmath$\sigma$}_{a}+\text{\boldmath$\sigma$}_{b} \right)\cdot  \left[\mathbf{q} \times\mathbf{k}_{1}  -\mathbf{q}_{} \times \mathbf{k}_{2}  \right] \, .
\end{eqnarray}
The constants $W_i^{(\alpha)}$ are combinations of NEDF coupling constants. They are given in the next Section for symmetric nuclear matter and pure neutron matter. 

The previous general expression for the residual interaction is the starting point for all the calculations and results presented in this paper. For each specific system (symmetric nuclear matter, asymmetric nuclear matter, pure neutron matter ...) one must determine of course the appropriate matrix elements in spin and isospin space. However, the method for obtaining the response function relies only on the dependence on momenta ${\mathbf k}_1, {\mathbf k}_2$, which is the same for all these systems. Based on this quadratic dependence, the induced interaction in the medium
was calculated in Ref. \cite{gar92} as an intermediate step to get the response function. Actually, this intermediate step is no longer required \cite{mar06,dav09} because the Bethe-Salpeter equation can be directly solved. This is the method used in the following.

\subsection{Solving the Bethe-Salpeter equation}
\label{solving-BS}
The Bethe-Salpeter equation (\ref{bethe-salpeter}) has to be solved in each channel. The strategy for obtaining the response function is very simple. The general structure in momentum space of the residual interaction (\ref{res})
contains only terms of the form
\begin{equation}
1 \, , \, {\bf k}_{12}^2 \, , (k_{12})^{(1)}_{\mu} \, , \, {\rm and} \, (k_{12})^{(1)}_{\mu} (k_{12})^{(1)}_{\mu'}\, .
\end{equation}
The idea is thus to multiply the equation with appropriate functions of $\mathbf{k}_{1}$ and integrate over momentum. Instead of the $ph$ propagator $G_{RPA}^{(\alpha)}$ one has to deal with its momentum averages. 
In general, a closed set of algebraic equations is obtained for the following quantities: $\langle G^{(\alpha)}_{RPA}\rangle$, 
$\langle k^2 G^{(\alpha)}_{RPA}\rangle$, $\langle k Y_1 ^0 G^{(\alpha)}_{RPA}\rangle$,
$\langle k^2 |Y_1 ^1| ^2 G^{(\alpha)}_{RPA}\rangle$ and $\langle k^2 |Y_1 ^0| ^2 G^{(\alpha)}_{RPA}\rangle$. 
Depending on the residual interaction, one can therefore obtain either a single equation or a set of equations coupling the different spin-isospin channels. Although in principle the system can be analytically solved, it is worth giving analytical expressions only for  symmetric nuclear matter and pure neutron matter. For other systems, the  expressions are cumbersome and not easily tractable. 

Let us illustrate the method assuming that particles and holes are restricted to be at the surface of their respective Fermi sphere. The $ph$ interaction adopts a form particularly simple, since it only depends on the relative angle $\theta$ between the two momenta ${\bf k}_1$ and ${\bf k}_2$ at the Fermi surface, besides the obvious dependence on the spin and isospin degrees of freedom. This is the well-known Landau approximation~\cite{mig67},
which is discussed in more details in Sec.~\ref{Sec:landau}. For the present illustrative purposes, it suffices to consider the case of channel $S=0, T=0$ of symmetric nuclear matter and assume a $ph$ interaction characterized by the monopole and dipole Landau parameters $f_0, f_1$ 
\begin{eqnarray}\label{Landau01}
n_{d}\left[f_{0}+f_{1}\frac{4\pi}{3} \sum_{\mu} Y^*_{1,\mu}( \mathbf{\hat k}_1) Y_{1,\mu}( \mathbf{\hat k}_2)\right]\delta_{S,0}\delta_{T,0}\, ,
\end{eqnarray}
where $\mathbf{\hat k}_{1,2}$ refer to the spherical angles of the hole momenta $\mathbf{k}_1$ and $\mathbf{k}_2$.

In this particular case, there is no mixing between the different spin/isospin channels, and the Bethe-Salpeter equation reads 
\begin{eqnarray}
G^{(0,0)}_{RPA}(\mathbf{k}_{1},\mathbf{q},\omega) &=& G_{HF}(\mathbf{k}_{1},\mathbf{q},\omega) 
+ n_d f_0 G_{HF}(\mathbf{k}_{1},\mathbf{q},\omega) \langle G_{RPA}^{(0,0)} \rangle \nnn
&+& n_d f_1 \frac{4 \pi}{3} \sum_{\mu}  Y^*_{1,\mu}(\mathbf{k}_{1})  
G_{HF}(\mathbf{k}_{1},\mathbf{q},\omega)  
\langle Y_{1,\mu} G_{RPA}^{(0,0)} \rangle \, .
\end{eqnarray}
Without loss of generality the vector $\mathbf{q}$ can be chosen along the $z$-axis. From (\ref{GHF}) one can see that $G_{HF}$ does not depend on the azimuthal angle $\phi$, so that the momentum integral $\langle Y^*_{1,\mu} G_{HF} \rangle$ vanishes unless $\mu=0$. The previous equation can thus be written as
\be
\langle G_{RPA}^{(0,0)}  \rangle = \alpha_0 + n_d f_0 \alpha_0 \langle G_{RPA}^{(0,0)} \rangle + n_d f_1  \alpha_1 \langle \cos \theta \; G_{RPA} ^{(0,0)} \rangle \, ,
\label{f01-1}
\ee
where we have defined the auxiliary functions
\be
\alpha_n(\mathbf{q},\omega) = \langle \cos^n \theta \, G_{HF} ^{(0,0)}  \rangle \, ,
\label{defalpha}
\ee
whose properties are given in \cite{pas13d}. 
One can see that the function $\langle G_{RPA}^{(0,0)} \rangle$ we are looking for is coupled to $\langle \cos \theta \; G_{RPA}^{(0,0)}  \rangle$, for which a second equation is required. It is obtained by multiplying the Bethe-Salpeter equation (\ref{bethe-salpeter}) with $\cos \theta_1$ and integrating over $\mathbf{k}_1$. The resulting equation is
\be
\langle \cos \theta \, G_{RPA} ^{(0,0)} \rangle = \alpha_1 + n_d f_0 \alpha_1 \langle G_{RPA}^{(0,0)} \rangle 
+ n_d f_1  \alpha_2 \langle \cos \theta \, G_{RPA} ^{(0,0)} \rangle\, .
\label{f01-2}
\ee
Eqs. (\ref{f01-1}) and (\ref{f01-2}) form a closed coupled system of algebraic equations for the quantities $\langle G_{RPA} ^{(0,0)}\rangle$ and $\langle \cos \theta \, G_{RPA} ^{(0,0)}\rangle$. 
The solution is immediate and we can finally write the response function as
\be
\chi_{RPA} ^{(0,0)}(\mathbf{q},\omega) = \chi_{HF} (\mathbf{q},\omega) \left\{ 1 - \left[ f_0 + f_1 \frac{\alpha_1^2/\alpha_0^2}{1-n_d f_1 (\alpha_0 \alpha_2 - \alpha_1^2)/\alpha_0} \right] 
\chi_{HF} (\mathbf{q},\omega) 
\right\}^{-1}\, .
\label{resp-f01}
\ee
This expression can be simplified if the $\alpha_i$'s functions in the brackets are calculated in the so-called
 Landau limit: $q \to 0$, but $\omega/q$ finite. In that case, one can replace $\alpha_1 \to k_F \nu \alpha_0$ and  $\alpha_2 \to k_F^2 \nu^2 \alpha_0 - N(0) / (3 n_d)$, where $\nu= \omega m^*/( q k_F)$ and 
 $N(0) = n_d \, \frac{k_F m^*}{2 \pi^2}$ is the density of states at the Fermi surface. We finally get the familiar expression for the Landau response function (see \emph{e.g.} \cite{lip08})
\be
\chi_{LAN}^{(0,0)}(\nu) = \chi_{HF}(\nu) \left\{ 1 - \left[ f_0 + \frac{f_1 \nu^2}{1+ F_1/3} \right] \chi_{HF}(\nu) 
\right\}^{-1} \, ,
\label{firstexample}
\ee
where $F_1=N(0) f_1$ is the dimensionless parameter. 
This example illustrates the method for obtaining the RPA response function from the Bethe-Salpeter equation. The solution using the complete residual interaction, or even a Landau interaction with tensor terms, is much more laborious. Nevertheless, in most cases it is possible to write analytic expressions for the response function.

\subsection{Sum rules and detection of instabilities}

The properties of the nucleus relevant to experimental measurements concerning inclusive processes can be related to energy-weighted integrals of the strength function 
\be
\int_{-\infty}^{\infty} {\rm d}\omega \, f(\omega, q) \, S^{(\alpha)}(q, \omega) \, .
\label{pepet}
\ee
The physical process considered defines the operator $Q$ and the weighting function $f\omega, q)$. The latter is completely determined by kinematics and does not involve any nuclear structure information, which is all contained in the strength function for the given operator $Q$. Energy-weighted integrals of the strength function can provide useful information of general applicability. Besides, closed expressions can be easily obtained in some cases. The $p$-th order sum rule is defined as
\be
M^{(\alpha)}_p(q) = \int_{-\infty}^{\infty} {\rm d}\omega \, \omega^p S^{(\alpha)}(q, \omega) \, .
\label{SR-numerical}
\ee
Besides a direct numerical integration of the strength function, it is also possible in principle to get analytical expressions for $M^{(\alpha)}_p$~\cite{boh79,lip89}. Replacing the definition (\ref{strength1}) of the strength function in the above integral one obtains 
\be
M^{(\alpha)}_p(q) = \sum_n E^p_{n0} \, | \langle n | \tilde{Q}^{(\alpha)} | 0 \rangle |^2 \, .
\ee
Assuming that $E_n$ and $|n\rangle$ are eigenvalues and eigenstates of the Hamiltonian $H$, it is then possible to write the sum rules with $p>0$ as the ground state expectation values of commutators and anticommutators of the Hamiltonian and the excitation operator $Q^{(\alpha)}$. In this way one can obtain useful expressions for the energy-weighted  and the cubic energy-weighted sum rules, $M^{(\alpha)}_1$ and $M^{(\alpha)}_3$ (in short EWSR and  CEWSR, respectively). Formally, sum rules with $p<0$ may also be expressed in terms of commutators and anticommutators of an operator $X^{(\alpha)}$ defined as the solution of $[H,X^{(\alpha)}]=Q^{(\alpha)}$ \cite{boh79,lip89}. However, this procedure requires the explicit knowledge of the Hamiltonian $H$, and thus is not applicable in the case of a general NEDF. 

At zero temperature sum rules can be defined as energy moments of either $S^{(\alpha)}({\bf q},\omega)$ or of
${\rm Im} \chi^{(\alpha)}({\bf q},\omega)$, as they coincide apart from a trivial factor. However this is not true at $T \neq 0$, as it has been discussed for instance in \cite{her96}. In that case, one should define the moments in the full energy interval, including negative values. The result is
\bea
\int_{-\infty}^{+\infty} {\rm d}\omega \, \omega^p S^{(\alpha)}(q, \omega) 
& = - \frac{1}{\pi} \int\limits_0^{\infty} {\rm d}\omega \, \omega^p {\rm Im}\chi^{(\alpha)}(q, \omega)& \quad , {\rm odd} \, p  \label{M1-3:num}\\
& =  - \frac{1}{\pi} \int\limits_0^{\infty} {\rm d}\omega \, \omega^p \coth \frac{\omega}{2T} {\rm Im}\chi^{(\alpha)}(q, \omega)& \quad , {\rm even} \, p \label{M-1:num}
\eea
Therefore, odd order sum rules can be calculated in the same way either at zero or finite temperature. 

A useful alternative way to obtain odd-order sum rules is to expand the response function in powers of $\omega$~  \cite{boh79}. In fact, as the strength and the response functions are defined per unit of volume one gets the sum rules per particle
\bea
{\omega \to \infty} \quad : \quad \chi^{(\alpha)}(q, \omega) &{\simeq}& 2 \rho \sum_{p=0}^{\infty} (\omega)^{-(2p+2)} M^{(\alpha)}_{2p+1}(q)/A \label{M1-3}\,, \\
{\omega \to 0} \quad : \quad \chi^{(\alpha)}(q, \omega) &{\simeq}& - 2 \rho \sum_{p=0}^{\infty} (\omega)^{2p} M^{(\alpha)}_{-(2p+1)}(q)/A \label{M-1} \, ,
\eea
where $\rho$ is the density of the system. Notice that, in the above equations, the imaginary part of
$\chi^{(\alpha)}(q, \omega)$ cancels out exactly in both limits. 

One can thus obtain some $M^{(\alpha)}_p$ in two ways: the first one is purely numerical 
and implies the whole response function; the second one is analytical and originates from 
the previous limits, which are a direct consequence of the dispersion relation satisfied by the response function. 
Comparing numerical and analytical sum rules is important for several reasons. It is a very good test about the reliability of the calculations. Besides, it provides us with an independent way to localize collective excitations. Indeed, 
in the presence of a pole the simplest form of the dispersion relations is no longer valid. 
And from the same token, the comparison also helps in detecting all instabilities, either physical or not.  
To this respect, the study of the inverse energy-weighted sum rule (IEWSR) $M_{-1}$ is particularly interesting because, apart from a trivial factor, it gives the susceptibility, or response function at zero energy. It is thus a very efficient tool to detect the instabilities related to a zero-energy mode~\cite{pas12}. 
Physically these instabilities can be related to a phase transition as the spinodal one (a mode which can be independently analyzed based on thermodynamical considerations) or to a non-physical instability, as the pion condensation largely discussed in the past \cite{alb80,bar73}. 


\section{Saturated systems}  
\label{sec:saturated}

We consider the two extreme cases of spin saturated nuclear matter, namely symmetric  nuclear matter (SNM) and pure neutron matter (PNM). From a technical point of view, both systems can be treated on the same footing, since in practice one has to deal with a single Fermi sea. In SNM the $ph$ excitations are characterized by the spin and isospin quantum numbers $(S, M, I, Q)$, where $M$ and $Q$ refer to the projections of the spin $S$ and isospin $I$, respectively. As long as we are not interested in charge-exchange processes, the isospin projection index $Q$ is irrelevant and will be ignored. In PNM only the two spin quantum numbers $(S,M)$ are required. All together, we shall employ a generic symbol $(\alpha)$ for these quantum numbers. It is possible to write the response functions and other magnitudes in a unified form for both systems assuming the following convention for the isospin index: $I=0, 1$ for SNM and $I=n$ for PNM.      

\subsection{ph interaction}

We turn now to the general form presented in Eq. (\ref{res}) for the $ph$ interaction. We write its matrix elements in the 
$(\alpha)$-space as
\begin{eqnarray}
V_{ph}^{(\alpha, \alpha')} &=& \delta(\alpha, \alpha') \bigg\{ W_{1}^{(\alpha)}(q) 
+  W_{2}^{(\alpha)} {\bf k}_{12}^2 + \delta_{S,1} \delta_{M,0}
W_{T2}^{(I)} \, q^2 
+ W_3^{(\alpha)} [ {\bf k}_1 ({\bf k}_2+{\bf q}) + {\bf k}_2 ({\bf k}_1+{\bf q}) ]
\bigg\} \nonumber\\ 
&+& \delta_{I,I'}  \, q W_{SO}^{(I)} \bigg( \delta_{S',0}\delta_{S,1} \, M(k_{12})_{-M} +
\delta_{S',1}\delta_{S,0} \, M'(k_{12})_{M'} \bigg) \nonumber \\
&+&  \delta_{I,I'}  \delta_{S,S'}\delta_{S,1} \, W_{T1}^{(I)} \, (-)^{M}(k_{12})_{-M}(k_{12})_{M'} \, . 
\label{ME-SNM}
\end{eqnarray}
The symbol $\delta(\alpha,\alpha')$ is a shorthand notation for the product of Kronecker delta for all quantum numbers. 
As compared to Eq. (\ref{res}) we have slightly modified the notation, using $W_{SO}^{(I)} = 4 C_{I}^{\nabla J}$, $W_{T1}^{(I)} = 4 C_I^F$, and $W_{T2}^{(I)}= 8 C_I^{\nabla s} - 2 C_I^F$, to stress the genuine spin-orbit and tensor origin of these terms. We have also included a new generic term $W_3^{(\alpha)}$, thus anticipating the generalization of the NEDF we shall discuss later on (Sec. \ref{sec:extra})  in connection to a general three-body effective interaction, which also contributes to $W_1^{(\alpha)}$ and $W_2^{(\alpha)}$.  Actually the momentum dependence associated to $W_3^{(\alpha)}$ was considered in \cite{gar92} in the framework of liquid $^3$He in connection to the density dependence of the  effective mass.

One can see in (\ref{ME-SNM}) that the spin-orbit term $W_{SO}^{(I)}$ mixes both spin channels $S=0$ and $S=1$. The tensor term $W_{T1}^{(I)}$ is only effective on the $S=1$ channel, but mixes the spin projection indices $M$. However,  tensor effects can influence also the $S=0$ channel, due to the spin coupling induced by the spin-orbit term $W_{SO}$. Notice that the matrix elements are diagonal in the isospin channel, which justifies the use of our isospin convention to simultaneously treat SNM and PNM.  

\begin{table}[H]
\caption{Definition of constants $W_i^{(\alpha)}$ entering (\ref{res}) and (\ref{ME-SNM}) for SNM.}
\label{table:W_i}
\begin{center}
\begin{tabular}{c|c|c}
\hline
\hline
$(\alpha)$ & $W_1^{(\alpha)}/4$ & $W_2^{(\alpha)}/4$  \\ 
 \hline 
(0,0) & $2 C_{0}^{\rho, 0} 
+ (2+\gamma)(1+\gamma) C_{0}^{\rho, \gamma} \rho^{\gamma} 
       -\left[ 2C_{0}^{\Delta \rho}+\frac{1}{2}C_{0}^{\tau}\right]q^{2}$ 
&  $ C_{0}^{\tau}$ \\
(0,1) & $2C^{\rho, 0}_{1}
+ 2C^{\rho,\gamma}_{1} \rho^{\gamma} 
-\left[2 C_{1}^{\Delta \rho}+\frac{1}{2}C_{1}^{\tau}\right]q^{2}$ &  $C_{1}^{\tau}$ \\
(1,0) & $2C_{0}^{s,0} 
+ 2C_{0}^{s,\gamma} \rho^{\gamma}
-\left[2C_{0}^{\Delta s}+\frac{1}{2}C_{0}^{T} \right]q^{2}$ &  $C_{0}^{T}$ \\
(1,1) & $2C_{1}^{s,0} 
+ 2C_{1}^{s\gamma} \rho^{\gamma}
- \left[2C_{1}^{\Delta s}+\frac{1}{2}C_{1}^{T} \right] q^{2}$ &  $C_{1}^{T}$ 
\\
\hline
\hline
\end{tabular}
\end{center}
\end{table}

\begin{table}[H]
\caption{Definition of constants $W_i^{(\alpha)}$ entering  (\ref{res}) and (\ref{ME-SNM}) for PNM.}
\label{table:W_i:PNM}
\begin{center}
\begin{tabular}{c|c|c}
\hline
\hline
$(\alpha)$ & $W_1^{(\alpha)}/2$ & $W_2^{(\alpha)}/2$  \\ 
\hline
(0,n) &$2\left( C_{0}^{\rho, 0} + C_{1}^{\rho, 0} \right) +(2+\gamma)(1+\gamma)\left[ C^{\rho\gamma}_{0}+C^{\rho\gamma}_{1}\right]\rho^{\gamma} $ & $C^{\tau}_{0}+C^{\tau}_{1}$\\
         &$-q^{2}\left[ 2C^{\Delta \rho}_{0}+2C^{\Delta \rho}_{1}+\frac{1}{2}C^{\tau}_{0}+\frac{1}{2}C^{\tau}_{1}\right]$ & \\
(1,n) &$ 2\left( C_{0}^{s, 0} + C_{1}^{s, 0}+ C^{s\gamma}_{0}+C^{s\gamma}_{1}\rho^{\gamma}  \right)$  & $ C^{T}_{0}+C^{T}_{1}$\\
         &$-q^{2}\left[ 2C^{\Delta s}_{0}+2C^{\Delta s}_{1}+\frac{1}{2}C^{T}_{0}+\frac{1}{2}C^{T}_{1}\right] $& \\
\hline
\hline
\end{tabular}
\end{center}
\end{table}

In Tables \ref{table:W_i}-\ref{table:W_i:PNM} are given the parameters $W_i^{(\alpha)}$ in terms of the coupling constants of the Skyrme functional (see Sec.\ref{EDF}) for SNM and PNM respectively.

\subsection{Explicit expressions for the response functions}  

The response function is proportional to the momentum average $\langle G^{(\alpha)}_{RPA} \rangle$ of the RPA $ph$ propagator. By inspecting the Bethe-Salpeter equation (\ref{bethe-salpeter}) one can see that due to the momentum structure of the $ph$ interaction (\ref{ME-SNM}), the response function is coupled to other momentum averages 
involving the weighting functions $k^2$, $k Y_{1,\mu}(\hat{k})$, $(k_{12})_{M'}$, and
$(k_{12})_{-M} (k_{12})_{M'}$. A closed algebraic system is obtained by multiplying the Bethe-Salpeter equation with these weighting functions and integrating over the momentum. We have explicitly given an example in Sec.~\ref{solving-BS} considering a simple $ph$ interaction described in terms of two Landau parameters $f_0$ and $f_1$. The algebraic system involves momentum averages of the HF propagator. One should keep in mind that the HF propagator 
(\ref{GHF}) is independent on the polar angle $\phi$. Therefore, integrals of the type 
$\langle f(k) Y_{l_1,m_1}(\hat{k}) Y_{l_2,m_2}(\hat{k}) \dots Y_{l_n,m_n}(\hat{k}) G_{HF} \rangle$ involving $n$ spherical harmonics vanish unless $m_1+m_2+ \dots +m_n=0$. This simplifies the system of equations by discarding the contribution of some averages.
 
Another simplification concerns the spin orbit terms. The form of the $ph$ interaction (\ref{ME-SNM}) shows a coupling between the functions $\langle G^{(0,0,I)}_{RPA} \rangle$ and $\langle k Y_{1,\pm M}(\hat{k}) G^{(1,M,I)}_{RPA} \rangle$. It has been shown \cite{mar06,dav09} that the equation for the latter function is only coupled to $\langle G^{(0,0,I)}_{RPA} \rangle$, and consequently this coupling can be absorbed into an effective coefficient $\widetilde{W}_1^{(\alpha)}$. 

We present now the explicit expressions of the SNM/PNM response functions in the different spin channels. 
They are given in terms of some averages of the Hartree-Fock $ph$ propagator (\ref{GHF}), labelled $\beta_i$ and $\chi_i$ and defined in \ref{app:beta}.  
For the sake of simplicity,  in the $S=0$ channel we omit the index $M=0$. The response can be written as 
\begin{eqnarray}
\frac{\chi_{HF}}{\chi_{RPA}^{(0,I)}}&=&1-{\widetilde{W}}_{1}^{(0,I)}\chi_{0}+\frac{1}{2}q^{2}{W}_{3}^{(0,I)}\chi_{0}+{W}_{2}^{(0,I)}\left( \frac{q^{2}}{2}\chi_{0}-2k_{F}^{2}\chi_{2}\right)\nonumber\\
&+&[{W}_{2}^{(0,I)}]^{2}k_{F}^{4}\left(-\chi_{0}\chi_{4}+\chi_{2}^{2}+\frac{m^{*2}\omega^{2}}{k_{F}^{4}} \chi_{0}^{2}-\frac{m^{*}q^{2}}{6\pi^{2}k_{F}}\chi_{0}\right)\nonumber\\
&+&\frac{2m^{*2}\omega^{2}}{q^{2}}\frac{{W}_{2}^{(0,I)}-{W}_{3}^{(0,I)}}{1-\frac{m^{*}k_F^3}{3\pi^{2}}({W}_{2}^{(0,I)}-{W}_{3}^{(0,I)})} \, \chi_{0}\, ,
\label{satu:chi0I}
\end{eqnarray}
where we have defined
\begin{eqnarray}
\label{tildeW}
\widetilde{W}_{1}^{(0,I)} &=& {W}_{1}^{(0,I)}(q) +\frac{q^4 [W_{SO}^{(I)}]^{2}(\beta_{2}-\beta_{3})}{1+q^{2}(\beta_{2}-\beta_{3})[{W}_{2}^{(1,I)}-{W}_{3}^{(1,I)}- \frac{1}{2} W_{T1}^{(I)}]} \, .
\end{eqnarray}
As mentioned above, the coupling between the $S$-channels induced by the spin-orbit interaction has been absorbed into the effective interaction coefficient $\widetilde{W}_1^{(0,I)}$. Notice that its second term is proportional to $[W_{SO}^{(I)}]^2$, and also depends on $W_2^{(1,I)}$, $W_3^{(1,I)}$ and $W_{T1}^{(1,I)}$.
Whereas $W_1^{(\alpha)}$ is real and independent of $\omega$ and $T$, the effective  $\widetilde{W}_1^{(0,I)}$ one is in fact a complex function of $q, \omega, T$, through the 
$\beta_2$ and $\beta_3$ functions. 
Since the tensor interaction only affects the $S=1$ channel one could expect the response function in the $S=0$ channel to be independent on the tensor parameter $W_{T1}^{(I)}$. This assertion should be considered more carefully because the spin-orbit interaction couples both $S=0$ and 1 channels and is reflected in the effective coefficient $\widetilde{W}_1^{(0,I)}$, defined above, which contains an explicit dependence on $W_{T1}^{(I)}$. However, this coupling term is proportional to $q^4$. We shall see that the spin-orbit interaction gives in general a noticeable contribution to the response function only for values of the transferred moment $q$ large enough as compared to $k_F$.   

For the channel $S=1$, we must distinguish the responses of the longitudinal ($M=0$) and transverse $(M=\pm1)$ channels. These are written as
\begin{eqnarray}
\frac{\chi_{HF}}{\chi_{RPA}^{(1,0,I)}}&=&\left(1+\frac{k_{F}^{3} m^{*} W_{T1}^{(I)}}{6\pi^{2}} \right)^{2} -\widetilde{W}_{1}^{(1,0,I)}\chi_{0}+\frac{1}{2}q^{2}W_{3}^{(1,I)}\chi_{0}\nonumber\\
&+& W_{2}^{(1,I)}\left[\frac{q^{2}}{2}\left( 1+\frac{W_{T1}^{(I)} k_{F}^{3}m^{*}}{3\pi^{2}}\right)\chi_{0}-2k_{F}^{2}\chi_{2}+\frac{k_{F}^{5}m^{*} W_{T1}^{(I)}}{3\pi^{2}}\left( \chi_{0}-\chi_{2}\right) \right]\nonumber\\
&+& [W_{2}^{(1,I)}]^{2}\left[k_{F}^{4}\chi_{2}^{2}-k_{F}^{4}\chi_{0}\chi_{4}+m^{*2}\omega^{2}\chi_{0}^{2}-\frac{k_{F}^{3}m^{*}q^{2}}{6\pi^{2}}\chi_{0} \right]\nonumber\\
&+& \frac{2m^{*2}\omega^{2}}{q^{2}}\frac{(W_{2}^{(1,I)}-W_{3}^{(1,I)}+W_{T1}^{(I)})\left[ 1+\frac{k_{F}^{3}m^{*}}{3\pi^{2}}X^{(1,0,I)}\right]}{1+\frac{k_{F}^{3}m^{*}}{3\pi^{2}}(X^{(1,0,I)}-W_{2}^{(1,I)}+W_{3}^{(1,I)}-W_{T1}^{(I)})} \, \chi_{0}\, ,
\label{SNM-S1-M0}
\end{eqnarray}
\begin{eqnarray}
\frac{\chi_{HF}}{\chi_{RPA}^{(1,\pm1,I)}}&=&\left[1- \frac{m^{*}k_{F}^{3} W_{T1}^{(I)}}{12\pi^{2}} \right]^{2}-{\widetilde{W}}_{1}^{(1,\pm1,I)}\chi_{0} +\frac{1}{2}q^{2}W_{3}^{(1,I)}\chi_{0}\nonumber\\
&+&\left[W_{2}^{(1,I)}+ \frac{1}{2} W_{T1}^{(I)} \right]\left\{ \frac{q^{2}}{2}\chi_{0}\left[1- 
\frac{m^{*}k_{F}^{3} W_{T1}^{(I)}}{6\pi^{2}}\right] -2k_{F}^{2}\chi_{2}- \frac{m^{*}k_{F}^{5} W_{T1}^{(I)}}{6\pi^{2}}(\chi_{0}-\chi_{2})\right\}\nonumber\\
&+&\left[W_{2}^{(1,I)}+ \frac{1}{2} W_{T1}^{(I)} \right]^{2}k_{F}^{4}\left\{\chi_{2}^{2}-\chi_{0}\chi_{4}+\left( \frac{m^{*}\omega}{k_{F}^{2}}\right)^{2} \chi_{0}^{2}-\frac{m^{*}}{6\pi^{2}k_{F}}q^{2}\chi_{0}\right\}\nonumber\\
&+&2\chi_{0}\left(\frac{m^{*}\omega}{q} \right)^{2}\frac{ (W_{2}^{(1,I)}-W_{3}^{(1,I)}) \left(1+\frac{m^{*}k_{F}^{3}}{6\pi^{2}}X^{(1,\pm1,I)} \right)}{1-\frac{m^{*}k_{F}^{3}}{6\pi^{2}}\left[2(W_{2}^{(1,I)}-W_{3}^{(1,I)})-X^{(1,\pm1,I)} \right]}\, ,
\label{SNM-S1-M1}
\end{eqnarray}
where we have defined
\begin{eqnarray}
\widetilde{W}_{1}^{(1,0,I)}&=& W_{1}^{(1,I)} + q^{2} W_{T2}^{(I)} + 
\frac{2m^{*2}\omega^{2}}{q^{2}} W_{T1}^{(I)} 
-  \left(\frac{k_{F}^{5}m^{*}}{6\pi^{2}}+\frac{k_{F}^{3}m^{*}q^{2}}{24\pi^{2}}-\frac{k_{F}^{3}m^{*3}\omega^{2}}{6\pi^{2}q^{2}} \right) [W_{T1}^{(I)}]^{2}\, ,
\end{eqnarray}
\begin{eqnarray}
{\widetilde{W}}_{1}^{(1,\pm1,I)} &=& W_{1}^{(1,I)} +\frac{ \frac{1}{2} q^{4} [W_{SO}^{(I)}]^{2}(\beta_{2}-\beta_{3})}{1+ (W_{2}^{0I}-W_{3}^{0I})q^{2}(\beta_{2}-\beta_{3})}
-\frac{m^{*2}\omega^{2}}{q^{2}} W_{T1}^{(I)}  \\
&+&  \frac{1}{16} [W_{T1}^{(I)}]^{2} \left[ \frac{2m^{*}k_{F}^{3}q^{2}}{3\pi^{2}}+\frac{1}{4}\left( q^{2}-4 \left( \frac{m^{*}\omega}{q}\right)^{2}\right)^{2} \chi_{0} 
-2k_{F}^{2}\left( q^{2}+4\left( \frac{m^{*}\omega}{q}\right)^{2}\right)\chi_{2}+4k_{F}^{4}\chi_{4}\right]\, ,
\nonumber
\end{eqnarray}
and
\begin{eqnarray}
X^{(1,0,I)}&=& \frac{\frac{1}{2} [W_{T1}^{(I)}]^{2} q^{2} (\beta_{2}-\beta_{3})}{1+q^{2}(\beta_{2}-\beta_{3}) {W}_{2}^{(1,I)}}\, , \\
X^{(1,\pm1,I)}&=&\frac{ \frac{1}{2} [W_{T1}^{(I)}]^{2} q^{2}(\beta_{2}-\beta_{3})}{1+(W_{2}^{(1,I)}-W_{3}^{(1,I)})q^{2}(\beta_{2}-\beta_{3})}\, .
\end{eqnarray}

Notice that the longitudinal channel $(1,0,I)$ does not depend on the spin-orbit parameter $W_{SO}^{(I)}$. Indeed, the spin-orbit contribution to the particle-hole interaction (\ref{ME-SNM}) is multiplied by the spin projection $M$, which is null in that channel. On the contrary, the transverse channel depends explicitly on the spin-orbit interaction, as reflected in the effective interaction coefficient 
$\widetilde{W}_1^{(0,I)}$ coefficient in the same way as in the $S=0$ channel. Analogously to this case, the spin-orbit interaction becomes noticeable only for large values of the transferred moment $q$, as compared to $k_F$.   

The tensor interaction couples the longitudinal and the transverse channels. In a similar way as for the spin-orbit coupling shown before, it is possible to reduce the algebraic system and obtain compact expressions for these responses \cite{dav09}. However the tensor coupling is only partly absorbed in an effective $\widetilde{W}_1^{(\alpha)}$ function as for the spin-orbit coupling, and appears also in the last terms of Eqs. (\ref{SNM-S1-M0}) and (\ref{SNM-S1-M1}), through the functions $X^{(\alpha)}$ defined above.

In conclusion, the response functions have been written in a compact and practical form, both for SNM and PNM, whose structure is the same whether or not the particle-hole interaction contains spin-orbit and tensor terms. Their inclusion induces a non-trivial coupling among all channels, which can be absorbed into effective interaction parameters that are in fact complex functions of the transferred momentum $q$, energy $\omega$, and temperature.

\subsection{Sum rules} \label{sub:SNM:sumrule} 

A detailed study of sum rules can shed some light on the contribution of the tensor for various physical situations (see Ref. \cite{boh79,lip89}). In this section we present explicit expressions for the odd power sum rules, according to the expansions (\ref{M1-3}, \ref{M-1}). 

The EWSR are written as
\begin{eqnarray}
\label{m100}
\frac{M^{(0,I)}_{1}}{A} &=& \frac{q^{2}}{2m^{*}}\left[1- \frac{1}{2} \left(W_{2}^{(0,I)}-W_{3}^{(0,I)} \right) m^* \rho \right]\, , \\
\frac{M^{(1,\pm1,I)}_{1}}{A} &=& \frac{q^{2}}{2m^{*}}\left[1- \frac{1}{2} \left(W_{2}^{(1,I)}-W_{3}^{(1,I)} \right) m^* \rho \right] \, ,\\
\frac{M^{(1,0,I)}_{1}}{A} &=& \frac{q^{2}}{2m^{*}}\left[1- \frac{1}{2} \left(W_{2}^{(1,I)}-W_{3}^{(1,I)}+ W_{T1}^{(I)} \right) m^* \rho \right]\, .
\end{eqnarray}
Due to Galilean invariance, the EWSR in channel (0,0) is equal to $q^2/2m$. That means that the term in squared brackets in (\ref{m100}) should be equal to $m^*/m$, as it can be easily checked. 
At first sight it seems that the $M=\pm 1$ channel is independent of the tensor since there is no explicit contribution of $W_{T1}^{(\alpha)}$. Actually the constant $W_2^{(\alpha)}$  depends on the tensor parameters, as shown in 
Tables \ref{table:W_i} and \ref{table:W_i:PNM}.  

For the CESWR we have
\begin{eqnarray}
\frac{M^{(0,I)}_{3}}{A}&=&\frac{k_{F}^{2} q^{4}}{2m^{*3}} \left[1- \frac{1}{2} \left(W_{2}^{(0,I)}-W_{3}^{(0I)} \right) 
m^* \rho \right]^{2} \nonumber\\
&& \times \left[\frac{3}{5}+k^{2}
+\frac{1}{4} \left(W_{1}^{(0,I)}+2k_F^2 W_{2}^{(0,I)} \right) \frac{m^{*}\rho}{k_F^2}
+\frac{1}{2} \left(W_{2}^{(0,I)}-W_{3}^{(0,I)} \right) m^* \rho k^2 \right] \, ,\\
\frac{M_3^{(1,\pm 1,I)}}{A} &=& \frac{k_F^2 q^4}{2 (m^*)^3}
\left[1- \frac{1}{2} \left( W_{2}^{(1,I)}-W_{3}^{(1,I)} \right) m^* \rho \right]^2 \nonumber \\
&& \times \left[ \frac{3}{5} + k^2 
+ \frac{1}{4} \left(  W_1^{(1,I)} + 2 k_F^2 W_2^{(1,I)} \right) \frac{m^* \rho}{k_F^2}
 + \frac{1}{2} \left( W_{2}^{(1,I)} -W_{3}^{(1,I)} \right) m^* \rho k^2 \right. \nonumber \\
&& \left. \quad + \frac{1}{10} W_{T1}^{(I)} m^* \rho- \frac{1}{160}\left[ W_{T1}^{(I)}  m^* \rho  \right]^2 \right]\, , \\
\frac{M_3^{(1,0,I)}}{A} &=& \frac{k_F^2 q^4}{2 (m^*)^3}
\left[1 - \frac{1}{2} ( W_{2}^{(1,I)} -W_{3}^{(1,I)} + W_{T1}^{(I)}) m^* \rho
\right]^2 \nonumber \\
&& \left[ \left( \frac{3}{5} + k^2 \right) \left( 1 + \frac{1}{2} m^* \rho W_{T1}^{(I)} \right) + \frac{1}{4}  \left( W_{1}^{(1,I)} + 2 k_F^2 W_{2}^{(1,I)} \right) \frac{m^* \rho}{k_F^2} 
 + m^* \rho k^2 W_{T2}^{(I)}  \right. \nonumber \\
&& \left. + \frac{1}{2}\left( W_2^{(1,I)} - W_3^{(1,I)} \right) m^* \rho k^2 
 - \frac{1}{40} \left[ W_{T1}^{(I)} m^* \rho \right]^2 
\right] \, ,
\end{eqnarray}
with $k=q/(2k_{F})$.

For IEWSR we have
\begin{eqnarray}
\frac{M^{(0,I)}_{-1}}{A}&=& \frac{3m^{*}}{2k_{F}^{2}} f(k) \bigg\{ 
 \left( 1 + \frac{3}{8} W_2^{(0,I)} m^* \rho \right)^2 
+ \frac{3}{4} \left( W_1^{(0,I)} + W_2^{(0,I)} k_F^2 (1-k^2) - 2 W_3^{(0,I)} k_F^2 k^2 \right) \frac{m^* \rho}{k_F^2} f(k) \nonumber \\
&&
- \frac{3}{64} \left[ W_2^{(0,I)} m^* \rho \right]^2 f(k) \left( 2+ \frac{26}{3} k^2 + (1-k^2) f(k) \right) \nonumber \\
&& - \frac{3}{8} \left[ W_{SO}^{(I)} m^* \rho \right]^2 
\frac{ k^2 f(k) \left[ 1+ 3(1 - k^2)f(k)\right]}{1-\frac{1}{8} \left[ W_{2}^{(1,I)} -W_{3}^{(1,I)}- \frac{1}{2} W_{T1}^{(I)} \right] 
m^* \rho \left[ +3(1 - k^2)f(k)\right]} \bigg\}^{-1} \quad \, ,
\end{eqnarray}
\begin{eqnarray}
\frac{M_{-1}^{(1,\pm1,I)}}{A}&=& \frac{3m^{*}}{2k_{F}^{2}} f(k)  \bigg\{ 
\left( 1+ \frac{1}{16} W_{T1}^{(I)} m^* \rho +\frac{3}{8} W_2^{(1,I)} m^* \rho \right)^{2} 
\nonumber \\
&&+ \frac{3}{4} m^* \rho f(k) \left[ \frac{1}{k_F^2} W_1^{(1,I)} + W_2^{(1,I)} 
- k^2 ( W_2^{(1,I)} + 2 W_3^{(1,I)} ) - \frac{1}{2} W_{T1}^{(I)} (1-k^2) \right. \nonumber \\
&& \left. -\frac{1}{8} m^* \rho [W_2^{(1,I)}]^2 (1+\frac{13}{3} k^2 ) - \frac{1}{6} m^* \rho W_2^{(1,I)} W_{T1}^{(I)} k^2
- \frac{1}{32} m^* \rho [W_{T1}^{(I)}]^2( 1- \frac{1}{3} k^2 ) \right] \nonumber \\
&& - \frac{3}{64} (m^* \rho)^2 f^2(k) [W_2^{(1,I)} + \frac{1}{2} W_{T1}^{(I)} ]^2 (1-k^2)^2 \nonumber \\
&&+ \frac{3}{8} \left[ W_{SO}^{(I)} m^* \rho \right]^2 
\frac{ k^2 f(k) \left[ 1+3 (1-k^2) f(k) \right]}{1- \frac{1}{8} m^* \rho (W_{2}^{(0,I)}-W_{3}^{(0,I)})  
\left[ 1+3 (1-k^{2}) f(k) \right] }  \bigg\}^{-1}\, ,
\end{eqnarray}
\begin{eqnarray}
\frac{M_{-1}^{(1,0,I)}}{A} &=& \frac{3m^{*}}{2k_{F}^{2}} f(k)  \bigg\{ \left[ 1+\frac{1}{4} W_{T1}^{(I)} m^* \rho + \frac{3}{8} W_{2}^{(1,I)} m^* \rho ) \right]^{2} \nonumber \\
&& + \frac{3}{4} m^* \rho f(k) \left( \frac{1}{k_F^2} W_1^{(1,0,I)} + W_2^{1,I} (1-k^2) - 2 k^2 
W_3^{(1,I)} + 4 k^2  W_{T2}^{(I)} \right. \nonumber \\
&& \left.  -\frac{1}{4} W_2^{1,I} W_{T1}^{(I)} m^* \rho ( 1+3k^2 ) 
- \frac{1}{4} \left[ W_{T1}^{(I)} \right]^2 m^* \rho (1+k^2)  
\right. \nonumber \\
&& \left. 
 -\frac{1}{16} [W_2^{(1,I)}]^2 m^* \rho \left[ 2 (1+\frac{13}{3} k^2) + (1-k^2)^2 f(k) \right] \right) 
 \bigg\}^{-1}\, ,
\end{eqnarray}
where the function
\begin{equation}
f(k)=\frac{1}{2}\left[ 1+\frac{1}{2k}(1-k^{2}) \log\left( \frac{k+1}{k-1}\right)\right]\, ,
\end{equation}
is related to the zero-frequency Lindhardt function. 

\subsection{Choice of interactions}
\label{choice}

We have calculated the strength functions in the different $(\alpha)$ channels  for several Skyrme sets which we present now.  
As an example of interaction without tensor terms we have chosen the set SLy5 \cite{cha97,cha98}. Its fit protocol includes, among other properties, a reasonable fit to the equation of state of pure neutron matter, as calculated by Wiringa {\em  et al.} \cite{wir88}, and it is a reliable interaction for astrophysical applications. 
Tensor terms have been added perturbatively to SLy5 by Col\`o {\em et al.} \cite{col07}, fitting experimental single-particle energies for the $N=82$ isotones and $Z=50$ isotopes, but keeping unchanged the remaining parameters. The resulting interaction is labelled SLy5-t in the following.  
We have also considered the T$IJ$ interaction family \cite{les07}, whose tensor parameters have been fixed following the restrictions found in Ref. \cite{sta77} to improve the spin-orbit splitting of some selected levels. Afterwards, a complete refit of the remaining parameters has been performed using the same procedure of \cite{cha97,cha98}. Among this family, we have chosen the sets T22 and T44. 
The peculiarity of the T22  functional is that the so-called $J^{2}$ term~\cite{les07} does not contribute to the ground state of even-even spherical nuclei, while T44 seems to be the preferable one according to properties of finite nuclei studied in ~\cite{les07}.
By construction, these four interactions give the same values for the energy per particle (-16 MeV), equilibrium density (0.16 fm$^{-3}$), effective mass (0.7) and incompressibility modulus (230 MeV) for symmetric nuclear matter. 

\begin{figure}[ht]
\begin{center}
  \includegraphics[clip,scale=0.4,angle=-90]{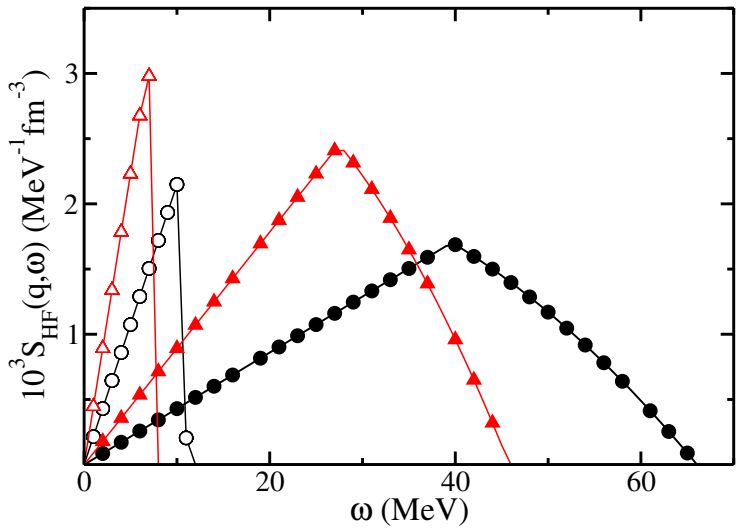}  
\caption{(Color online) HF strength functions associated to SLy and Skx sets (circles and triangles, respectively) 
for transferred momenta $q=0.1k_F$ and $0.5k_F$ (open and filled symbols, respectively) at saturation density in SNM.}
\label{SNM-HF}
\end{center}
\end{figure}

We have also considered the Skx interaction family \cite{bro06}, whose tensor parameters have been first calibrated with results from a finite-range $G$-matrix calculation, and afterwards all parameters have been varied to a best fit of data. Among this family, we have selected the Skxta and Skxtb sets. They give the same values for energy per particle (-16 MeV), equilibrium density (0.156 fm$^{-3}$), effective mass (1.0) and incompressibility modulus (313 MeV) for symmetric nuclear matter. 
The main differences between SLy and Skx sets concern the effective masses and the incompressibility modulus. 

The effective mass is the only dynamical ingredient relevant for the uncorrelated HF strength function $S_{HF}(q,\omega)$. In Fig. \ref{SNM-HF} are plotted the $S_{HF}(q,\omega)$ associated to SLy and Skx sets at saturation density and two transferred momenta. They display the familiar pattern of a Fermi gas (see Fig. \ref{reponselibre}), with the $ph$ continuum edge being inversely proportional to the effective mass. One should keep in mind this figure as a reference for the coming discussion concerning the effects of the residual interaction.

\subsection{Symmetric nuclear matter results}

\begin{figure}
\begin{center}
  \includegraphics[width=0.6\textwidth,angle=-90]{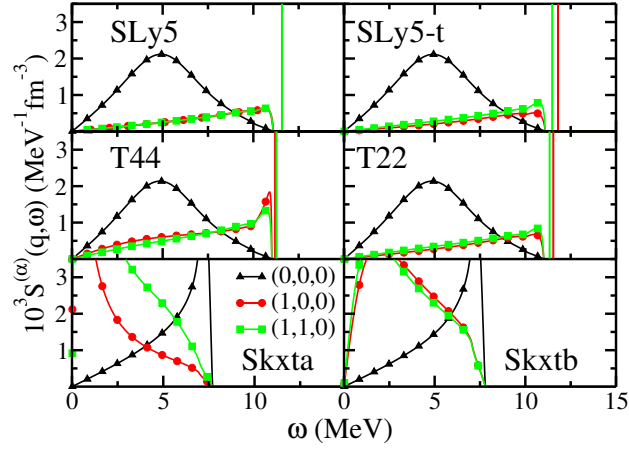}  
    \includegraphics[width=0.6\textwidth,angle=-90]{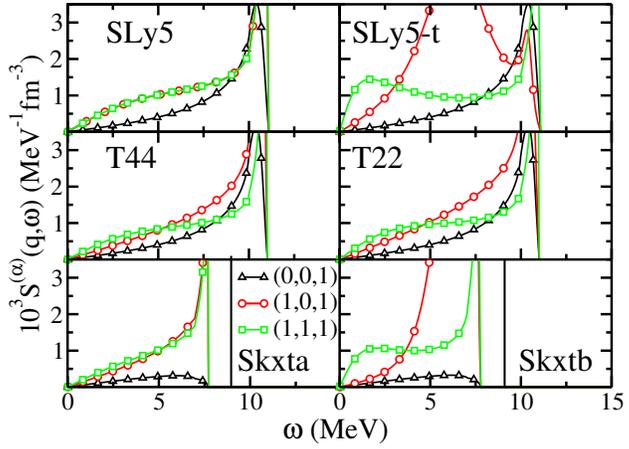}  
\mbox{}
\caption{(Color online) SNM strength functions obtained from several interactions at their respective saturation density and transfer momentum $q=0.1 k_F$ and $T=0$. Isospin $I=0$ ($I=1$) channels are given in top (bottom) panels.}
\label{SNM-ALL}
\end{center}
\end{figure}

We discuss here the general features of the SNM strength functions calculated with the selected interactions. In Fig. \ref{SNM-ALL} are plotted the strengths as a function of $\omega$ at the saturation density $\rho_0$, for a transferred momentum $q=0.1 k_F$ and zero temperature.    
The SLy5 interaction has no tensor terms, and thus the $S=1$ strength functions are essentially the same whatever the value of $M$, the small differences being related to the spin-orbit contribution. The isoscalar strength has a large peak at around 5 MeV in the $S=0$ channel, and the $S=1$ intensity is very reduced. The isovector strength is largely concentrated in two peaks near the $ph$ edge, associated with two collective isovector states.  
One can see that the $S=0$ results obtained with interactions SLy5-t, T22 and T44 are barely distinguishable from those obtained with SLy5. We have seen that the tensor terms affect the $S=0$ channels through the spin-orbit interaction, which couples both $S=0$ and $S=1$ channels. However, this coupling is only relevant for relatively large values of the transferred momentum. We have checked that in practice it is negligible for values of $q$ below $\simeq k_F$.   
The SLy5-t results show the main effect of tensor terms: as compared to SLy5 results, there is a strong splitting of the isovector $S=1$ strength into the $M=0$ and $M=1$ channels, with two visible bumps at lower energies. The splitting is however reduced for T44, which has fully fitted parameters.
The effect on the isoscalar $S=1$ is very small; we shall see that this is related to the very small transferred momentum considered in this specific case.
As regarding the Skxta and Skxtb results one can clearly see the effect of the different effective mass, as compared to the other results, reducing the $ph$ continuum edge in about 30\%. Both interactions predict  resonances in both isoscalar and isovector $S=0$ channels, with a small strength in the latter. The isoscalar $S=1$ strength is concentrated in the low energy part, with a divergence in the isoscalar channel.

\begin{figure}
\begin{center}
\includegraphics[width=0.6\textwidth,angle=-90]{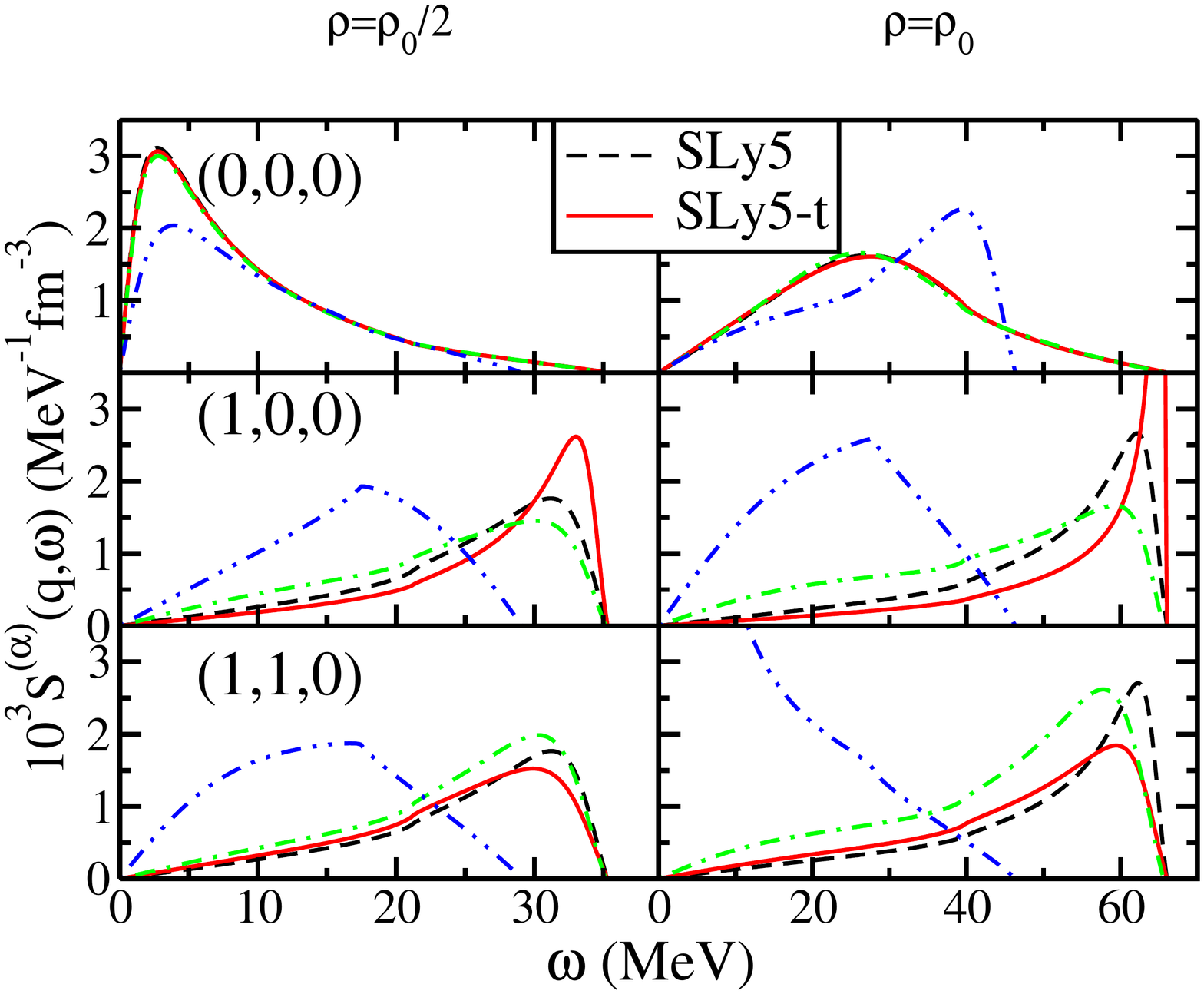}  
\includegraphics[width=0.6\textwidth,angle=-90]{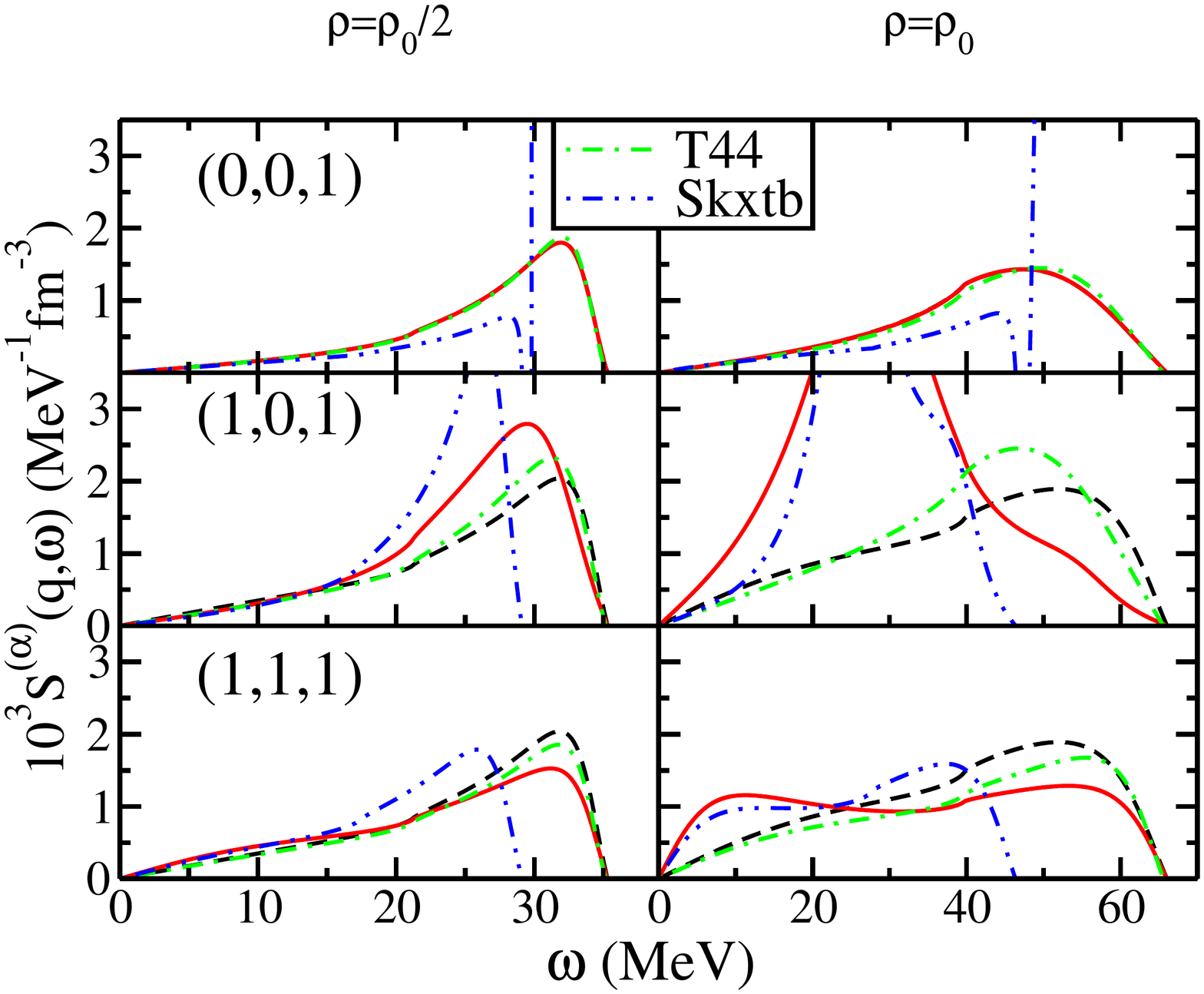}  
\caption{(Color online) SNM RPA strength function obtained from several interactions for density values $\rho=\rho_0/2$ (left panels) and $\rho=\rho_0$ (right panels), where $\rho_0$ is the respective SNM saturation density. In both cases the transfer momentum is $q=0.5k_{F}$.  Isospin $I=0$ ($I=1$) channels are given in top (bottom) panels.}
\label{RESP-qkF05}
\end{center}
\end{figure}

\begin{figure}[H]
\begin{center}
\includegraphics[width=0.49\textwidth,angle=0]{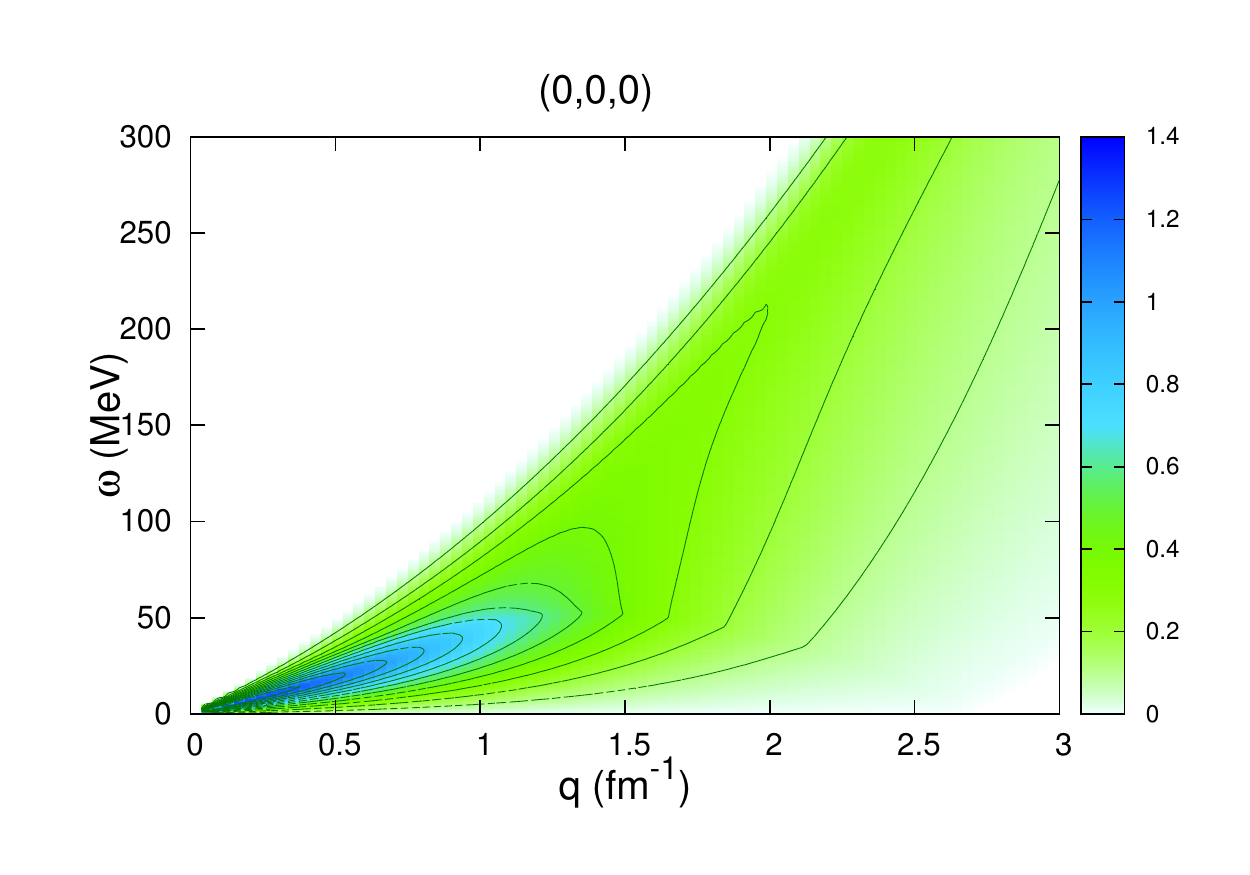}  
\includegraphics[width=0.49\textwidth,angle=0]{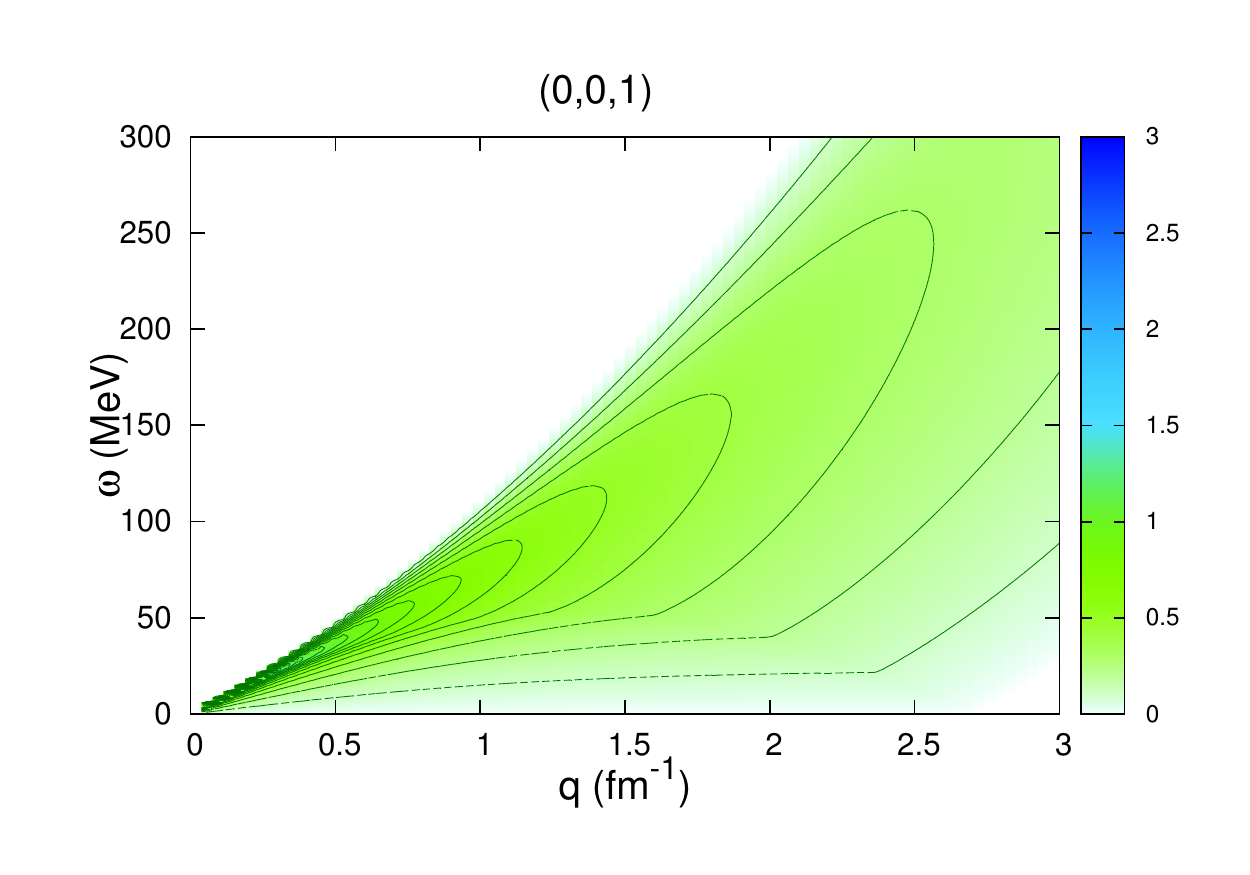}\\
\includegraphics[width=0.49\textwidth,angle=0]{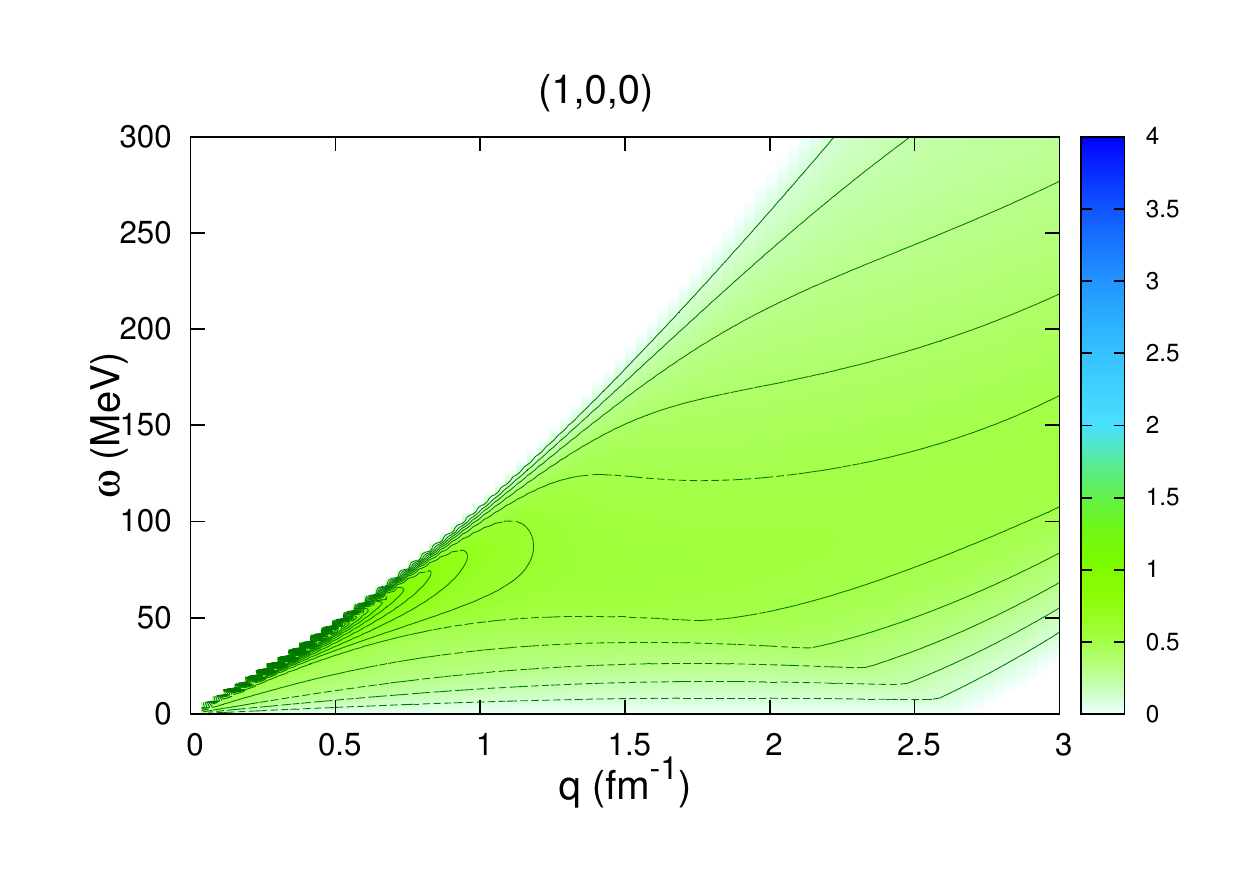}  
\includegraphics[width=0.49\textwidth,angle=0]{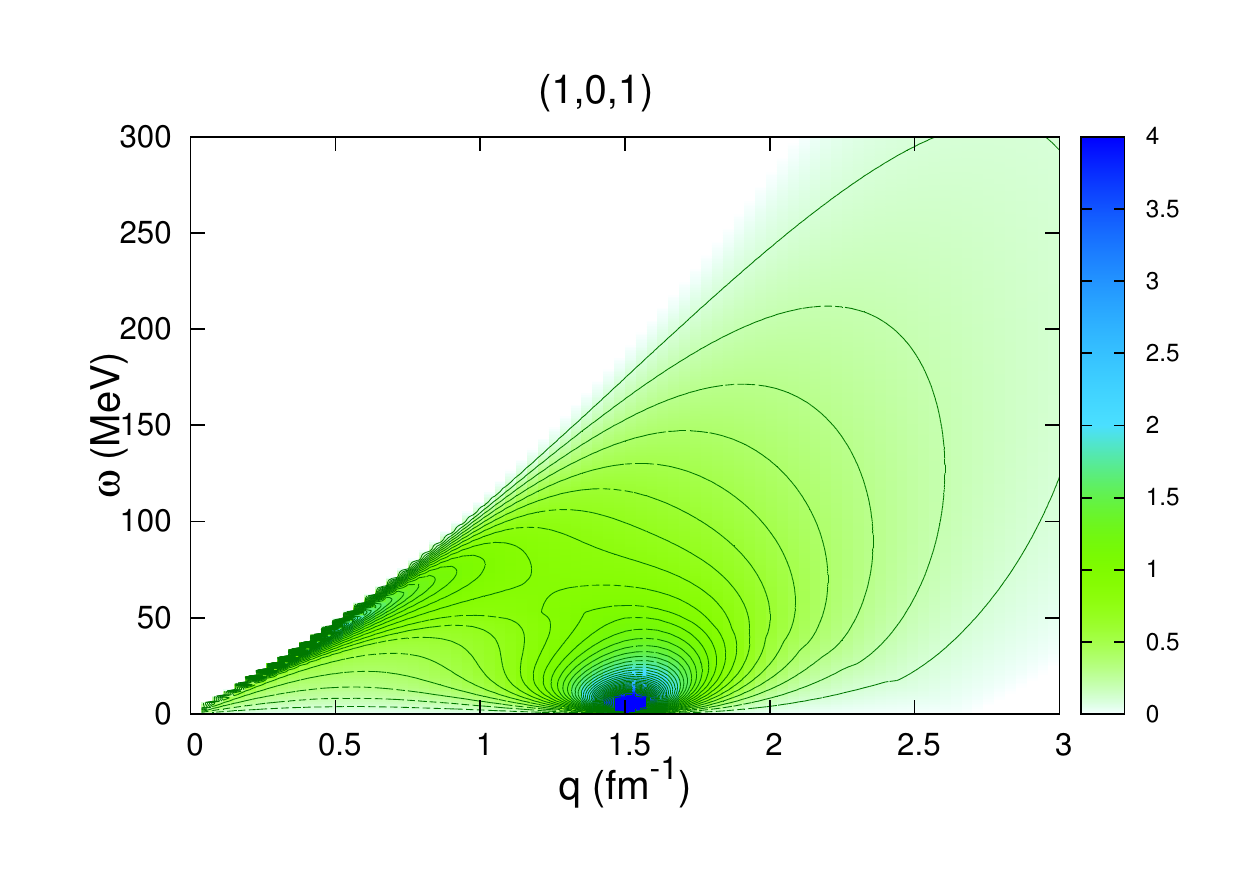}\\  
\includegraphics[width=0.49\textwidth,angle=0]{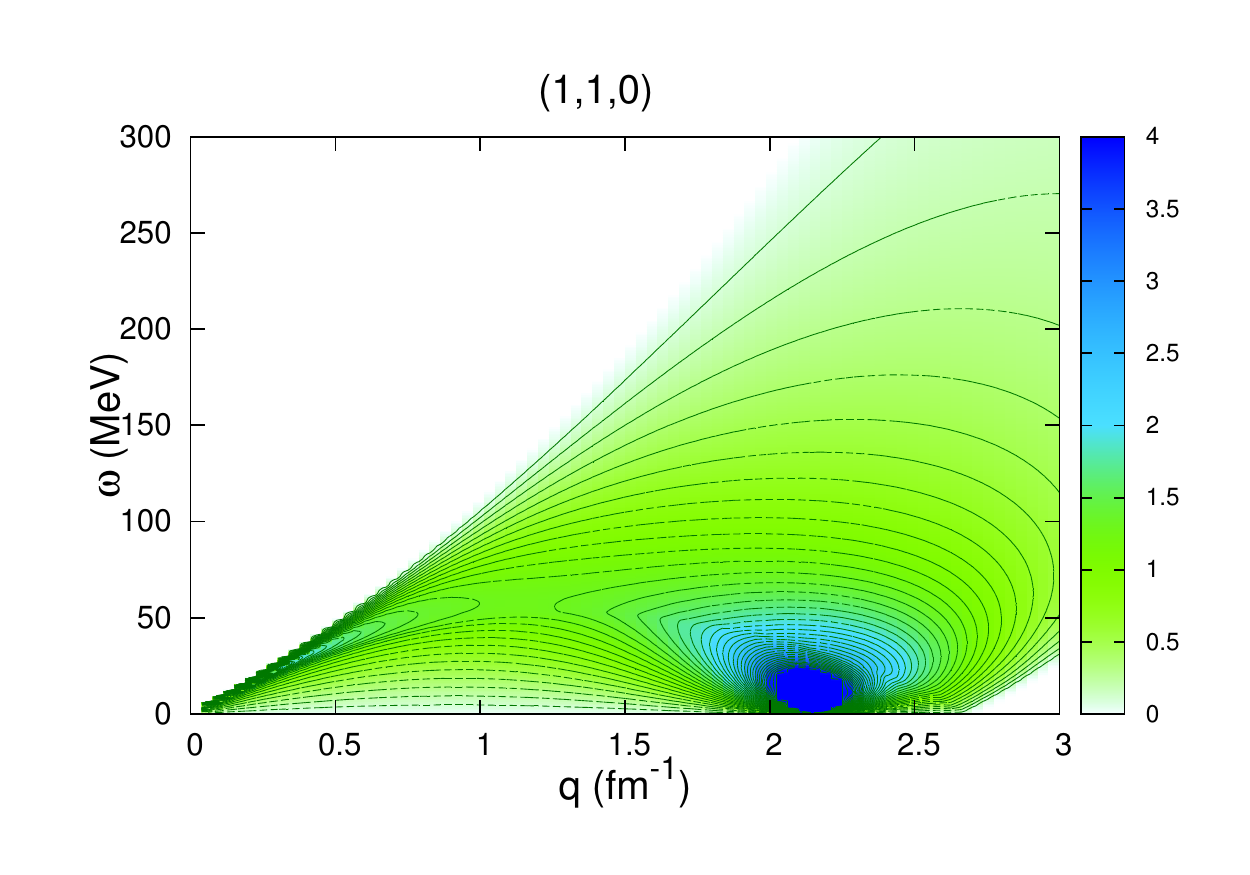}  
\includegraphics[width=0.49\textwidth,angle=0]{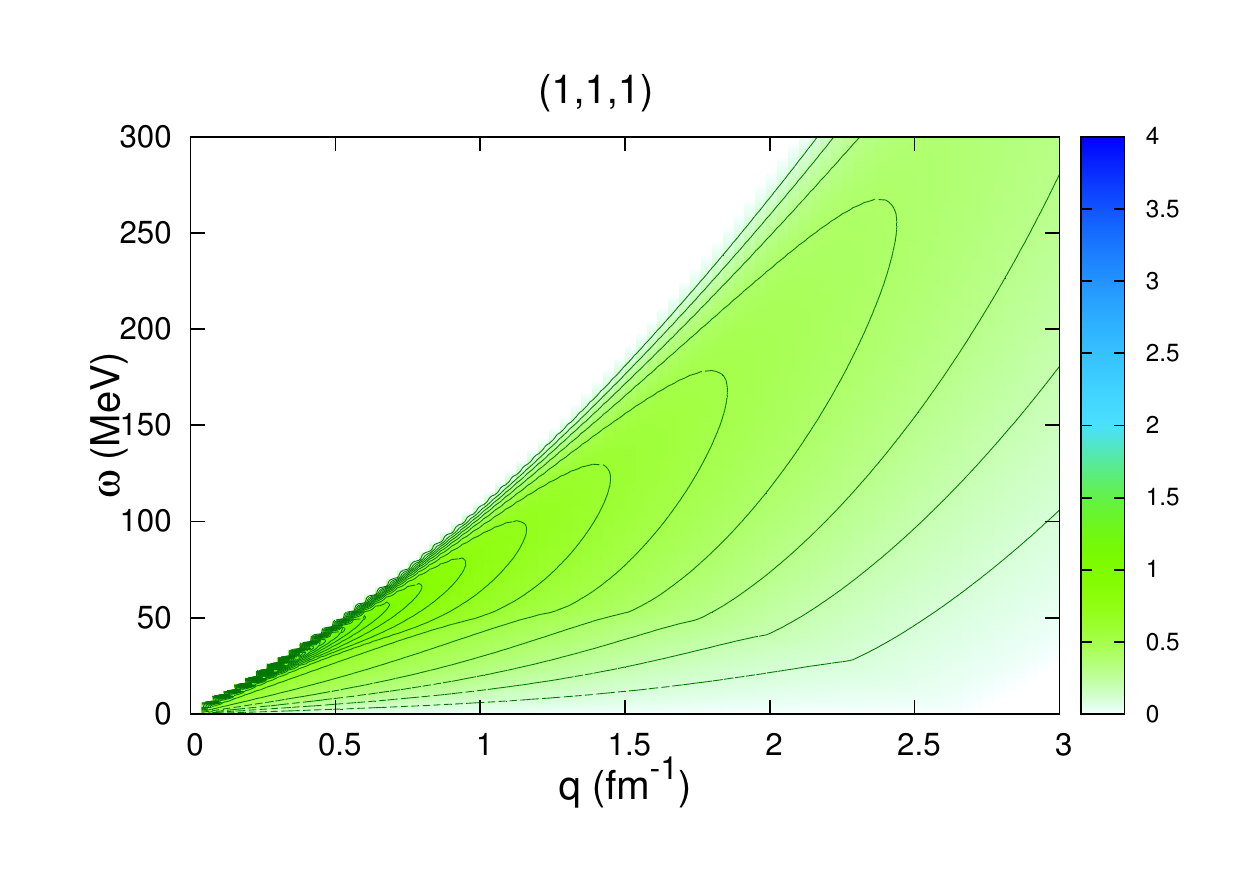}  
\caption{(Color online) SNM RPA strength function T44 calculated at zero temperature and at saturation density. For graphical reasons, the plotted function is $S^{(\alpha)}(q,\omega)/\rho$ in fm units.}
\label{RESP-3D-16}
\end{center}
\end{figure}

\begin{figure}[H]
\begin{center}
\includegraphics[width=0.49\textwidth,angle=0]{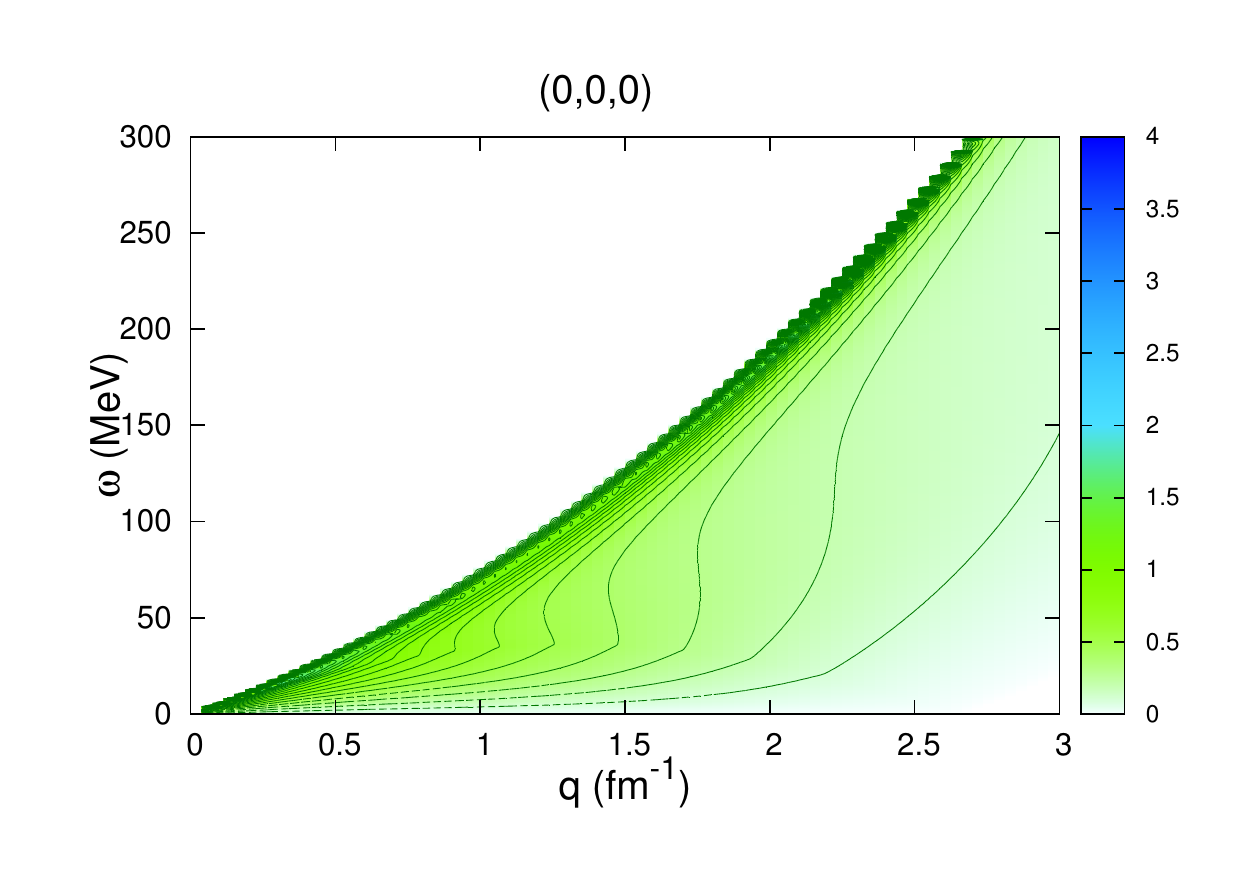}  
\includegraphics[width=0.49\textwidth,angle=0]{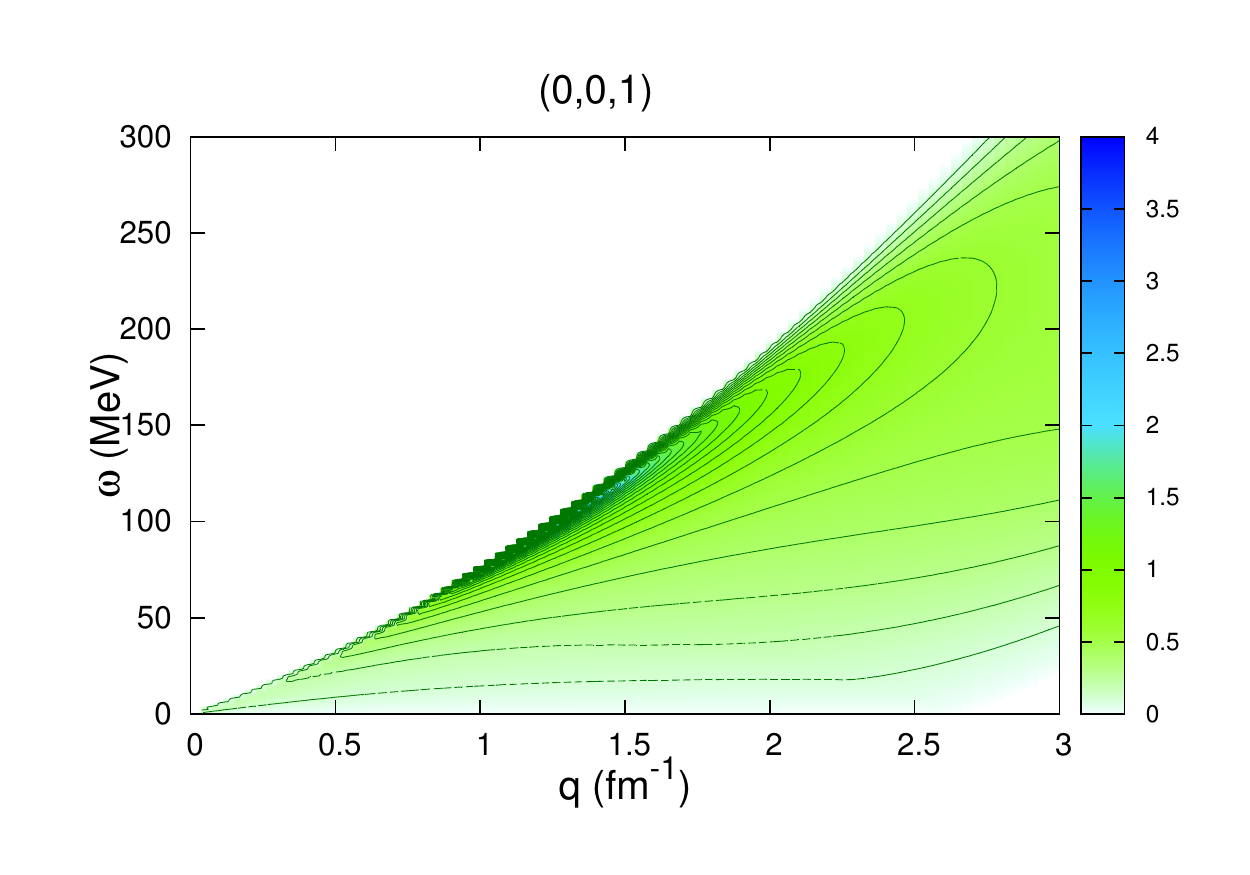}\\
\includegraphics[width=0.49\textwidth,angle=0]{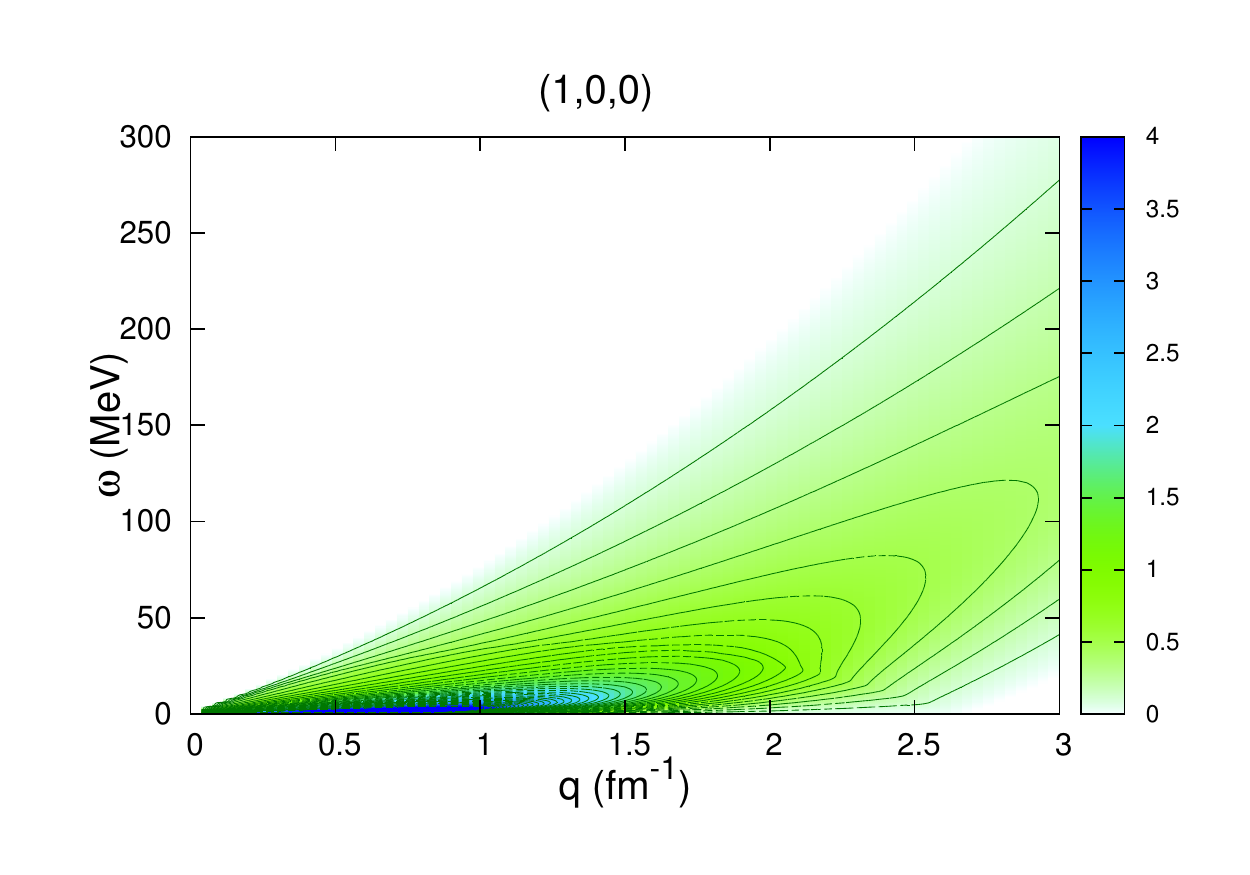}  
\includegraphics[width=0.49\textwidth,angle=0]{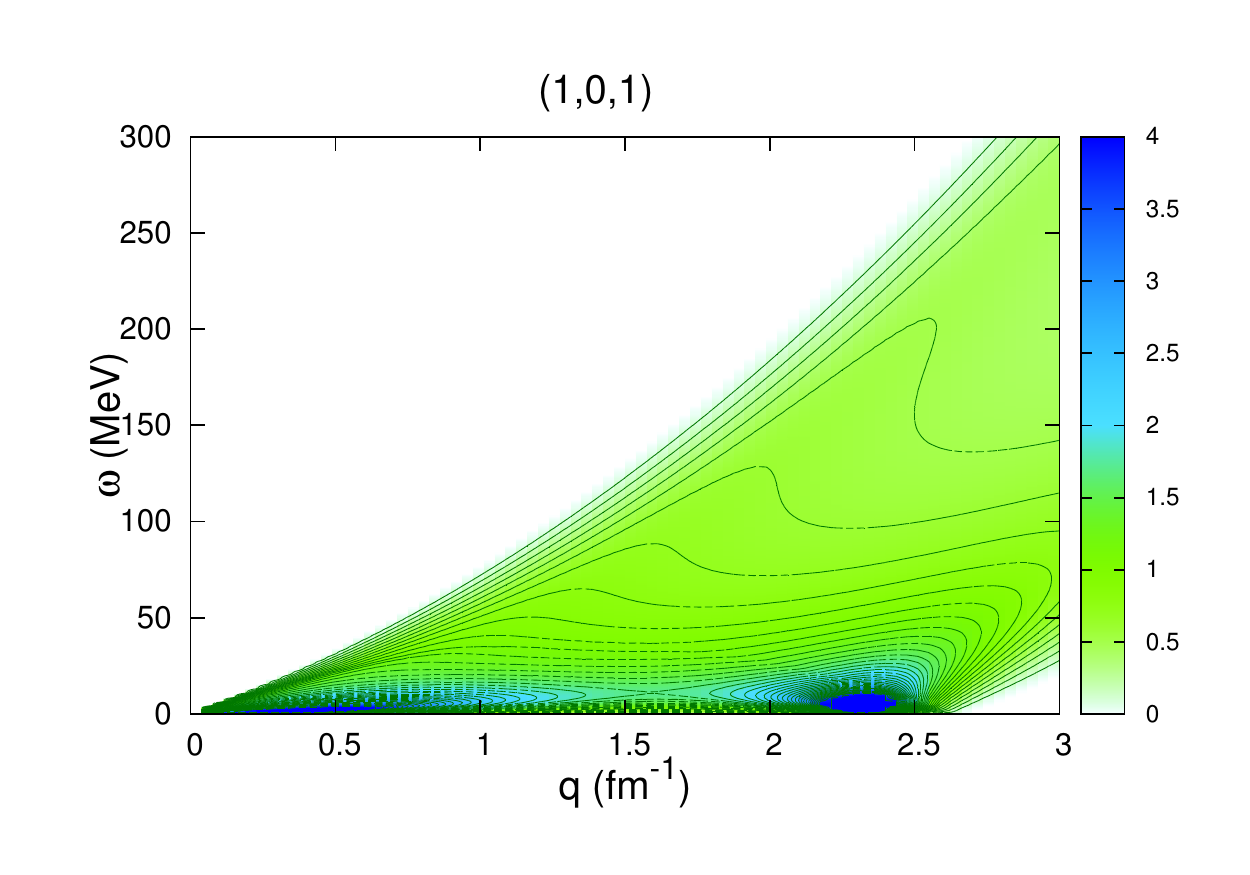}\\  
\includegraphics[width=0.49\textwidth,angle=0]{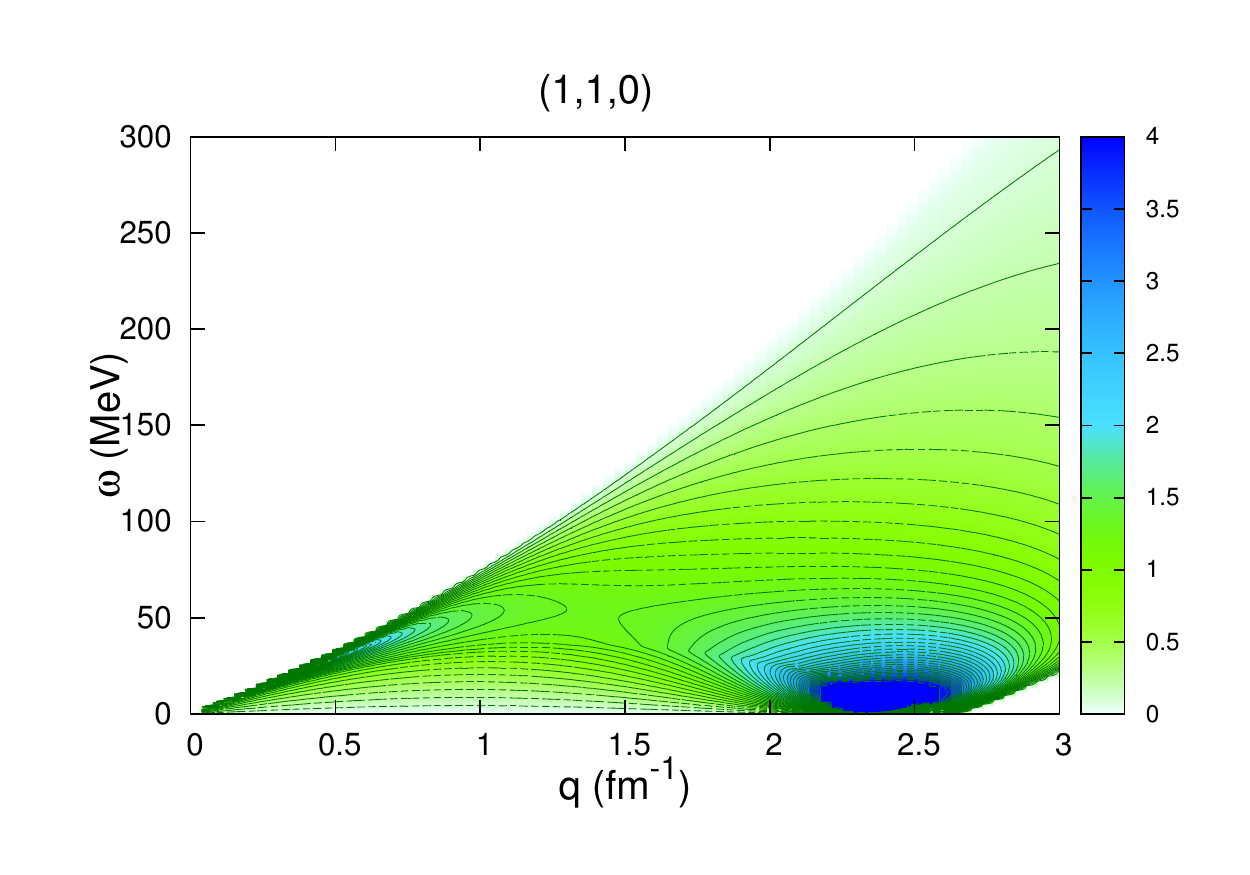}  
\includegraphics[width=0.49\textwidth,angle=0]{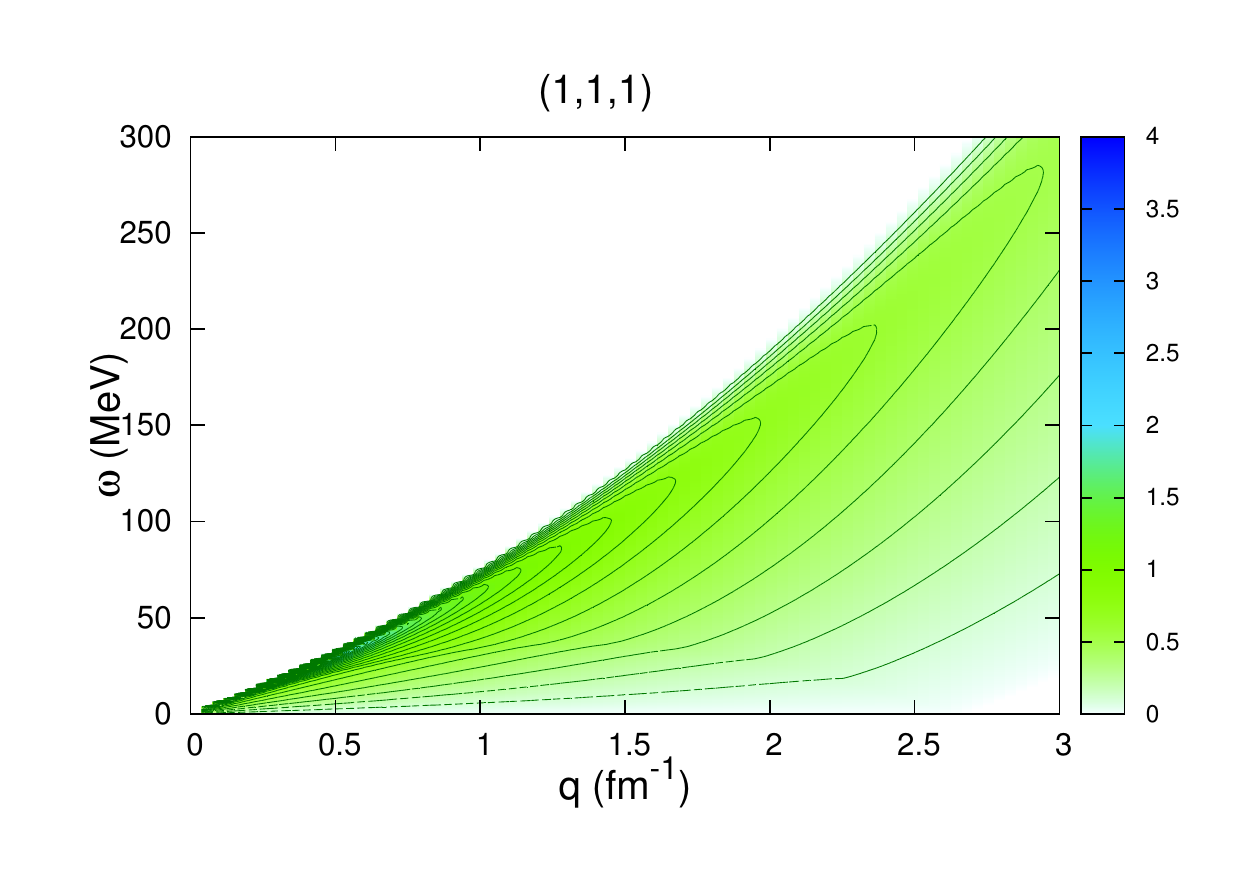}  
\caption{(Color online) Same as Fig. \ref{RESP-3D-16} for interaction Skxta.}
\label{RESP-3D-16:skxta}
\end{center}
\end{figure}

To see the effect of increasing the value of the transferred momentum, we have plotted in Fig.  \ref{RESP-qkF05} the strength function for $q=0.5k_F$ and two density values.  
All in all, the residual interaction produces a redistribution of the strength in all channels, and the most remarkable effect is the presence of huge peaks at very low energies in some cases. 
Divergences at zero energy reveal the presence of instabilities observed in nuclei \cite{les06}, with the appearance of domains with typical size of the order of $2 \pi /q$. To have a global view of the instabilities, we have plotted the projection of the strength function onto the $(q,\omega)$-plane in Figs. \ref{RESP-3D-16} and \ref{RESP-3D-16:skxta} for interactions T44 and Skxta, respectively. The calculations have been done at their respective saturation density $\rho_0$. 

In that way one can easily detect collective modes and singularities. 
We observe that the interaction T44 presents two singularities at nearly zero energy in the (1,0,1) and (1,1,0) channels at transferred momentum  $q\approx1.5$~fm$^{-1}$ and 2.1~fm$^{-1}$ respectively. In comparison, the
Skxta interaction is more repulsive in the S=0 channels as it can be seen by a strong accumulation of the response function at high values of transfered energy $\omega$.

\begin{figure}[H]
\begin{center}
\includegraphics[width=0.4\textwidth,angle=-90]{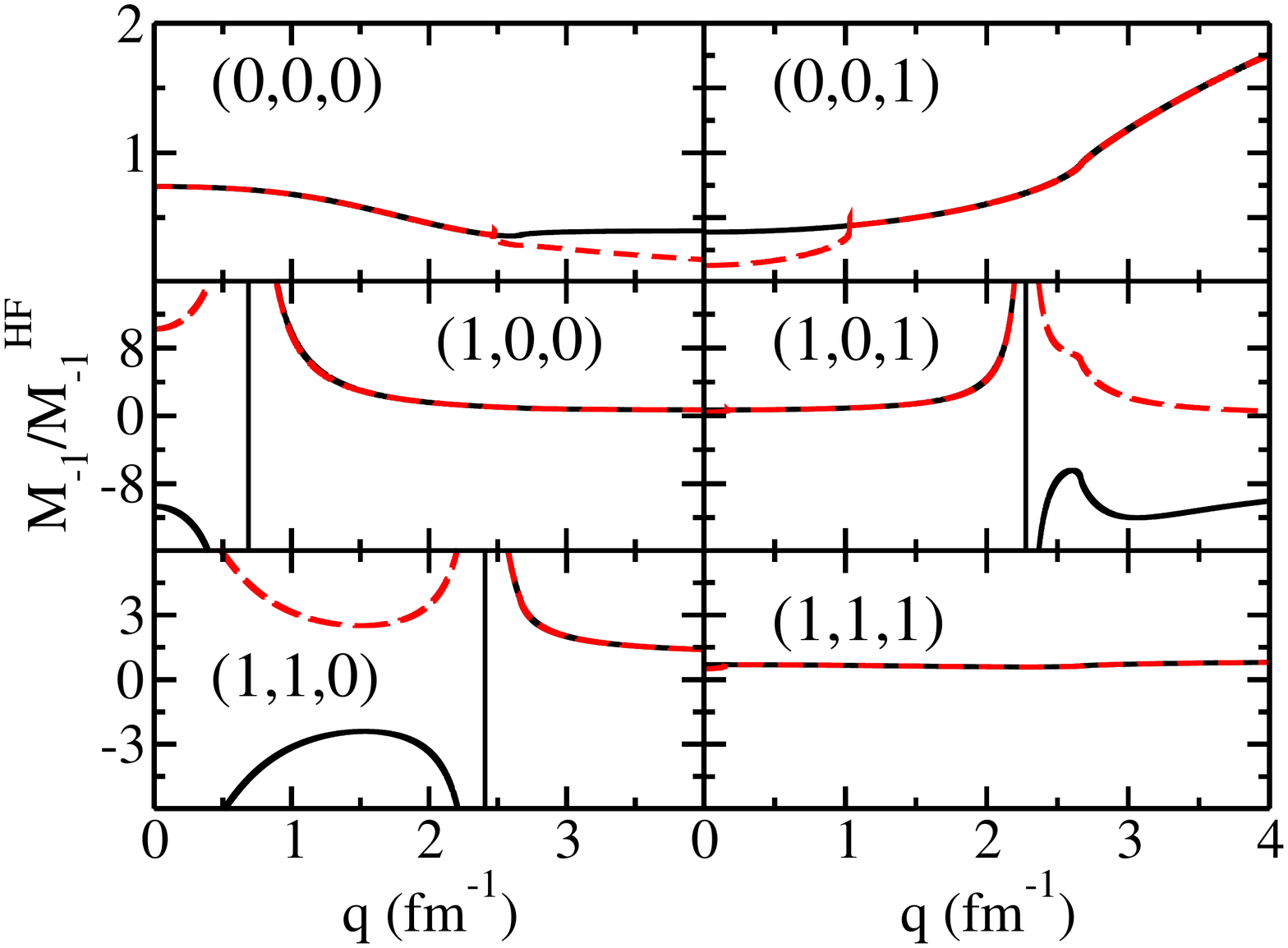}  
\hspace{-0.8cm}  
\includegraphics[width=0.40\textwidth,angle=-90]{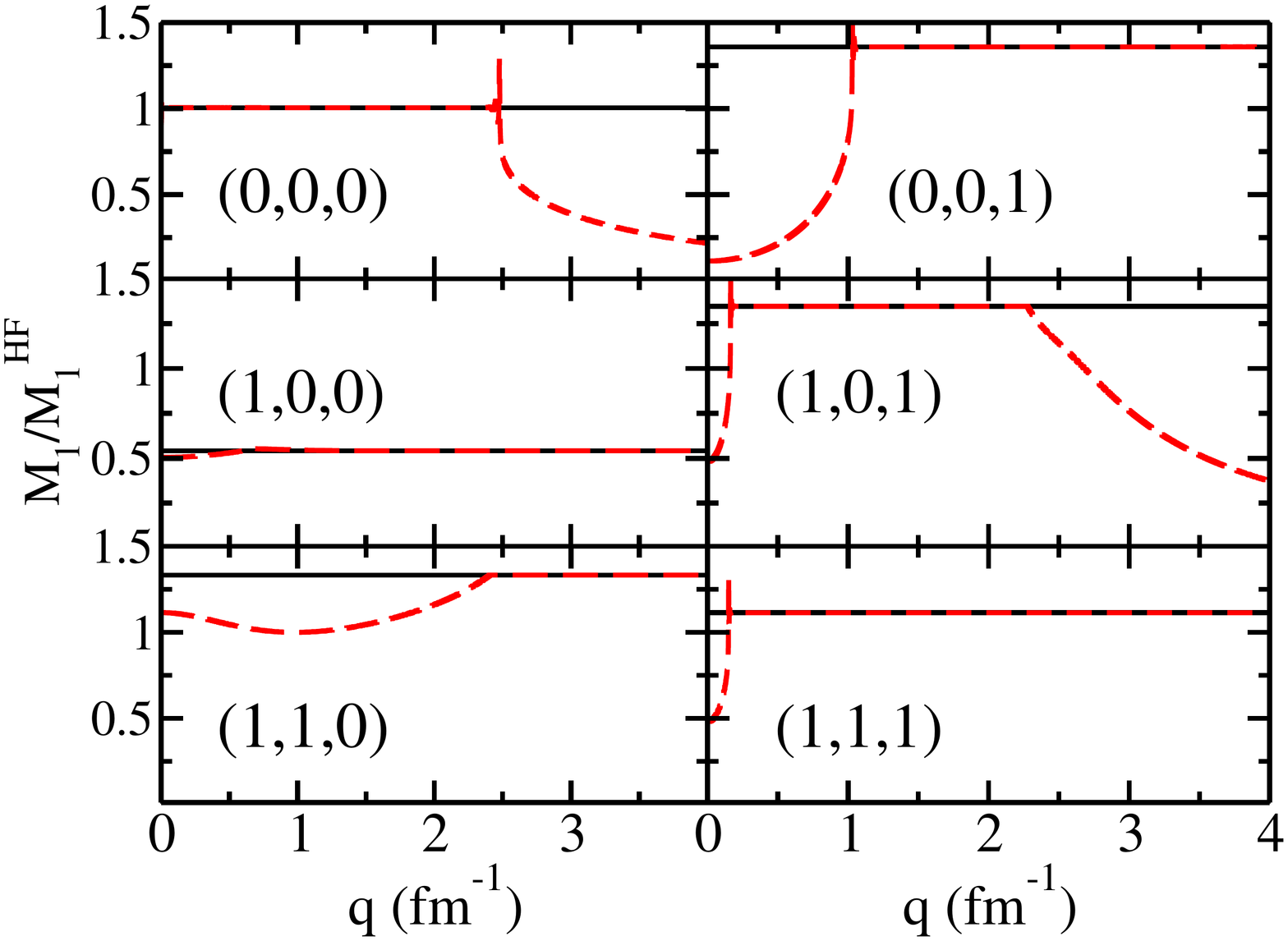}
\caption{(Color online) RPA and HF ratios of  IEWSR (left panel) and EWSR (right panel) for the Skxta interaction at saturation density. The solid and dashed lines represent analytic and numerical RPA sum rules, respectively.}
\label{ewsr:skxta}
\end{center}
\end{figure}

A much simpler way to identify singularities is provided by the sum rules. Indeed, the simplest form of the dispersion relations is no longer available in the presence of a pole, which shows up by comparing the sum rules calculated 
analytically (Sec. \ref{sub:SNM:sumrule}) and numerically (Eq. \ref{SR-numerical}). The value of $q$ at which both calculations differ indicates the existence of a pole, and this value depends of course on the interaction. 
To highlight the connection of the $ph$ interaction and the singularities, it is convenient to plot the ratio of RPA and HF sum rules. The latter are obtained from the former by simply switching all interaction parameters to zero, but keeping of course the HF effective mass. In Fig.~\ref{ewsr:skxta} are plotted these ratios as a function of $q$ for the $M_1^{(\alpha)}$ and $M_{-1}^{(\alpha)}$ sum rules, using Sktxa interaction at saturation density. One can see that analytic (solid lines) and numerical (dashed lines) results coincide except in the proximity of either a singularity or a zero-sound mode. These two cases can be easily distinguished. A singularity at zero transferred energy and with infinite strength appears as a divergence in the sum rules, while the zero sound mode shows up with a loss of sum rule, but with finite value.

\subsection{Spinodal and finite size instabilities}
The IEWSR magnifies the presence of singularities in the response function at low values of the transferred energy. 
In fact, as shown in Eq.~(\ref{M-1}), the IEWSR is proportional to the response function at zero transferred energy (or dynamic susceptibility). It is thus the tool of choice for studying zero-energy instabilities, because the detailed knowledge of the response function is not necessary and the calculations are much simpler. 

\begin{figure}[H]
\begin{center}
\includegraphics[width=0.4\textwidth,angle=-90]{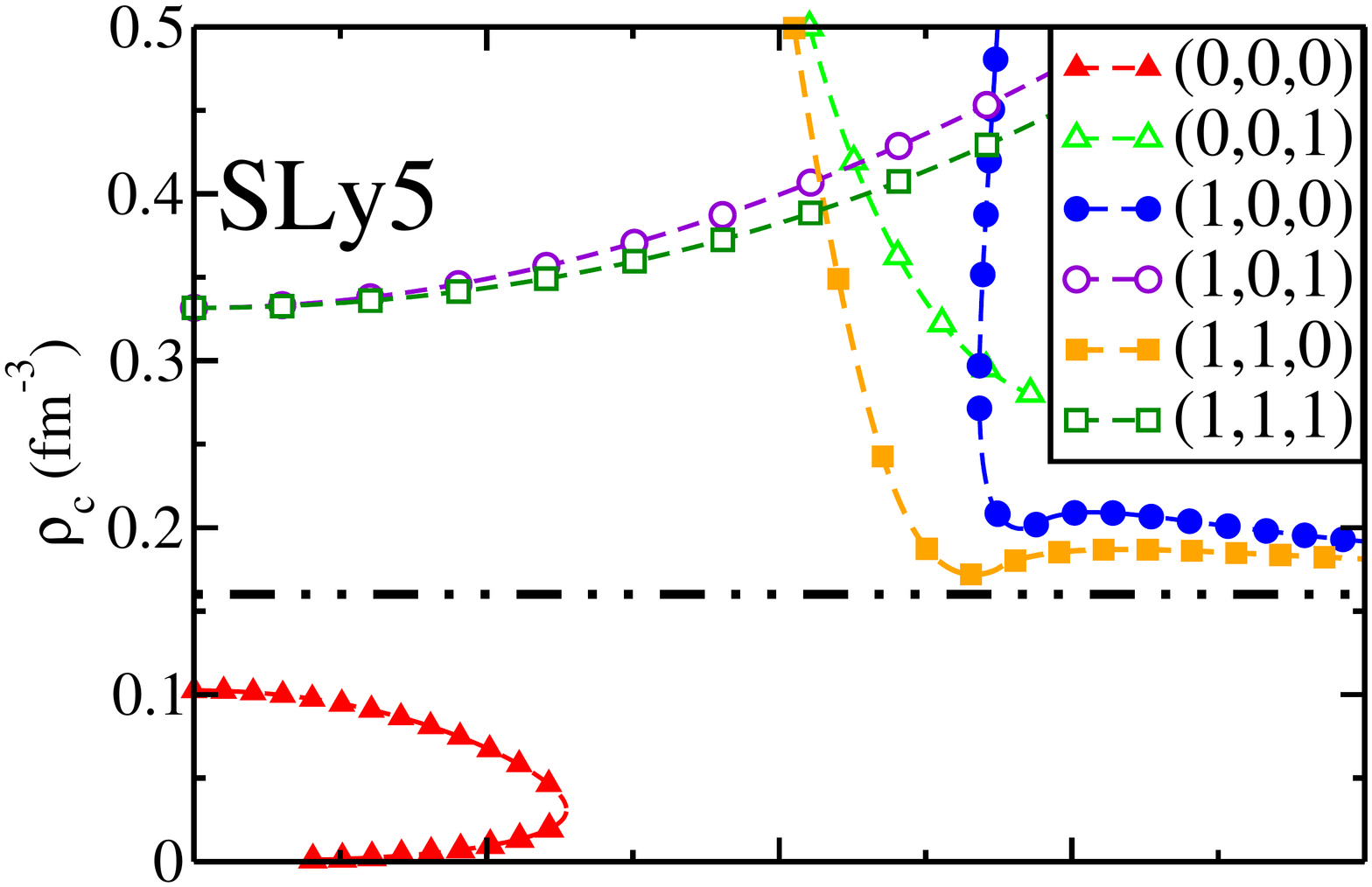}
\hspace{-1.773cm}
\includegraphics[width=0.4\textwidth,angle=-90]{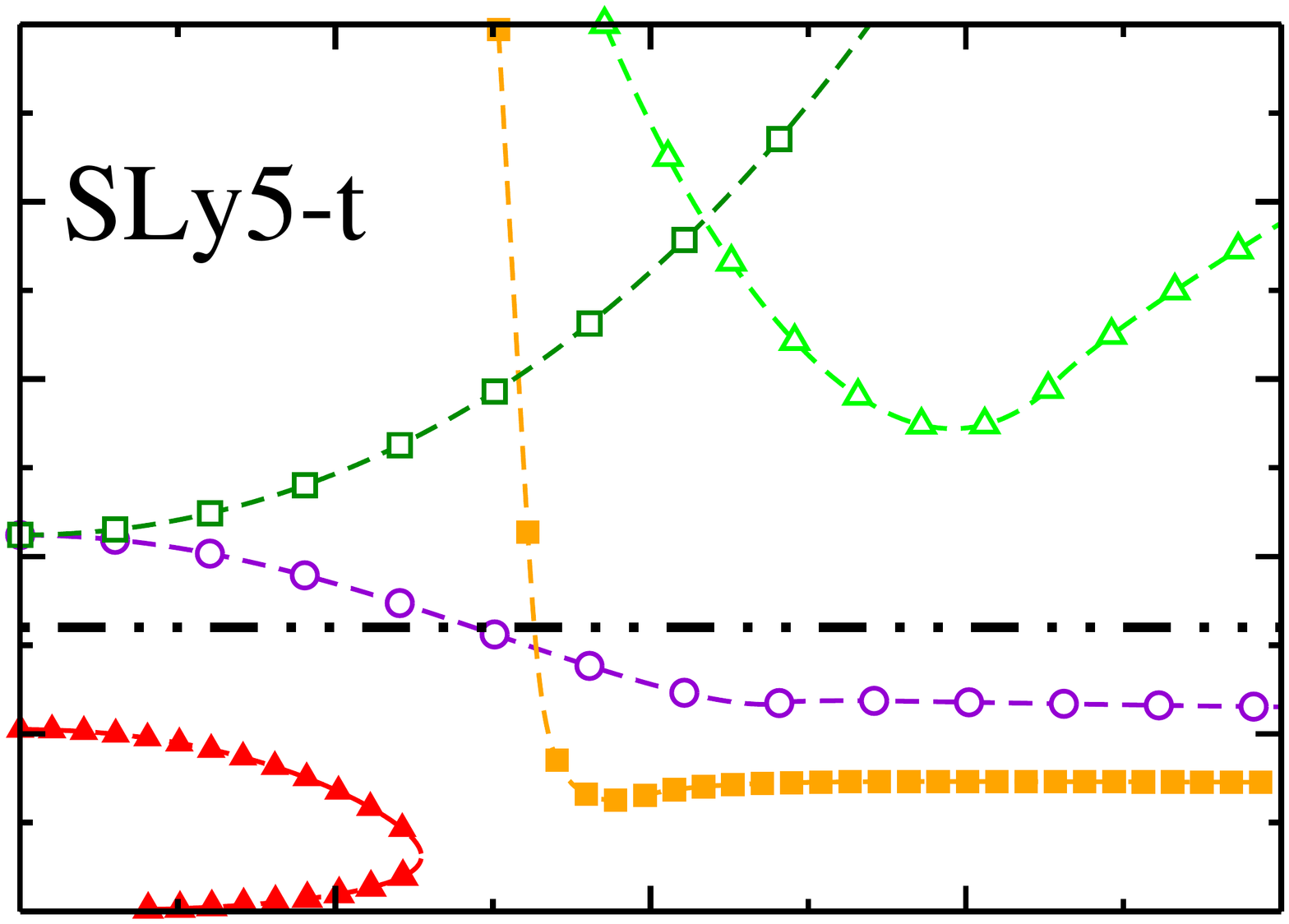}  \\  
\vspace{-1.67cm}
\includegraphics[width=0.4\textwidth,angle=-90]{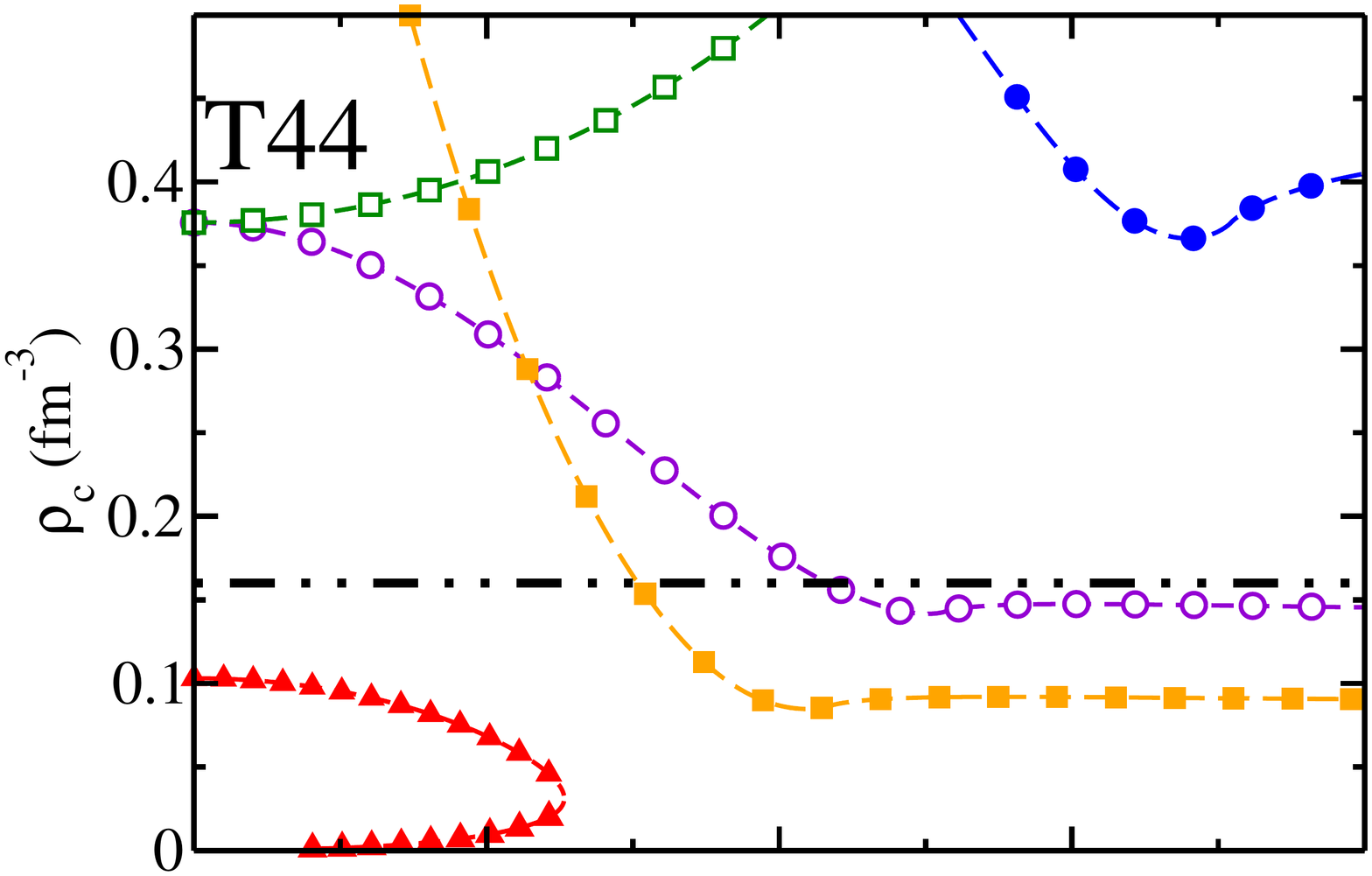}
\hspace{-1.773cm}
\includegraphics[width=0.4\textwidth,angle=-90]{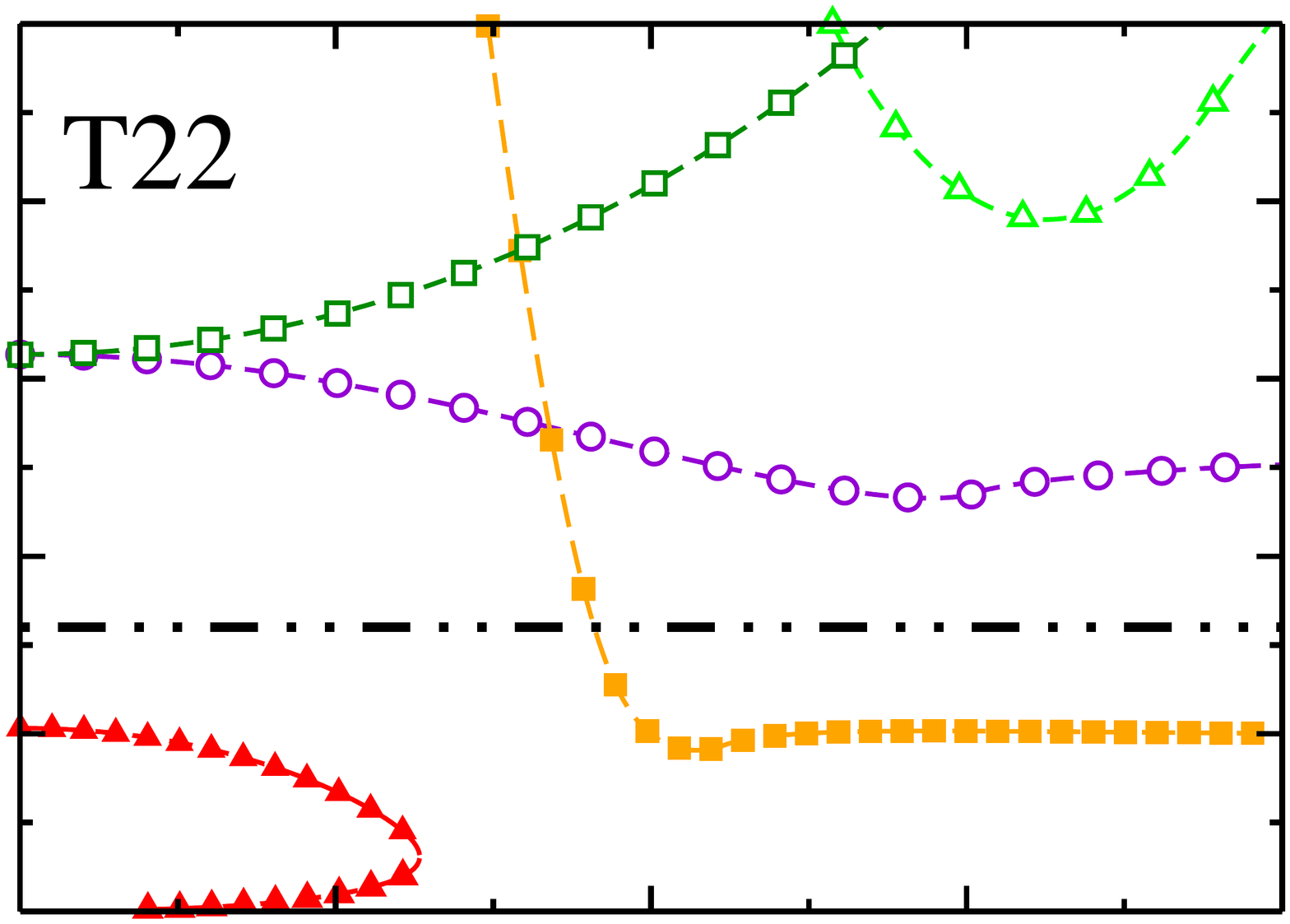}\\
\vspace{-1.67cm}
\includegraphics[width=0.4\textwidth,angle=-90]{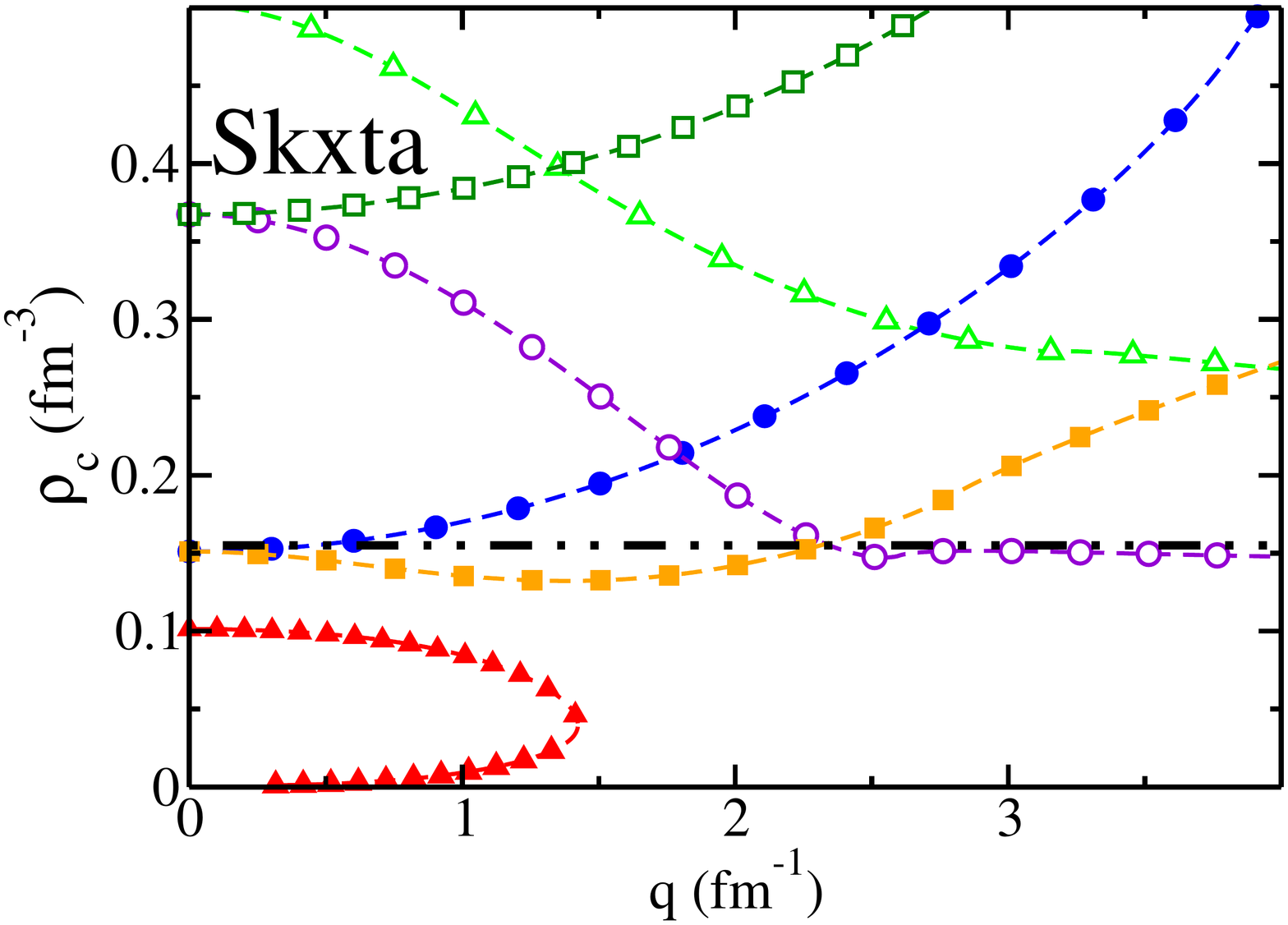} 
\hspace{-1.773cm} 
\includegraphics[width=0.4\textwidth,angle=-90]{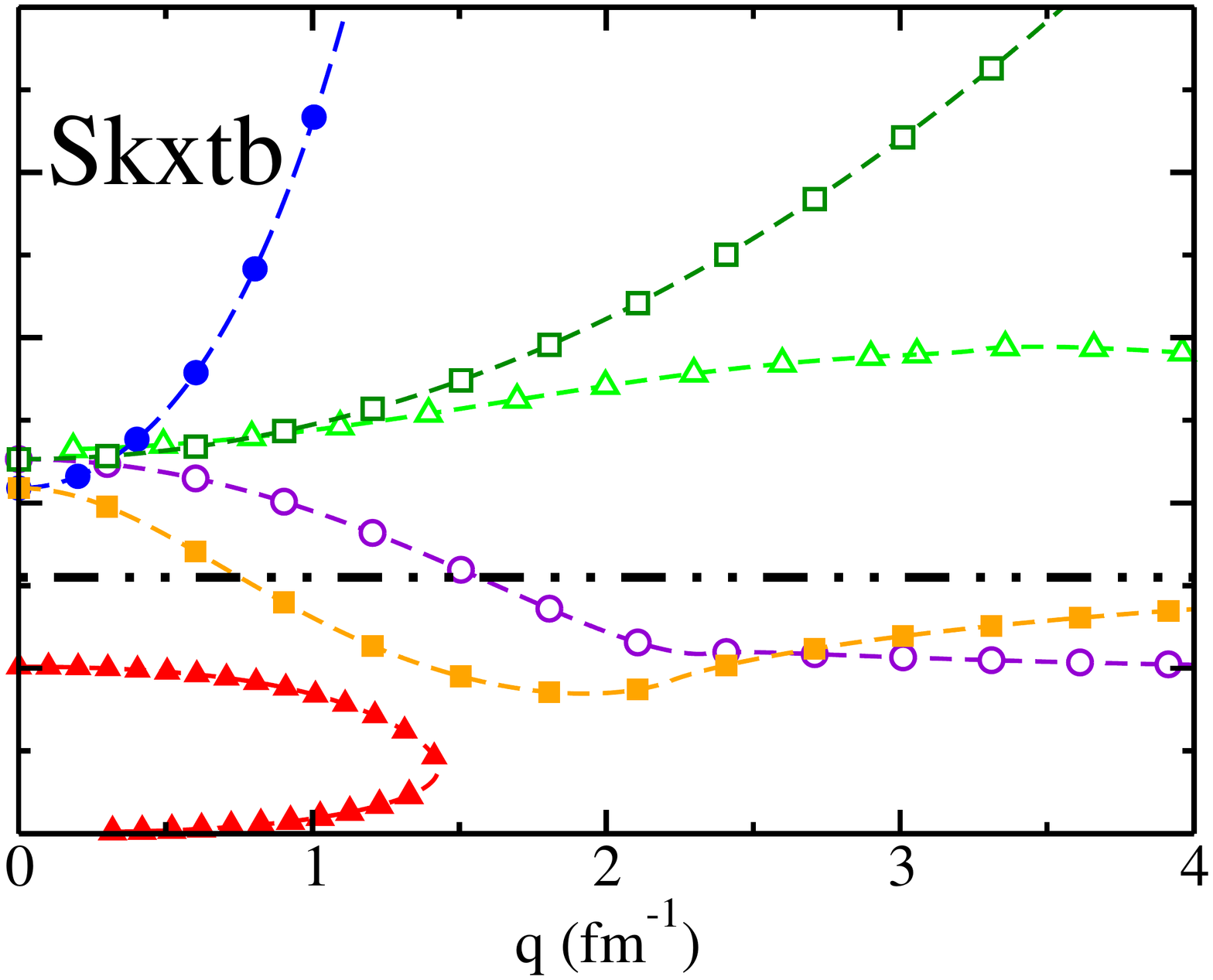}  
\caption{(Color online) Critical densities in SNM for different Skyrme interactions.
The horizontal dashed-dotted-dotted line represents the density value 0.16~fm$^{-3}$}
\label{Critical:SNM:pole}
\end{center}
\end{figure}

A thorough study of instabilities requires varying also the value of the density. 
 To this end, we define a critical density $\rho_c$ as the density at which an instability appears at a given value of $q$. The system becomes unstable for densities beyond the critical value, and this is an important information in the process of fixing NEDF parameters. This information is summarized by plotting the critical density $\rho_c$ as a function of $q$ for the different channels, as shown in Fig.~\ref{Critical:SNM:pole} for the six chosen interactions. As a reference, we have also drawn the value 0.16~fm$^{-3}$ as a dot-dot-dash line.  
One can observe that all the interactions present an instability in the $(0,0,0)$ channel, which corresponds to the so-called spinodal instability~\cite{duc08}. This is a transition of homogeneous matter, where density fluctuations induce a decrease of the total free energy and are thus amplified until a separation in two distinct stable phases, liquid and gas.
A more general discussion about it will be given in Sec.~\ref{spinod:asym}, to include also the isospin asymmetry parameter.

The spinodal is the only physical instability displayed in Fig.~\ref{Critical:SNM:pole}, all the others being unphysical.
Among them, those appearing at zero momentum transfer are best studied in terms of the Landau parameters, related to the specific interaction as will be shown in Sec.~\ref{Sec:landau}. 
It is thus useful to control these instabilities within the optimization procedure~\cite{cha97,cha10c}, by checking 
that the corresponding Landau parameters fulfill some inequalities. 
A systematic study of these instabilities have been done in \cite{mar02,cao10}, thus giving some bounds to the reliability of the interactions. 

However, not all the NEDF terms do contribute to the Landau parameters, and cannot be checked in that way. For instance,  the coupling constants $C^{\Delta \rho}_{I},C^{\nabla s}_{I},C^{\Delta s}_{I}$ entering the general NEDF (\ref{res}) are multiplied by $q^{2}$. The use of a $(\rho_c,q)$ plot is thus the only way to study the possible existence of finite-size instabilities that manifest at finite momentum transfer. These instabilities can affect the description of finite nuclei properties, if they appear at density values around the saturation density. 

\begin{figure}[H]
\begin{center}
\includegraphics[width=0.36\textwidth,angle=-90]{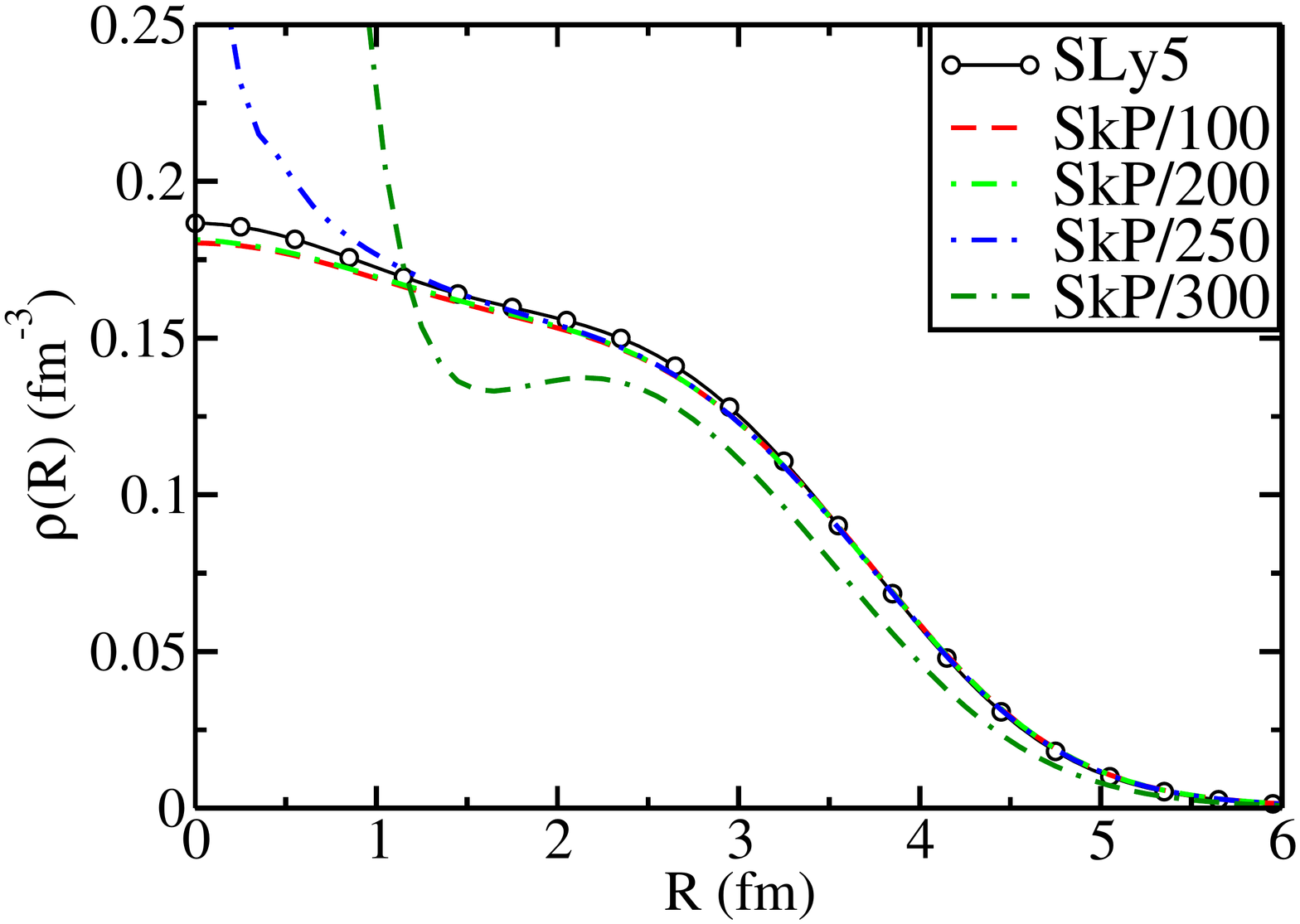}
\hspace{-0.1cm}
\includegraphics[width=0.36\textwidth,angle=-90]{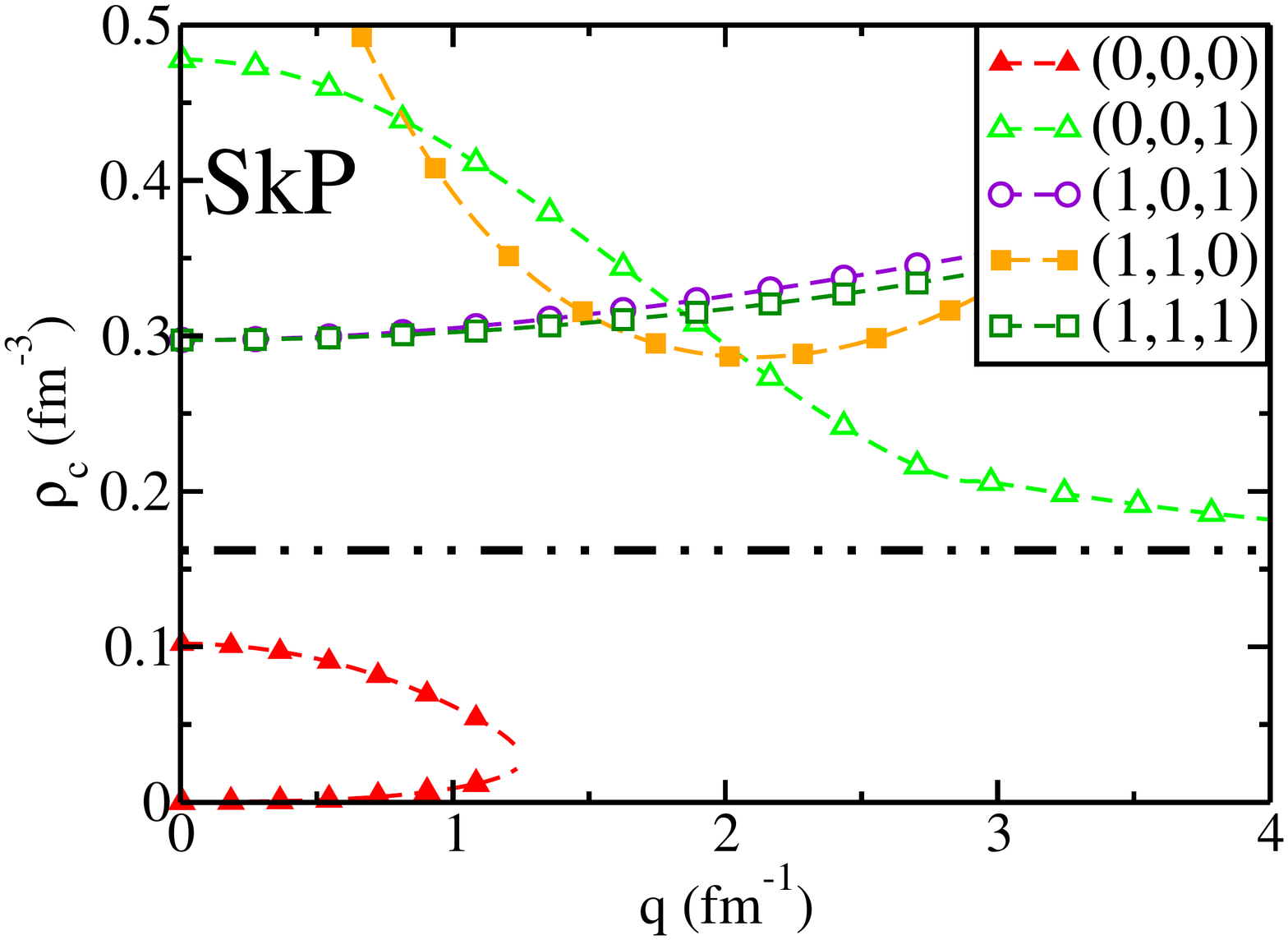} 
\caption{(Color online) On the left panel, we show the total density profile of $^{40}$Ca obtained for a fully converged result obtained with SLy5 functional and the results obtained using the SkP functional for different values of HF iterations done to find the solution. On the right panel, we show the position of the instabilities in SNM for SkP functional. The horizontal line represents the saturation density of the system.}
\label{Critical:instability:skp}
\end{center}
\end{figure}

As an example, we refer to the calculations for $^{40}$Ca performed in \cite{les06} using an accurate spherical HF code  HFBRAD~\cite{ben05} with the SkP pseudo-potential~\cite{dob84}. 
Since time-reversal invariance has been imposed, the conclusions affect only  the time-even part of the functional~\cite{dob95}.  
In the left panel of Fig.~\ref{Critical:instability:skp} are plotted the density profiles obtained when the number of iterations in the HF calculation of the ground state energy is increased from 100 to 300. The profiles depend of course on the number of iterations, but not in the expected way: the solution worsens as the number is increased. 
As a reference, the density profile obtained from a fully converged HF calculation with SLy5 interaction is also plotted. 
To understand the origin of the problem we must have a look to the instabilities in the $(\rho_c,q)$ plane in $S=0$ channels at densities around the saturation value, 
which excludes the spinodal. As shown in Fig.\ref{Critical:SNM:pole} for SLy5, there is a pole in the $S=0,I=1$ channel  
at $q\approx2.7$ fm$^{-1}$ and $\rho_{c}\approx0.3$ fm$^{-3}$. We don't expect this density value be relevant for finite nuclei. The right panel in Fig.~\ref{Critical:instability:skp} displays the $(\rho_c,q)$ plane for SkP interaction. 
There is a pole in the $S=0,I=1$ channel at $q\approx4$ fm$^{-1}$ and $\rho_{c}\approx0.18$ fm$^{-3}$, which should manifest in the central region of finite nuclei. 
 
Of course a clear one-to-one correspondence between SNM and $^{40}$Ca cannot be done. A systematic analysis of scalar/isovector instabilities have been performed in Ref.~\cite{hel13}, paying particular attention to the numerical uncertainties. The main outcome is that a functional is stable if the lowest critical density at which a pole occurs in SNM is larger than the central density of $^{40}$Ca, in practice around 1.2 times the saturation density. In addition, one has to also verify that this pole represents a distinct global minimum in the $(\rho_{c},q)$-plane and not a monotonously decreasing function of transferred momenta.

A similar analysis for the time-odd part of the functional has not yet been done. Some authors have noticed that finite size instabilities in the $S=1$ channel can also affect the calculations of finite nuclei~\cite{fra12,hel12,sch10}. In this case, it is necessary to break time-reversal invariance, thus making the calculations more time consuming and small basis size are usually employed. That could hide the development of the instabilities~\cite{hel13}.


\subsection{Thermal effects}

We consider now the effect of the temperature. A redistribution of the strength is to be expected, spreading the $ph$ band and giving non-null contribution to the negative energy region. The natural variable to analyze thermal effects in the response function is the dimensionless temperature $T / \varepsilon_F$. However, it is worthwhile noticing that when results from different interactions are compared, the absolute temperatures can be different due to the different values of the Fermi energy $\varepsilon_F$. 

\begin{figure}[H]
\begin{center}
  \includegraphics[width=0.6\textwidth,angle=-90]{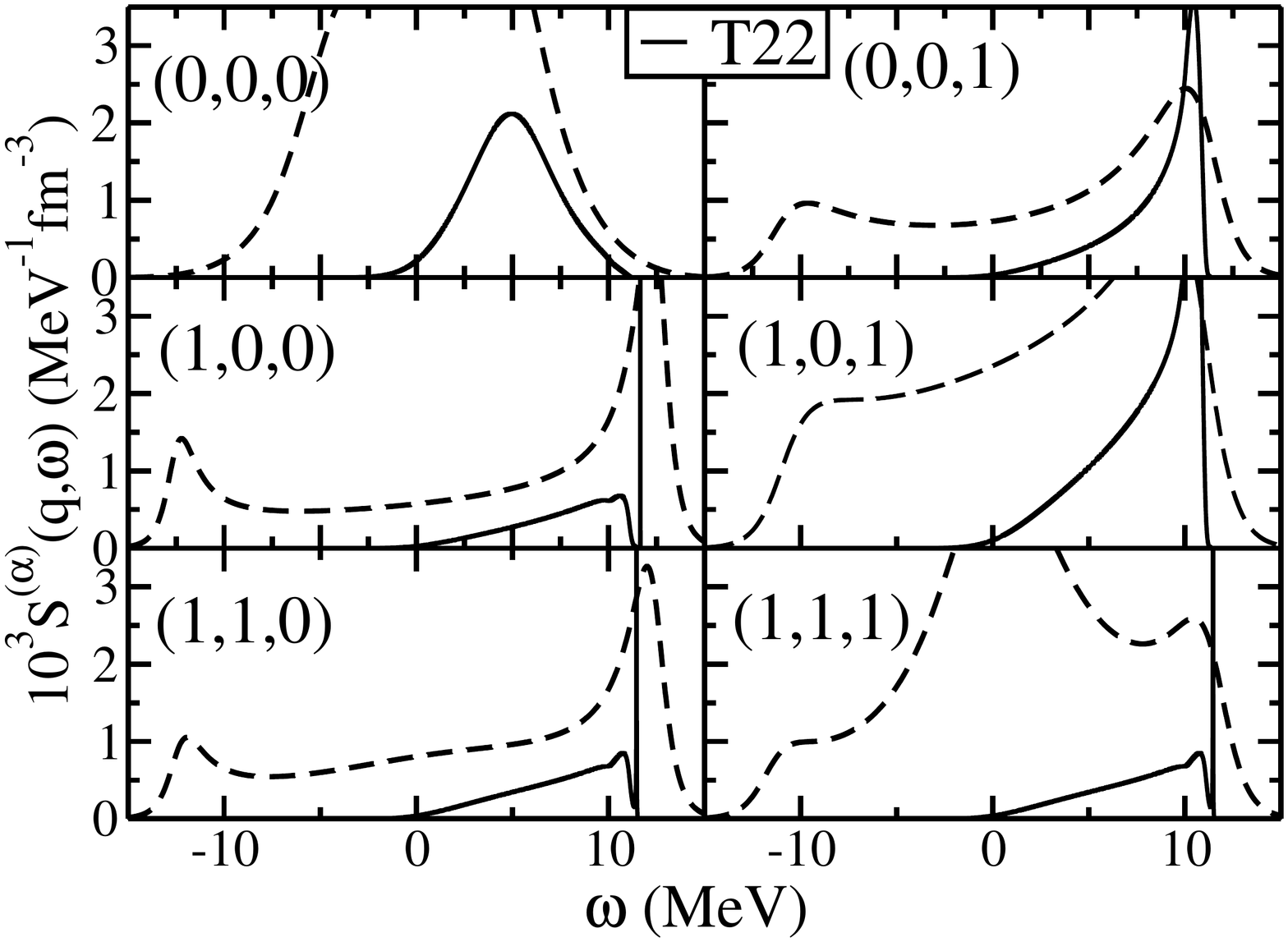}  \\
    \includegraphics[width=0.6\textwidth,angle=-90]{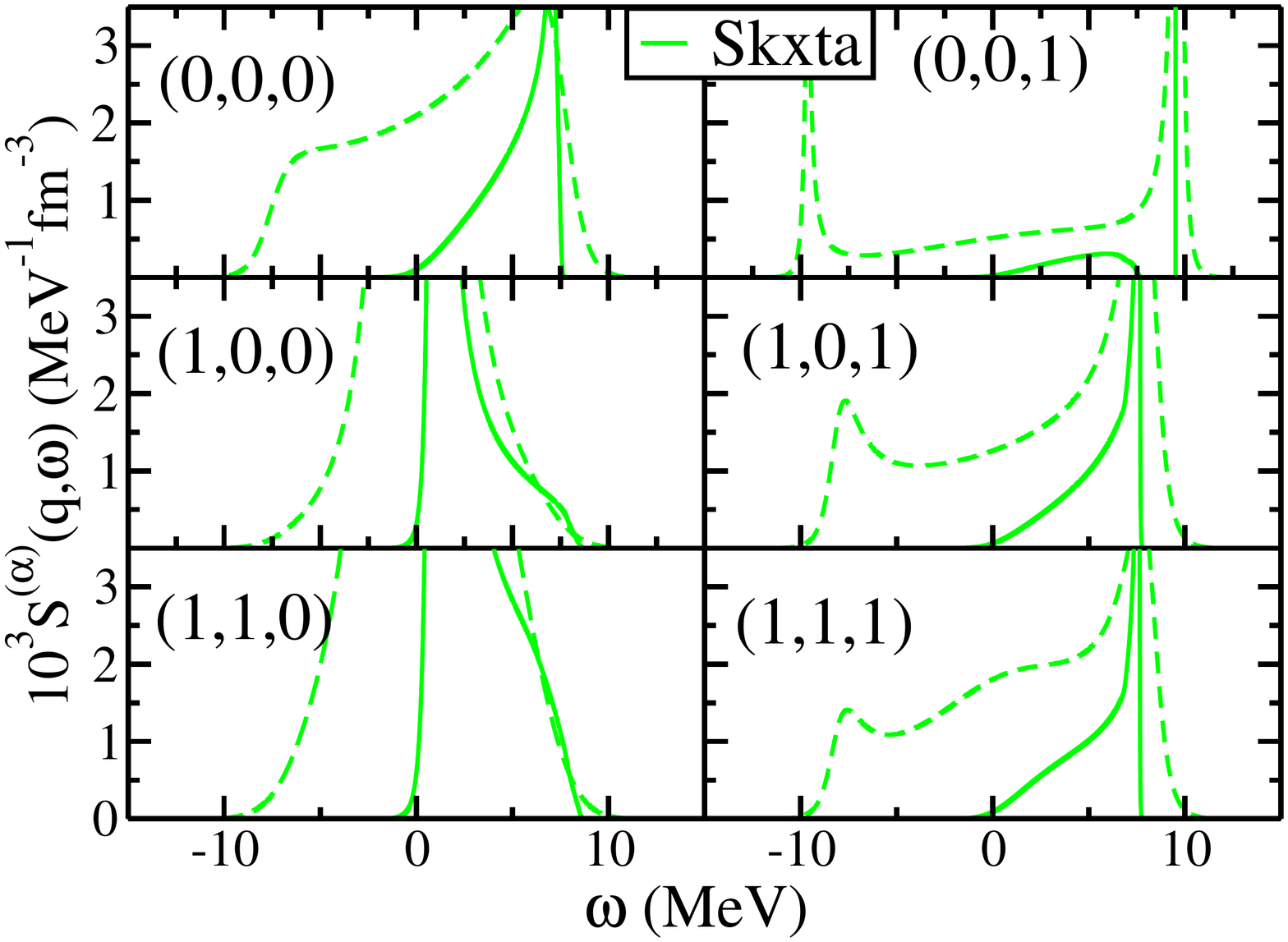}  
\caption{(Color online) SNM thermal effects on the strength function obtained using T22 and Skxta. The calculations have been performed at their respective saturation density and for a value of the transferred momentum $q=0.1k_{F}$. The solid line represent the calculations at $T=0$ and the dashed line $T=0.2\varepsilon_{F}$.}
\label{RESP-thermal}
\end{center}
\end{figure}

In Fig. \ref{RESP-thermal} are plotted the results obtained with interactions T22 and Skxta for temperatures 0 and 0.2$\varepsilon_F$. The latter values correspond to absolute temperatures of about 10 and 7 MeV, respectively for T22 and Skxta. One can see that at non-null temperature a weak ``image peak" appears in the strength at negative energies, which corresponds to the desexcitation of the heated system. One may additionally recognize that the larger the effective mass --that is, the smaller the kinetic energy per particle-- the more concentrated remains the strength in the low-energy region.
Both interactions predict sizable zero-sound modes in the (1,1,0) and (1,1,1) channels at $T=0$.
At finite temperature, these modes are absorbed into the thermally extended $ph$ continuum and become subject to Landau damping. Mirror peaks of such modes appear on the negative energy axis. 

Finally, we have checked that the position of finite size instabilities in the $(\rho_c,q)$ plane
is not very sensitive to temperature: a variation around 1-2\% for values of $T$ up to $\approx0.5\varepsilon_{F}$ is observed. 
The influence of $T$ on the spinodal is discussed in the more general context of Subsec. \ref{spinod:asym}.

\subsection{Pure neutron matter results}

\begin{figure}[H]
\begin{center}
  \includegraphics[width=0.5\textwidth,angle=-90]{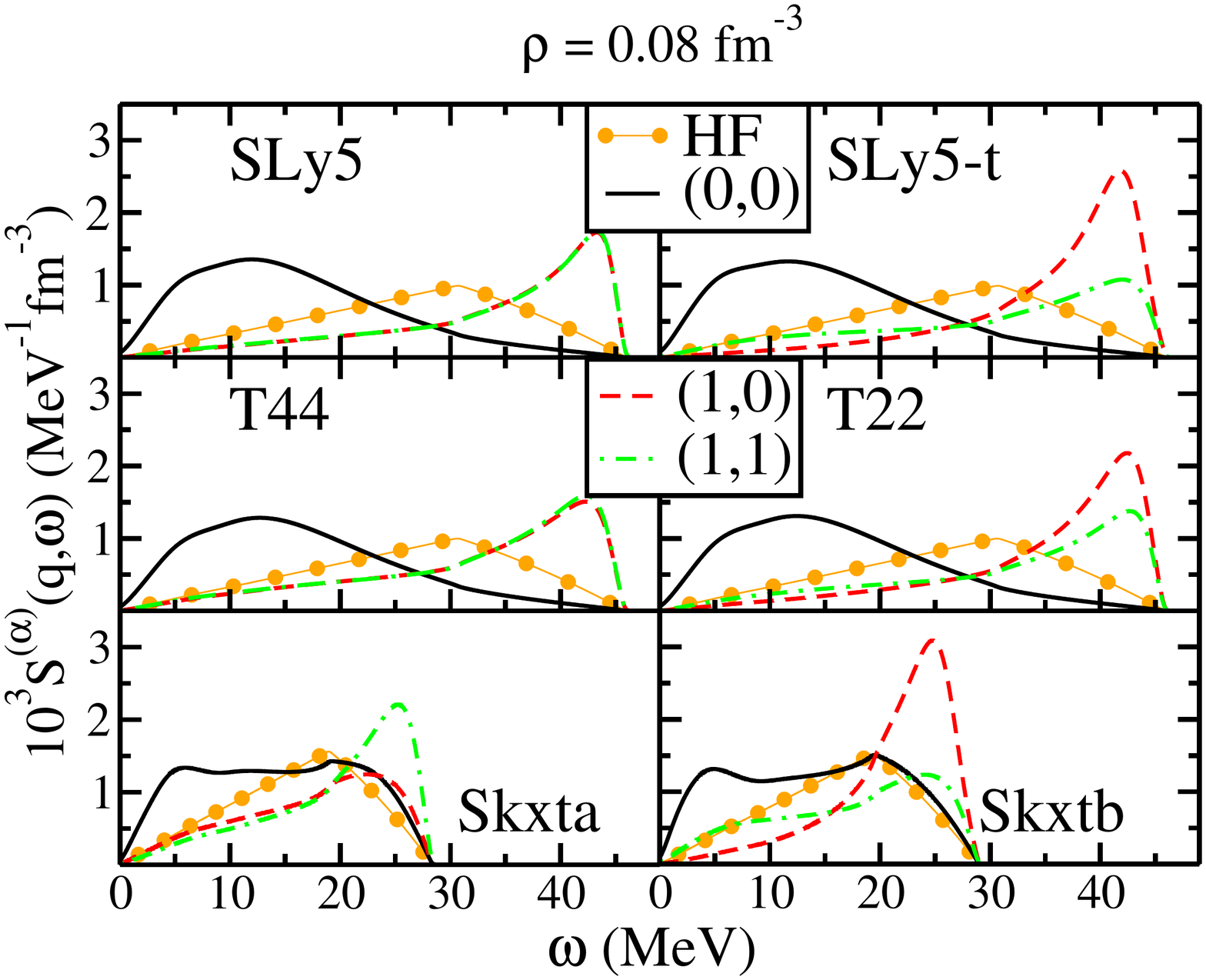}  
  \\
    \includegraphics[width=0.5\textwidth,angle=-90]{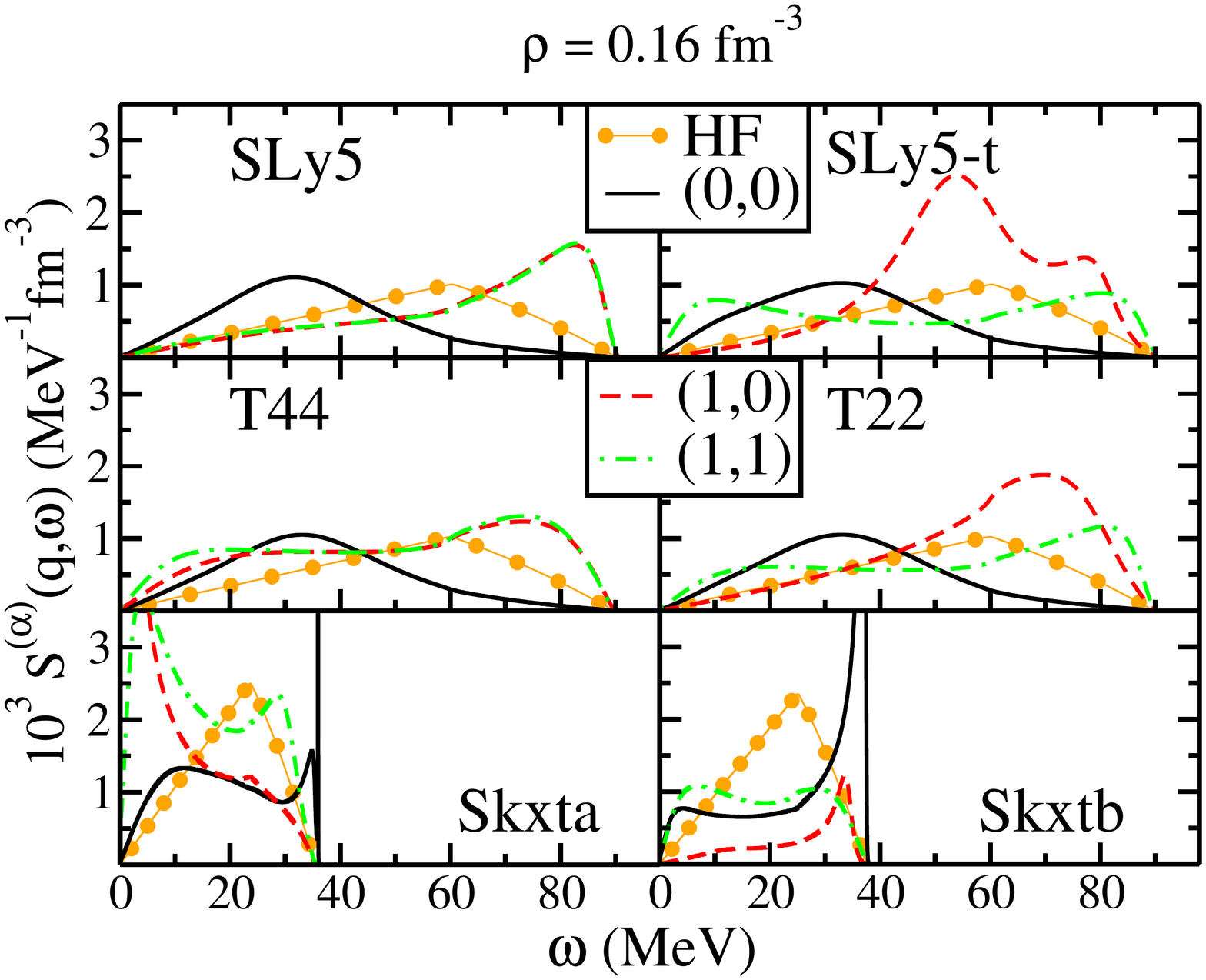}  
\caption{(Color online) RPA strength functions in PNM calculated at $\rho=0.08$~fm$^{-3}$ (top panel) and $\rho=0.16$~fm$^{-3}$ (bottom panel) and a momentum transfer $q=0.5k_F$.}
\label{PNM-1-05}
\end{center}
\end{figure}

The response function is shown in Fig.~\ref{PNM-1-05} at $T=0$, $q=0.5 k_F$ and densities $\rho=0.08$ and $0.16$ fm$^{-3}$. One can observe that the SLy family of interactions produces nearly the same response in the $S=0$ channel, but different shapes in the $S=1$ channels due to the tensor terms. The reason for this behavior has been discussed in the SNM case: for this relatively low value of the momentum transfer, the spin-orbit contribution to the residual interaction is negligible. Therefore, the tensor acts only in the $S=1$ channel and the $S=0$ responses are essentially the same for these SLy interactions.

\begin{figure}[H]
\begin{center}
\includegraphics[width=0.4\textwidth,angle=-90]{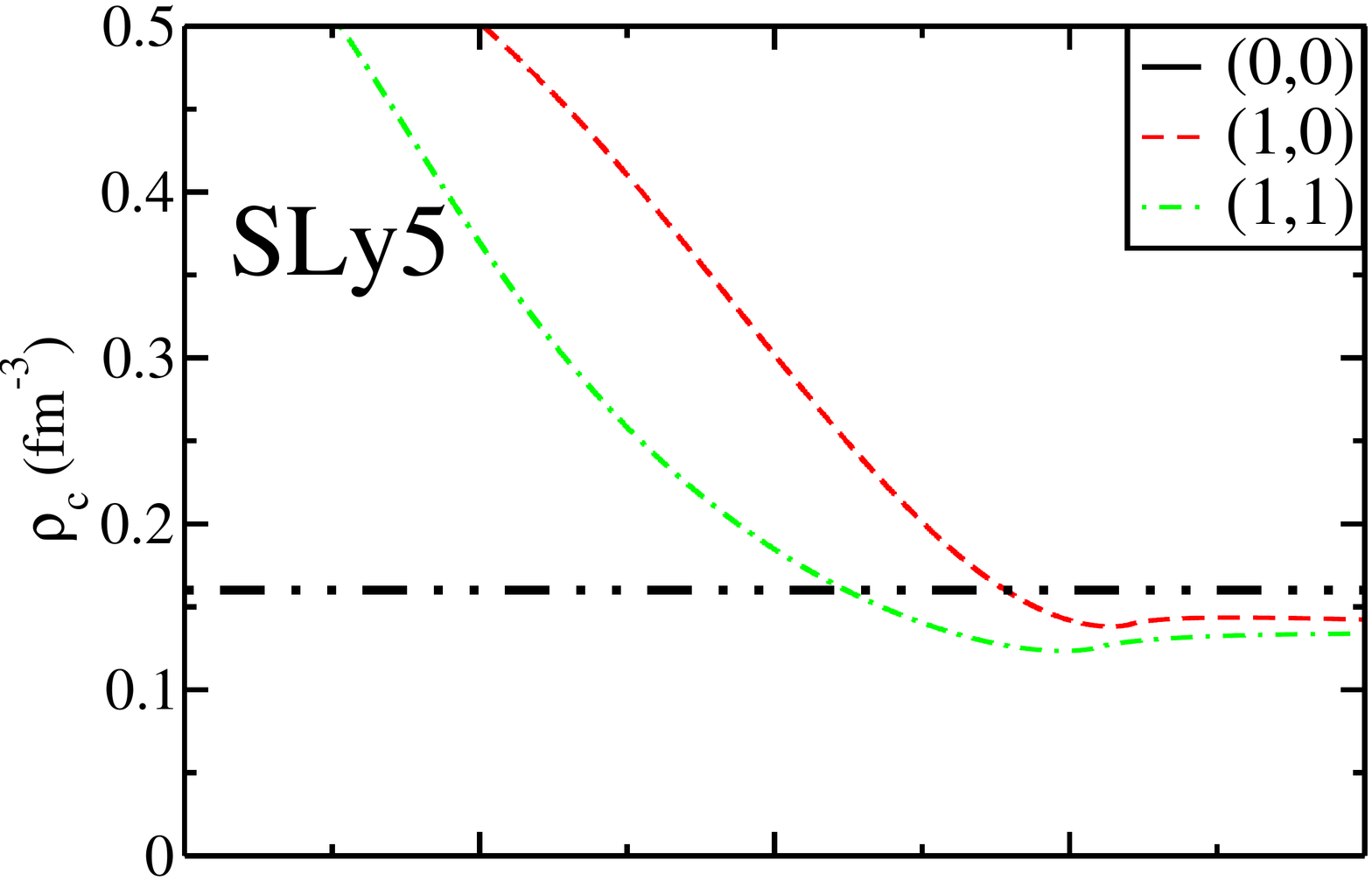}
\hspace{-1.773cm}
\includegraphics[width=0.4\textwidth,angle=-90]{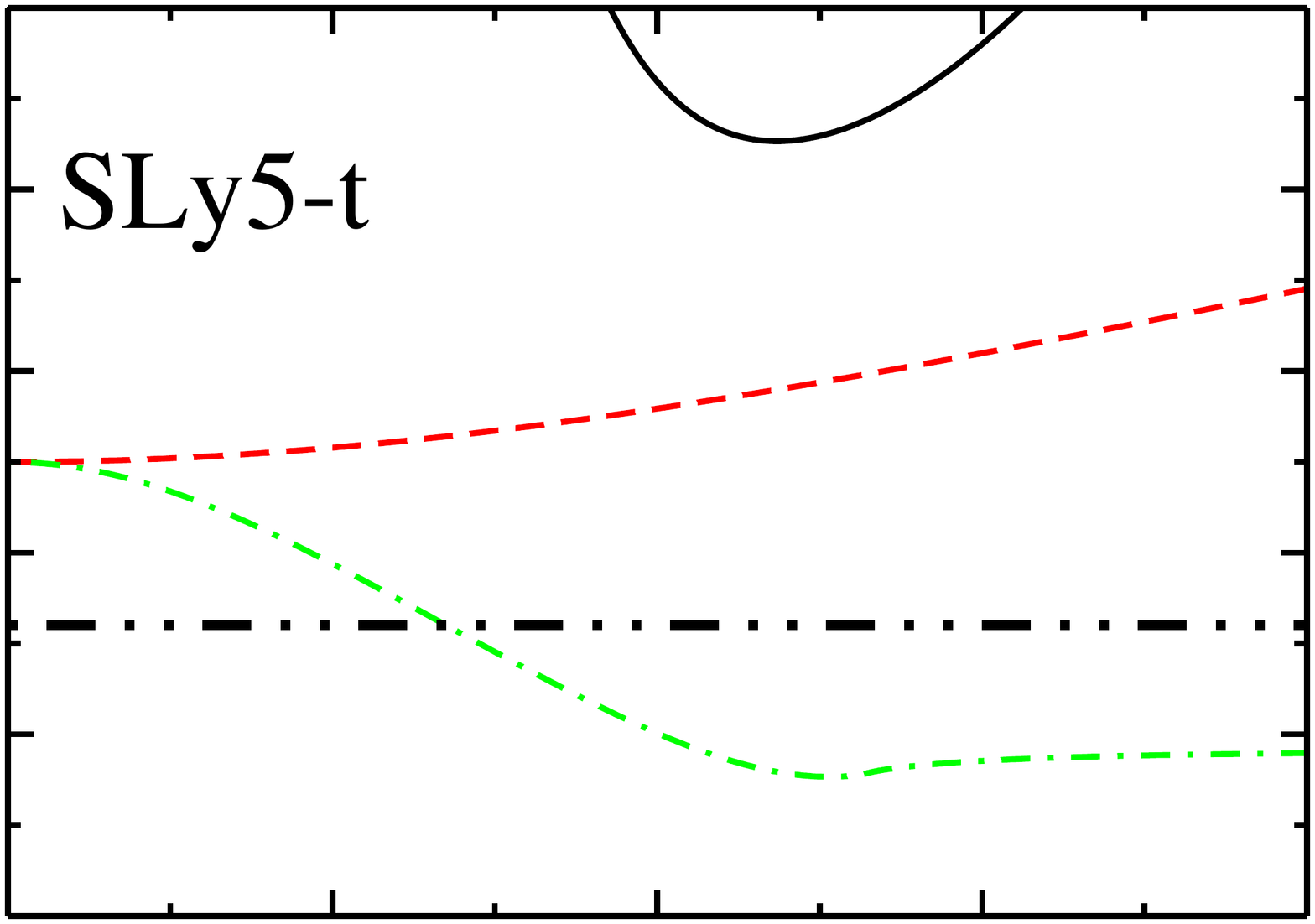}  \\  
\vspace{-1.67cm}
\includegraphics[width=0.4\textwidth,angle=-90]{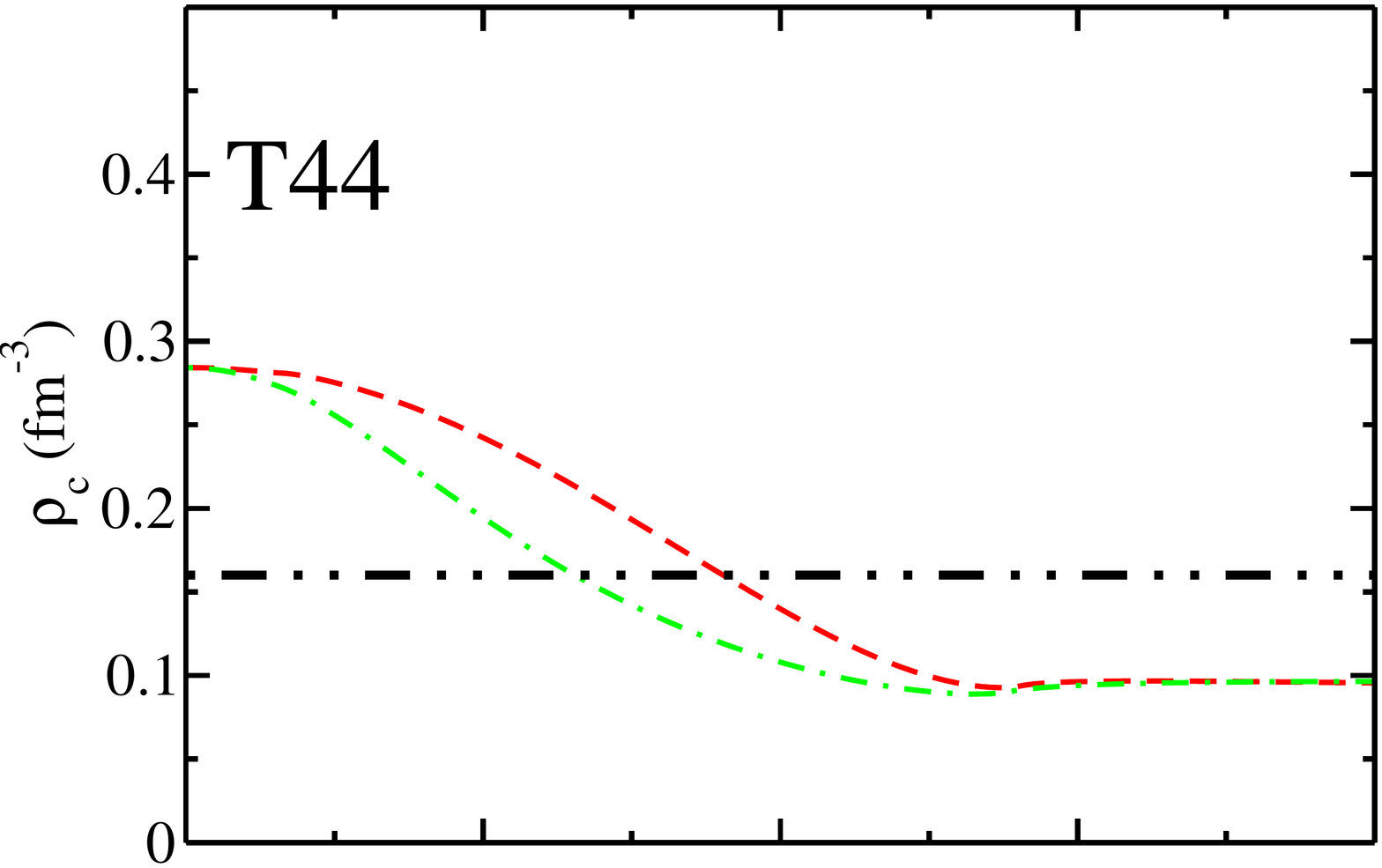}
\hspace{-1.773cm}
\includegraphics[width=0.4\textwidth,angle=-90]{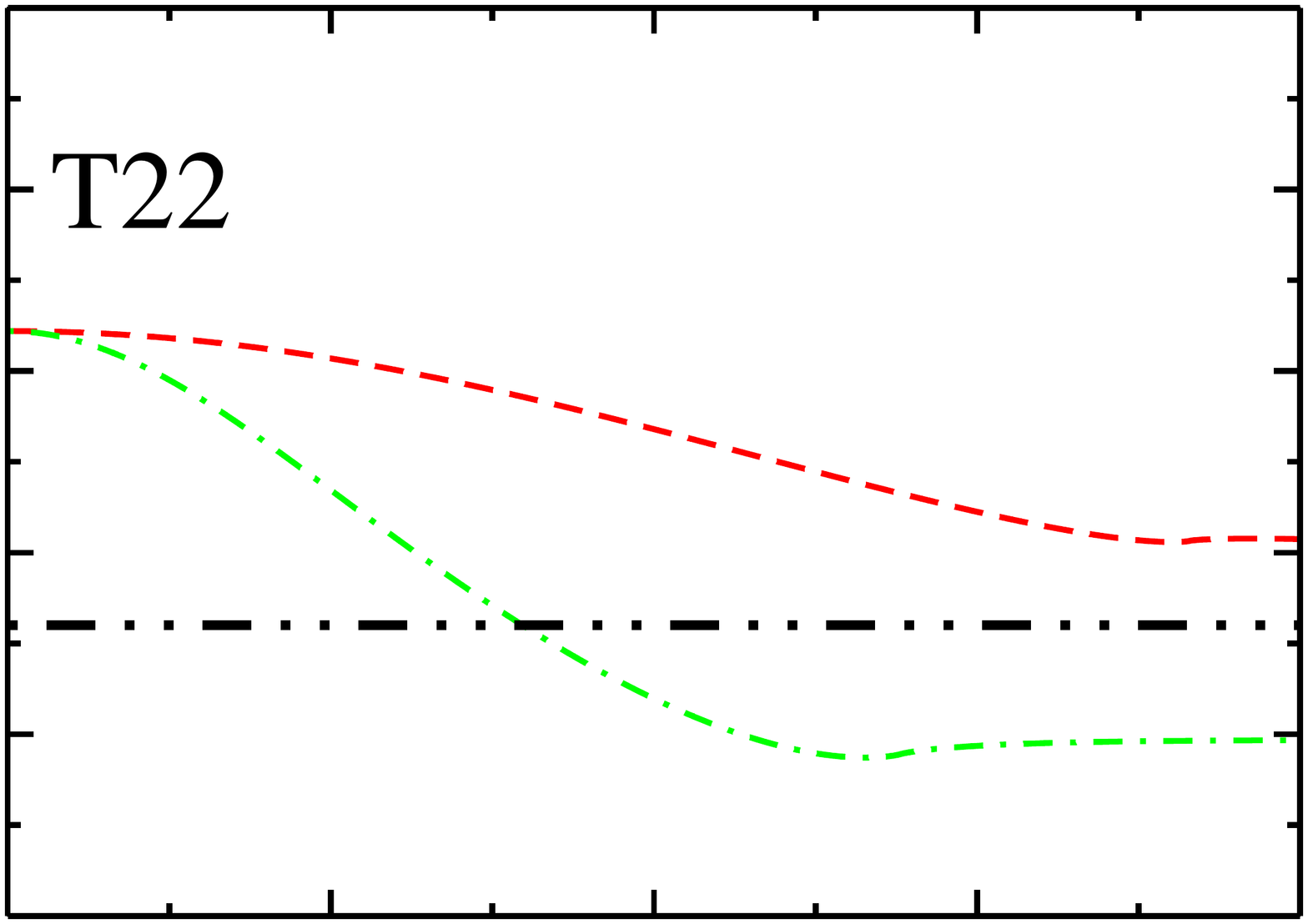}\\
\vspace{-1.67cm}
\includegraphics[width=0.4\textwidth,angle=-90]{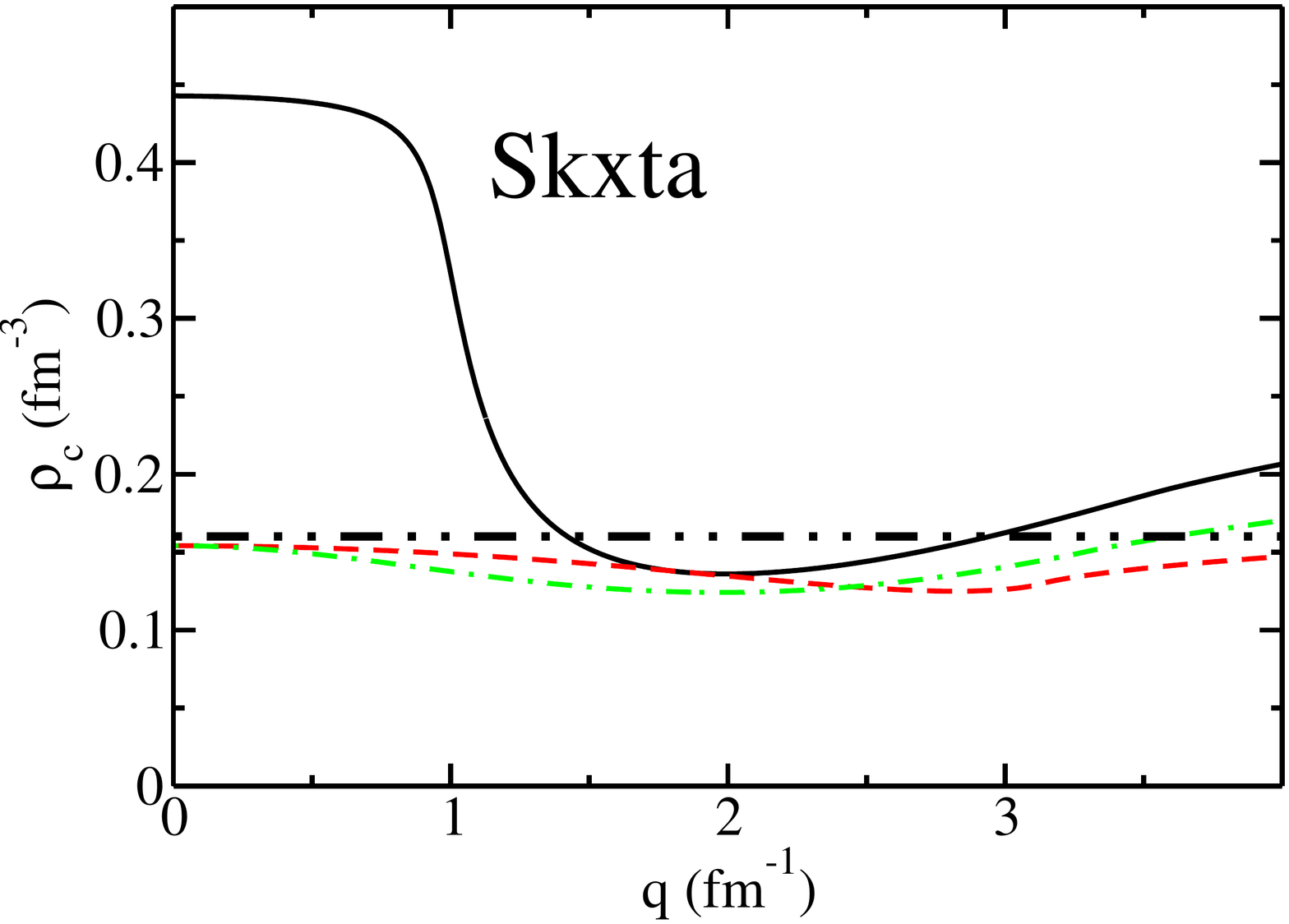} 
\hspace{-1.773cm} 
\includegraphics[width=0.4\textwidth,angle=-90]{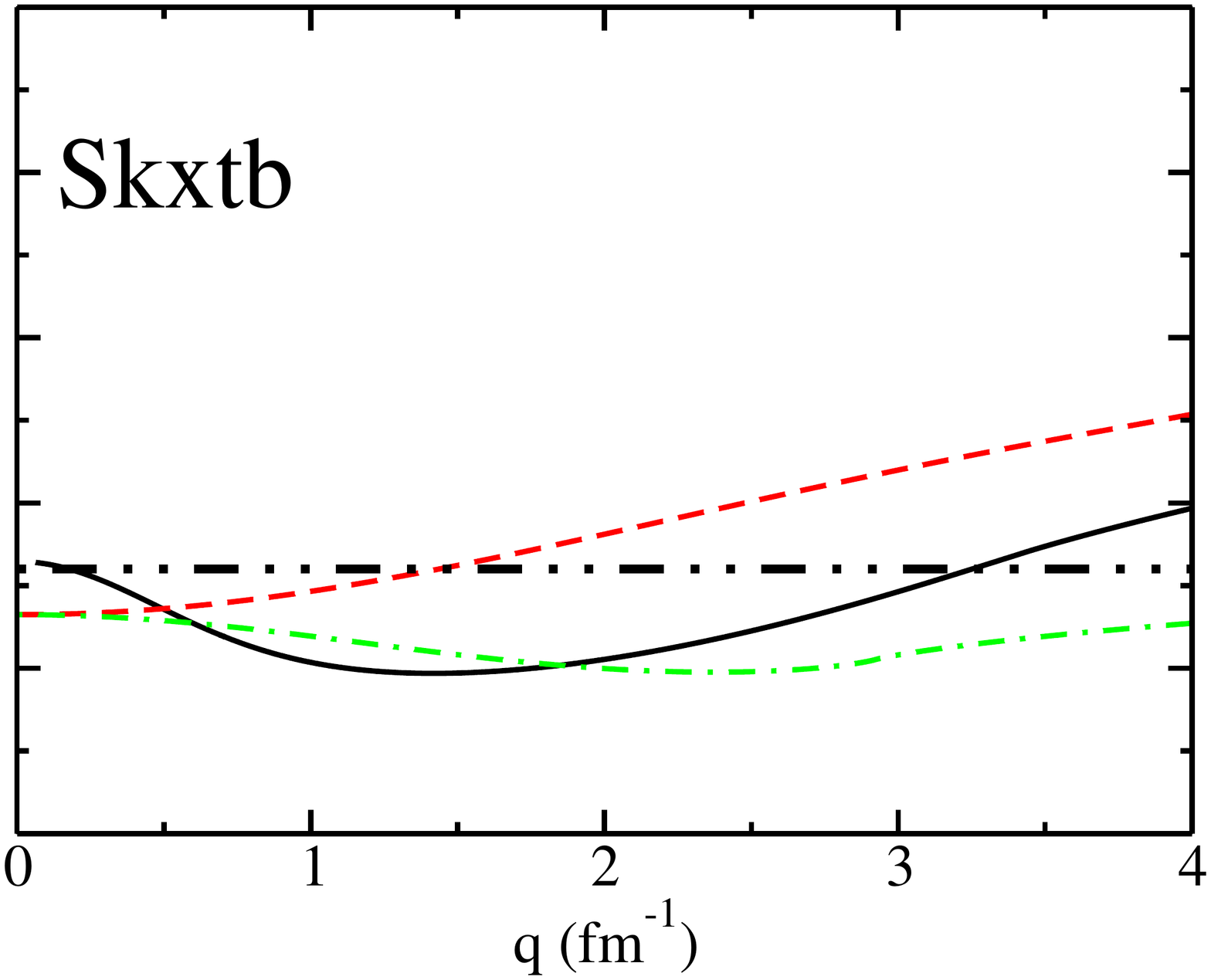}  
\caption{(Color online) Critical densities in PNM for different Skyrme interactions. The horizontal dashed-dotted-dotted line represents the density value 0.16~fm$^{-3}$.}
\label{Critical:PNM:pole}
\end{center}
\end{figure}

In contrast, the Skx interactions show a different behavior in both channels. In particular a very huge singularity appears in the $S=1$ channel for Skxta at the density $\rho=0.16$~fm$^{-3}$. As discussed in the SNM case, 
these instabilities are best analyzed in the $(\rho_c,q)$ plane. In Fig.~\ref{Critical:PNM:pole} are plotted the position of the instabilities for the six interactions.

As PNM is not a self-bound system there is no a direct connection with finite systems. Calculations of neutron droplets require confining them in an external potential, and trapped neutron droplets have been recently used to test various \emph{ab-initio} approaches against NEDF calculations with phenomenological functionals~\cite{pie03,gan11}. 
A new parameterization of the Skyrme NEDF, dubbed UNEDF2, has been proposed in \cite{kor13}. In that reference, neutron droplets have been confined by a spherical harmonic oscillator potential, and HF-Bogoliubov calculations have been performed. It has been shown that the functional UNEDF2 presents an instability in PNM at $\rho_{c}\approx0.16$ fm$^{-3}$ and $q\approx0$. It turns out that  the central neutron density exceeds this critical value for more than 38 neutrons and an external trap with frequency $\omega=10$ MeV. Consequently, the calculations fail to converge, much in the same way as we have previously discussed for $^{40}$Ca.

\subsection{Neutrino mean-free path in pure neutron matter} 
As an application, let us present calculations of the neutrino mean-free path (NMFP) in PNM, which requires the detailed knowledge of the strength functions $S^{(\alpha)}$ in all channels.
 During the gravitational collapse of a massive star, the energy excess is dissipated by the emission of neutrinos. In this process, neutrinos are scattered by the nucleons, loosing energy and making thus the cooling process less efficient \cite{iwa82}. Consider the scattering process $n+\nu \rightarrow n'+\nu'$ assuming non-degenerate neutrinos, thus ignoring Pauli blocking effects. Denoting with $\mathbf{k}_{\nu(\nu')}$ and  $E_{\nu(\nu')}$ the momentum and energy of incoming (outcoming) neutrino, respectively, the momentum-energy conservation gives for the energy and momentum transferred to the nucleon medium the relations
$\omega=E_{\nu'}-E_{\nu}$, and $q^{2}=(\mathbf{k}_{\nu}-\mathbf{k}_{\nu'})^{2}$. 
With these notations, the double differential cross-section is written as \cite{mar09,mar10}
\begin{eqnarray}\label{cross:nu}
\frac{d^{2}\sigma (E_{\nu})}{d\Omega_{k'}d\omega}
&=&\frac{G_{F}^{2}E_{\nu}^{2}}{16\pi^{2} \rho}\left\{ g_{V}^{2}(1+\cos\theta )S^{(0,0)}(q,\omega) \right. \nonumber \\
&& \left. \quad +g_{A}^{2}\left[ \frac{2(E_{\nu'}\cos\theta-E_{\nu})(E_{\nu'}-E_{\nu}\cos\theta)}{q^{2}}+1-\cos\theta\right]S^{(1,0)}(q,\omega)\right.\nonumber\\
&& \quad + 2g^{2}_{A} \left. \left[ \frac{E_{\nu}E_{\nu'}}{q^{2}}\sin^{2}\theta+1-\cos\theta\right]S^{(1,1)}(q,\omega)\right\}\,.
\end{eqnarray}
This expression neglects corrections of order $E_{\nu}/m$ arising from weak magnetism and other effects~\cite{hor02} as the finite size of the nucleon~\cite{che09}. We refer to \cite{mar09} for a more detailed discussion.
Neglecting tensor contributions to the neutron-neutron interaction, it can be further simplified because the spin longitudinal and transversal channels  responses are identical. 
Within the literature it is possible to find several examples of more recent calculations of NMFP using different models based on Skyrme effective interaction ~\cite{nav99,mar06,pas12a,red98,pas14}, or \emph{ab-initio} methods ~\cite{lov14,cow04,zha10,she03,bac09} among others, and relativistic calculations~\cite{sul05,hut04,nie01} as well.

\begin{figure}[H]
\begin{center}
\includegraphics[width=0.4\textwidth,angle=-90]{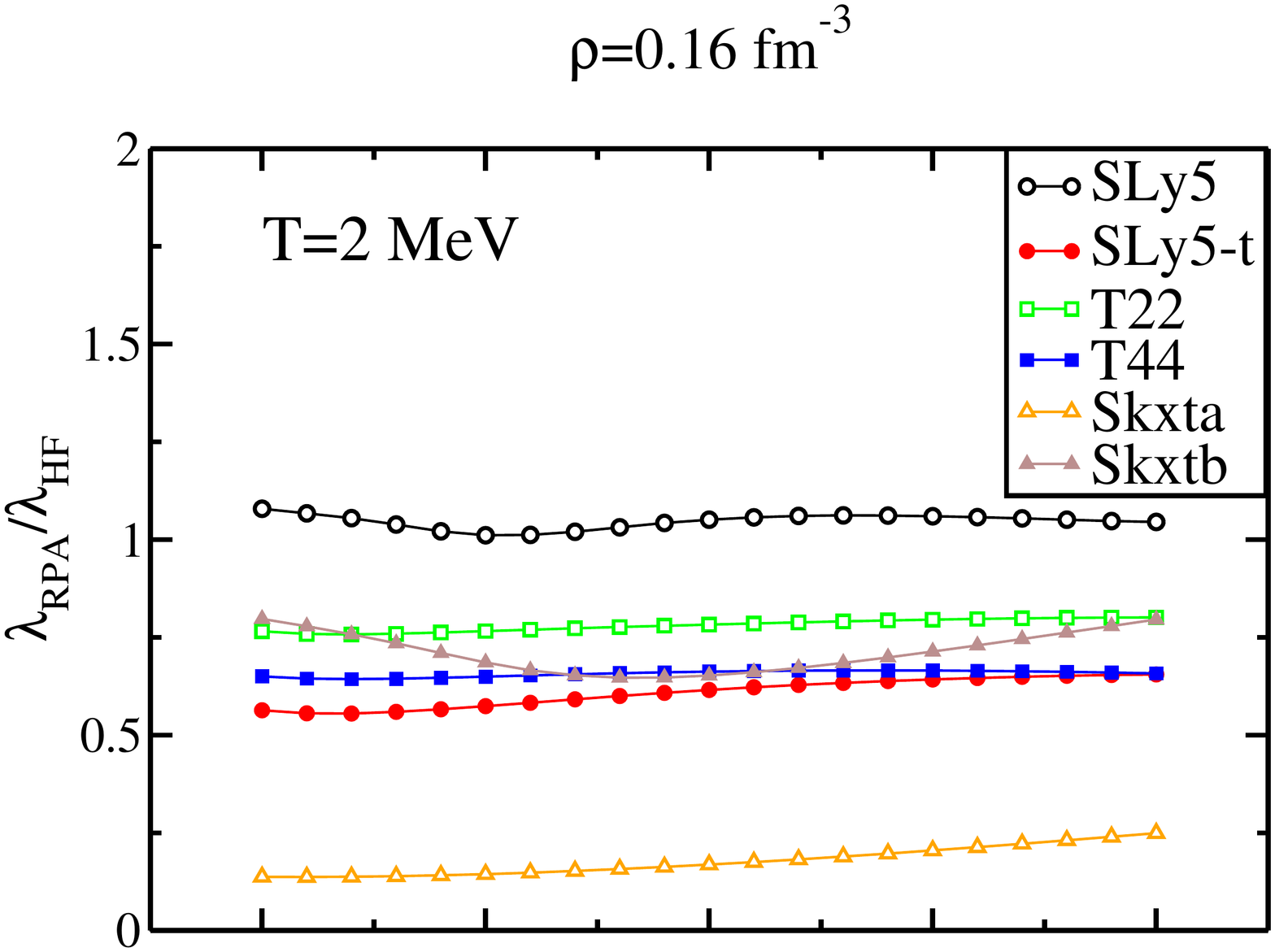}
\hspace{-1.773cm}
\includegraphics[width=0.4\textwidth,angle=-90]{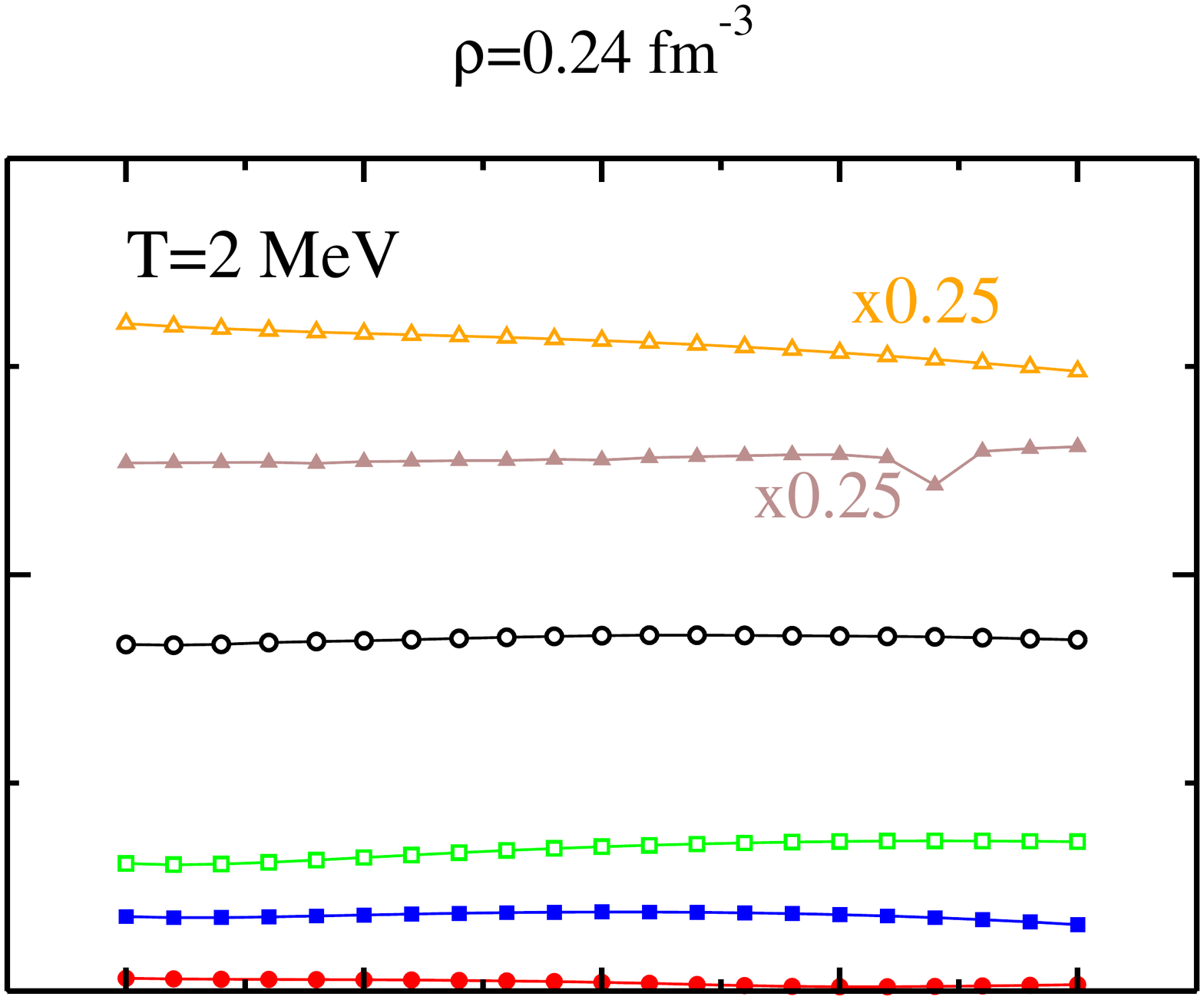}  \\  
\vspace{-1.67cm}
\includegraphics[width=0.4\textwidth,angle=-90]{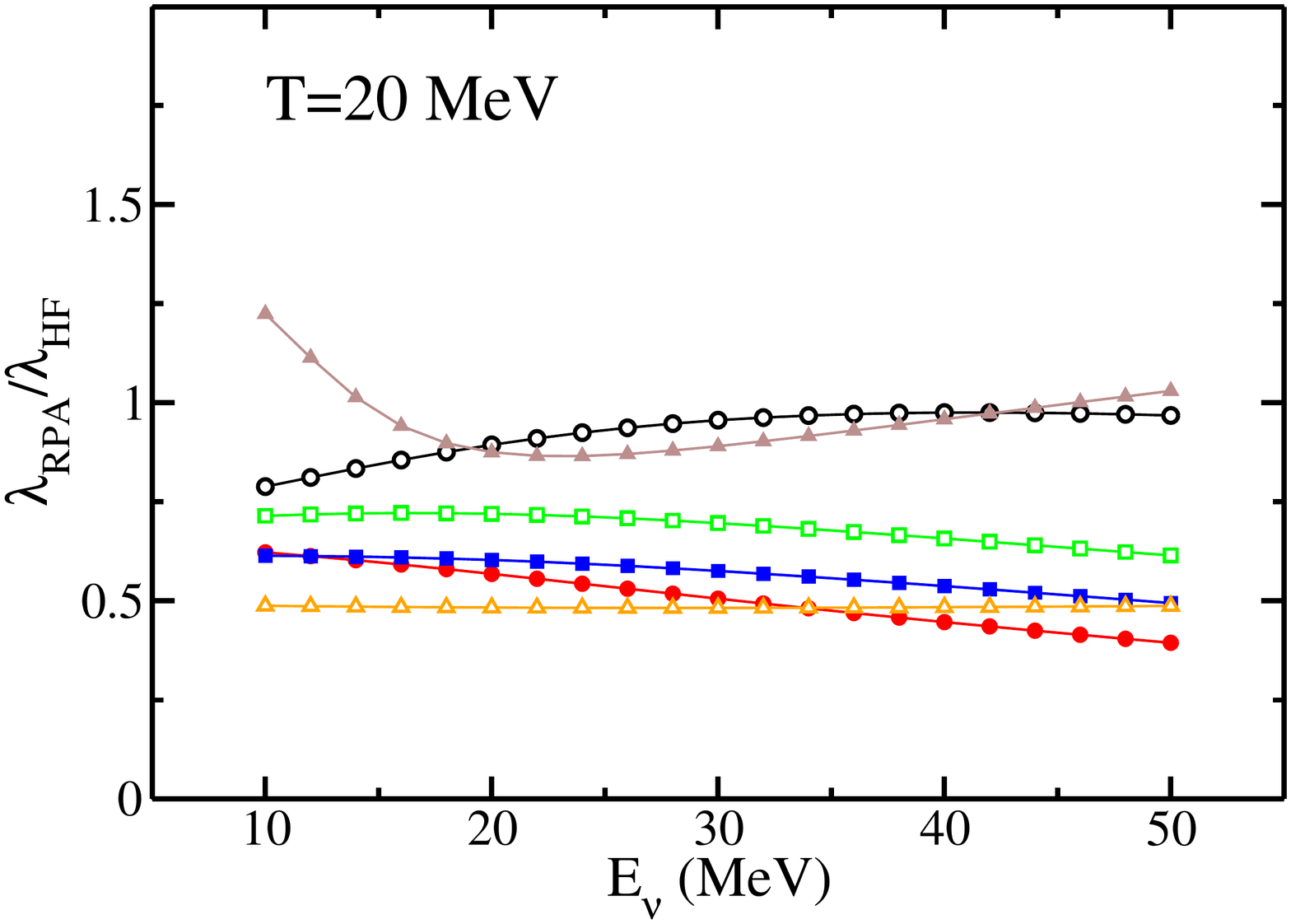}
\hspace{-1.773cm}
\includegraphics[width=0.4\textwidth,angle=-90]{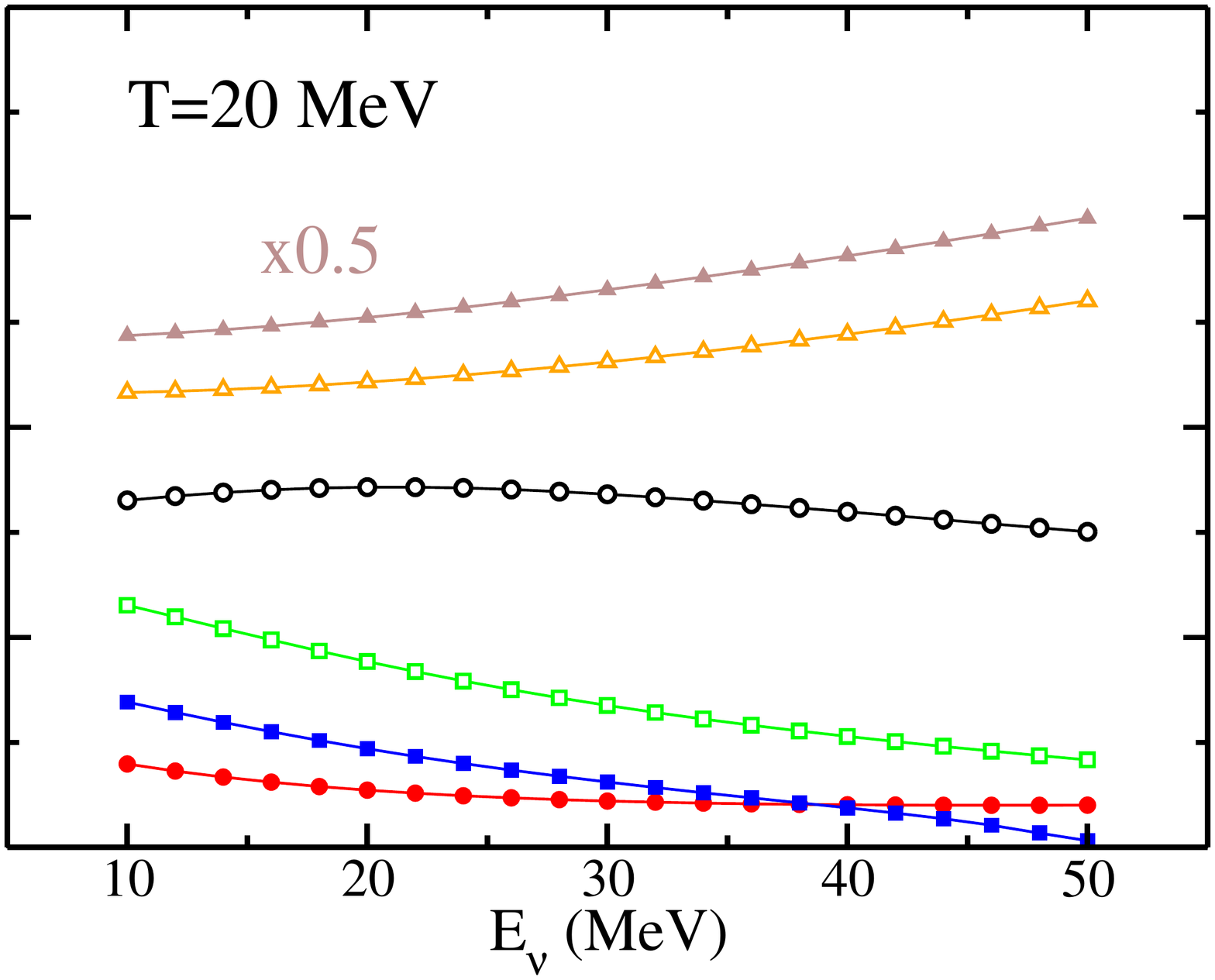}  
\caption{(Color online) The ratio of calculated RPA and HF neutrino mean free paths as a function of the neutrino energy, calculated for two values of temperature and density. Some curves have been multiplied by the displayed scaling factor.}
\label{nmfp:ratio:T2}
\end{center}
\end{figure}

To properly highlight interaction effects other than those contained in the effective mass, we have plotted in Fig. \ref{nmfp:ratio:T2} the ratio  $\lambda_{RPA}/\lambda_{HF}$ for different functionals and two values of temperature and density. We notice that in the low density region the presence of a residual ph-interaction does not strongly affect the value of $\lambda_{RPA}$, which is reduced by a factor $\simeq 0.5-1$ for all functionals, except Skxtb at $T=2$ MeV. At $\rho = 0.24$~fm$^{-3}$, the results strongly depend on the specific interaction.
We clearly see that the functionals Skxta-b and SLy5-t present low density instabilities that strongly affects the results of mean free path giving unrealistic suppressions.
For that reason, we will not consider them in our analysis.
Concerning the differences among T22,T44 and SLy5 functionals they are essentially related to the tensor term. As already discussed in \cite{pas12a}, the cross section given in Eq.\ref{cross:nu}, is mainly dominated by the spin transverse ($S=1,M=\pm1$) response function, which is very sensitive to the properties of the tensor.
 
The net effect of the temperature is to increase collisions. It also modifies the response function and change the integration domains. 
As discussed in \cite{iwa82}, for non-degenerated neutrinos there is no Pauli blocking, so that there is no lower limit. Due to the temperature, negative energies excitations are allowed, thus the available phase space in Eq. (\ref{cross:nu}) is increased.
Since $S^{(S,M)}(q,\omega,T)$ is always positive, we conclude that the cross section can only increase, and thus the mean free path will decrease. 

\begin{figure}[H]
\begin{center}
    \includegraphics[width=0.5\textwidth,angle=-90]{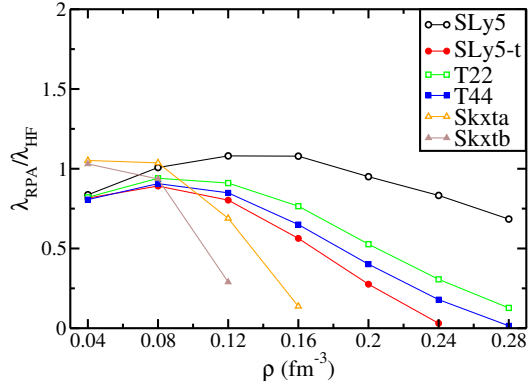}
\caption{(Color online) Evolution of the mean free path at temperature of $T=2$ MeV  and at $E_{\nu}=10$ MeV as a function of the density of the system. }
\label{nmfp:ratio:T0:rho}
\end{center}
\end{figure}

To test the effect of the tensor interaction, we compare the neutrino mean free path obtained with our selected interactions. We recall that these functionals give similar SNM and PNM EoS, except for significant differences in the PNM effective mass as a function of density. By plotting the ratio  $\lambda_{RPA}/\lambda_{HF}$ the mean field effects are smoothed, in particular the effective mass, and the remaining differences among the functionals can be highlighted. In Fig. \ref{nmfp:ratio:T0:rho} the calculated mean free path are plotted as a function of density for an incoming neutrino energy $E_{\nu}=10$ MeV at $T=0$.
As already discussed in \cite{pas12a}, the mean free path is strongly affected by the tensor. 

Taking SLy5 results as the reference for the SLy family, one can clearly see that even a small tensor contribution induces significant variations of the NMFP. The Skx interactions give a more rapid reduction of the NMFP as a function of the density, being zero at around 0.16 fm$^{-3}$. This behavior can be grasped by inspecting Figs.\ref{PNM-1-05} and \ref{Critical:PNM:pole}. Firstly, there is a concentration of strength near the origin of energy for the $S=1$ channel. Secondly, the critical densities are close to the value 0.16 fm$^{-3}$. The presence of these divergences shows up in the quick vanishing of the NMPF.


\section{RPA response functions with Landau $\mathbf{ph}$ interactions}
\label{Sec:landau}

Landau's theory \cite{pin66,mig67,bay91} encompasses the basic properties of Fermi liquids. The particle-hole excitations in a normal Fermi system are described in terms of weakly interacting quasiparticles, which are long-lived only near the Fermi surface. A general $ph$ interaction is thus approximated by assuming the interacting particles and holes on the Fermi surface, that is $k_1 = k_F$, $k_2=k_F$, $q=0$. The only variable is thus the relative angle $({\mathbf {\hat k}_1} \cdot {\mathbf {\hat k}_2})$ between the initial and final hole momenta. 
The response to an external perturbation is obtained by solving the linearized Boltzmann transport equation, in the so-called Landau limit, by taking long-wavelengths $q \to 0$ but keeping $\omega/q$ fixed. In principle, this approximation is improved by using the RPA response, which is a function of the two variables $\omega$ and $q$. There exists an intermediate possibility \cite{pas13d}, which consists in describing the $ph$ interaction in terms of Landau parameters and calculating the RPA response function with no further approximations. In this section, we discuss and compare these possibilities. It is worth mentioning that the use of Landau parameters also allows us to be in touch with finite-range interactions.

\subsection{The response functions}
\label{LAN:respfunc}
As the Landau $ph$ interaction depends only on the relative angle between the hole momenta, it is conveniently
expanded in Legendre polynomials with argument $({\mathbf {\hat k}_1} \cdot {\mathbf {\hat k}_2})$, and it is solely characterized by a set of parameters. For symmetric nuclear matter it is written as
\begin{eqnarray}
\label{landau-Vph}
 V^{(\alpha)}_{ph}  &=&  \sum_{\ell} \,
\bigg\{ f_{\ell} + f_{\ell}' \, (\tau_1 \cdot \tau_2 ) 
+ \left[ g_{\ell}+ g_{\ell}' \, (\tau_1 \cdot \tau_2) 
\right] (\sigma_1 \cdot \sigma_2)  \\
&& \quad  + \left[ h_{\ell} + h_{\ell}' \, (\tau_1 \cdot \tau_2 ) \right]  \frac{{\mathbf k}_{12}^{2}}{k_{F}^{2}} \, S_{12}
\bigg\} \, P_{\ell} ( {\mathbf {\hat k}_1} \cdot {\mathbf {\hat k}_2}  ) \nonumber\,,
\end{eqnarray}
where $f_{\ell}, f'_{\ell}, \dots$ are the Landau parameters, ${\mathbf k}_{12} = {\mathbf k}_1 - {\mathbf k}_2$, and $S_{12}= 3( \hat{\mathbf{k}}_{12} \cdot \sigma_1)( \hat{\mathbf{k}}_{12} \cdot \sigma_2 )- ( \sigma_1 \cdot \sigma_2) $ is the tensor operator. The factor ${\mathbf k}_{12}^{2}/k_F^2$ in the tensor term should be properly written in its Landau limit as $2 [1 - ( {\mathbf {\hat k}_1} \cdot {\mathbf {\hat k}_2}  )]$. We have kept the improper writing 
just for a short discussion in this subsection.  

In the following, all the expressions for SNM and PNM will be cast in a single formal expression by using the symbol $(\alpha)$ to indicate the relevant spin-isospin quantum numbers, in the same way used in the previous Section. For instance, the Landau parameters $f_{\ell}, f'_{\ell}, g_{\ell}, g'_{\ell}$ will be written as $f_{\ell}^{(\alpha)}$, with $(\alpha) =$ (0,0), (0,1), (1,0), (1,1), respectively. Notice that the $ph$ interaction has been defined such that the Landau parameters are independent of the spin projection $M$.  Dimensionless Landau parameters $F_{\ell}^{(\alpha)}$ are defined by multiplying $f_{\ell}^{(\alpha)}$ with the density of states at the Fermi surface
\begin{equation}
N(0) = n_d \, \frac{k_F m^*}{2 \pi^2} \, .
\end{equation}
They provide a dimensionless measure of the $ph$ interaction strength on the Fermi surface. 
The stability  of the spherical Fermi surface of nuclear matter against small deformations can be expressed in terms of the Landau parameters which should fulfill some stability criteria. For the central parameters these are simply written as the inequality $F_{\ell}^{(\alpha)} > - (2 \ell +1)$. The tensor ones are coupled to the central spin-dependent parameters, and the criteria are much more involved, as shown in \cite{bac79}. 

The Landau parameters are obtained by taking the Landau limit of the full $ph$ interaction, which in turn can be 
determined from phenomenological or microscopic interactions. For the general residual interaction  given in Eq.
(\ref{res}), the only non vanishing Landau parameters correspond to the central $\ell=0, 1$ and tensor $\ell =0$ multipolarities. 

Two comments are in order concerning the tensor part. Firstly, we have not included other non-central components considered by other authors (see for instance \cite{ols04,sch04}), as the center-of-mass tensor and cross-vector interactions. Secondly, the tensor part has been written according to the conventional definition~\cite{dab76,bac79}, which is well adapted to the present method to calculate the response function. Some authors \cite{ols04,sch04,ben13,hol13} have defined it without the factor ${\bf k}_{12}^2/k_F^2$, because it leads to a faster convergence \cite{ols04}, in the sense that the absolute value of parameters $h_{\ell}, h'_{\ell}$ decreases as $\ell$ increases. Although the physical information contained in the $ph$ interaction is the same in both cases, the Landau parameters are different, because of the extra $2 [1 - ( {\mathbf {\hat k}_1} \cdot {\mathbf {\hat k}_2}  )]$ factor entering the conventional definition. A recurrence relation \cite{ols04} connects both sets of parameters.

The matrix elements of the Landau $ph$ interaction between spin-isospin states are written as
\begin{eqnarray}
\label{LAN-Vph-ME}
 V_{ph}^{(\alpha,\alpha')}/n_d &=& \delta(M,M') \sum_{\ell}   f^{(\alpha)}_{\ell} 
P_{\ell}( {\mathbf {\hat k}_1} \cdot {\mathbf {\hat k}_2}  ) 
 + \delta(S,1) \sum_{\ell} h^{(\alpha)}_{\ell}  
P_{\ell}( {\mathbf {\hat k}_1} \cdot {\mathbf {\hat k}_2}  ) 
S_T^{(M,M')}( {\mathbf {\hat k}_1} , {\mathbf {\hat k}_2}  ) \, , \quad 
\end{eqnarray}
where 
\begin{eqnarray}
S_T^{(M,M')}( {\mathbf {\hat k}_1} , {\mathbf {\hat k}_2}  ) =  
3  (-)^M \left( k_{12}\right)^{(1)}_{-M} \left( k_{12}\right)^{(1)}_{M'}
-  2 \, \left[ 1- ( {\mathbf {\hat k}_1} \cdot {\mathbf {\hat k}_2}  ) \right] \, \delta(M,M')\,.
\end{eqnarray}
An implicit product $\delta(S,S') \delta(I,I')$ in the r.h.s. of Eq.~(\ref{LAN-Vph-ME}) is to be assumed. 

We have considered in Sec. \ref{solving-BS} the first non-trivial case of a Landau $ph$ interaction characterized by the $\ell=0, 1$ central parameters, and given the response function for the SNM $(S=0, T=0)$ channel. It is trivially extended to the SNM $(S=0, T=1)$ and the PNM $S=0$ channels, by using the appropriate Landau parameters. The response functions have been written in terms of auxiliary $\alpha_i$ functions (\ref{defalpha}), which play the analogous role of the $\beta_i$ functions entering Eqs.~(\ref{satu:chi0I}, \ref{SNM-S1-M0}, \ref{SNM-S1-M1}). These functions are momentum averages of the HF 1p-1h propagator, with different weighting functions.

Since it is possible to get analytical expressions for the RPA response function with the full $ph$ interaction, including tensor terms, one could expect that using a simpler $ph$ interaction would lead to simpler formulae, but it is not so. 
The $\beta_i$ functions satisfy some symmetry properties which help a lot in simplifying the final expressions. These properties are not satisfied by the analogous $\alpha_i$ functions. Therefore the coupled equations for a Landau $ph$ interaction become in comparison much more cumbersome, except in two cases. The first one is when the $ph$ interaction is described with only two central parameters, as it has been considered in Eq. (\ref{resp-f01}). The second one corresponds to the limit of zero-frequency and long-wavelength. As shown in \cite{nav13}, this method allows to write the static susceptibility in terms of the Landau parameters. The final expression is in agreement with the results obtained in \cite{ols04,fuj87} by solving the linearized Boltzmann equation in the same limit.

In general, the equations can be obtained using a symbolic code, but the solution must be numerically obtained, being unpractical to write analytical expressions for the response functions in terms of more Landau parameters.  
Interestingly, the method is applicable when the $ph$ interaction contains an arbitrary number of Landau parameters, the only limitation being the memory size. This allows one to obtain response functions connected to finite-range interactions, either phenomenological or realistic, based on the related Landau $ph$ interaction.

\subsection{Comparing responses in the Landau limit and in the RPA}
The response functions in the Landau limit depend on the single dynamic variable $\omega/q$,
and it is often used beyond the strict long-wavelength limit because of its simplicity. 
In contrast, the RPA response depends on the variables $\omega$ and $q$, and it is worth to compare both responses for non-vanishing values of the transferred momentum $q$. The dimensionless variable $\nu=\omega m^*/(q k_F)$ will be used for a proper comparison. We restrict ourselves to the case of a $ph$ interaction with two central $\ell=0, 1$ and a single tensor $\ell=0$ parameters, which is the natural approximation to the standard Skyrme interaction. In that case, it is possible to solve analytically the algebraic system in the Landau limit. The $S=0$ response has been given in Eq.~(\ref{firstexample}). We give now the expressions for the $S=1$ responses in the longitudinal channel 
\begin{eqnarray}\label{LAN-S10}
 \frac{\chi_{HF}(\nu)}{\chi^{(1,0)}_{LAN}(\nu)} 
&=& (1+H_0)^2  \\
 &-& \left( g_0 - 2 h_0 (1-  3 \nu^2) - 3 H_0 h_0 (1- \nu^2)  
+ \frac{ \nu^2 (g_1 - 4 h_0) (1+X^{(0)}) }{1+X^{(0)} + 
\frac{1}{3} ( G_1- 4 H_0)} \right) \chi_{HF}(\nu)\,, \nonumber
\end{eqnarray}
with
\begin{equation}
 X^{(0)} = \frac{ 3 H_0^2 (\nu^2 -2/3)}
{1- \frac{1}{2} (G_1 -7 H_0) (\nu^2 -2/3)} \, ,
\end{equation}
and in the transverse channel  
\begin{eqnarray}
 \frac{\chi_{HF}(\nu)}{\chi^{(1,1)}_{LAN}(\nu)} 
&=& (1 - \frac{1}{2} H_0)^2 \\
 &-& \left( g_0 + h_0 (1- 3 \nu^2) - \frac{9}{4} H_0 h_0 ( \nu^2 -1)^2   
+ \frac{\nu^2 (g_1 + 2 h_0)(1 +  X^{(1)})}{1 +  X^{(1)} + 
\frac{1}{3} (G_1 + 2 H_0)} \right) \chi_{HF}(\nu)\,, \nonumber 
\label{LAN-S11}
\end{eqnarray}
with
\begin{equation}
 X^{(1)} = \frac{ \frac{3}{2} H_0^2 (\nu^2 -2/3)}
{1- \frac{1}{2} (G_1 + 2 H_0) (\nu^2 -2/3)} \, .
\end{equation}

\begin{figure}[H]
\begin{center}
\includegraphics[width=0.65\textwidth,angle=-90]{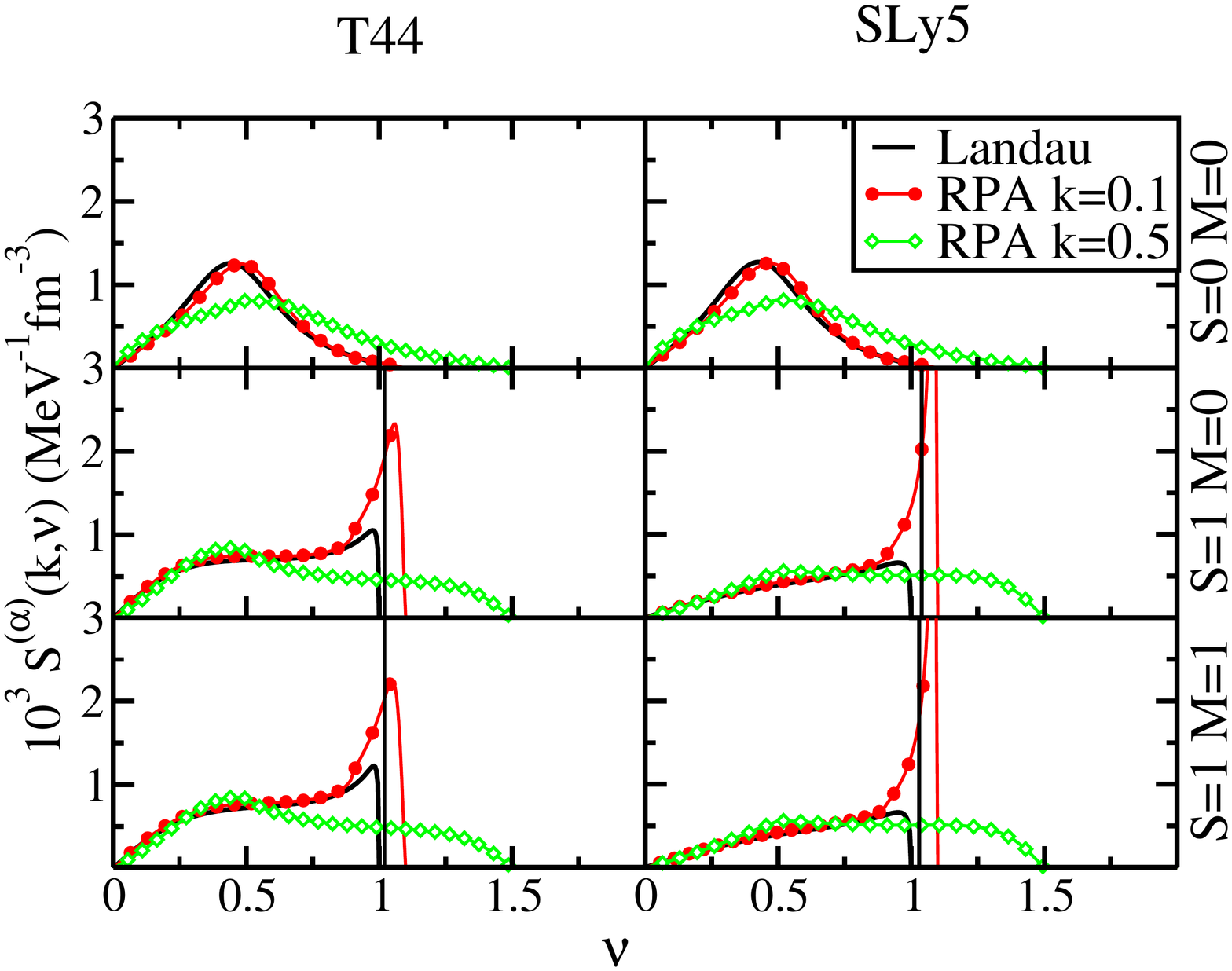} \\
\includegraphics[width=0.65\textwidth,angle=-90]{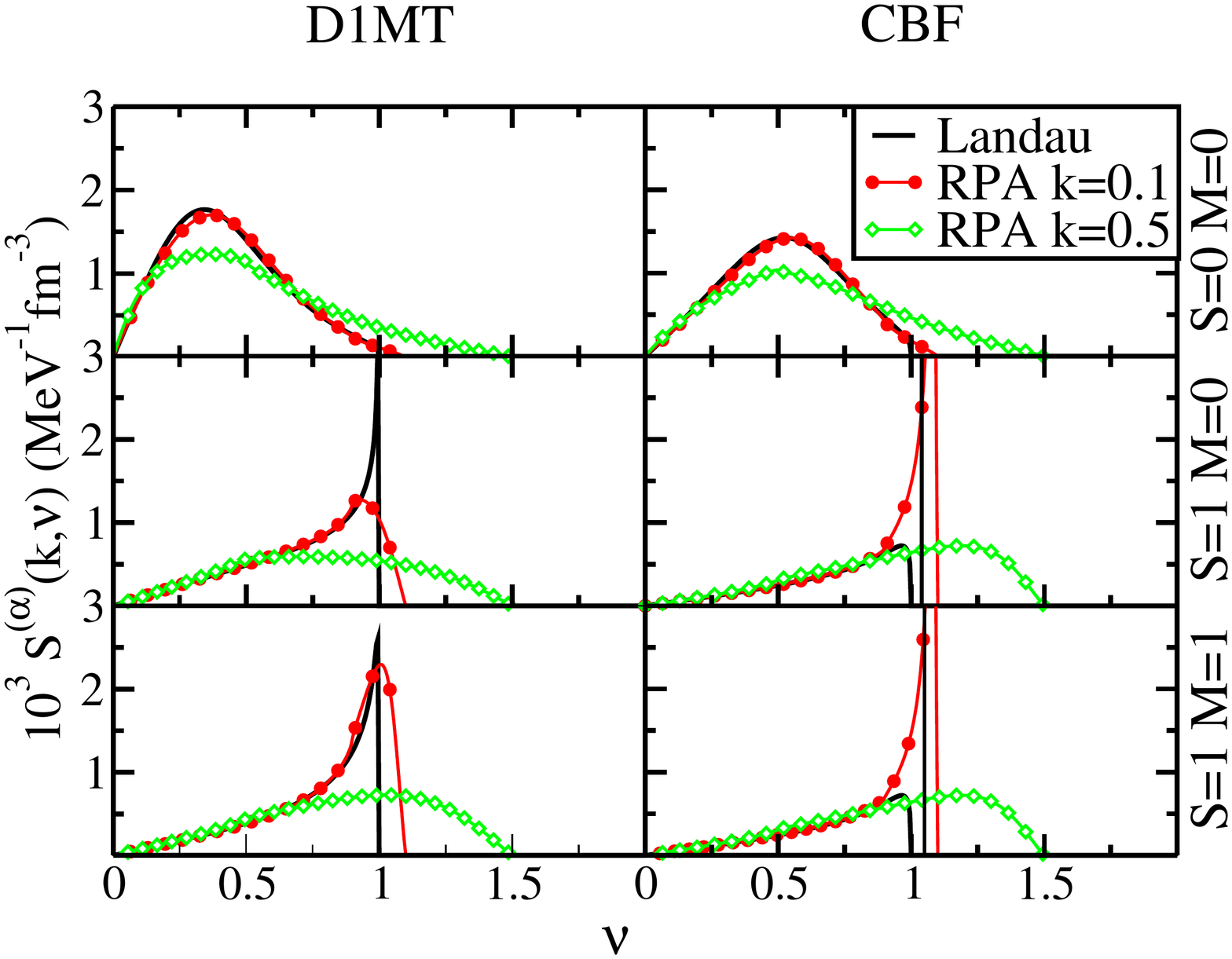} 
\end{center}
\caption{(Color online) PNM strength functions  at $\rho=0.16$~fm$^{-3}$ calculated in the Landau limit (Eqs. \ref{LAN-S10}, \ref{LAN-S11}) and in the RPA (Eqs. \ref{SNM-S1-M0}, \ref{SNM-S1-M1}) with Landau parameters obtained from Skyrme functionals T44 and SLy5, and from finite-range interactions D1MT and CBF. The Landau limit is represented with solid lines, while RPA results calculated at $k=0.1$ and $k=0.5$ correspond to solid lines with circles and triangles, respectively.}
\label{fig:proba1}
\end{figure}

To compare the responses obtained in the Landau limit and in the RPA, we choose the PNM $S=1$ channel.
In Figs.~\ref{fig:proba1} are plotted both strength functions as a function of the dimensionless parameter $\nu$. Two values of the transferred momentum $q$ have been considered in the RPA case, 
$k=q/(2 k_F)= 0.1$ and 0.5. We have employed the Landau parameters calculated from interactions T44 and SLy5, and also from finite-range interactions which will be detailed in Subsec. \ref{convergence}.   

As expected, both responses coincide for low values of the transferred momentum (typically $q \leq 0.2 k_F$) and low values of $\nu$. However neat differences are visible when $q$ increases. The RPA strength displays an important reshaping, in particular the absorption of the zero sound peak into the continuum. This conclusion is particularly interesting if one considers quantities such as the neutrino mean free path in dense matter, since it requires the knowledge of the response function at transferred momentum values which are not small.

One can also see in these plots that the responses in channels $(S=1,M=0)$ and $(S=1,M=1)$ look quite similar, thus indicating that the tensor plays no significant role. Actually, the SLy5 interaction has no tensor terms, and so these responses are identical. The other interactions have indeed very small tensor parameters, and differences between those channels are only visible for the D1MT interaction.

\subsection{Landau and full Skyrme $ph$ interactions}
\begin{figure}[H]
\begin{center}
\includegraphics[width=0.58\textwidth,angle=-90]{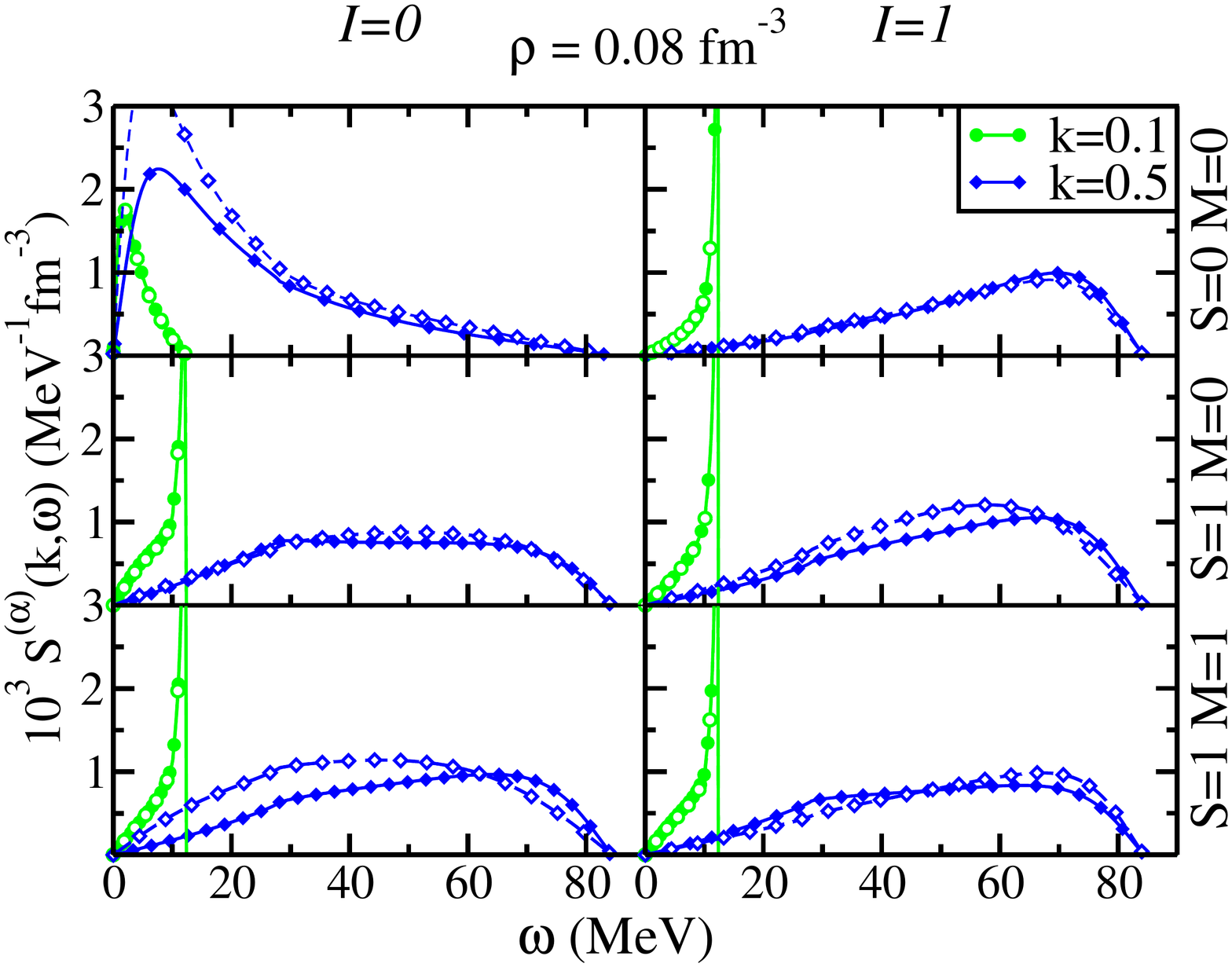} \\
\includegraphics[width=0.58\textwidth,angle=-90]{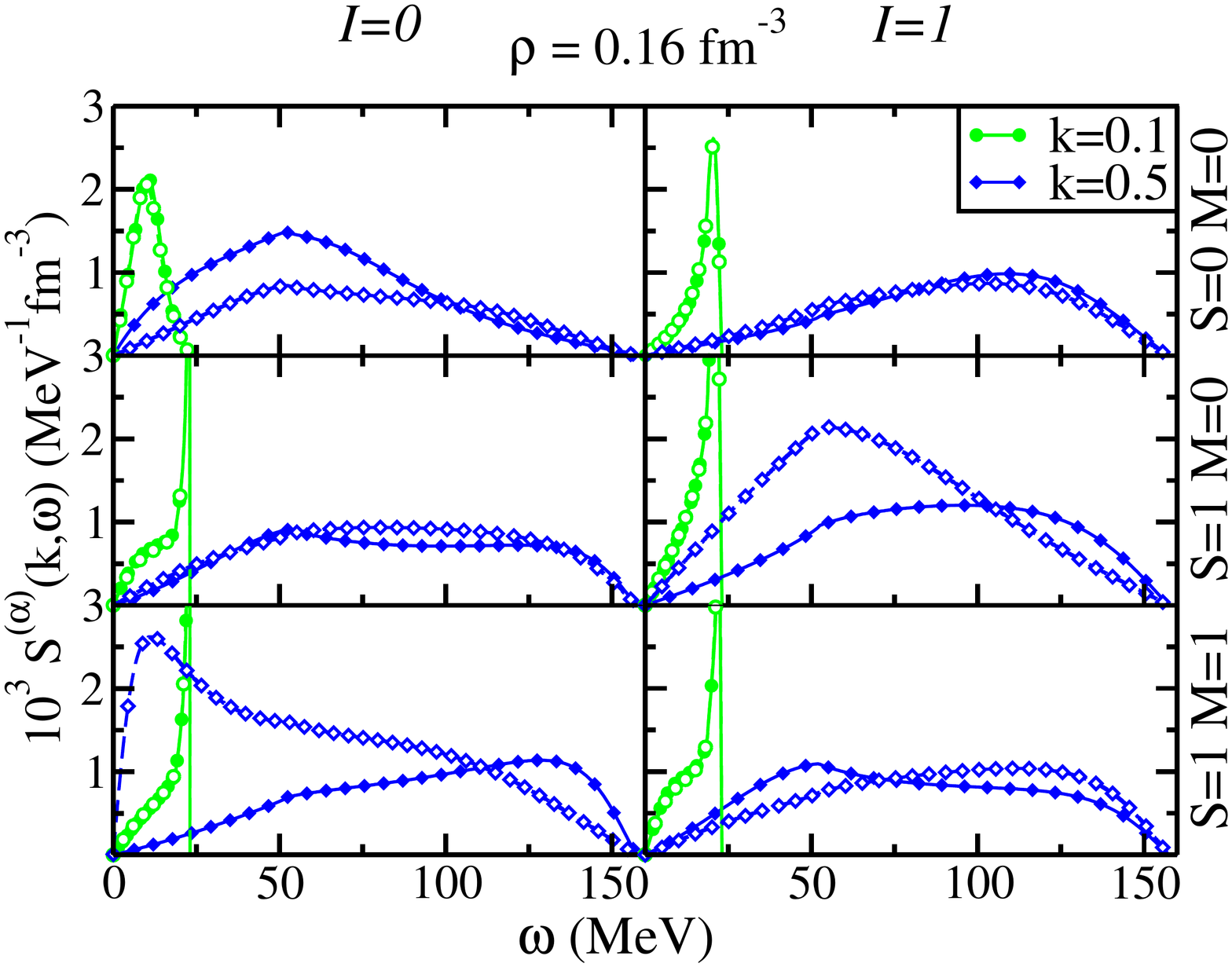}
\end{center}
\caption{(Color online) RPA strength functions in SNM using the full T44 Skyrme $ph$ interaction (open symbols) and its Landau approximation (full symbols) at densities $\rho=0.08$~fm$^{-3}$ (top panels) and 0.16~fm$^{-3}$ (bottom panels), and two values of the momentum transfer.}
\label{fig:landau-rpa-skyrme}
\end{figure}

We consider now the differences in the RPA responses calculated using either the full $ph$ interaction given in (\ref{ME-SNM}) or its Landau approximation ($k_i=k_F, q=0$). To have a rough estimate of the  range of validity of the Landau approximation, 
we have chosen the interaction T44 and two transferred momentum values $k=0.1$ and 0.5. The responses are plotted in Fig.~\ref{fig:landau-rpa-skyrme}, for density values 0.08 and 0.16 fm$^{-3}$ (top and bottom panels, respectively).

As expected, both responses are in good agreement for $k=0.1$. 
At the lower density value (0.08~fm$^{-3}$) they are clearly distinguished for $k=0.5$, but still in reasonably agreement, except for the $S=0, I=0$ channel. That means that the spinodal instability cannot be well reproduced within the Landau approximation for the $ph$ interaction. This is not surprising because the momentum dependence is mandatory for a correct description of this instability. At the higher density value (0.16~fm$^{-3}$) neat differences are 
visible in several channels. For instance, with the full Skyrme $ph$ interaction, the strength in channel $(1,1,0)$ exhibits a peak at low energy, which points towards a finite-size instability related to the specific parameterization. Such an instability will not be located using the corresponding Landau approximation.
In conclusion, the use of such an approximated $ph$ interaction should be reasonably restricted to momentum transfer values $k \le 0.1$ and density values $\rho \le 0.08$~fm$^{-3}$.  

\subsection{Convergence in terms of $\ell_{max}$}
\label{convergence}

The Landau approximation for the $ph$ interaction allows us to consider finite-range interactions in a simple way. 
In principle, any finite-range interaction gives non-vanishing Landau parameters for any value of $\ell$, so that Eq. (\ref{LAN-Vph-ME}) contains an infinite number of terms. However, for practical use, such an expansion has to be truncated up to a maximum index $\ell_{max}$, and the convergence of the results should be analyzed in terms of it. Our choice for the truncation is guided by the developments of the Skyrme pseudo-potential at the next-to-next-to-next-to leading order (N3LO) \cite{car08,rai11,dav14c}. Following the method presented in \cite{dav13}, one can show that a given N$\ell$LO pseudo-potential contributes to the central Landau parameters up to the multipolarity $\ell$, and to the tensor ones up to $(\ell-1)$. For instance, the standard Skyrme interaction is the N1LO pseudo-potential. We shall show that for the finite range interactions considered here, the results are fully converged for $\ell_{max}=3$.

We have calculated the Landau parameters using two types of phenomenological effective interactions. The first one is the familiar Gogny interaction, supplemented with a tensor term as done in \cite{ang11}. These authors have 
started from the tensor part of the Argonne Av8' interaction \cite{wir02}, and have softened its isospin component with  a factor $\left(1-\exp (-b r_{12}^2)\right)$, adjusting the parameter $b$ to fit the lowest $0^-$ states of some selected nuclei in a HF plus RPA description.  In that way they provided two new interactions: D1ST  and D1MT \cite{ang11}, corresponding to the Gogny interactions D1S \cite{ber91}  and D1M~\cite{gor09}, respectively.  

The second type of phenomenological interaction has been constructed by Nakada~\cite{nak03}. It is a modification of the M3Y interaction, that was originally derived from a bare NN interaction by fitting Yukawa functions to a G-matrix \cite{ber77}, including tensor and spin-orbit components, to which an effective density-dependent finite-range part has been added. All the parameters are obtained through a complete fitting procedure including doubly magic nuclei and nuclear matter properties. We choose the M3Y-P2 parameterization \cite{nak03}. 

We have also considered Landau parameters obtained from two recent microscopic calculations for PNM, which we shall indicate as CBF and CEFT. The former refers to the results given in \cite{ben13}, where the formalism of the correlated basis function (CBF) was applied using the Argonne V18 interaction \cite{wir95}. The latter refers to the parameters deduced in  the framework of chiral effective-field theory (CEFT) including two- and three-nucleon interactions \cite{hol13}. As mentioned above, these tensor parameters are based on a different definition than ours. Both sets of parameters are related by a recurrence relation \cite{ols04}, from which one can deduce the values required for the present formalism.  

\begin{figure}[H]
\begin{center}
\includegraphics[width=0.7\textwidth,angle=-90]{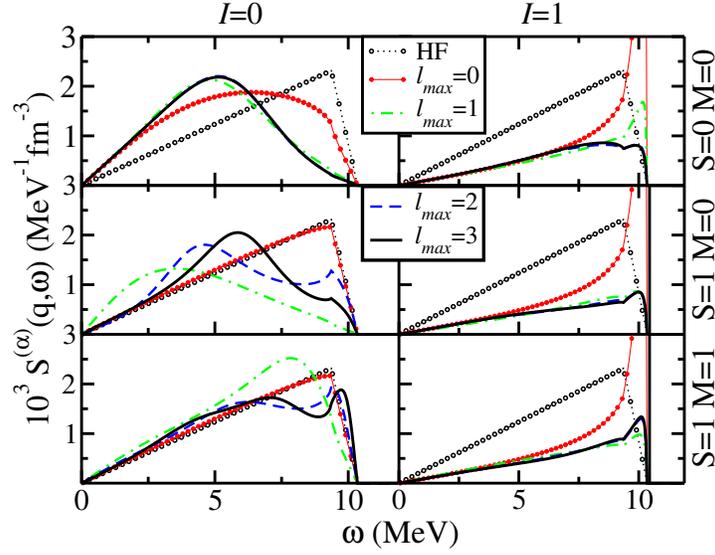}
\end{center}
\caption{(Color online) RPA strength functions in symmetric nuclear matter at density $\rho=0.16$ fm$^{-3}$ and $q=0.1 k_F$, calculated using D1MT parameters. Left and right panels display the isospin channels $I=0$ and 1, respectively. The spin channels $(S,M)$ are plotted in the different rows.}
\label{fig:cov:gognyD1MT}
\end{figure}

In Fig. \ref{fig:cov:gognyD1MT} are plotted the SNM strength functions for the D1MT interaction calculated for a low value of transferred momentum  ($q/k_{F}=0.1$). To have a better insight of the effect of the $ph$ interaction, the HF strength function with the same effective mass is also plotted in the figures. One can see that a good convergence is obtained at $\ell_{max}=2$ in the $S=0$ channel. Since there is no tensor contribution to the $S=0$ channel, this convergence only affects the central Landau parameters. We should keep in mind that using the full Skyrme interaction, the tensor term contributes to both spin channels, due to the presence of a residual spin-orbit interaction. This coupling is suppressed in the Landau limit since there is no spin-orbit contribution to the $ph$ interaction (\ref{landau-Vph}). The shape of $S=1$ strengths is more complex due to the explicit tensor contribution, but one can see that 
 the truncation at $\ell_{max}=2$ insures also a good convergence, except in the channel $S=1, M=0, I=0$. The same conclusions about the convergence hold for higher values of the transferred momentum \cite{pas13d}. The slow convergence in the channel $S=1, M=0, I=0$ depends on the particular interaction considered. For instance, a fast convergence in all channels is found for M3Y-P2. 
 
\begin{figure}[H]
\begin{center}
\includegraphics[width=0.7\textwidth,angle=-90]{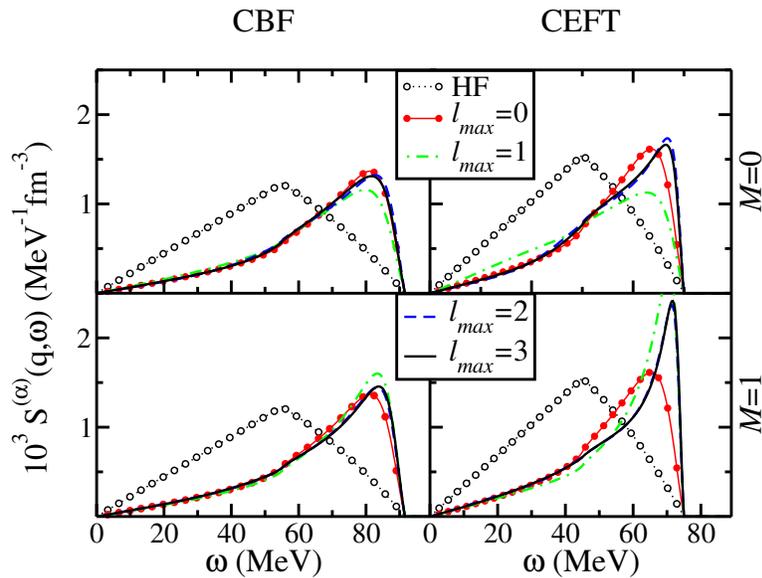}
\end{center}
\caption{(Color online) Strength function in pure neutron matter for $S=1$ channel at $q=0.5 k_F$ for interactions CBF \cite{ben13} at density $\rho=0.16$ fm$^{-3}$ and CEFT \cite{hol13}  at $k_F=1.7$ fm$^{-1}$.}
\label{fig:cov:cbf}
\end{figure}

To see the convergence at a higher momentum transfer, we have plotted in Fig. \ref{fig:cov:cbf} the PNM strength functions in the $S=1$ channels calculated at $q=0.5 k_F$ with the Landau parameters from interactions CBF and CEFT, whose effective masses at the considered density are $m^{*}_{n}/m=0.798$ and $m^{*}_{n}/m=0.995$, respectively. These values are noticeably different, and are reflected in the energy interval where the response functions are defined. In general, similar differences are also to be expected for other microscopic calculations. We refer to ~\cite{bal12} for a critical comparison of symmetric nuclear matter properties with different microscopic methods. There are still uncertainties to unambiguously extract from them a general pattern for the Landau parameters. We remind that we are using approximated values for tensor parameters, as they have been obtained truncating a recurrence relation. These interactions produce very different responses in both $(S=1, M)$ channels. In particular, CEFT predicts a narrow resonance, whereas CBF displays a broad structure. In both cases, a good convergence is attained for $\ell_{max}=2$. 

Notice that tensor effects are not very significant, since the response functions are quite similar for $M=0$ and $M=1$. It has been shown \cite{pas13c}, that tensor terms are responsible of the slow convergence observed in the 
the channel $S=1, M=0, I=0$ with D1MT interaction. One should keep in mind that the D1MT tensor terms have not been derived from a complete re-fit of the parameters. However, as the effect is highly non linear, the convergence of the strength function cannot be guessed directly from the numerical values of the tensor Landau parameters.


\section{Asymmetric nuclear matter}
\label{sect:asym}

The previous studies on SNM and PNM provide a first physical insight of the bulk properties of nuclei. However, isospin asymmetry ($Y=(\rho_n-\rho_p)/\rho$) is the most common situation encountered along the drip-lines or in astrophysics. Thus a systematic study of the influence of asymmetry is clearly required. In particular one may wonder whether the critical densities for unphysical instabilities are located between those of SNM and PNM, and how the spinodal region evolves when one  continuously varies the asymmetry.
In this Section, we adapt the previous formalism to asymmetric nuclear matter (ANM). This has already been done for the central part of a Skyrme interaction in \cite{her97,her97a}, and in \cite{dav14b} for the complete standard interaction, but limited to the particular T44 case. A systematic study of asymmetry effects is presented here, including several relevant Skyrme NEDF. We restrict ourselves to isospin excitations induced by the operator 
$\hat{\text{\boldmath{$\tau$}}}^z$, that is to processes not involving charge exchange.  

\subsection{The residual particle-hole interaction}\label{matrix:el}

In ANM, it is more convenient to work in the proton/neutron formalism. Dropping, for the sake of clarity, momenta and spin dependence, the matrix elements of the $ph$  interaction are written as 
$\langle \tau_1, \tau_3^{-1} | V_{ph} | \tau_4, \tau_2^{-1} \rangle$, where the indices $\tau_i$ refer to protons ($p$) or neutrons ($n$). Because of charge conservation and since charge exchange processes are neglected, there are only four remaining matrix elements, namely 
$$\langle p, p^{-1} | V_{ph} | p, p^{-1} \rangle, \langle n, n^{-1} | V_{ph} | n, n^{-1} \rangle, 
\langle p, p^{-1} | V_{ph} | n, n^{-1} \rangle, \langle n, n^{-1} | V_{ph} | p, p^{-1} \rangle.$$
Actually, since $\langle p, p^{-1} | V_{ph} | n, n^{-1} \rangle = \langle n, n^{-1} | V_{ph} | p, p^{-1} \rangle$, the number actually reduces to three. To avoid repetition of indices they will be written as
\begin{equation}
V_{ph}^{(\tau SM;\tau' S'M')}({\mathbf k}_1, {\mathbf k}_2)
= \langle \tau, \tau, SM | V_{ph}({\mathbf k}_1, {\mathbf k}_2)| \tau',\tau' S'M'\rangle\,.
\end{equation}

From the general residual interaction Eq.~(\ref{res}), we can straightforwardly deduce
\begin{eqnarray}\label{mel:nn}
V_{ph}^{(\tau SM;\tau' S'M')}({\mathbf k}_1, {\mathbf k}_2)&=&
\frac{1}{2}  \delta_{SS'}\delta_{MM'} 
\left( W_1^{(\tau,\tau',S)} + W_2^{(\tau,\tau',S)} {\mathbf k}_{12}^2  \right) 
\nonumber \\
&+& \frac{1}{2}  \delta_{SS'} \delta_{S1}  W_{T1}^{(\tau,\tau')} \, (-)^M (k_{12})^{(1)}_{-M} (k_{12})^{(1)}_{M'}  \nonumber \\
&+& \frac{1}{2} W_{SO}^{(\tau,\tau')} 
\left( \delta_{S'0}\delta_{S1}M(k_{12})^{(1)}_{-M}+\delta_{S'1}\delta_{S0}M'(k_{12})^{(1)}_{M'}\right)\, ,
\end{eqnarray}
which in momentum space is identical to the matrix elements previously determined for SNM and PNM. 
From a technical point of view, the Bethe-Salpeter equations will have the same structure, but due to the isospin asymmetry the number of coupled equations is doubled. The coefficients $W_{i}^{(\tau,\tau',S)}$ can be written in terms of the constants $W_i^{(\alpha)}$ given in Table \ref{table:W_i} as
\begin{eqnarray}
W_1^{(\tau,\tau,S)} &=& W_1^{(S,0)} + W_1^{(S,1)} + 8 b(\tau) C_1^{\rho,\gamma} \rho_1 \gamma \rho^{\gamma-1}
+ W_{T2}^{(0)} + W_{T2}^{(1)} \,,\\
W_1^{(\tau,-\tau,S)} &=& W_1^{(S,0)} - W_1^{(S,1)} + W_{T2}^{(0)} - W_{T2}^{(1)}\,,
\end{eqnarray}
with $b(n)=1$ and $b(p)=-1$. For the remaining indices ($a= 2, T1, SO$) we have
\begin{eqnarray}
W_a^{(\tau,\tau,S)} &=& W_a^{(S,0)} + W_a^{(S,1)} \,, \\
W_a^{(\tau,-\tau,S)} &=& W_a^{(S,0)} - W_a^{(S,1)}\,.
\label{Wa}
\end{eqnarray}
Notice that some of them explicitly depend on the isovector density $\rho_1$. 

\subsection{Response functions}\label{sec:form}

The first required ingredient is the Hartree-Fock retarded propagator of the non-interacting $ph$  pair, whose expression is given in Eq. (\ref{app:beta:betafunct}). Since charge-exchange processes are not considered here, the particle and the hole in a given pair share the same isospin quantum number $\tau$ (either $p$ or $n$). General expressions for the momentum integral of $G^{(\tau)}_{HF}(\mathbf{k}, q,\omega)$ and the moments required for the calculation are given in \ref{app:beta}, together with typical figures for different asymmetries and temperatures. Since the HF propagator does not carry spin quantum numbers, the Bethe-Salpeter equation simply writes as 
\begin{eqnarray}
\label{BSeq}
G^{(\tau \tau' SM)}(\mathbf{k}_1,q,\omega) &=& 
\delta(\tau,\tau') G_{HF}^{(\tau)}(\mathbf{k}_1,q,\omega) \\
&+& 
G^{(\tau)}_{HF}(\mathbf{k}_1,q,\omega) 
\sum_{(\tau'',S''M'')} \int \frac{d^{3}\mathbf{k}_2}{(2 \pi)^3} \,
V_{ph}^{(\tau S M;\tau'' S''M'')}({\mathbf k}_1, {\mathbf k}_2)
G^{(\tau'' \tau' S''M'')}(\mathbf{k}_2,q,\omega) \, . \nonumber 
\end{eqnarray}
The form of the residual interaction (\ref{mel:nn}) allows one to write a closed set of algebraic equations by multiplying Eq. (\ref{BSeq}) successively with the functions $1$, $k^2$, $k Y_{1,0}$, $k^2 |Y_{1,\pm m}| ^2$ and integrating over the momentum $\mathbf{k}_1$. This is analogous to the previously discussed SNM and PNM cases, but because of the isospin indices $\tau,\tau'$ the number of equations is doubled in the ANM case. However, due to isospin properties of the residual interaction, the set of equations can be actually decoupled in two subsystems in each $(S,M)$ channel. One subsystem is for the couple $(nn)-(pn)$ and the other is for the couple $(pp)-(np)$, which is obtained from the previous one by simply exchanging $n \leftrightarrow p$ (see \cite{dav14b} for details). These subsystems can be found in \cite{dav14b}. Getting the analytic expressions for ANM response functions is rather cumbersome, and not particularly enlightening for the physics. It is preferable to solve numerically the subsystems in the $(q,\omega)$-space. Their analytical expressions have been used within a formal computation code just to derive their associated sum rules (see next part).
The response function for each channel reads
\begin{equation} 
\chi^{(\tau \tau' SM)}(q,\omega) = n_d \langle G^{(\tau \tau' SM)} \rangle  \, ,
\label{chittRPA}
\end{equation} 
with the spin degeneracy $n_d = 2$ for ANM. 

Actually, the relevant spin-isospin responses are linear combinations of $\chi^{(\tau \tau' SM)}(q,\omega)$. 
Let us first consider the spin channels $S=0$. The general excitation operator can be written in the form
\begin{eqnarray}
 \alpha \sum_i {\rm e}^{i {\mathbf q} \cdot {\mathbf r}_i} 
+ \beta  \sum_i {\rm e}^{i {\mathbf q} \cdot {\mathbf r}_i} \hat{\text{\boldmath{$\tau$}}}^z_i
= (\alpha + \beta) \sum_i {\rm e}^{i {\mathbf q} \cdot {\mathbf r}_i} \hat{\mathbf p}_i +
(\alpha - \beta) \sum_i {\rm e}^{i {\mathbf q} \cdot {\mathbf r}_i} \hat{\bf n}_i \, ,
\end{eqnarray}
where $\alpha, \beta$ are arbitrary coefficients, and the operators $\hat{\bf p}=(1+ \hat{\text{\boldmath{$\tau$}}}^z)/2$ and $\hat{\bf n}=(1- \hat{\text{\boldmath{$\tau$}}}^z)/2$ project upon proton or neutron spaces, respectively. 
Therefore, the response in the $ph$  spin channels $S=0$ can be written as
\begin{eqnarray}
 | \alpha + \beta |^2 \chi^{(pp 0)}(q, \omega) + 
(\alpha + \beta) (\alpha - \beta)  \chi^{(pn0)}(q, \omega) 
+ (\alpha - \beta) (\alpha + \beta)\chi^{(np0)}(q, \omega) 
+| \alpha - \beta |^2 \chi^{(pp0)}(q, \omega) \nonumber\,.
\end{eqnarray}
In particular, taking $\beta=0$ or $\alpha=0$ the excitation operator respectively acts on the isoscalar or the isovector  channels. The antisymmetric spin channels $S=1$ correspond to including 
$\hat{\text{\boldmath{$\sigma$}}}_i$ in the excitation operator. Finally, the 
 relevant spin-isospin responses are given by the combinations
\begin{eqnarray}
\chi^{(SM;I=0)}(q, \omega) &=& 
\chi^{(nnSM)}(q, \omega) + \chi^{(pnSM)}(q, \omega) 
+ \chi^{(ppSM)}(q, \omega) + \chi^{(npSM)}(q, \omega)\,, \\
\chi^{(SM;I=1)}(q, \omega) &=& 
\chi^{(nnSM)}(q, \omega) - \chi^{(pnSM)}(q, \omega) 
+ \chi^{(ppSM)}(q, \omega) - \chi^{(npSM)}(q, \omega) \,.
\label{isospinchi}
\end{eqnarray}

\subsection{Results}\label{sec:res}

We shall discuss the strength functions $S^{(SM;I)}(q, \omega)$ for some selected values of the asymmetry parameter. 
Besides the extreme cases $Y=0$ (SNM) and $Y=1$ (PNM), we have chosen $Y=0.21$  (isospin asymmetry of $^{208}$Pb) and $Y=0.5$  ({\it i.e.} $Z =N/3$, which roughly corresponds to infinite matter in $\beta$-equilibrium). Furthermore, we have performed calculations at densities 0.16 and 0.08 fm$^{-3}$.

\begin{figure}[H]
\begin{center}
\includegraphics[width=0.6\textwidth,angle=-90]{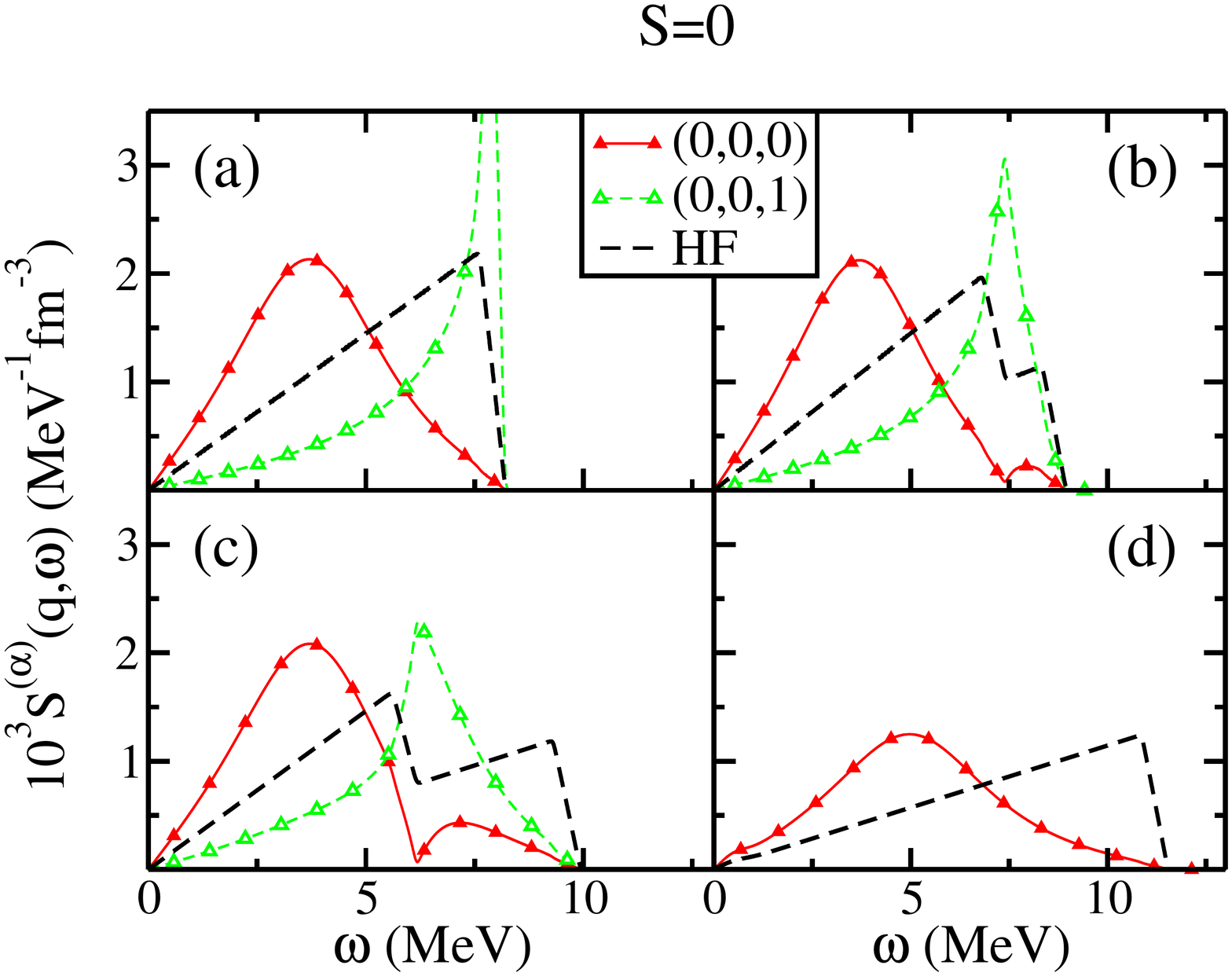}\\
\includegraphics[width=0.6\textwidth,angle=-90]{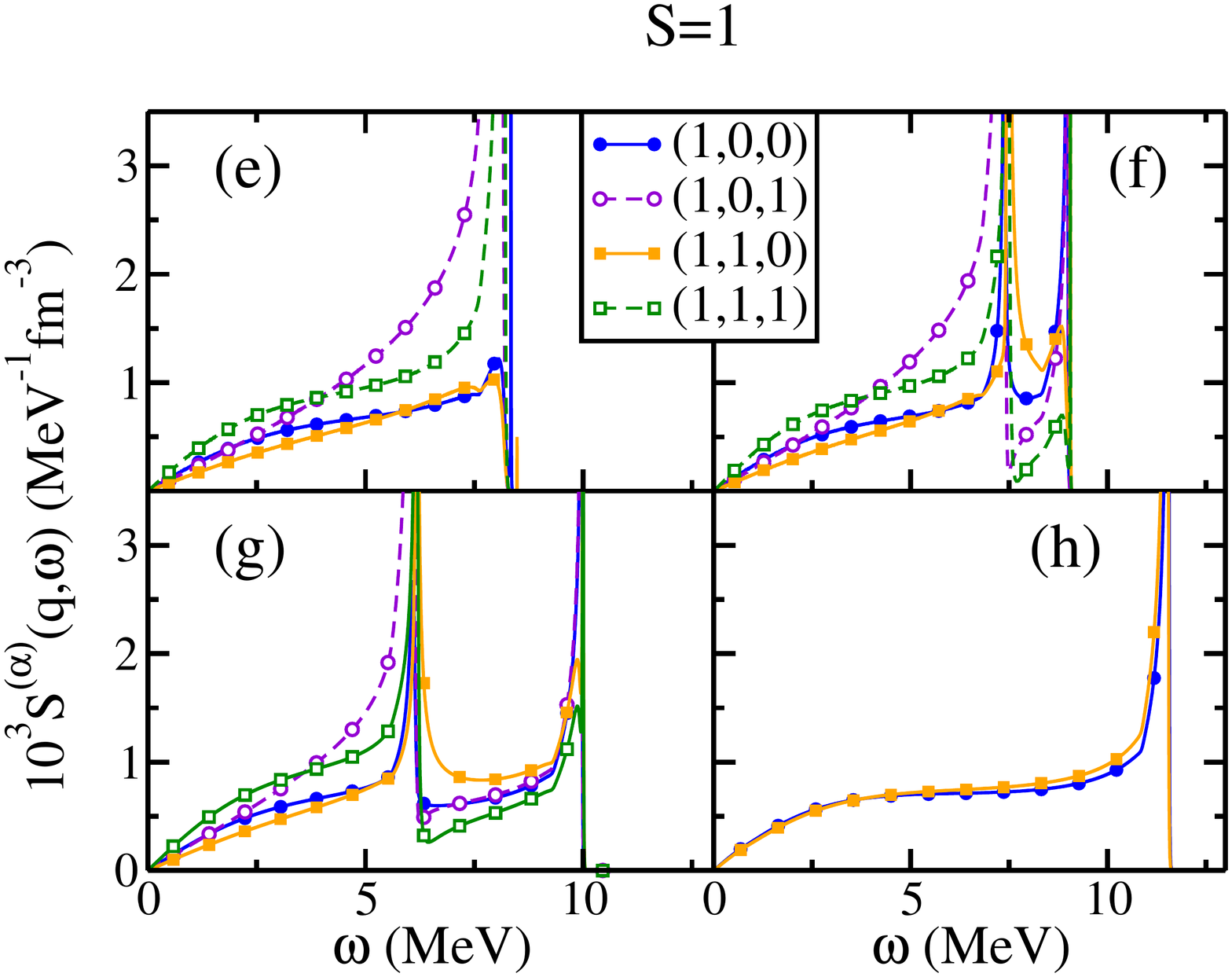}
\end{center}
\caption{(Color online) Strength functions $S^{(\alpha)}(q,\omega)$ for the spin-isospin channels $(\alpha)=(S,M,I)$ in asymmetric nuclear matter, calculated with Skyrme interaction T44 at density $\rho=0.16$ fm$^{-3}$ and momentum transfer $q=0.1$~fm$^{-1}$.  Panels (a) and (e) correspond to the asymmetry parameter $Y=0$ (SNM). Panels (b) and (f) to $Y=0.21$. Panels (c) and (g) to $Y=0.5$. Panels (d) and (h) to $Y=1$ (PNM). Only channels $(S,M)$ are relevant in PNM.}
\label{response:q01:T44}
\end{figure}

We have performed calculations for the interactions discussed in subsection \ref{choice}. Let us start presenting results for interaction T44. In Fig.~\ref{response:q01:T44}, strength functions at $\rho=0.16$ fm$^{-3}$ and momentum transfer $q=0.1$ fm$^{-1}$ are displayed for different asymmetries and for all $(S,M,I)$ channels. Panels (a) to (d) show results for channels $S=0$. The HF strength function has also been plotted in these panels to make more visible the effect of the residual interaction. Panels (e) to (h) show results for channels $S=1$. 

The HF strength function exhibits a two-peak structure for intermediate values of asymmetry ($Y=0.21$ and $Y=0.5$ in our case), related to the different characteristics of the proton and neutron Fermi seas. 
For instance, at $Y=0.5$ and saturation density the effective masses are $m^*_p/m=0.75$ and $m^*_n/m=0.66$,  as compared to the SNM value 0.7, leading thus to two different Fermi seas.
This two-peak structure is conserved by RPA in the $(S,I)=(0,0)$ channel, but not in the $(S,I)=(0,1)$ one where the collective state is shifted to low energy as $Y$ increases. For $S=1$, one can observe the presence of  a collective state for both SNM and PNM systems (panel (e) and (h)). Between these two extreme cases, the contribution of protons and neutrons is clearly distinguished. On the contrary, the tensor effect which manifests itself by a splitting between $M=1$ and $M=0$ is relatively small for important asymmetries and negligible for PNM.

\begin{figure}[H]
\begin{center}
\includegraphics[width=0.62\textwidth,angle=-90]{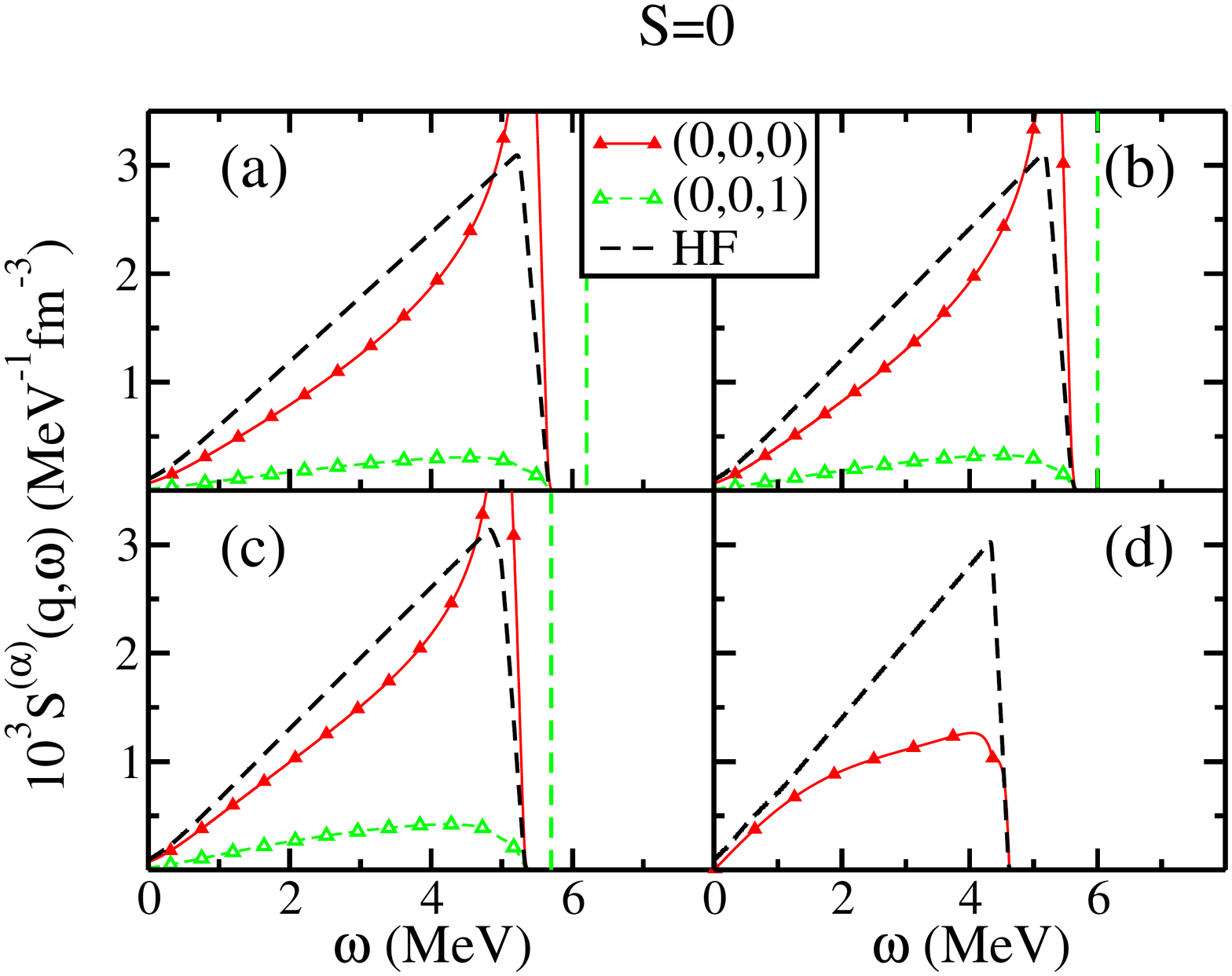}\\
\includegraphics[width=0.62\textwidth,angle=-90]{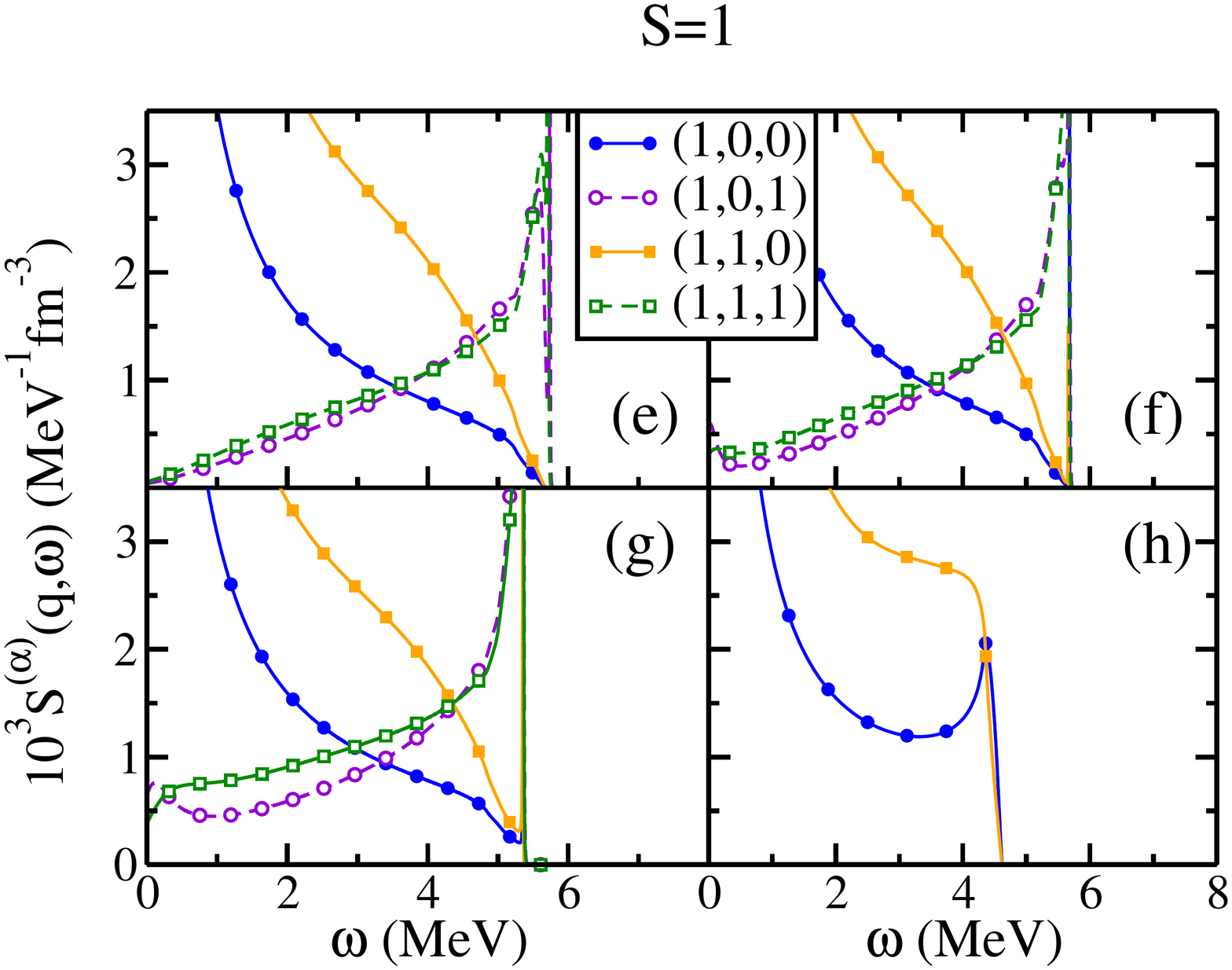}\\
\end{center}
\caption{(Color online) Same as Fig.\ref{response:q01:T44}, but for Skxta interaction.}
\label{response:q01:Skxa}
\end{figure}

The situation is different for interaction Skxta. As shown in Fig. \ref{response:q01:Skxa}, the two-peak structure is absent, even at the HF level. 
This is related to the particular behavior of the proton and neutron effective masses as the asymmetry is increased. 
For instance, at $Y=0.5$ and saturation density one has $m^*_p/m=0.86$ and $m^*_n/m=1.22$, as compared to the SNM value 1. Not only the variation goes in opposite direction, but also the absolute value of the difference is larger for Skxta as compared to T44. These differences manifest in conditions \ref{effmcon1}-\ref{effmcon2} by suppressing the imaginary part of the functions $\beta_i^{n,p}$. Consequently, with interaction Skxta, we deal in practice with a single Fermi sea.
The residual interaction does not induce any sizable modification in the isoscalar channel, whereas we can observe a huge strength depletion in the isovector one. For vector channels we can see that the effect of the tensor (manifested through the difference between spin projections) is negligible for $I=1$, but very important for $I=0$. In the limiting case of PNM, the degeneracy of both isospin channels is in favor of a strong tensor effect. Moreover, independently of the asymmetry, we have a huge accumulation of strength at zero energy in the isovector channel, which is the prelude to a physical instability as we shall discuss later on in this Section. 

\begin{figure}[H]
\begin{center}
\includegraphics[width=0.62\textwidth,angle=-90]{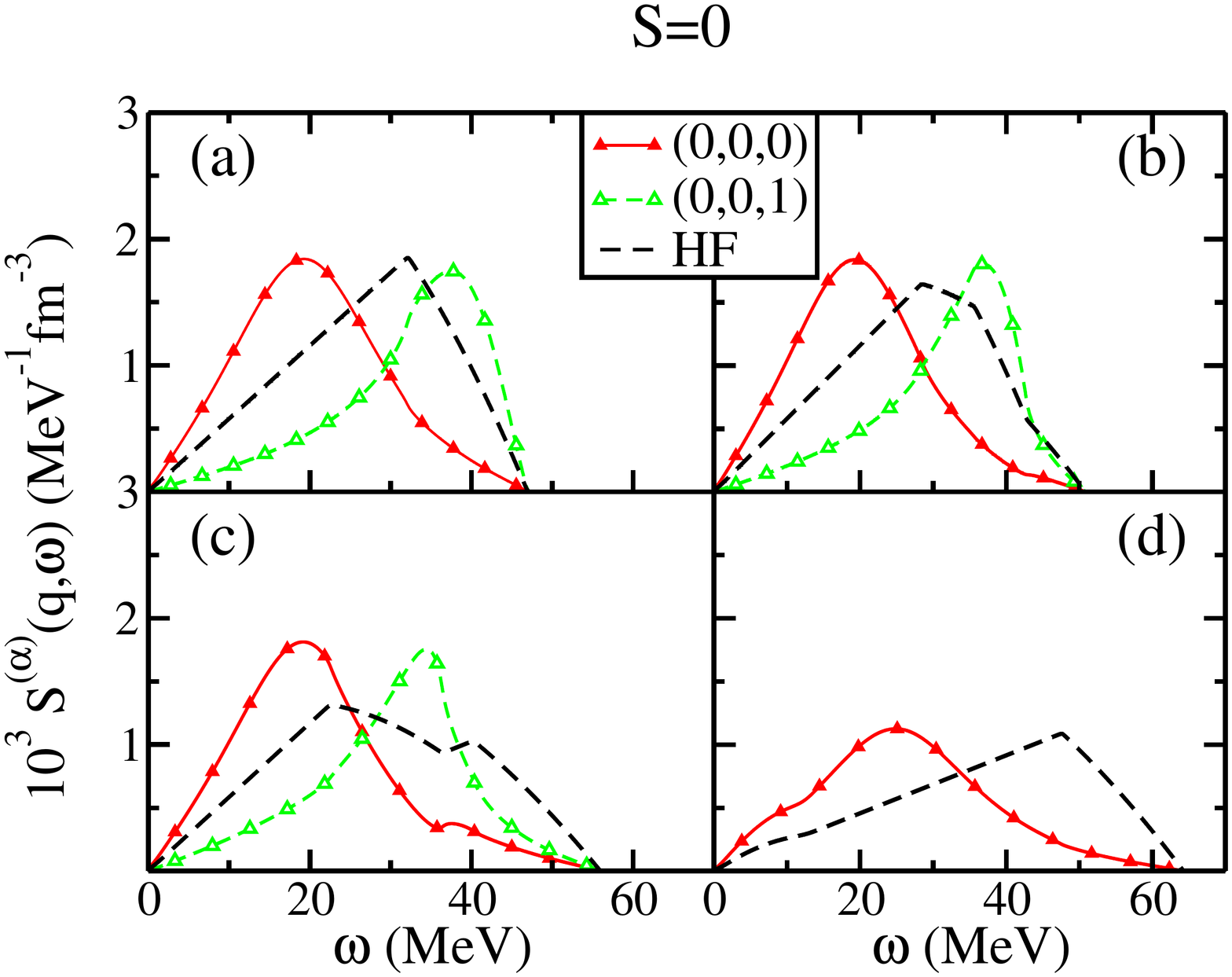}
\includegraphics[width=0.62\textwidth,angle=-90]{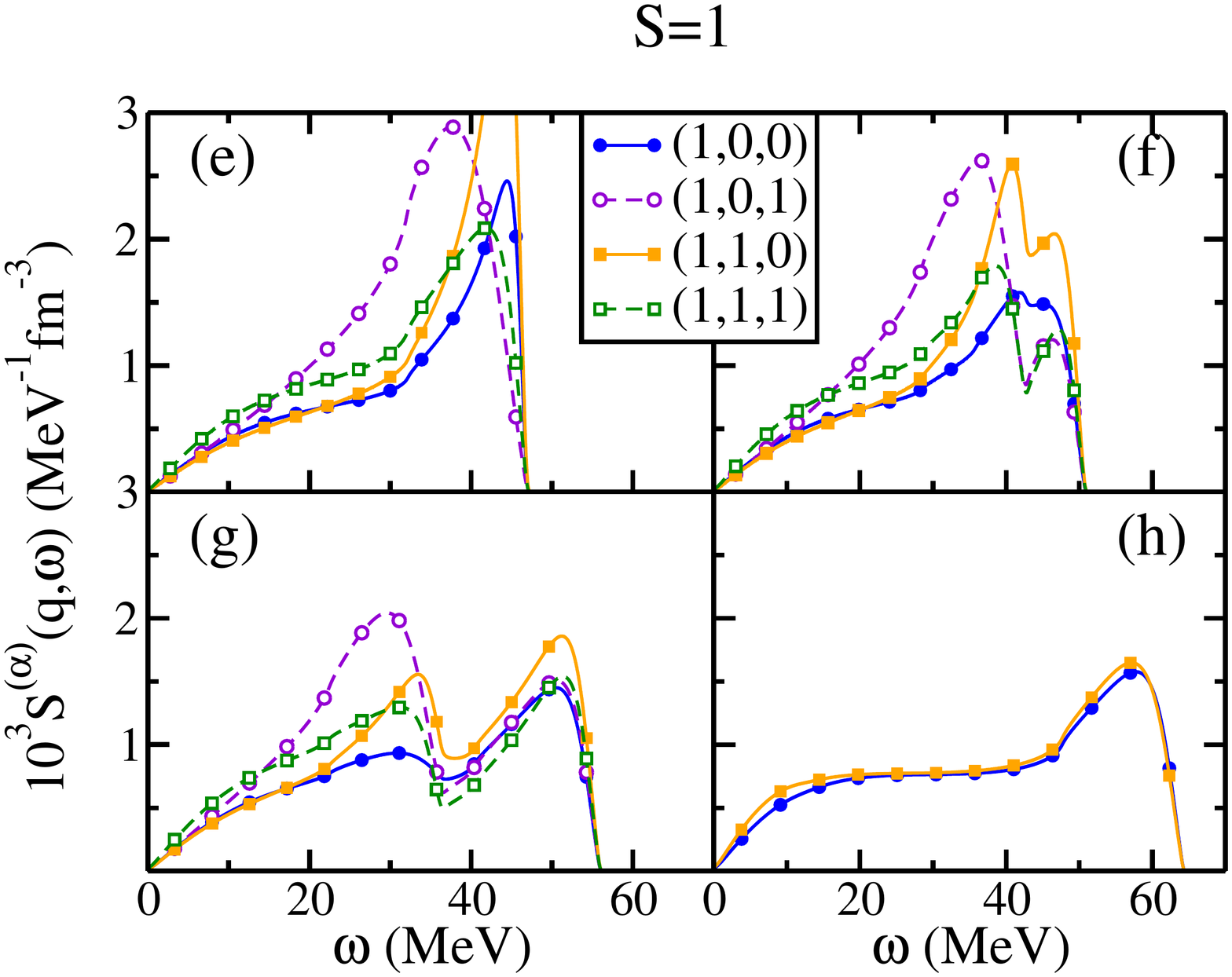}\\
\end{center}
\caption{(Color online) Strength functions $S^{(\alpha)}(q,\omega)$ for the spin-isospin channels $(\alpha)=(S,M,I)$ in asymmetric nuclear matter, calculated with Skyrme interaction T44 at density $\rho=0.16$ fm$^{-3}$ and momentum transfer $q=0.5$~fm$^{-1}$.  Panels (a) and (e) correspond to the asymmetry parameter $Y=0$ (SNM). Panels (b) and (f) to $Y=0.21$. Panels (c) and (g) to $Y=0.5$. Panels (d) and (h) to $Y=1$ (PNM). Only channels $(S,M)$ are relevant in PNM.}
\label{response:q05:T44}
\end{figure}

\begin{figure}[H]
\begin{center}
\includegraphics[width=0.62\textwidth,angle=-90]{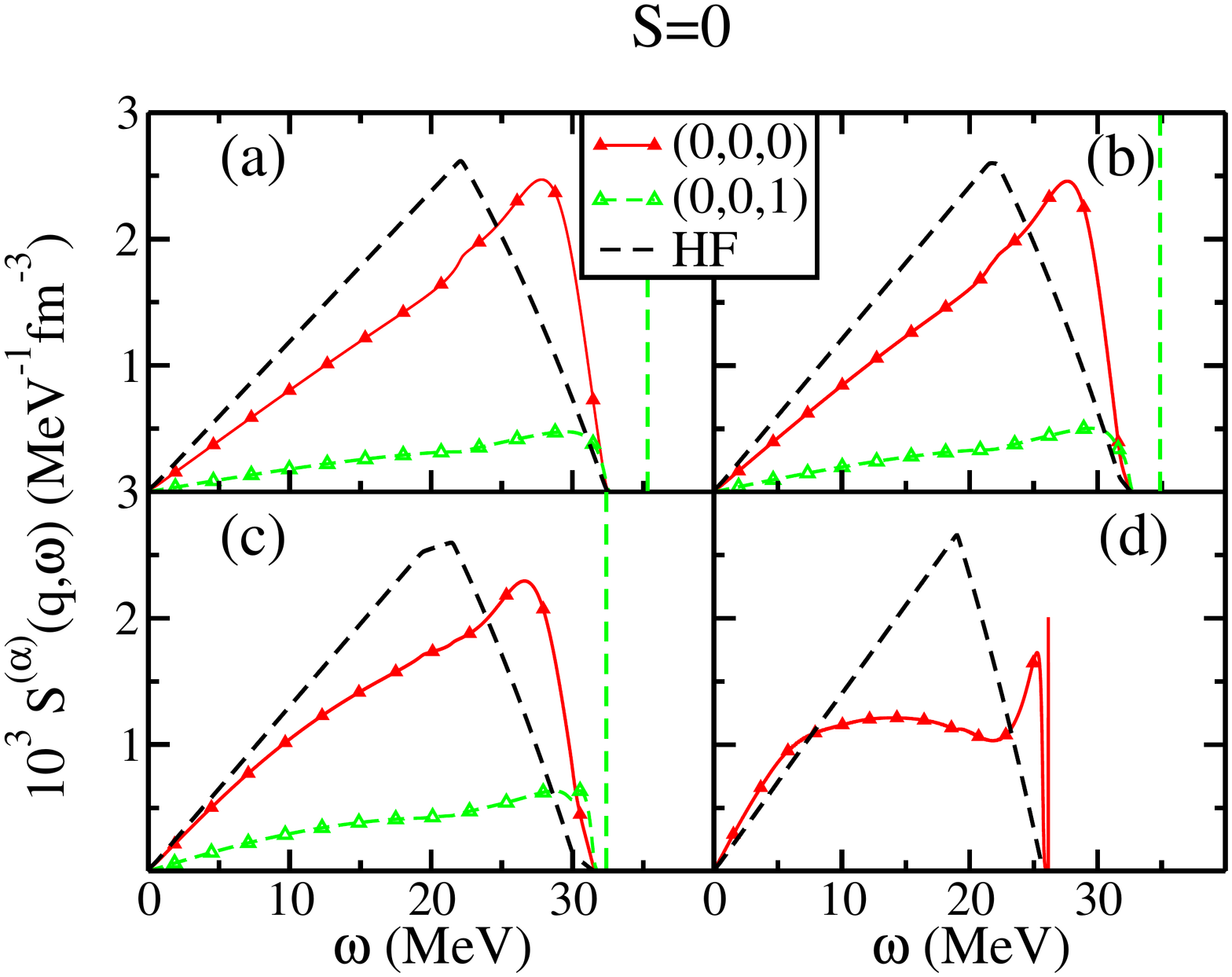}
\includegraphics[width=0.62\textwidth,angle=-90]{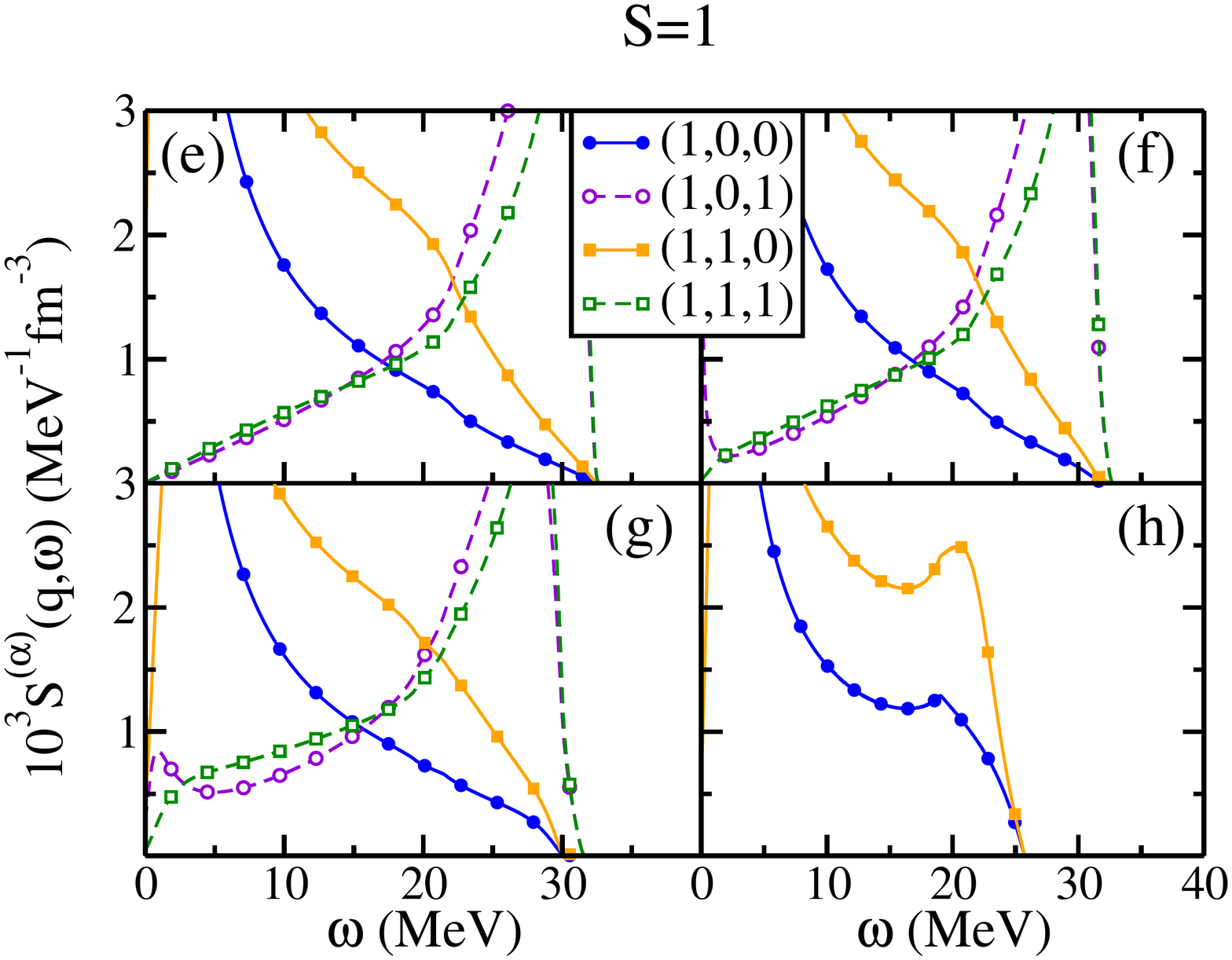}\\
\end{center}
\caption{(Color online) Same as Fig.\ref{response:q05:T44}, but got the Skxta  interaction.}
\label{response:q05:SKx}
\end{figure}

\begin{figure}[H]
\begin{center}
\includegraphics[width=0.62\textwidth,angle=-90]{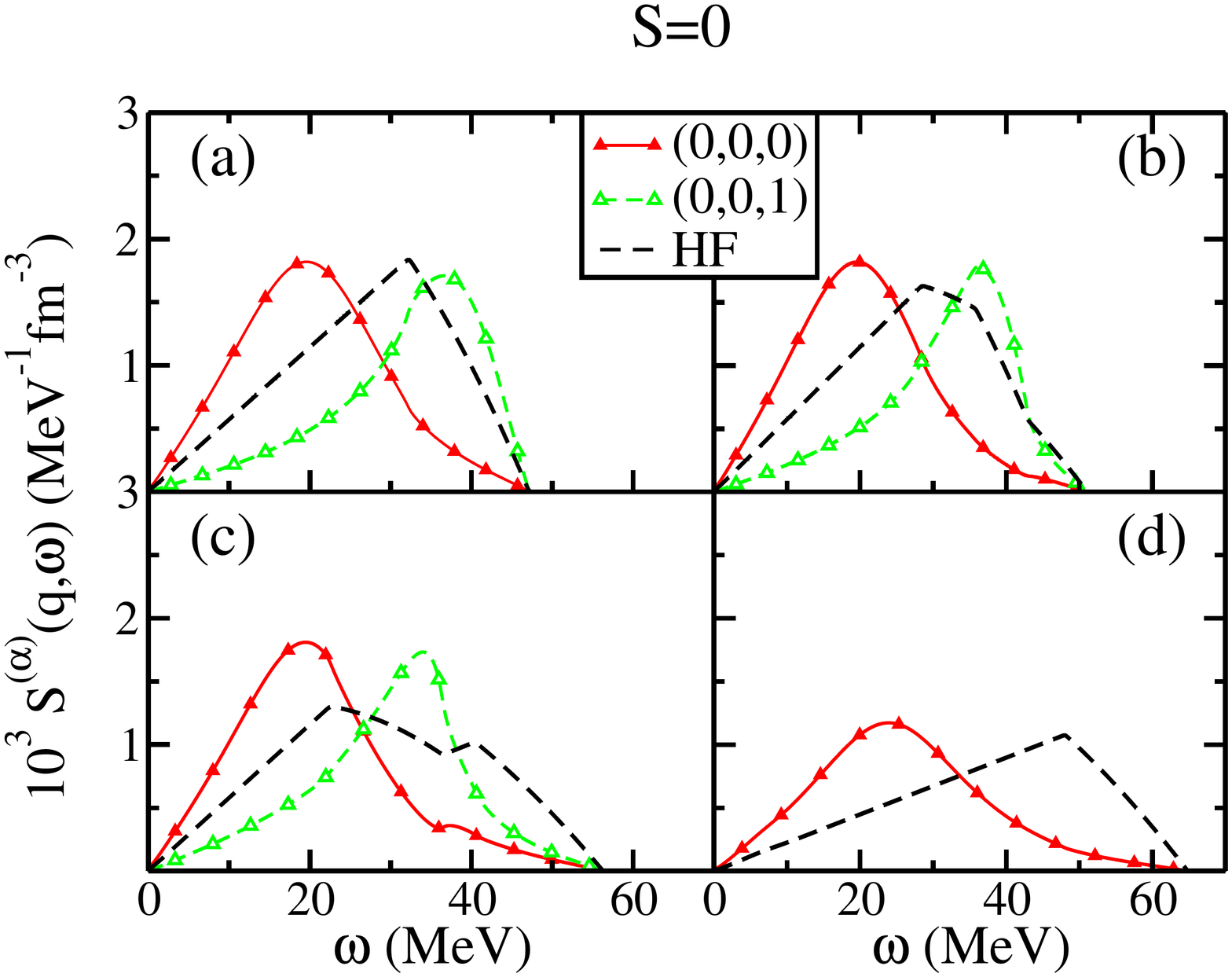}\\
\includegraphics[width=0.62\textwidth,angle=-90]{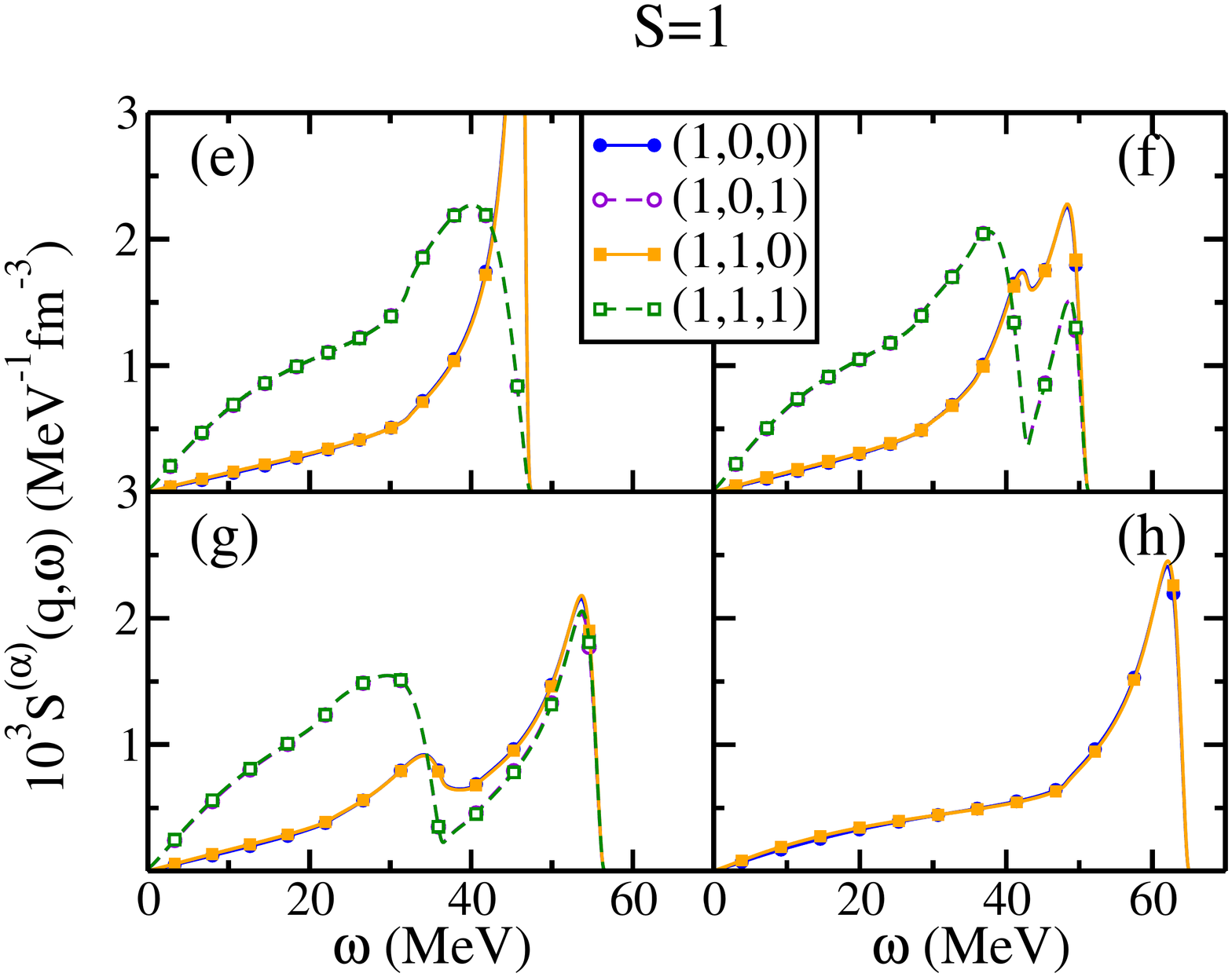}\\
\end{center}
\caption{(Color online) Strength functions $S^{(\alpha)}(q,\omega)$ for the spin-isospin channels $(\alpha)=(S,M,I)$ in asymmetric nuclear matter, calculated with Skyrme interaction SLy5 at density $\rho=0.16$ fm$^{-3}$ and momentum transfer $q=0.5$~fm$^{-1}$.  Panels (a) and (e) correspond to the asymmetry parameter $Y=0$ (SNM). Panels (b) and (f) to $Y=0.21$. Panels (c) and (g) to $Y=0.5$. Panels (d) and (h) to $Y=1$ (PNM). Only channels $(S,M)$ are relevant in PNM.}
\label{response:q05:SLy5}
\end{figure}

\begin{figure}[H]
\begin{center}
  \includegraphics[width=0.43\textwidth,angle=-90]{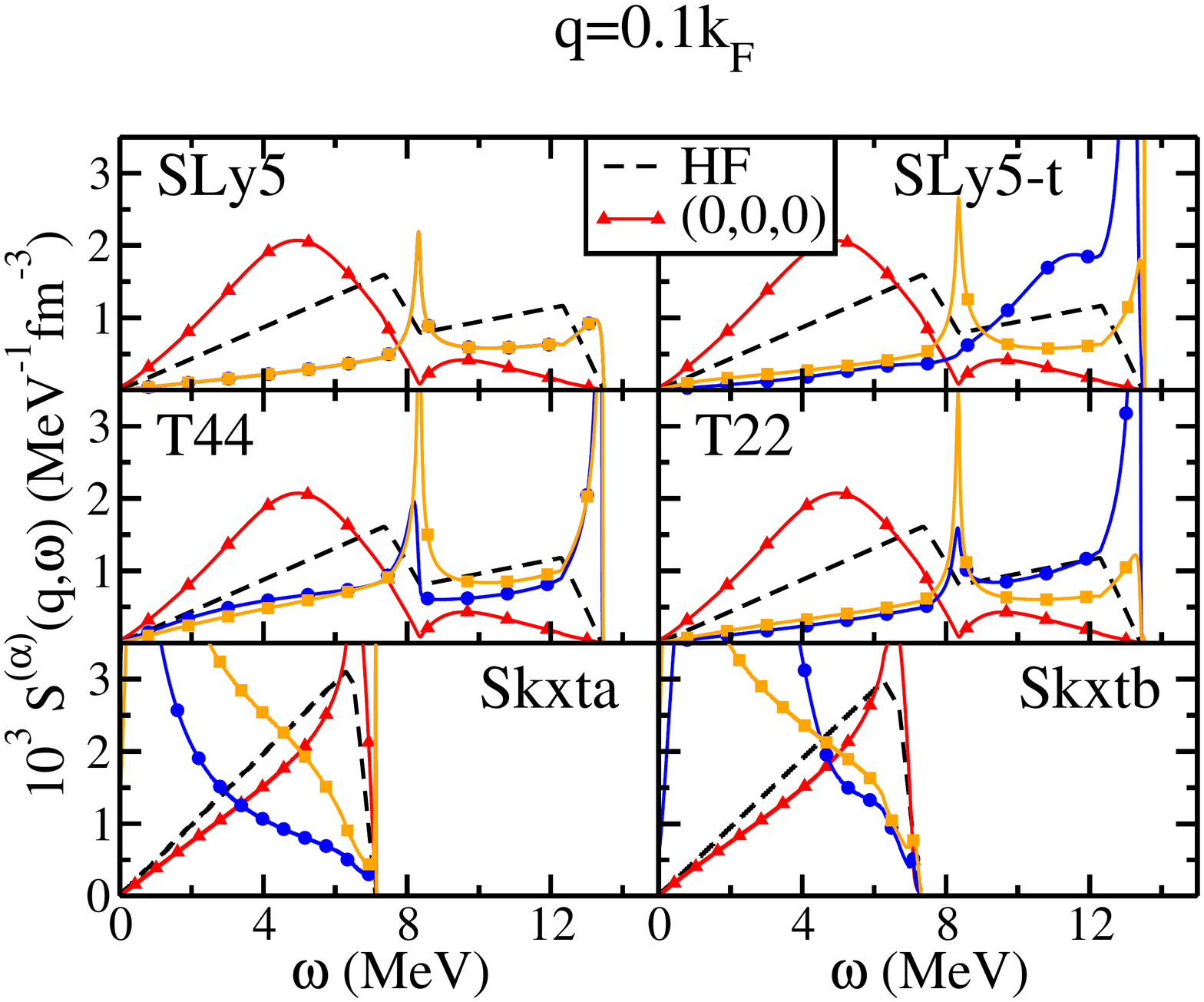} 
      \hspace{-1.9cm}
    \includegraphics[width=0.43\textwidth,angle=-90]{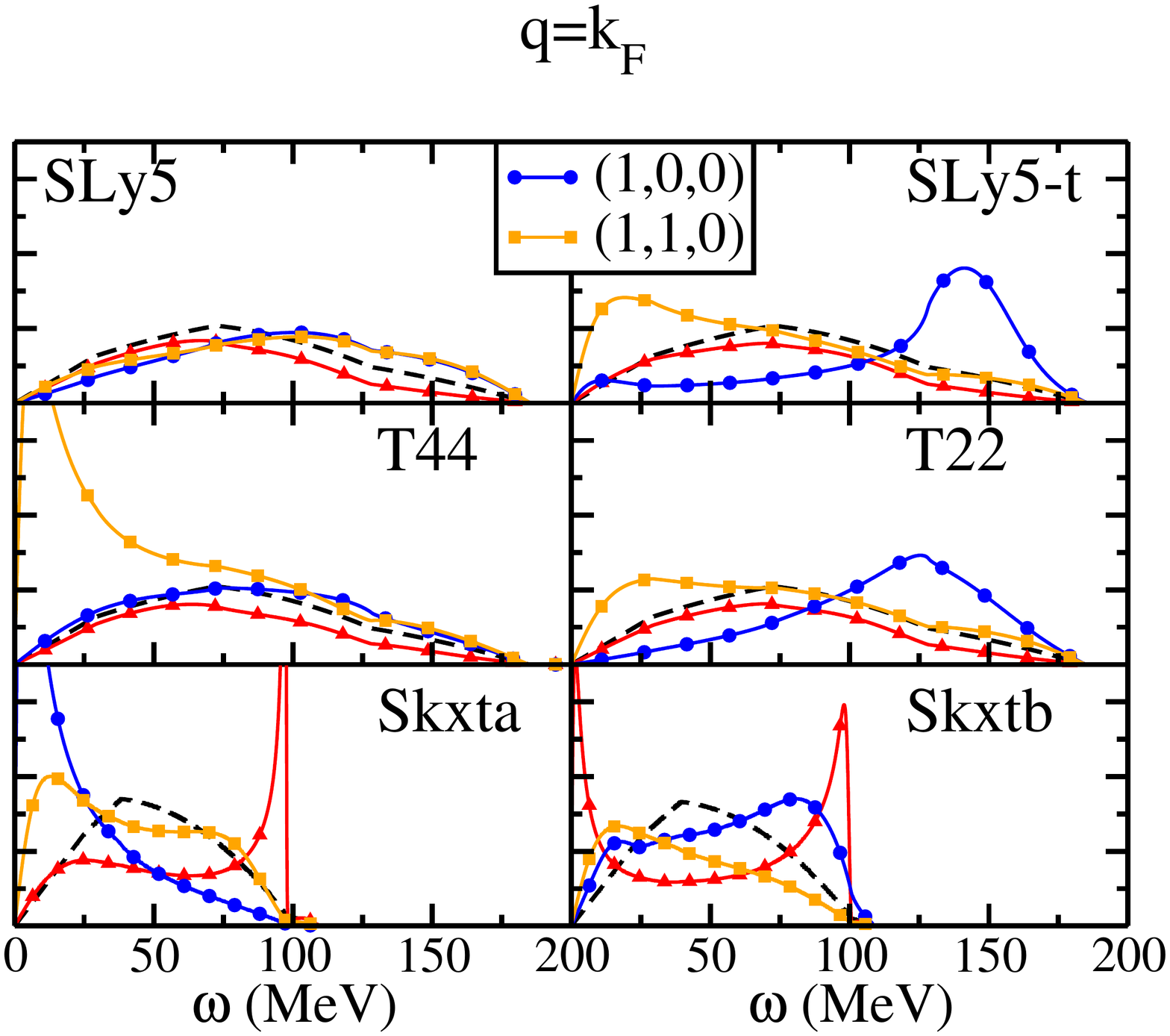} \\
    %
   %
  \includegraphics[width=0.43\textwidth,angle=-90]{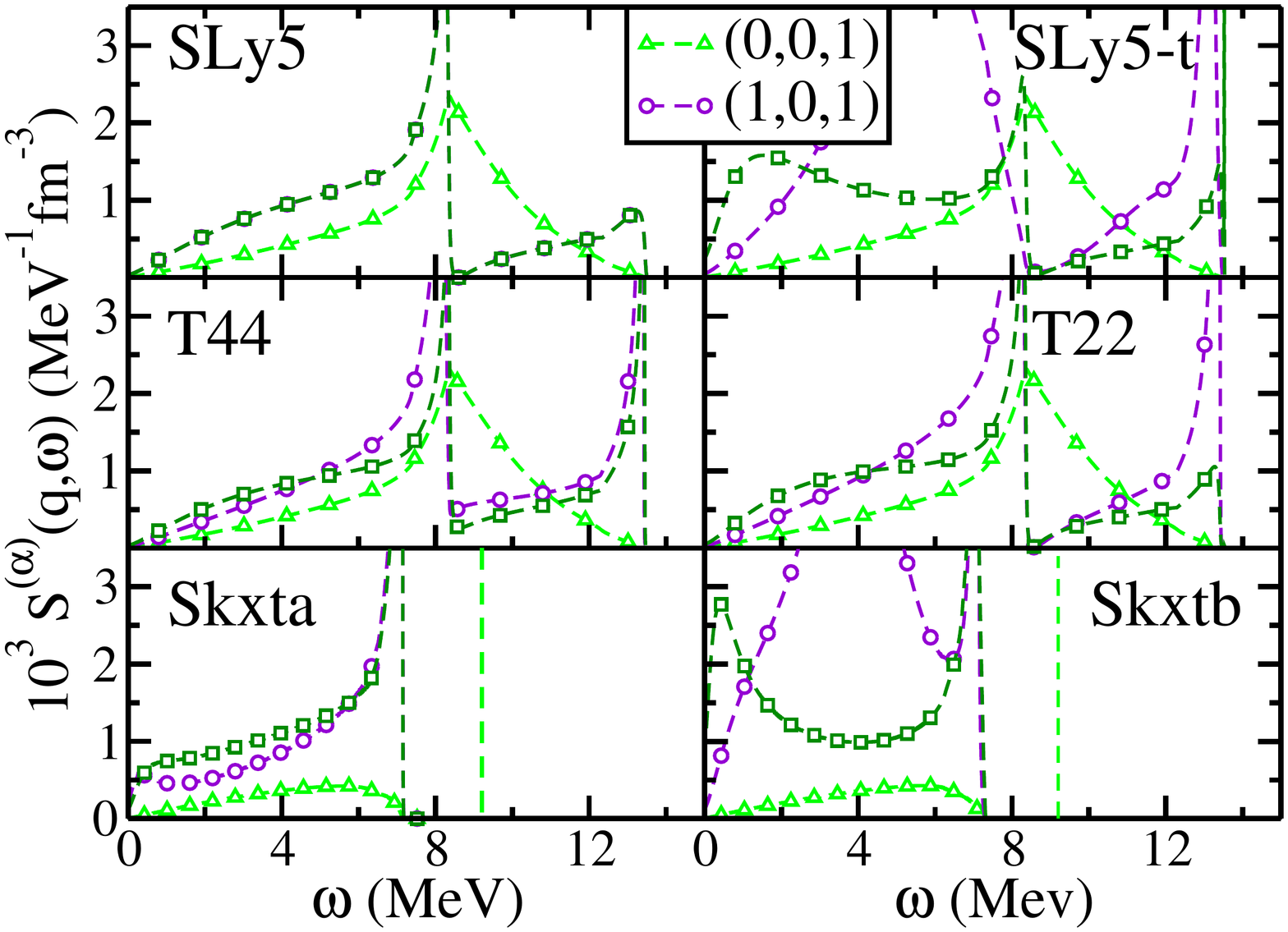} 
      \hspace{-1.9cm}
    \includegraphics[width=0.43\textwidth,angle=-90]{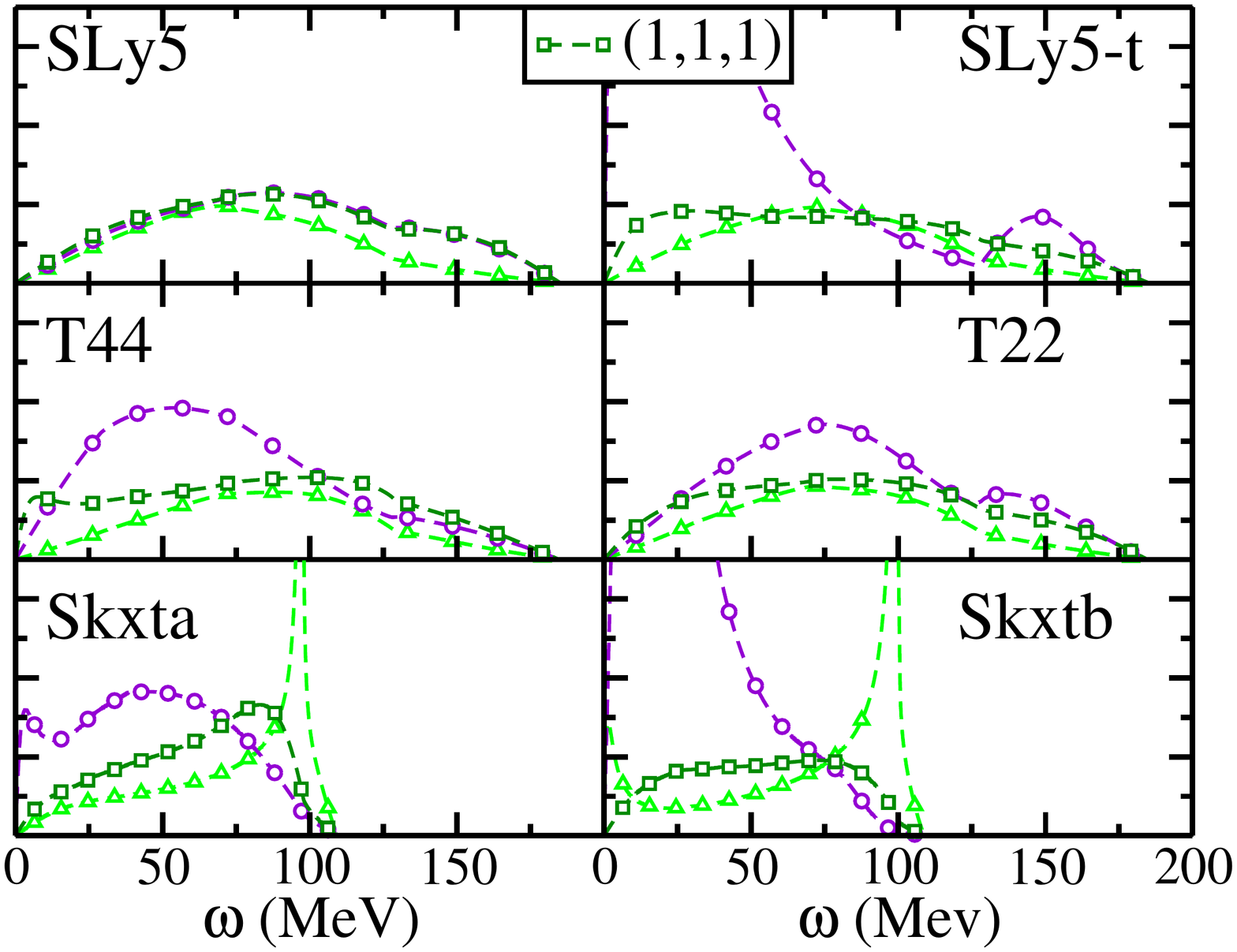} 
\caption{(Color online) Asymmetric RPA strength function, $Y=0.5$, obtained from several interactions at $\rho=\rho_0$, $q=0.1 k_F$ and $q=k_{F}$  and $T=0$. The upper panels represent the channel $I=0$, while in the lower panels we show $I=1$.}
\label{ASYM-ALL:q}
\end{center}
\end{figure}

\begin{figure}[H]
\begin{center}
   \includegraphics[width=0.43\textwidth,angle=-90]{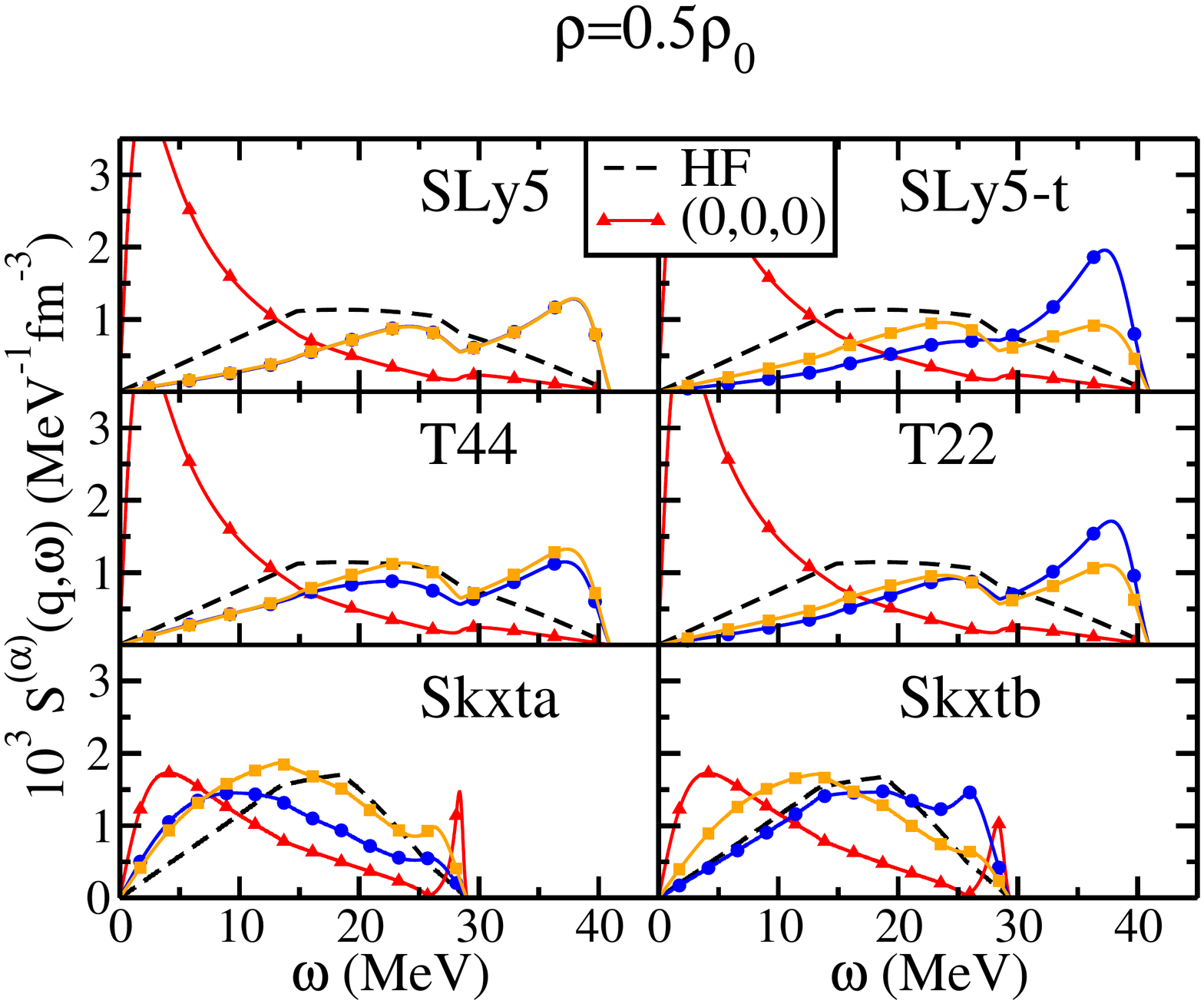} 
      \hspace{-1.9cm}
    \includegraphics[width=0.43\textwidth,angle=-90]{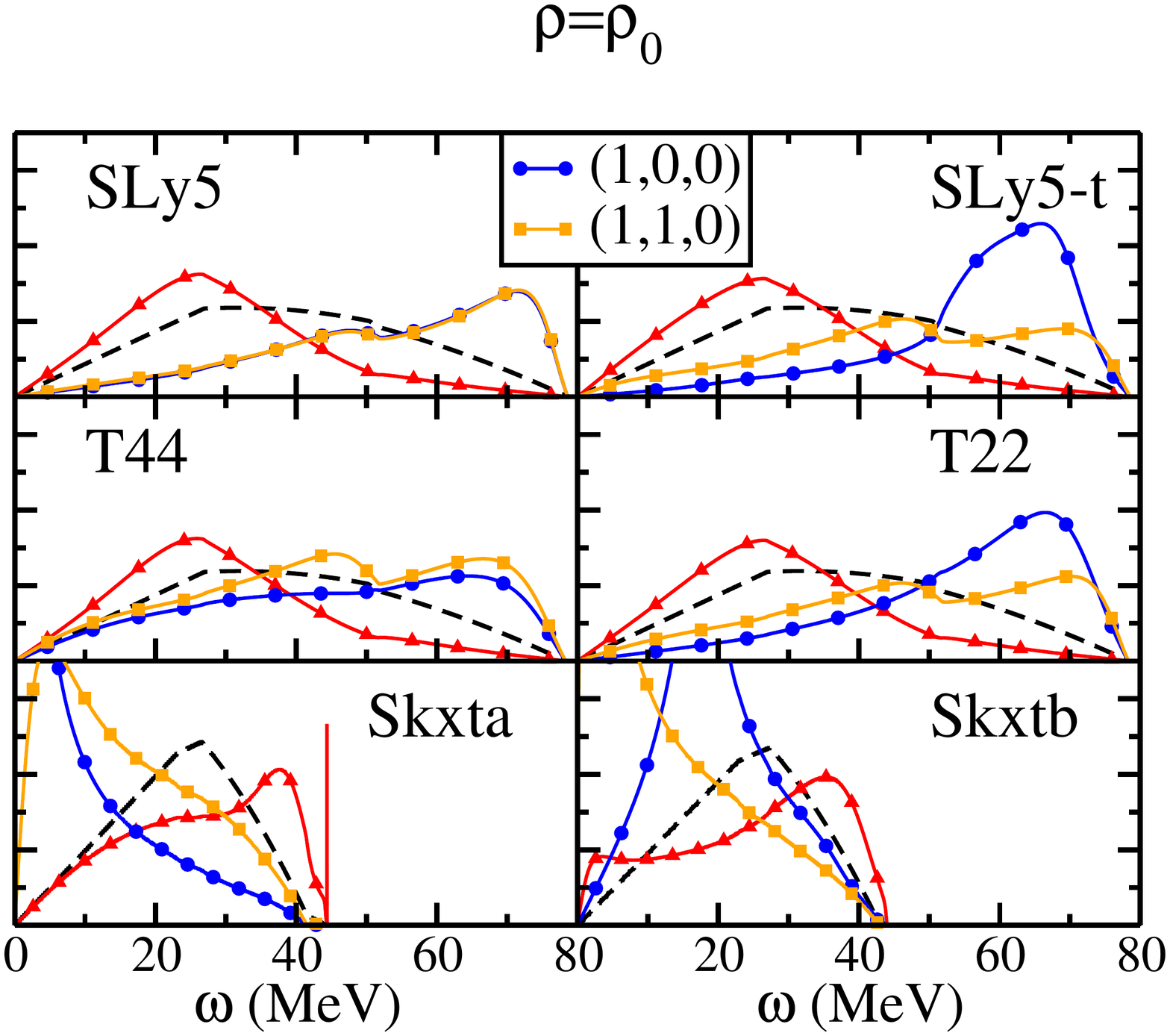} \\
%
%
  \includegraphics[width=0.43\textwidth,angle=-90]{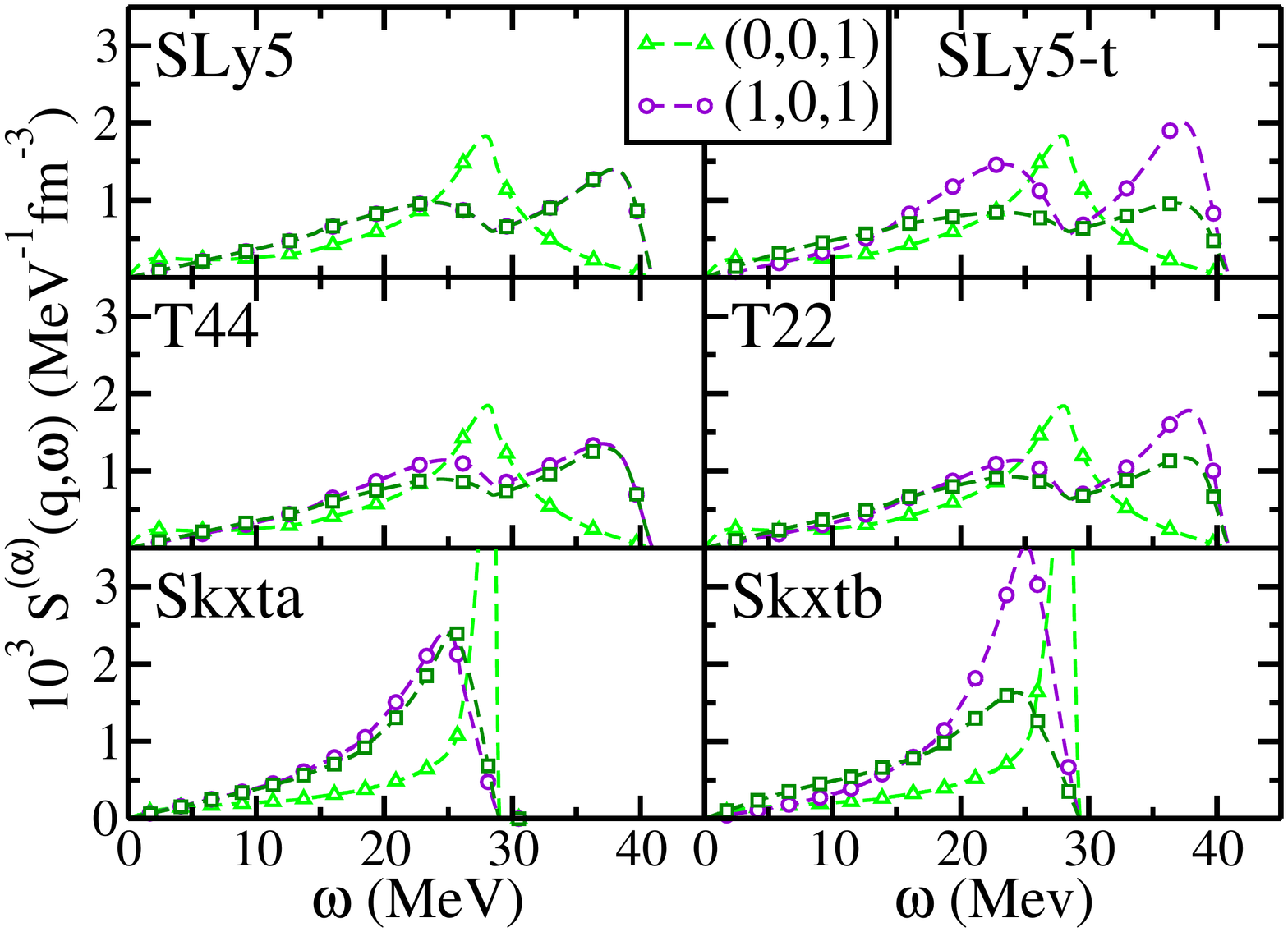} 
      \hspace{-1.9cm}
    \includegraphics[width=0.43\textwidth,angle=-90]{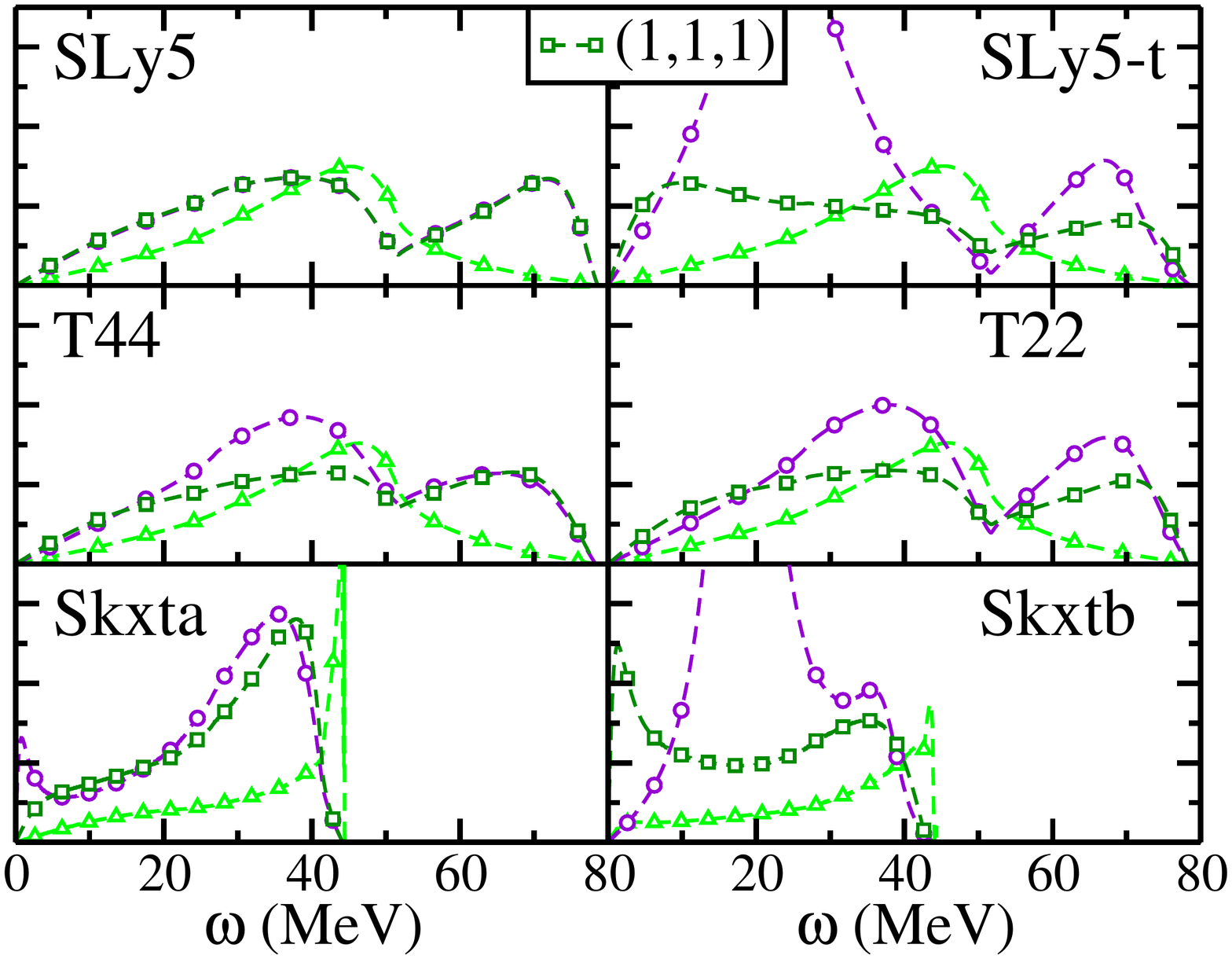} 
 \caption{(Color online) Asymmetric RPA strength function, $Y=0.5$, obtained from several interactions at $\rho=0.5 \rho_0$ and $\rho=\rho_0$, $q=0.5 k_F$ and $T=0$.}
\label{ASYM-ALL:qrho}
\end{center}
\end{figure}

In Fig. \ref{response:q05:T44} and Fig. \ref{response:q05:SKx} are displayed strength functions for T44 and Skxta in the conditions as the previous figures, but for a transferred momentum of $q=0.5$~fm$^{-1}$. A first common feature between these two interactions is that the two-peak structure is wiped out when the momentum transfer is increased, the only reminiscence being in $S=1$ channels for intermediate asymmetries. Moreover, one can observe the tensor effect is intensified for both interactions: the splitting between $M=0$ and $M=1$ is more pronounced. Concerning Skxta specifically, one can observe that for $(S,I)=(0,0)$ no collective state is present for SNM, but develops itself when the asymmetry increases: as we will see in the next section, this phenomenon can be interpreted in terms of physical instabilities. For $(S,I)=(0,1)$, on the contrary, the collective state becomes closer and closer to the continuum as $Y$ increases.
In Fig.~\ref{response:q05:SLy5} are plotted the strength functions calculated with the same conditions as before, but for SLy5. As expected we observe no difference between the different spin projections since there is no tensor. In addition, there is no two-peak structure for $S=0$ but only for $S=1$, the one in the isovector channel being far more pronounced.
Finally, we have plotted strength functions for our set of interactions for an intermediate asymmetry $Y=0.5$, $\rho=\rho_0$ and for $q/k_F=0.1, 0.5$ and $1$ (see respectively Fig.~\ref{ASYM-ALL:q}, six left panels of Fig.~\ref{ASYM-ALL:qrho} and Fig.~\ref{ASYM-ALL:q} again). We see that the general tendency when $q$ increases, is to wash out the characteristic two-peak structure of strength functions. We see by instance that for SLy5 this particular structure is present for $q=0.1 k_F$, attenuated for $q=0.5 k_F$ and absent for $q=k_F$. The 
inclusion of the tensor (SLy5-t) induces an enhancement of all these effects, but also the appearance of a pole at zero energy. SLy5-t, T22 and T44 behave similarly at small and intermediate values of $q$, but one can see an accumulation of strength for T44 in the $(S,M,I)=(1,1,0)$ channel. Obviously, one observes here the sole effect of the tensor. Finally, Skxta and Skxtb have a large splitting between the different spin projections and present some poles at zero energies, but for different channels.
\begin{figure}[H]
\begin{center}
\includegraphics[width=0.43\textwidth,angle=-90]{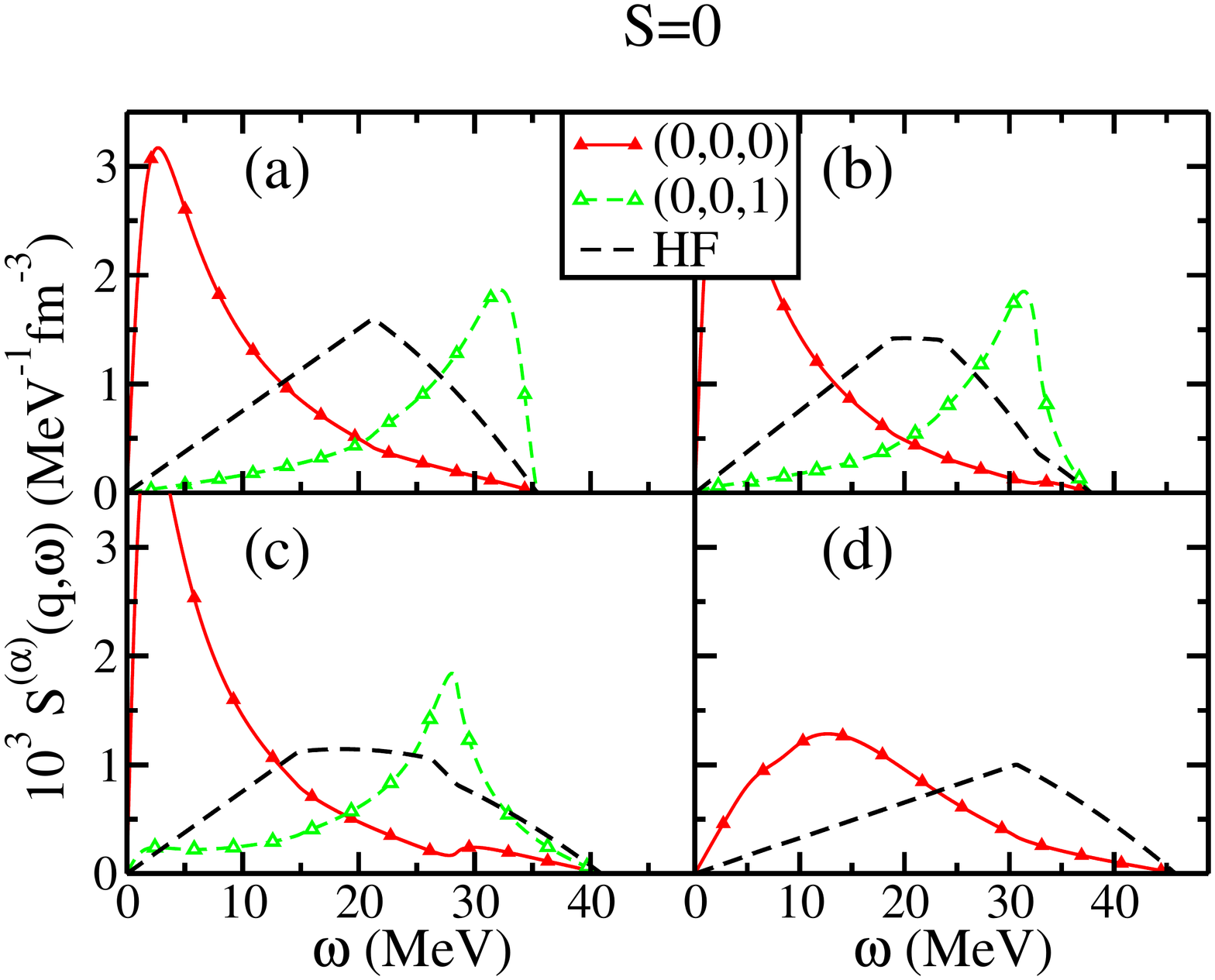}
\hspace{-1.9cm}
\includegraphics[width=0.43\textwidth,angle=-90]{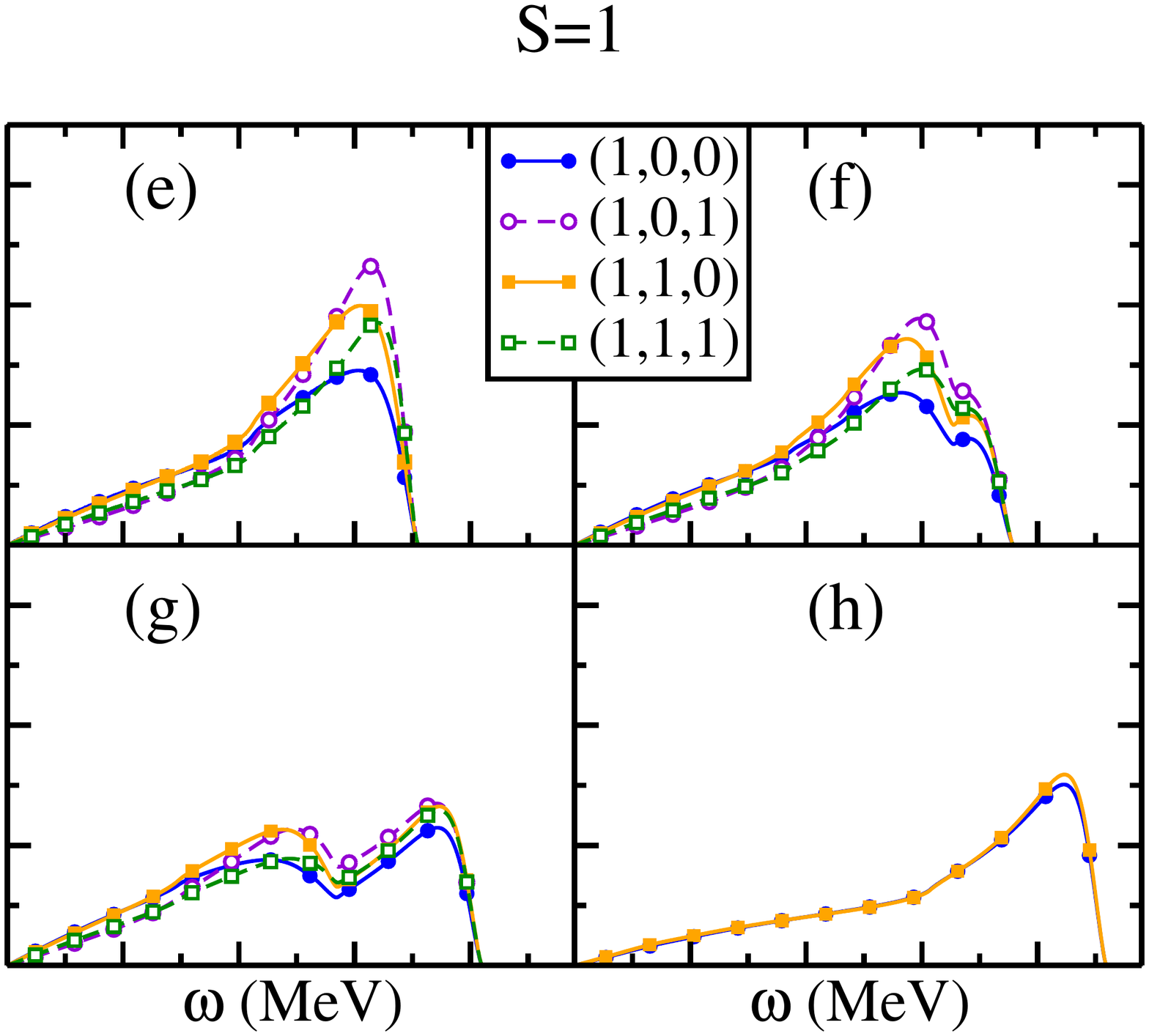}\\
\end{center}
\caption{(Color online) Strength functions $S^{(\alpha)}(q,\omega)$ for the spin-isospin channels $(\alpha)=(S,M,I)$ in asymmetric nuclear matter, calculated with Skyrme interaction T44 at density $\rho=0.08$ fm$^{-3}$ and momentum transfer $q=0.5$~fm$^{-1}$.  Panels (a) and (e) correspond to the asymmetry parameter $Y=0$ (SNM). Panels (b) and (f) to $Y=0.21$. Panels (c) and (g) to $Y=0.5$. Panels (d) and (h) to $Y=1$ (PNM). Only channels $(S,M)$ are relevant in PNM.}
\label{SlySKx:rho}
\end{figure}

\begin{figure}[H]
\begin{center}
\includegraphics[ width=0.43\textwidth,angle=-90]{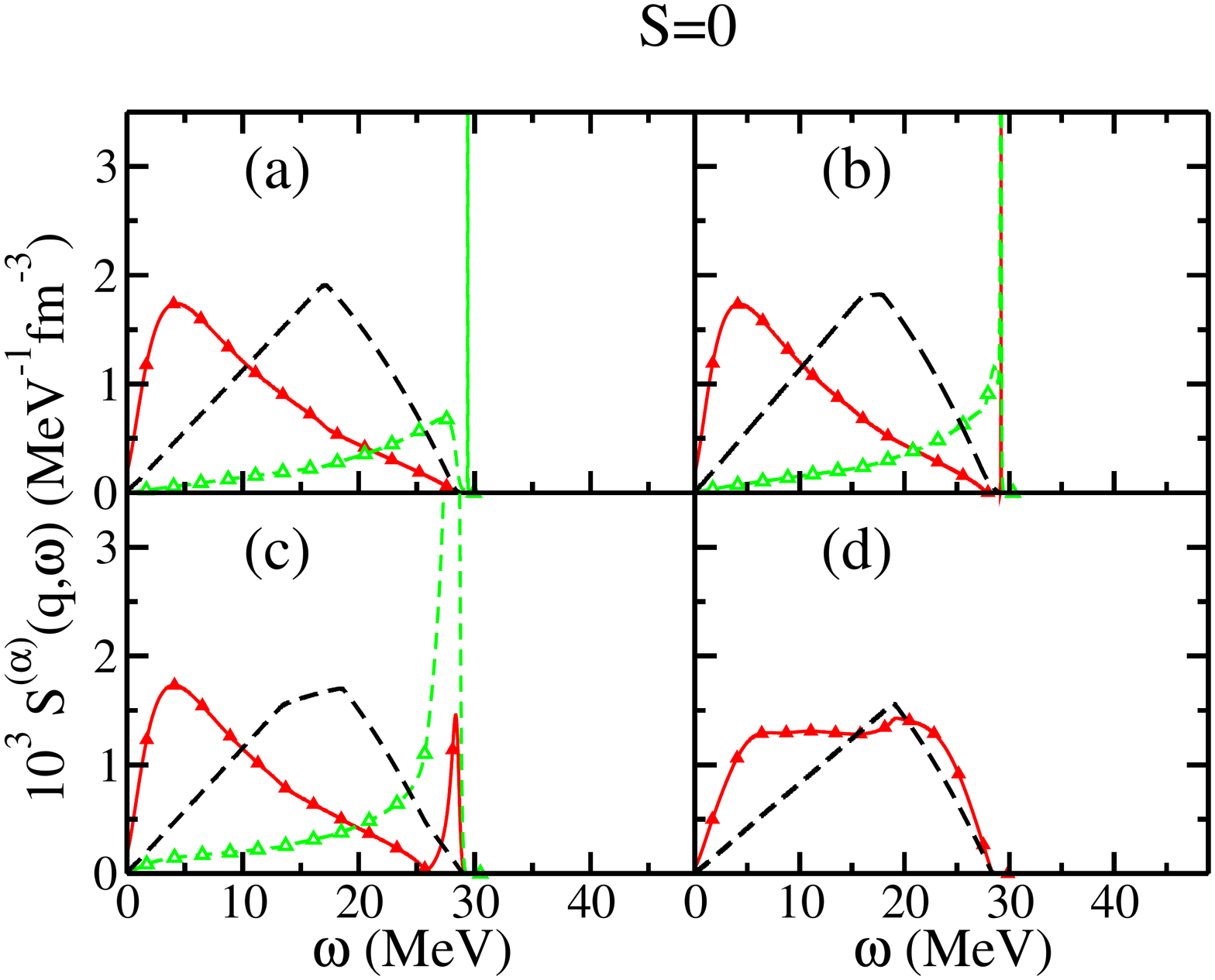}
\hspace{-1.9cm}
\includegraphics[width=0.43\textwidth,angle=-90]{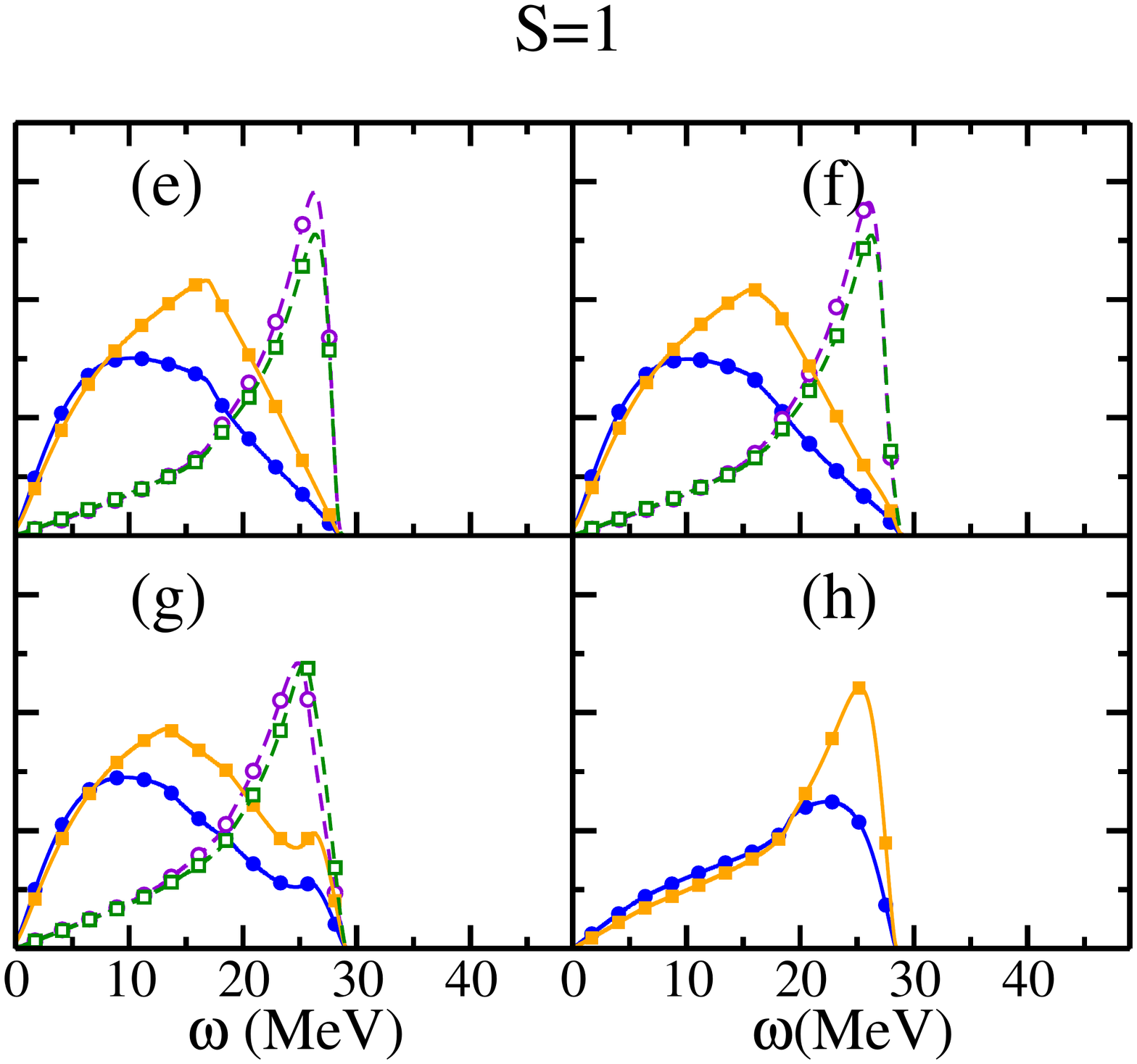}
\end{center}
\caption{(Color online) Same as Fig.\ref{SlySKx:rho}, but for Skxta.}
\label{SlySKx:rho2}
\end{figure}

Another important parameter regarding its influence on strength functions is the density. On Fig.~\ref{SlySKx:rho} we plotted strength functions for T44 calculated for $q=0.5$ fm$^{-1}$, the same different asymmetries considered previously and for $\rho=0.08$ fm$^{-3}$. As mentioned earlier, this density gives some insight of what happens at the surface of a nucleus. A first common feature is that the lowering of the density induces a global shift of the strength towards the low-energy region. Second, we can notice that there are precursor signs of a pole at zero energy in the scalar-isoscalar channel. This will be interpreted in the next part as the presence of the so-called spinodal instability. We can furthermore notice that this instability seems to disappear for PNM : we will see that the dependence of the spinodal with asymmetry is coherent with this result. We can also remark in Fig.\ref{SlySKx:rho2}, that a collective state appears in the scalar-isovector channel for Skxta functional, but this state disappears when $Y$ increases : asymmetry has an influence on the presence of a zero sound. Neutrino mean-free path, by instance, can thus be altered, and actually be very different, in pure neutron matter or in asymmetric matter.

Let us briefly comment on thermal effects. In practice one has to modify the expressions of auxiliary functions $\beta^{(\tau,\tau')}_{i=0,8}(q,\omega,T)$ (see \ref{app:beta}), so that the generalization to finite temperature of the Bethe-Salpeter equation (\ref{BSeq}) and the related algebraic systems of equations is straightforward \cite{dav14b}. 
The effects of temperature on the strength functions are similar to the SNM and PNM cases, discussed in Sec. \ref{sec:saturated}. The general trend is to spread strength functions and wipe out their structure. A striking feature is that negative energies now receive an important part of the strength. Moreover, the general structure reflects the new allowed physical processes: the excitation of the system (\emph{i.e.} a peak at positive energies) has a corresponding image in negative energies (\emph{i.e.} the desexcitation of the heated system). An important accumulation of strength of zero energy is observed in some channels. This is related to the presence of instabilities, as we shall immediately discuss.

\subsection{Instabilities}

\begin{figure}[H]
\begin{center}
\includegraphics[width=0.41\textwidth,angle=-90]{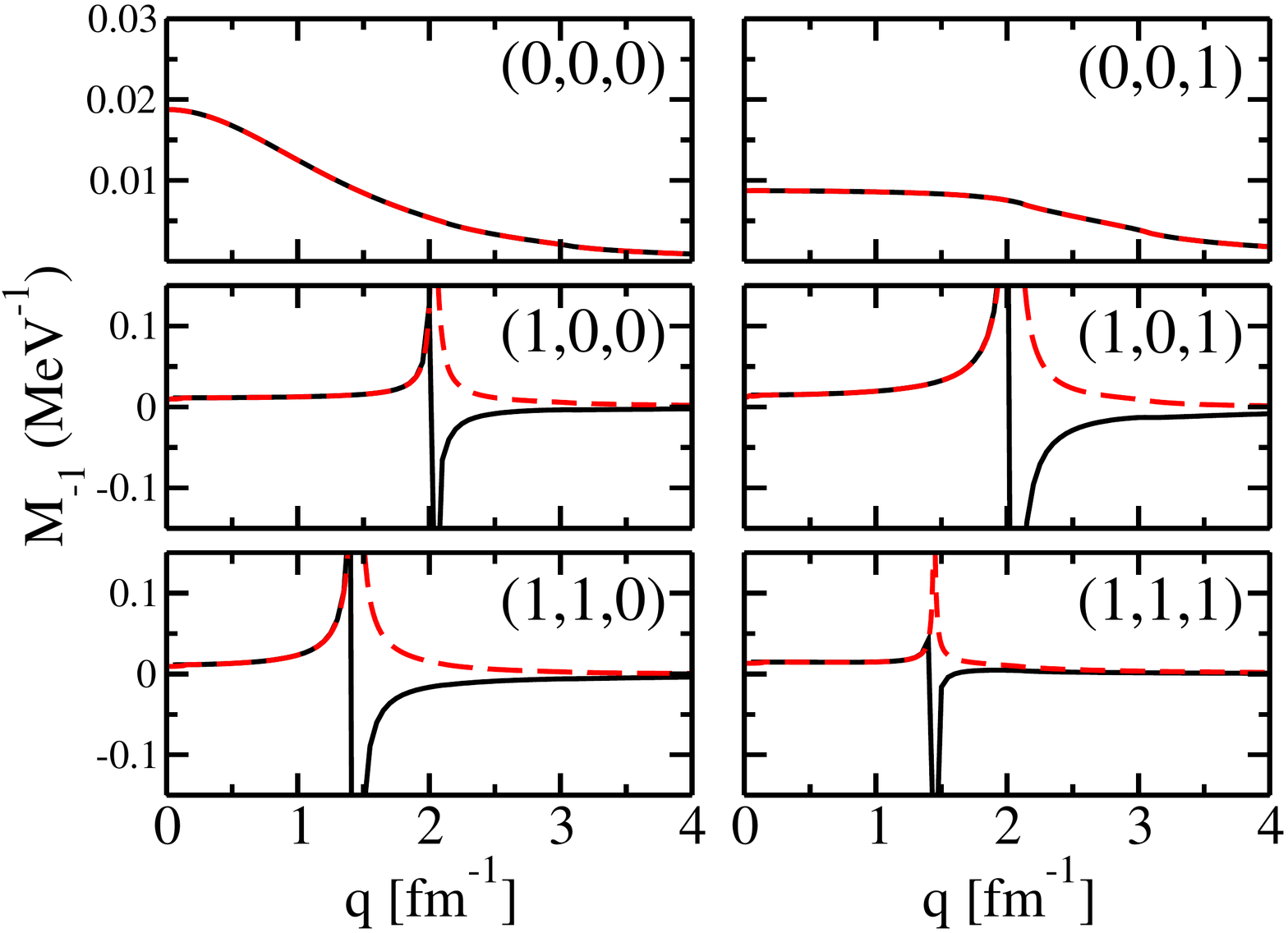}
\hspace{-1.6cm}
\includegraphics[width=0.41\textwidth,angle=-90]{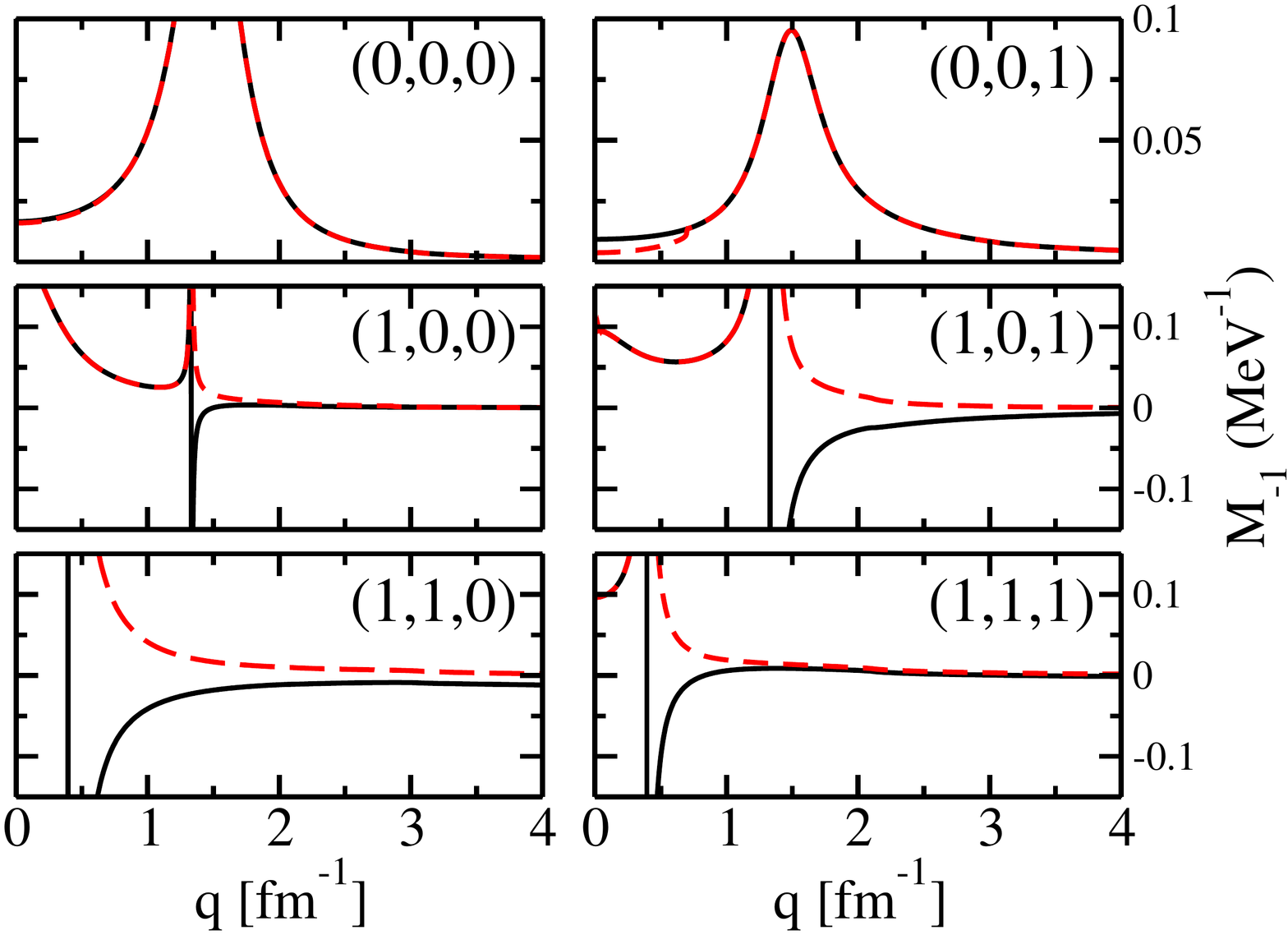}\\
\end{center}
\caption{(Color online) IEWSR $M_{-1}$ for spin-isospin channels $(S, M, I)$ at asymmetry $Y = 0.5$ and $\rho=0.16$~fm$^{-3}$ as calculated analytically (solid line) and numerically (dashed line). The results for T44 and Skxtb are displayed on the left and right panels, respectively.}
\label{asym:mm1}
\end{figure}

As discussed in Sec. \ref{sec:saturated}, the inverse energy-weighted sum rule $M_{-1}$ is the tool of choice for detecting finite size instabilities. As compared to SNM or PNM, the analytical expression is now much more cumbersome, and we do not give it explicitly. 
The two ways of calculating $M_{-1}$ are plotted in Fig.~\ref{asym:mm1} 
for $Y=0.5$ and $\rho=0.16$~fm$^{-3}$ and interactions T44 and Skxtb.
Solid and dashed curves correspond to analytical and numerical calculations, respectively.
One can see that both curves are on top of each other for $S=0$, except in channel $(0,0,1)$ and interaction Skxtb.
For the other case, no instabilities are present at this density and for this asymmetry. 
As already discussed in \cite{dav14b} for the particular interaction T44, 
asymmetry has a sizable effect on the location of instabilities. 
However, it turns out that in the $S=0$ channel the density is the most important parameter to this respect.
Indeed, at low density values one reaches the spinodal instability, which will 
be specifically discussed in the next subsection. The small discrepancy in channel $(0,0,1)$ at small values of $q$ appearing for Skxtb indicates a deficit of strength in the scalar-isovector channel. This is due to the presence of a collective state (zero sound) as we can see in Fig.~\ref{ASYM-ALL:qrho} which survives up to $q\simeq 0.8$ fm$^{-1}$ before being absorbed in the continuum.

For $S=1$, we can observe the presence of a finite-size instability in all 
channels.
Indeed, both analytical and numerical curves coincide for values of $q$ smaller than some specific value which depends on the interaction. The sum rule $M_{-1}$ changes of sign at this specific value of $q$, indicating the presence of the instability. Strictly, the plotted results are meaningless beyond such a value of $q$. The curves are nevertheless plotted in that region to visually show the instability.

\begin{figure}[H]
\begin{center}
\includegraphics[width=0.43\textwidth,angle=-90]{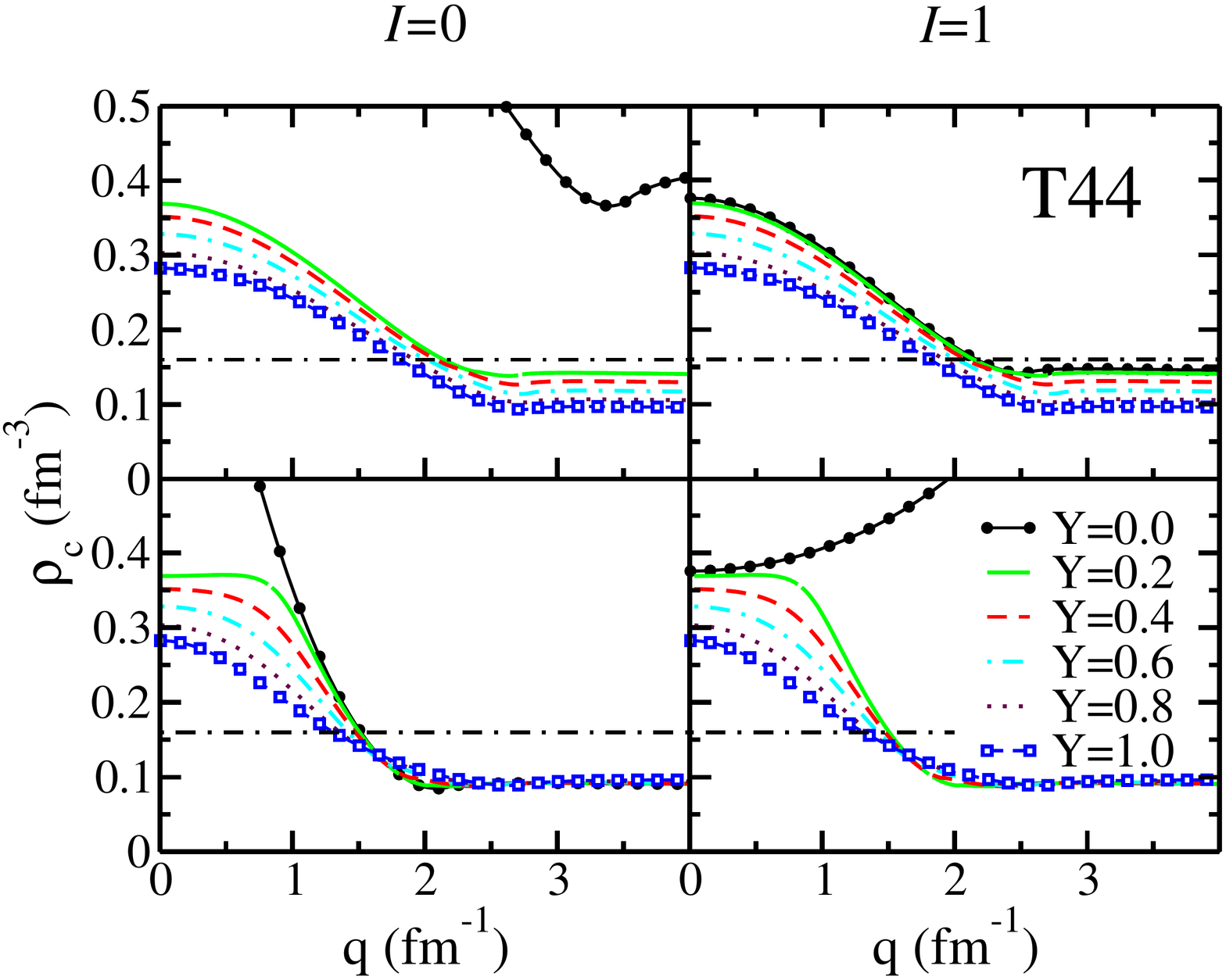}
\hspace{-1.92cm}
\includegraphics[width=0.43\textwidth,angle=-90]{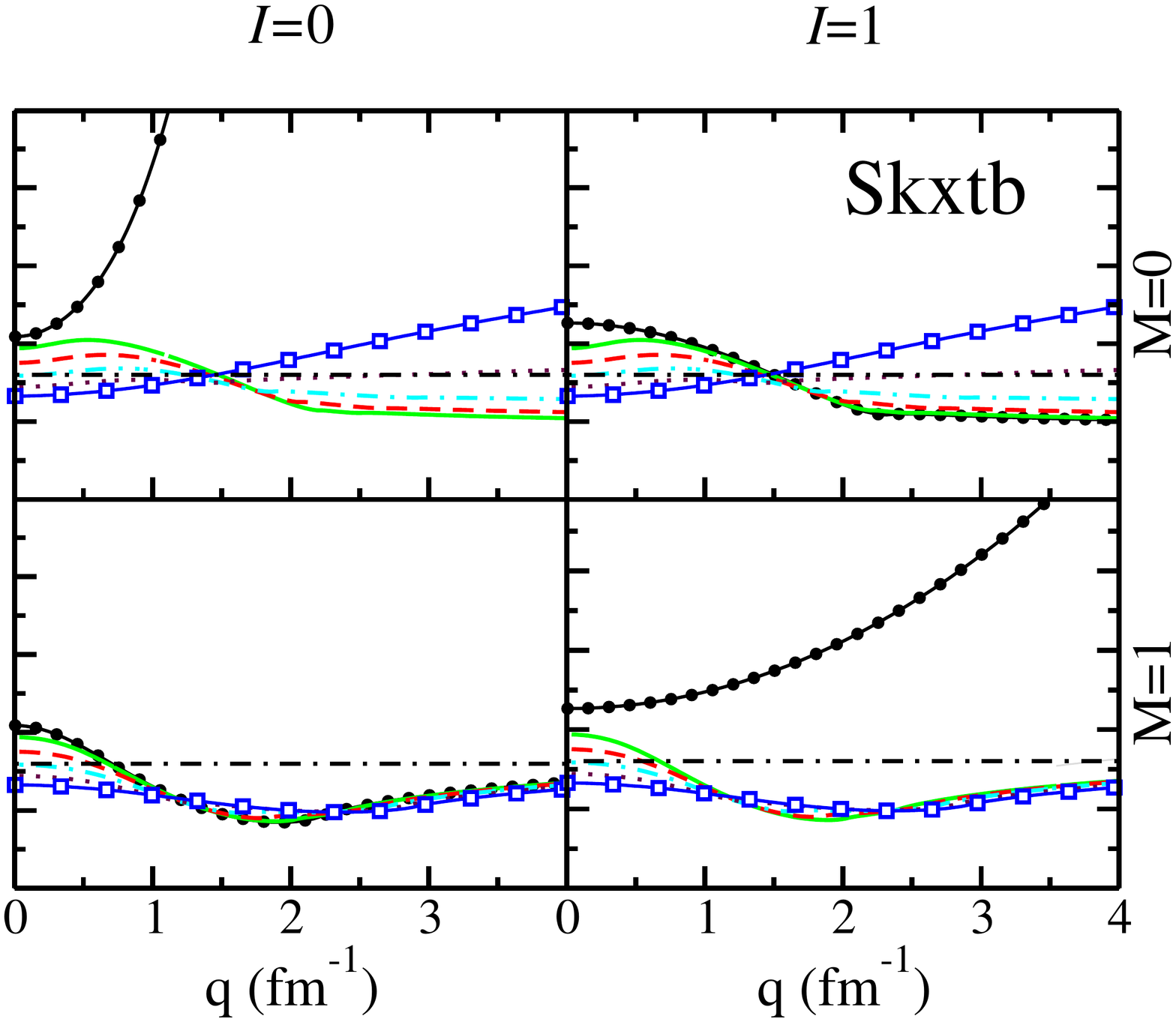}
\end{center}
\caption{(Color online)  Critical density $\rho_{c}$ for the T44 and Skxtb. In the different panels we represent the different channels $S=1,M=0,1,I=0,1$ for different values of the asymmetry parameter Y and at zero temperature. The horizontal dashed line corresponds to $\rho=0.16$~fm$^{-3}$.}
\label{asym:rhocrit}
\end{figure}

The previous results have been obtained for a single couple of density and asymmetry values. A thorough analysis requires an extended exploration. To this end, we first define the critical density $\rho_c$ as the value of $\rho$ at which an instability appears for given $q$ and $Y$. For density values higher than $\rho_c$ the system is unstable. The detection of instabilities can be done by calculating the critical density varying the transferred momentum and the asymmetry parameter. We shall concentrate now on the $S=1$ channels, leaving the $S=0$ for the next subsection. In Fig.~\ref{asym:rhocrit} is plotted $\rho_c$ as a function of $q$ for the interactions T44 and Skxtb. The different curves correspond to values of the asymmetry parameter $Y$ from 0 to 1 in steps of 0.2. The variation of $\rho_c$ with $Y$ sensibly depends on the employed interaction, as it could be expected from the discussion in the previous paragraph. Consider for instance the region of small $q$ values. With the interaction T44 the critical density in the isoscalar channels strongly varies with $Y$. In particular, there is an important gap when $Y=0$ is slightly modified from the SNM case. In contrast, a smooth variation is found in the isovector channels, as well as with the interaction Skxtb in all channels. To understand these discontinuities, also observed in a different combination for other values of $q$, one should keep in mind that the response in the channel $(S,M,I)$ is actually a linear combination of the $(nn)$, $(np)$, $(pn)$, and $(pp)$ responses. 
For small values of the asymmetry, some of the latter response functions may 
suddenly acquire a non-zero value which can strongly influence the whole combination. 

The knowledge of these critical regions could in principle be of some help to establish physical bounds to the interaction parameters. To this respect, the previous figures show that SNM and PNM critical densities provide reasonable accurate limits for any asymmetry value only in the case of interaction T44. This is not true for Skxtb, where we can observe that the critical densities reach their lowest values for a small asymmetry (see isovector channel with $M=0$). Therefore, it seems that no general conclusions can be drawn for the finite-size instabilities, neither for their location nor their evolution with asymmetry. When included in a complete fitting procedure, it is thus not sufficient to consider SNM and PNM to avoid instabilities at small densities (empirically $\rho_c \simeq 1.2 \rho_0$ \cite{hel13}).

\begin{figure}[H]
\begin{center}
\includegraphics[width=0.43\textwidth,angle=-90]{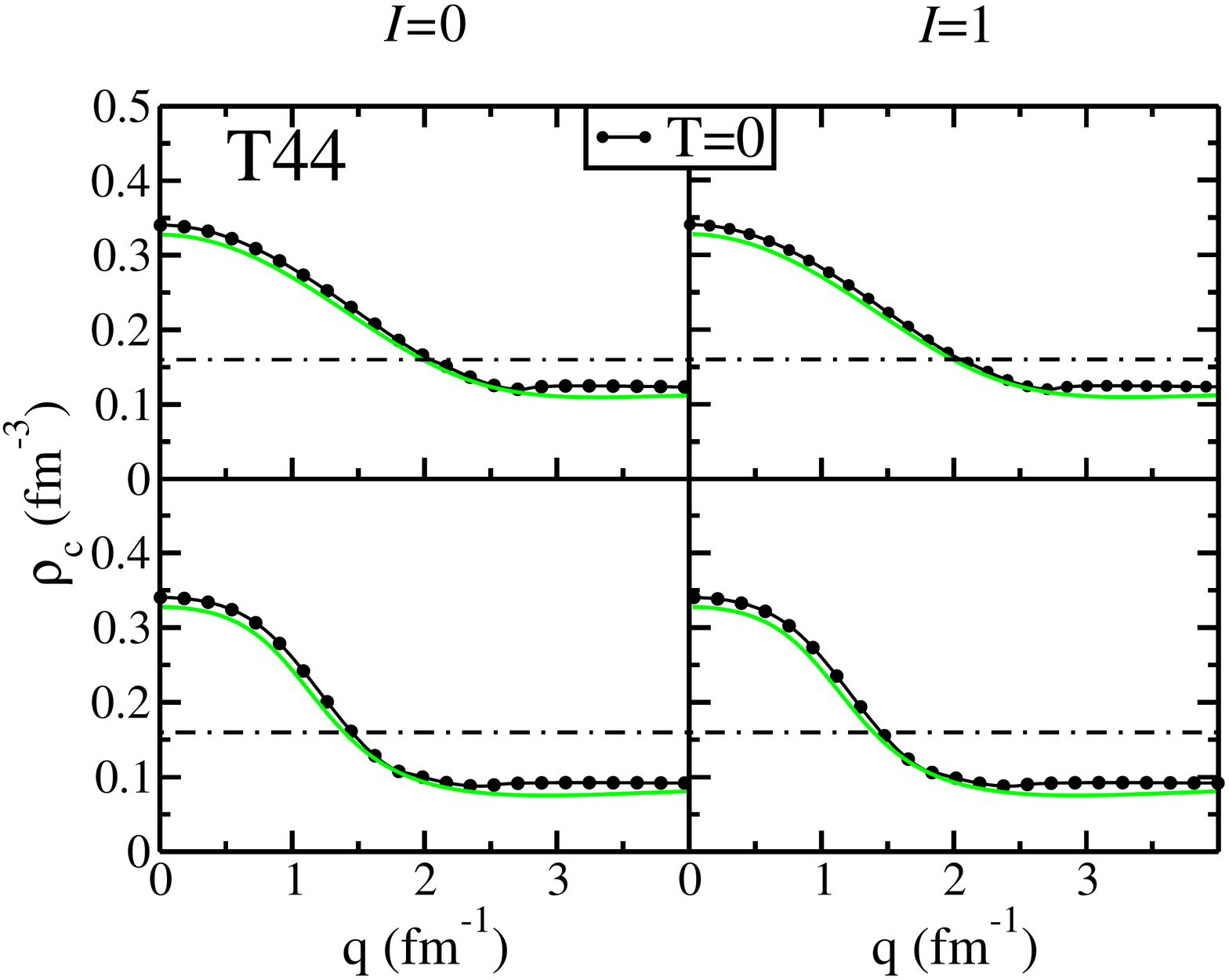}
\hspace{-1.92cm}
\includegraphics[width=0.43\textwidth,angle=-90]{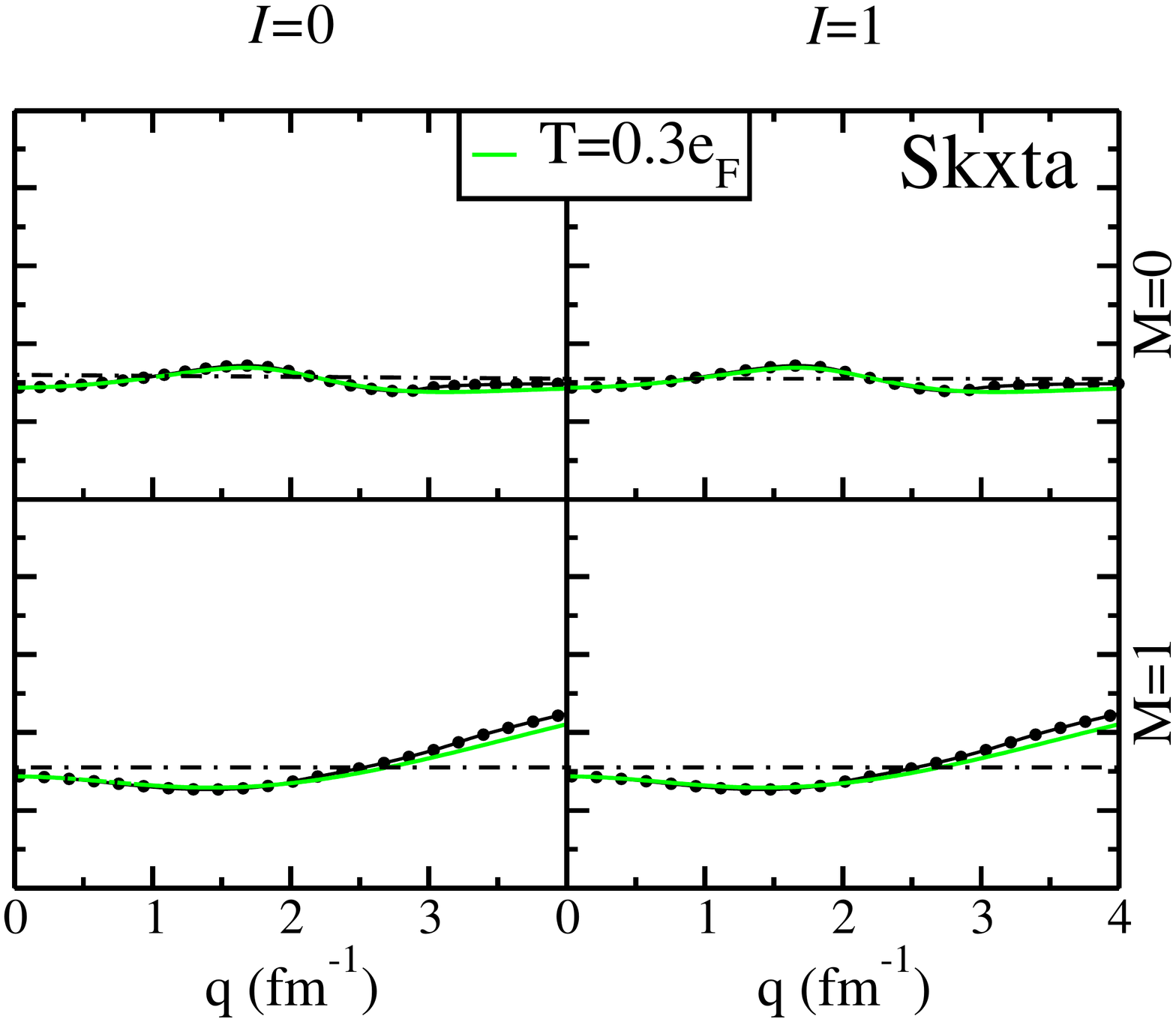}
\end{center}
\caption{(Color online) Critical densities of T44 (left) and Skxta (right) functional for Y=0.5 and S=1 for different values of the temperature.}
\label{asym:rhocritT}
\end{figure}

In principle, the instabilities and the critical densities are influenced by the temperature. However, as we can see on 
Fig.~\ref{asym:rhocritT}, even a high temperature does not modify significantly the critical densities compared to the $T=0$ case.

\subsection{The spinodal}
\label{spinod:asym}
Let us now consider the instabilities in the $S=0, I=0$ channel at low densities. These instabilities are physical, contrarily to the previously discussed for $S=1$. They are related to the thermodynamic spinodal transition of homogeneous matter, where density fluctuations induce a decrease of the total free energy which is amplified until a separation in two distinct stable phases, liquid and gas, is reached. As discussed in \cite{duc08}, the curvature of the free energy which governs the stability of the whole system contains two antagonistic terms. One is originated by the density gradient terms of the functional and is proportional to $q^2$. The other one is related to the Coulomb interaction, and is proportional to $1/q^2$. Therefore, for small values of $q$ the energy cost due to Coulomb interaction dominates, which implies a sensitive reduction of the spinodal region for proton-rich systems. In the case of a neutron star where a background of electrons is present, the net effect of the Coulomb interaction in the residual interaction is also to reduce the region of the spinodal \cite{duc08,bal09}. 
\begin{figure}[H]
\begin{center}
\includegraphics[width=0.4\textwidth,angle=-90]{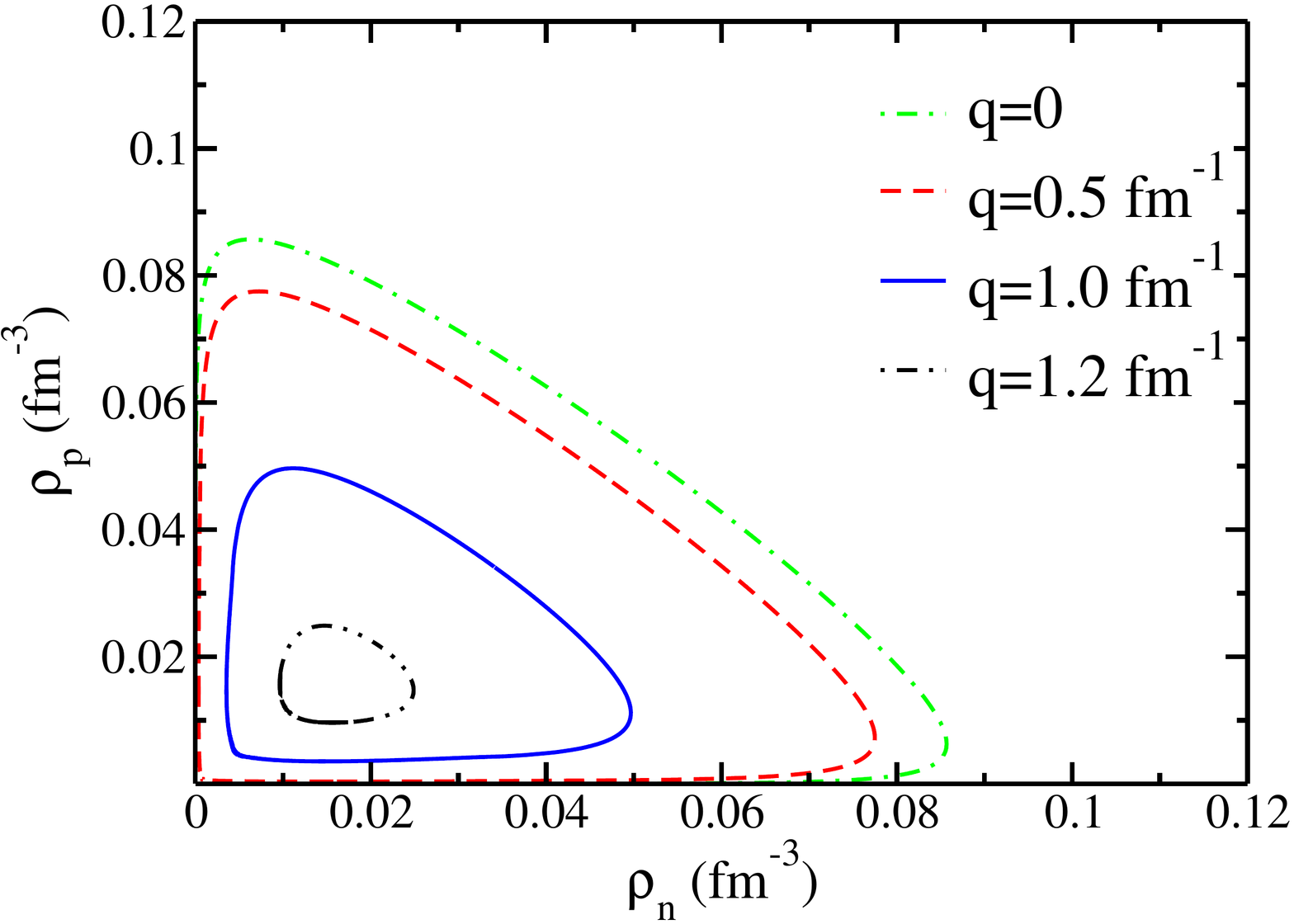}
\hspace{-0.8cm}
\includegraphics[width=0.4\textwidth,angle=-90]{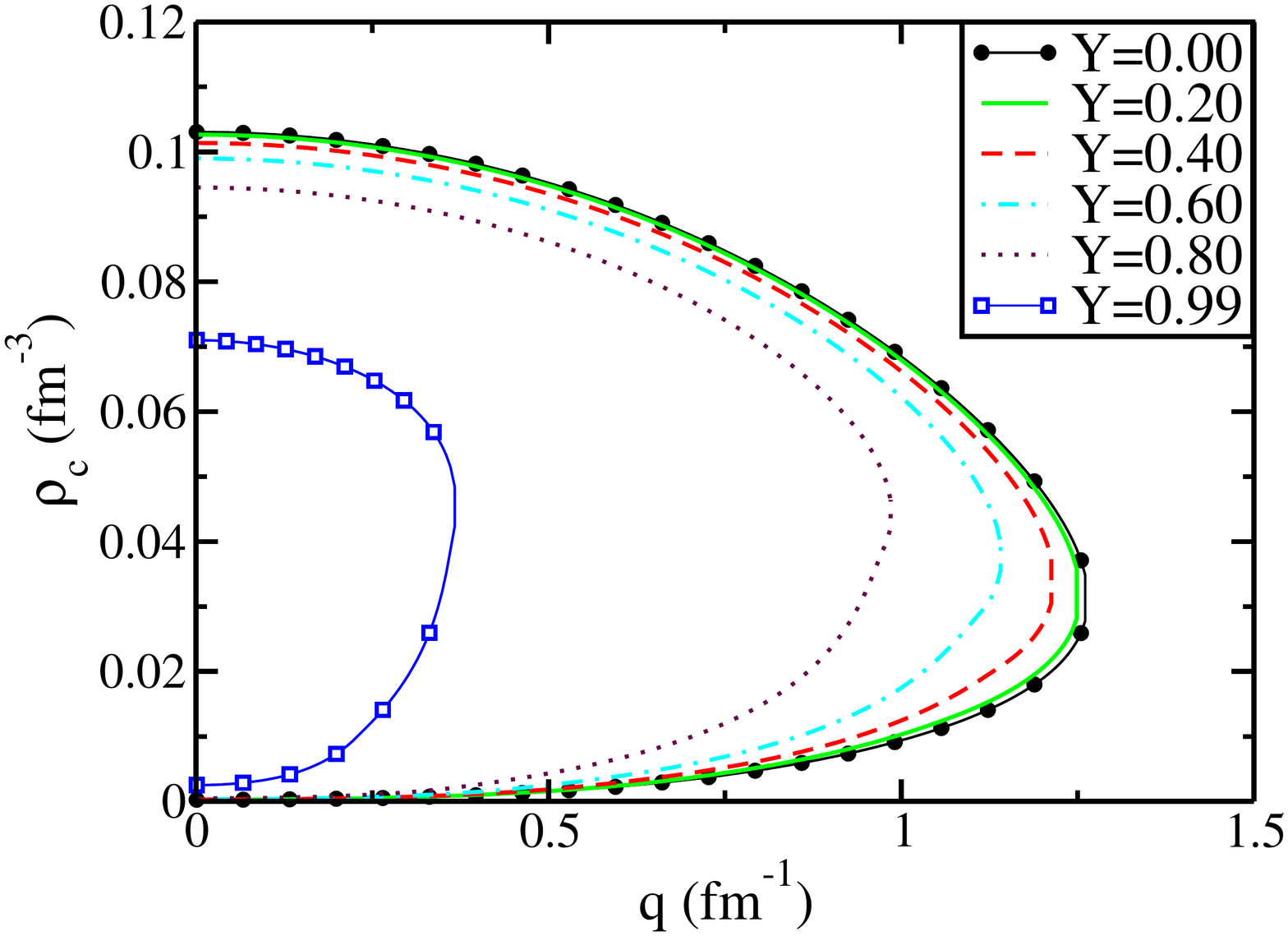}
\end{center}
\caption{(Color online) We show the spinodal instability (S=0,M=0,I=0) for different values of the neutron $\rho_{n}$ and proton $\rho_{p}$ density. On the left panel the T44 Skyrme functional for several values of $q$. On the right panel several interactions for $q=0$.}
\label{spinodal:t44}
\end{figure}
As discussed in \cite{dav14b} and illustrated on the left panel of Fig.~\ref{spinodal:t44}, the spinodal contours decrease when the transferred momentum is increased, the largest area being obtained for $q=0$.  Moreover this diminution is not linear with respect to $q$ and since Coulomb interaction is not taken into account, the spinodal is symmetric in the $(\rho_p,\rho_n)$  proton/neutron density plane. With the help of this figure we can now understand the shape of the spinodal critical lines displayed on the right panel of Fig.~\ref{spinodal:t44}. Let us start with SNM. On Fig.~\ref{spinodal:t44}, left panel, this corresponds to a straight line $\rho_n=\rho_p$. In that case we can follow the evolution of the two intersection points between this line and the spinodal contour, recovering the shape of Fig.~\ref{Critical:SNM:pole}. Following the same approach we can have an insight of the evolution of the spinodal region with asymmetry: by instance, when the slope becomes less and less important (neutron rich systems), the intersection points are only for small values of $q$, leading to a smaller (and even non existing for PNM) spinodal region. Finally, it is worth recalling \cite{duc08} that the spinodal is quite "universal" in the sense that its shape is globally the same, independently of the interaction. This can be understood because a reasonable interaction must reproduce the equation of state ($E/A$): the special combination of Skyrme parameters entering in $E/A$ must be approximately the same. On the contrary one may wonder how the tensor, which is very different for each interaction induces some differences between two different parametrizations. Actually, the tensor acts in the scalar-isoscalar channel only via the spin-orbit which mixes $S=0,1$ channels. However, similarly to the SNM case, this mixing can be absorbed into effective coefficients which contain a term proportional to $q^4$ (see Eq. \ref{tildeW} and Ref. \cite{dav14b}). Since the spinodal implies relatively small values of the transferred momentum, the tensor interaction marginally affects the spinodal and can be safely neglected in a first approximation. 

\begin{figure}[H]
\begin{center}
\includegraphics[width=0.6\textwidth,angle=-90]{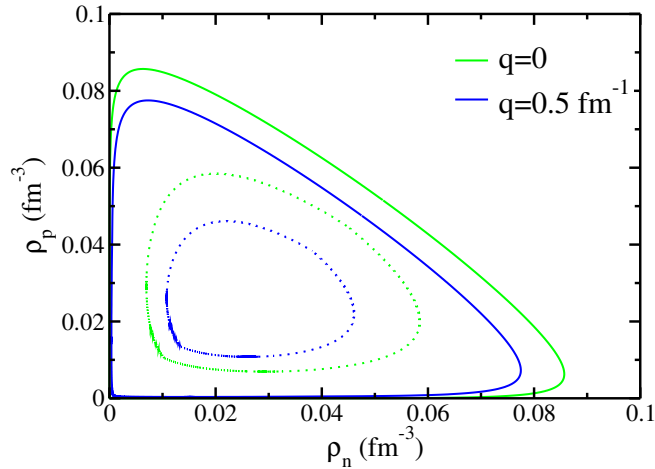}
\end{center}
\caption{(Color online) We show the spinodal instability (S=0,M=0,I=0) in the $(\rho_{n}, \rho_{p})$ plane for the T44 Skyrme functional at $T=0$ (solid line) and $T=10$~MeV (dots).}
\label{spinodal_T:t44}
\end{figure}

On Fig.~\ref{spinodal_T:t44} one can observe the evolution of the spinodal contours for two different values of $q$ when the temperature is not equal to zero. The net effect of the temperature is to smooth the fluctuations. For that reason, one expects a sizable reduction of the spinodal region, which is actually observed as compared to the corresponding curves at $T=0$ displayed in Fig.~\ref{spinodal:t44}. The temperature is rather high ($T=10$ MeV) for illustrative purposes, but the effect is quite important even for small temperatures. From $T=0$ to $T=10$ MeV, the typical range observed with T44 for the spinodal instability goes from $0-0.08$ fm$^{-3}$ to $0.01-0.06$ fm$^{-3}$


\section{Extra terms in Skyrme functionals}
\label{sec:extra}
All the results described in previous sections have been obtained with standard Skyrme functionals.
However, as recently discussed in Ref.~\cite{kor13}, the standard form of these functionals appears to be not flexible enough to satisfy all the constraints required by high precision calculations.
Two major avenues are thus currently explored. The first one aims at including correlations into the density functional; the second one is much in the spirit of the self-consistent mean field theory, where the building block is an effective pseudo-potential and the correlations are added afterwards~\cite{ben03}.
Several groups have investigated possible extensions of Skyrme functionals either increasing the number of terms or adding \emph{ad hoc} density dependent terms. In the following we will briefly illustrate, for the case of symmetric nuclear matter only, how our formalism can be extended and modified to properly include these extra terms.

\subsection{Brussels-Montreal Skyrme potentials}
\label{BMSpot}

Among the family of Skyrme functionals, the ones proposed by the Brussels-Montreal collaboration and named BSk~\cite{gor09a,gor13b} have the interesting characteristics of being fitted over the 2200 known experimental masses with an accuracy of $\simeq 500$ KeV in addition to the usual infinite matter results.
These functionals have not been designed for finite nuclei only, but also to determine some important properties of neutron stars~\cite{pea12,cha11c,pea11}. However, in a recent series of articles~\cite{cha10c,gor10}, the authors detected the presence of ferromagnetic instabilities in some of their functionals. In order to avoid such a problem, but also to have more flexibility during the fitting procedure, they introduced on top of the standard Skyrme interaction two extra density-dependent terms namely
\begin{eqnarray}\label{extra:bsk}
V^{Sk}_{ex}&=&\frac{1}{2}t_{1b}(1+x_{1b}P_{\sigma})\left( \mathbf{k}^{'2}\rho^{\beta_{1}}+\rho^{\beta_{1}} \mathbf{k}^{2}\right)+t_{2b}(1+x_{2b}P_{\sigma}) \mathbf{k}^{'}\rho^{\beta_{2}} \mathbf{k}.
\end{eqnarray}
The corresponding functional reads 
%
\begin{eqnarray}\label{extra:bsk:func}
E^{ex}[\rho]&=& \int  \sum_{t=0,1}\left\{C_{t}^{\Delta\rho,\beta_{1}}\rho^{\beta_{1}} \rho_{t}\Delta \rho_{t}+C^{\tau,\beta_{1}}_{t}\rho^{\beta_{1}}\left[ \rho_{t} \tau_{t} -\mathbf{j}_{t}^{2} \right] +C^{\tau,\beta_{2}}_{t}\rho^{\beta_{2}}\left[ \rho_{t} \tau_{t} -\mathbf{j}_{t}^{2} \right] \right.\nonumber\\
&+&\left.C^{\nabla \rho,\beta_{1}}_{t}\rho^{\beta_{1}}\nabla^{2}\rho_{t}+C^{\nabla \rho,\beta_{2}}_{t}\rho^{\beta_{2}}\nabla^{2}\rho_{t} +C^{\Delta s,\beta_{1}}_{t}\rho^{\beta_{1}}s_{t}\Delta s_{t}+C^{T,\beta_{1}}_{t}\rho^{\beta_{1}}(s_{t} T_{t} -J^{2}_{t})\right. \nonumber\\
&+&\left.C^{T,\beta_{2}}_{t}\rho^{\beta_{2}}\left[ s_{t} T_{t} -J^{2}_{t}\right]+C^{\nabla s,\beta_{1}}_{t} \rho^{\beta_{1}}  
| \nabla s_{t}|^2 +C^{\nabla s,\beta_{2}}_{t} \rho^{\beta_{2}}  | \nabla s_{t}|^2 \right\}.
\end{eqnarray}
The connection between the interaction parameters $t_{ib}, x_{ib}$ and the functional coupling constants is given in Eqs. \ref{app:bru1}-\ref{app:bru2}. Performing a second functional derivative, we can easily obtain the contribution to the residual interaction, which 
gives additional contributions $W_{i,ex}^{(\alpha)}$ to the general form given in Eq. (\ref{ME-SNM}). In Table \ref{table:W_iex} are given the terms $W_{1,ex}^{(\alpha)}$ and 
$W_{2,ex}^{(\alpha)}$ as a function of the coupling constants of the functional (\ref{extra:bsk:func}) in SNM. 
The term $W_{3,ex}^{(\alpha)}$ does not appears using the standard Skyrme functional, and in the present case it only
appears in the $(0,0)$ channel
\begin{equation}
\frac{1}{4}W_{3,ex}^{(0,0)} = C^{\tau,\beta_{1}}_{0}\beta_{1} \rho_{0}^{\beta_{1}} +C^{\tau,\beta_{2}}_{0}\beta_{2} \rho_{0}^{\beta_{2}}.
\end{equation}
As shown in Eq. (\ref{ME-SNM}) this term is associated to a momentum dependence ${\bf k}_1 ({\bf k}_2+{\bf q}) + {\bf k}_2 ({\bf k}_1+{\bf q})$, which has been previously included to derive the response function.   
This dependence is not limited to the present extra-terms, we will also encounter it in the Section \ref{sec:3b} dedicated to the three-body interaction. 
Actually, it appeared in Ref.~\cite{gar92} for Skyrme-like functionals used to describe liquid $^{3}$He properties~\cite{str84}, in particular to reproduce the experimental effective mass. In the present case, the effective mass is written as
\begin{equation}
\frac{\hbar^2}{2m^*} = \frac{\hbar^2}{2m}+C^{\tau}_0 \rho +C^{\tau,\beta_1}_0  \rho_0^{\beta_1+1} +
C^{\tau,\beta_2}_0  \rho_0^{\beta_2+1} \, .
\end{equation}

\begin{table}[H]
\caption{Definition of constants $W_{i,ex}^{(\alpha)}$ entering (\ref{ME-SNM}).}
\label{table:W_iex}
\begin{tabular}{c|c|c}
\hline
\hline
$(\alpha)$ & $W_{1,ex}^{(\alpha)}/4$ & $W_{2,ex}^{(\alpha)}/4$  \\ 
 \hline 
(0,0) & $C^{\tau,\beta_1}_0 (\beta_1+1)\beta_1 \rho^{\beta_1-1} \tau +C^{\tau,\beta_2}_0 (\beta_2+1) \beta_2 \rho^{\beta_2-1}\tau$ & $C^{\tau,\beta_1}_0 (1+\beta_1) \rho^{\beta_1}$ \\ 
& $-2 {\bf q}^2 \left[ C^{\Delta \rho,\beta_1}_0 (\beta_1+1) \rho^{\beta_1} - C^{\nabla \rho,\beta_1}_0 \rho^{\beta_1} -C^{\nabla \rho,\beta_2}_0 \rho^{\beta_2} \right]$ & $+C^{\tau,\beta_2}_0 (1+\beta_2) \rho^{\beta_2}$ \\
&- $\frac{1}{2} {\bf q}^2 \left[ C^{\tau,\beta_1}_0 \rho^{\beta_1} + C^{\tau,\beta_2}_0 \rho^{\beta_2}  \right]$ &  \\
(0,1) & $- 2 {\bf q}^2 \left[ C_1^{\Delta\rho,\beta_1} \rho^{\beta_1} - C^{\nabla \rho,\beta_1}_1 \rho^{\beta_1} 
- C^{\nabla \rho,\beta_2}_1 \rho^{\beta_2} \right]$ & 
$C^{\tau,\beta_1}_1 \rho^{\beta_1} +C^{\tau,\beta_2}_1 \rho^{\beta_2}$ \\
& $ -\frac{1}{2} {\bf q}^2 \left[ C^{\tau,\beta_1}_1 \rho^{\beta_1} + C^{\tau,\beta_2}_1 \rho^{\beta_2}  \right]$
& \\
(1,0) & $- 2 {\bf q}^2 \left[ C^{\Delta s,\beta_1}_0 \rho^{\beta_1} -  C^{\nabla s,\beta_1}_0 \rho^{\beta_1}
-  C^{\nabla s,\beta_2}_0 \rho^{\beta_2}   \right]$ 
& $C^{T,\beta_1}_0 \rho^{\beta_1} +C^{T,\beta_2}_0 \rho^{\beta_2}$ \\
& $ -\frac{1}{2}  {\bf q}^2 \left[ C^{T,\beta_1}_0 \rho^{\beta_1} + C^{T,\beta_2}_0 \rho^{\beta_2}  \right]$ & \\
(1,1) & $- 2 {\bf q}^2 \left[ C^{\Delta s,\beta_1}_1 \rho^{\beta_1} - C^{\nabla s,\beta_1}_1 \rho^{\beta_1}
- C^{\nabla s,\beta_2}_1 \rho^{\beta_2} \right]$ & $C^{T,\beta_1}_1 \rho^{\beta_1} +C^{T,\beta_2}_1 \rho^{\beta_2}$
\\
& $-\frac{1}{2}  {\bf q}^2 \left[ C^{T,\beta_1}_1 \rho^{\beta_1} +  C^{T,\beta_2}_1 \rho^{\beta_2}  \right]$ & \\
\hline
\hline
\end{tabular}
\end{table}
For our analysis we have selected  the functionals BSk19, BSk20 and BSk21~\cite{gor10}, which  have been designed to avoid some instabilities without spoiling the good general description achieved with the previous parameterizations of the same family. The three functionals give the same value for the SNM effective mass at saturation, and their main difference is the stiffness of the EoS in PNM, BSk19 being the softest and BSk21 the hardest.

\begin{figure}[H]
\begin{center}
  \includegraphics[width=0.6\textwidth,angle=-90]{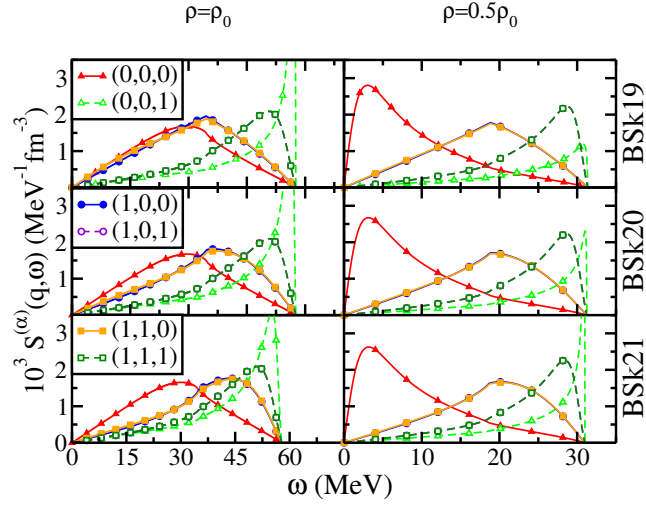}
\caption{(Color online) SNM Strength functions at $q=0.5 k_F$ and density values $\rho = 0.08$ (right panels) and $0.16$ fm$^{-3}$ (left panels) for the three selected BSk functionals.}
\label{Resp:bsk21}
\end{center}
\end{figure}

The calculated SNM strength functions $S^{(S,M,I)}(q,\omega)$ at zero temperature are plotted in 
Fig.~\ref{Resp:bsk21} for densities $\rho = 0.08$ and $0.16$ fm$^{-3}$ and momentum transfer $q=0.5 k_F$. We notice that due to absence of a tensor term, the two spin projections are essentially degenerate because the momentum transfer is not high enough for the spin-orbit coupling to produce visible results. 
As compared to other interactions previously studied (see Fig. \ref{RESP-qkF05}), there are two visible differences. The first one is that channels with isospin $I=1$ display a peak near the upper edge of the $ph$-band. The second one is the strong accumulation of strength in the channel (0,0,0) at low transferred energy at density $\rho=0.08$ fm$^{-3}$, which is of course the manifestation of the spinodal instability. 

\begin{figure}[H]
\begin{center}
  \includegraphics[width=0.6\textwidth,angle=-90]{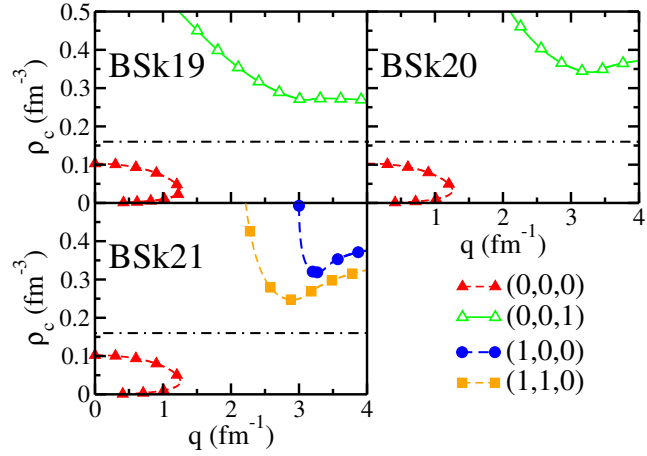}
\caption{(Color online) Critical densities for the three selected BSk functionals in SNM.}
\label{crit:bsk}
\end{center}
\end{figure}

In Fig.~\ref{crit:bsk} are plotted the critical densities $\rho_{c}$ as a function of $q$. We notice that, apart from the spinodal, BSk functionals do not present instabilities in the low $q$ region. This is consistent with the fitting procedure of Refs.~\cite{cha10c,gor10}, where it has been imposed that Landau inequalities~\cite{nav13} are fulfilled up to several times saturation density.
Actually we checked that the BSk20-21 functionals respect the empirical criterion defined in Ref.~\cite{hel13} so that we do not expect to find finite size-instabilities in nuclei.

%

Concerning BSk19, we have observed that the scalar-isovector instability decreases slowly when we increase the transfer momentum, the minimum being at $\rho_{c}=0.24$~fm$^{-3}$ for $q=15$~fm$^{-1}$. As discussed in Ref.~\cite{hel13}, values of $\rho_c \simeq 0.16-0.24$ fm$^{-3}$ found even at very high values of $q$ can play an important role concerning instabilities in finite nuclei. Hence, BSk19 must be tested in nuclei before any firm conclusion can be drawn. Clearly, since our formalism can be easily extended, it could be included directly in future fitting procedure.

\subsection{Extended $t_{3}$ with spin-density dependence}

In refs.~\cite{mar09c,mar09d}, the authors discussed the inclusion of two extra density dependent terms similar to the usual $t_{3}$ term \cite{cha97}. The main motivation was to improve the properties of the functional in the time odd part, but also to avoid ferromagnetic instabilities in infinite matter in the long-wavelength limit. Furthermore, they also improved the agreement between Landau parameters $G$ (vector channel), in both SNM and PNM with the Brueckner-Hartree-Fock results obtained in ref.~\cite{zuo03}. These extra terms read
\begin{eqnarray}\label{extra:dd}
V^{Sk}_{dd}&=&\frac{1}{6}t_{3}^{s}\left(1+x_{3}^{s}P_{\sigma}\right)|\mathbf{s}_{0}|^{2}+\frac{1}{6}t_{3}^{st}\left(1+x_{3}^{st}P_{\sigma}\right)|\mathbf{s}_{1}|^{2} \, , 
\end{eqnarray}
where $\mathbf{s}_{0},\mathbf{s}_{1}$ are the isoscalar and isovector spin densities.
This can be easily generalized to even exponents of the spin densities. It is worth noticing that as an attempt to describe magnetic properties of liquid $^3$He, in Refs. \cite{stri86,wei88,bar96,gat98} spin-density and current dependent terms were considered. 

The corresponding functional reads
\begin{eqnarray}
E^{dd}&=&\int d\vec{r} \,\left\{ C^{s\rho}_{0}\mathbf{s}_{0}^{2}\rho_{0}^{2}+C^{s\rho}_{1}\mathbf{s}_{1}^{2}\rho_{1}^{2}+C^{ss}_{0}\mathbf{s}_{0}^{4}+C^{ss}_{1}\mathbf{s}_{1}^{4} + C^{s\rho}_{01}\mathbf{s}_{0}^{2}\rho_{1}^{2}+C^{s\rho}_{10}\mathbf{s}_{1}^{2}{\rho}^{2}_{0}+(C^{ss}_{10}+C^{ss}_{01})\mathbf{s}_{0}^{2}\mathbf{s}_{1}^{2} \right\}.\nonumber\\
\end{eqnarray}
The connection between the interaction parameters $t_3^{s,st}, x_3^{s,st}$ and the functional coupling constants is given in Eqs. \ref{app:spi1}-\ref{app:spi2}. The residual interaction is deduced as usual by the second functional derivative with respect to the density. The momentum dependence of the general $ph$ interaction (\ref{ME-SNM}) is preserved, with additional contributions to the $W_{1}$ coefficients in the $(S,M)=(1,0)$ and $(1,1)$ channels
\begin{eqnarray}
\frac{1}{4}W_{1,dd}^{(1,0)}&=&2\rho_{0}^{2}C^{s\rho}_{0},\\
\frac{1}{4}W_{1,dd}^{(1,1)}&=&2\rho_{0}^{2}\left[ C^{s\rho}_{1}+C^{s\rho}_{10}\right].
\end{eqnarray}

\begin{figure}[H]
\begin{center}
  \includegraphics[width=0.43\textwidth,angle=-90]{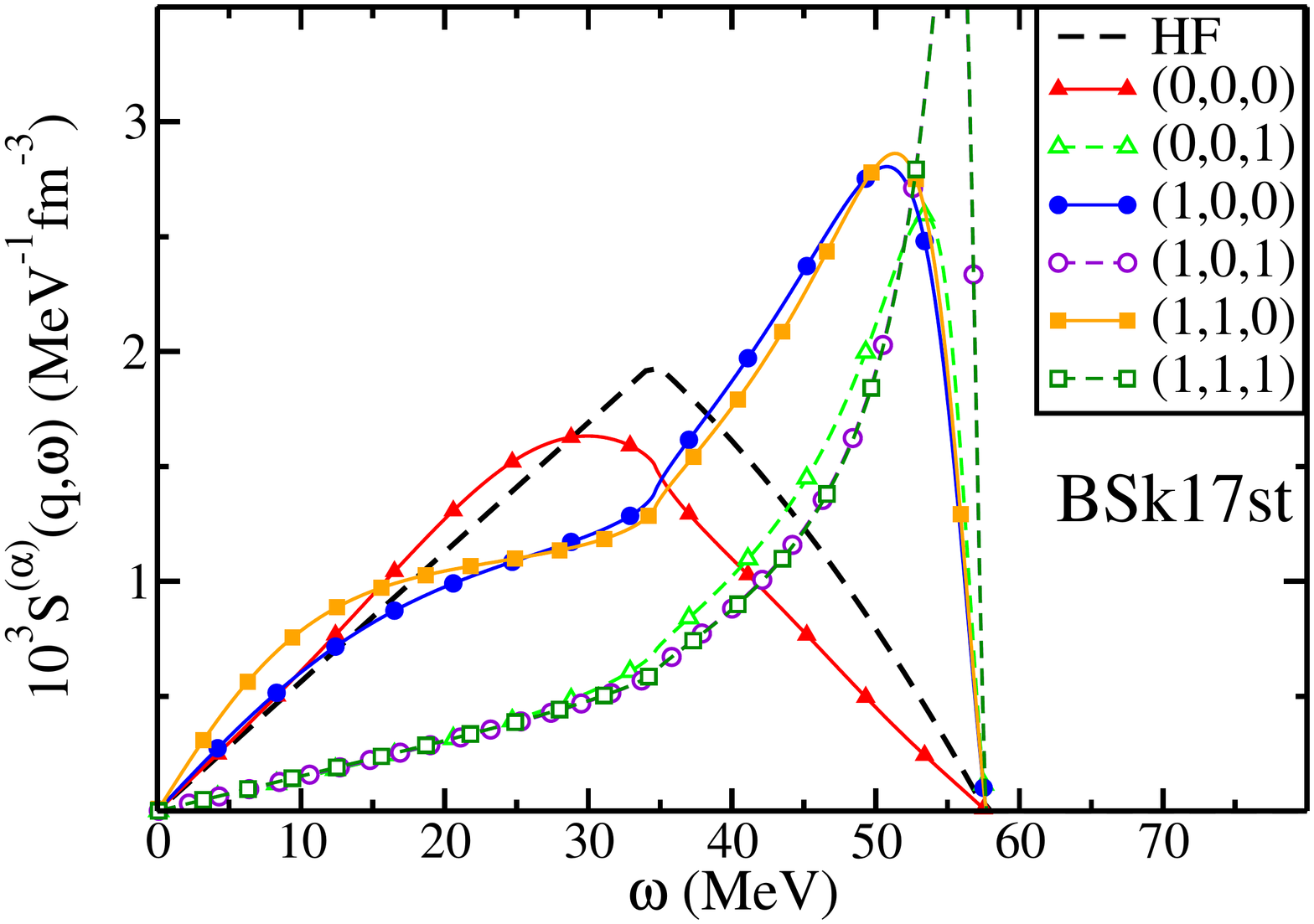}
\hspace{-1.89cm}
    \includegraphics[width=0.43\textwidth,angle=-90]{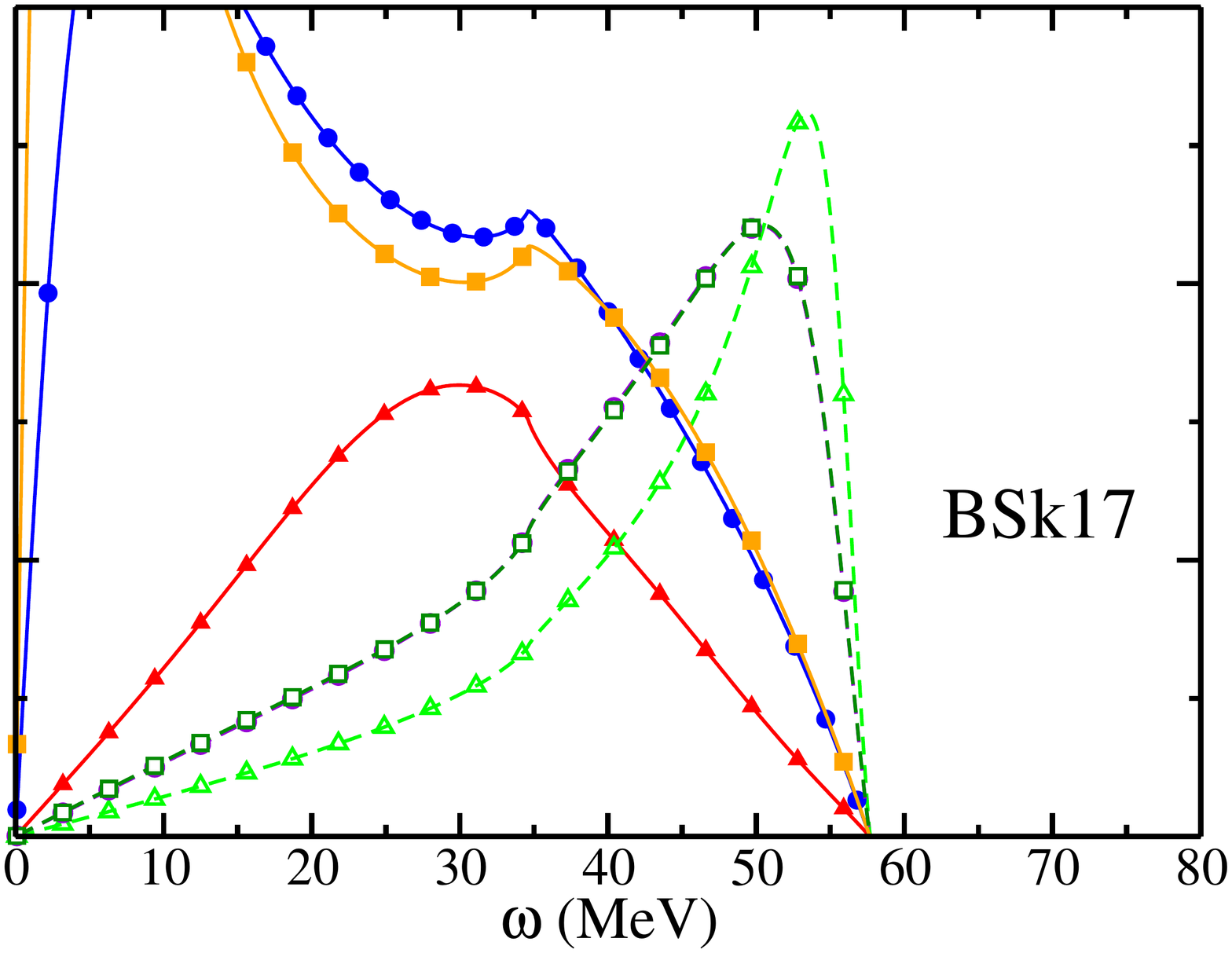}
\caption{(Color online) Response function at saturation density for $q=0.5k_{F}$ for the functionals BSk17st and BSk17.}
\label{Resp:bsk17st}
\end{center}
\end{figure}

The parameterization  BSk17st was obtained in \cite{mar09d} adding these new terms to BSk17, and fitting them so as to improve some infinite matter properties. In Fig.~\ref{Resp:bsk17st}, we compare the SNM strength functions obtained for both functionals at $\rho=0.16$~fm$^{-3}$ and $q=0.5k_{F}$. 
We see that the net effect of the extra terms is to make the residual interaction more repulsive in $(S,M,I)=(1,0,0)$ and $(1,1,0)$ channels : the accumulation of strength at low energy present for BSk17 disappears completely for BSk17st. This is in agreement with recent RPA calculations in finite nuclei~\cite{pei14} done with these functionals.
In all other channels, the extra terms play no significant role.

\begin{figure}[H]
\begin{center}
  \includegraphics[width=0.43\textwidth,angle=-90]{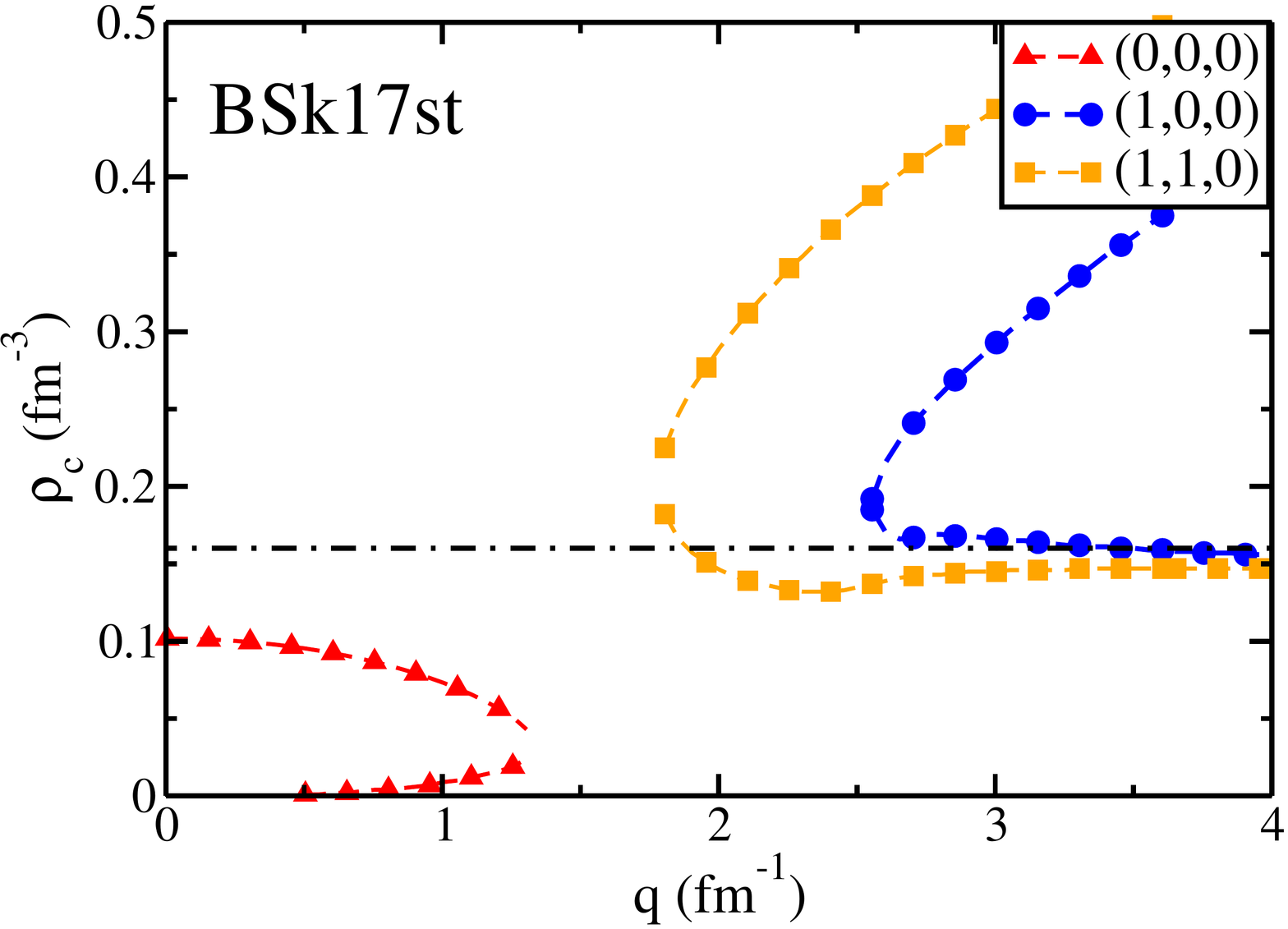}
\hspace{-1.89cm}
    \includegraphics[width=0.43\textwidth,angle=-90]{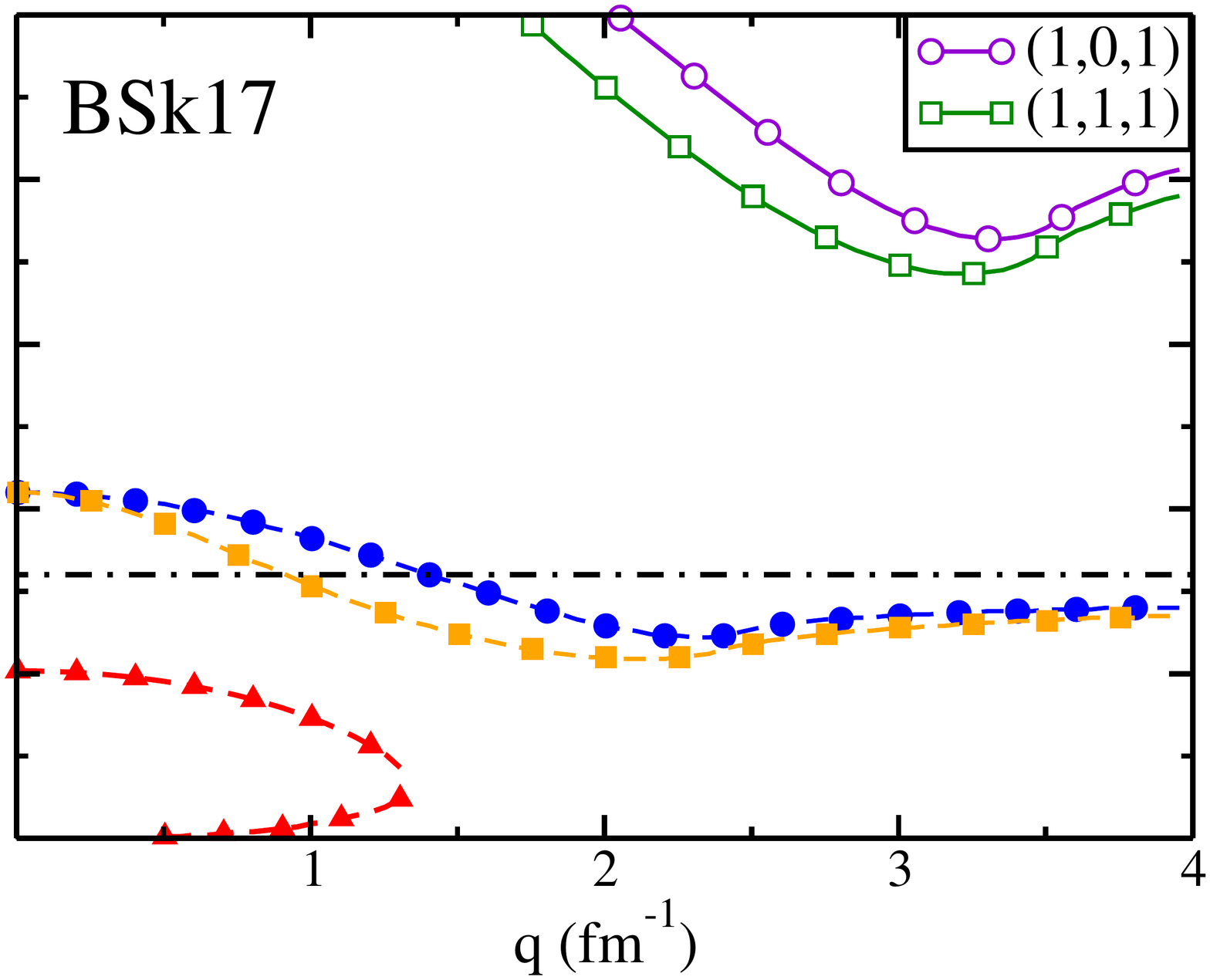}
\caption{(Color online) Critical densities as a function of the transferred momentum $q$ for the functionals BSk17 and BSk17st. The horizontal dot-dashed line represents the density 0.16 fm$^{-3}$.}
\label{crit:bsk17st}
\end{center}
\end{figure}
The instabilities of both functionals in the $(\rho_c,q)$ plane are compared in Fig.~\ref{crit:bsk17st}. As expected, the effect of the extra terms is to push to higher transferred momentum values the spin instabilities given by the BSk17 functional~\cite{cha10c}. It is interesting to note that these spin-density dependent terms do not change the quality of the finite nuclei adopted in the fit, but considerably improve the spin channels properties.

\subsection{Three- and four-body terms}\label{sec:3b}

There is a well-known problem met in calculating SNM saturation properties using realistic two-body interactions within a Brueckner-Hartree-Fock framework, which is summarized in the so-called Coester band \cite{coe70}. More precisely, even if the predicted phase-shifts agree well with experiments, the calculated saturation points are located on a narrow band in the $(E/A,\rho)$ plane which does not cross the empirical point. A similar Coester line is also obtained in Dirac-Brueckner-Hartree-Fock calculations based on OBE (one boson exchange) potentials \cite{bro90}. It has also been pointed out that two-body interactions cannot simultaneously satisfactorily reproduce ground state properties of $^3$He and $^4$He nuclei, the energies being correlated along the so-called Tjon band \cite{tjo75}, which does not include the experimental point. Actually, three-body interactions have been identified as the responsible for SNM saturation \cite{gra89,zuo02,heb11,lov12,car13} and
their importance in describing light nuclei properties has been highlighted in recent \emph{ab-initio} calculations \cite{pie01,hag07}. It has also received an interest in light hypernuclei~\cite{ric14,hiy14,gal14,gar14}. Generally speaking, the three-body interaction is thus a subject of fundamental importance in physics \cite{ham13}. 

Concerning the response function calculations, it has been shown that for normal Fermi liquids the contributions of three-quasiparticle interactions are suppressed at low excitation energies \cite{fri11}. However, recent microscopic calculations of the quasiparticle interactions within many-body perturbation theory~\cite{hol11B,hol12B} have been used for an accurate determination of Landau parameters thus showing the impact of three-body corrections~\cite{hol13} within the nuclear medium. 

T.H.R. Skyrme~\cite{sky59} already suggested in his original paper that his effective two-body interaction should also be equipped with an effective three-body term. However, in the first applications~\cite{vau72} it was shown that considering HF calculations of even-even nuclei, the central contact three-body term was equivalent to a zero-range interaction with a linear density dependence. Since then, a density dependent term has been added to the effective interaction, with a real exponent to have more flexibility during the fitting procedure~\cite{cha97}. Unfortunately, it has been realized a long time after that non integer exponents can lead to divergences in multi-reference calculations~\cite{ben09,lac09,dug09}. The possibility of adding an effective three-body interaction with the same form of the two-body part, $i.e.$ that includes up to second order gradients have therefore been examined very recently \cite{sad11,sad13}.
Following \cite{sad13}, we write the most general central three-body contact interaction as
\begin{align}\label{veff:3b}
V^{Sk}_{3b} = &
3
P^{\{x^0\}} 
\delta (\vec{r}_1 - \vec{r}_2) \delta (\vec{r}_1 - \vec{r}_3)
\nonumber \\
&
+ \frac{3}{2}
\bigg[
P^{\{x^1\}} \Big(
k^\prime_{12,u} k^\prime_{12,u} 
\Big)
 + \Big(
k_{12,u} k_{12,u} 
\Big) P^{\{x^1\}}
\bigg] \delta (\vec{r}_1 - \vec{r}_2) \delta (\vec{r}_1 - \vec{r}_3)
\nonumber 
\\
&
+ \frac{3}{2}
\bigg[
P^{\{x^2\}} \Big(
k_{12,u} k^\prime_{12,u} 
\Big)
 + \Big(
k^\prime_{12,u} k_{12,u} 
\Big) P^{\{x^2\}}
\bigg] \delta (\vec{r}_1 - \vec{r}_2) \delta (\vec{r}_1 - \vec{r}_3) \, ,
\end{align}
\noindent where the exchange operators $P^{\{x\}}$ are defined as \cite{sad11}
\begin{align}
P^{\{x^0\}} = \nonumber &
u_0\,,
\\
P^{\{x^1\}} = \nonumber &
u_1
+u_1y_1P^{\sigma}_{12}\,,
\\
P^{\{x^2\}} = &
u_2
+u_2y_{21}P^{\sigma}_{12}
+u_2y_{22}P^{\sigma}_{13} 
+u_2y_{22}P^{\sigma}_{23} \,.
\end{align}
Starting from the pseudo-potential given above, one can derive the energy density functional (see Ref.~\cite{sad13} for the explicit expressions) and then derive the residual interaction. The resulting momentum structure is included in the general form described earlier in Section \ref{BMSpot}. One has simply to add extra contributions from the three-body part to the coefficients $W_i^{(\alpha)}$. They are given in Table \ref{table:W_i3b} in terms of combinations of the interaction parameters, which in turn are given  in Table \ref{tab:Skyrme_int:3bodyEDF:coeff}, extracted from \cite{sad13}.

\begin{table}[H]
\caption{Definition of constants $W_{i,3b}^{(\alpha)}$ entering (\ref{ME-SNM}).}
\label{table:W_i3b}
\begin{tabular}{c|c|c|c}
\hline
\hline
$(\alpha)$ & $W_{1,3b}^{(\alpha)}/4$ & $W_{2,3b}^{(\alpha)}/4$  & $W_{3,3b}^{(\alpha)}/4$ \\ 
 \hline 
(0,0) & $6B_0^{\rho} \rho  + 2B_0^{\tau} \tau + {\bf q}^2 \left(\frac{1}{2}B_0^{j} \rho+ 2B_0^{\nabla \rho}\rho\right)$
& $2B_0^{\tau}  \rho  $ & $2 B_0^{\tau} \rho + B_0^{j} \rho$ \\
(0,1) & $2 B_1^{\rho} \rho + 2B_{10}^{\tau} \tau + {\bf q}^2 \left( \frac{1}{2} B_{1}^{j} \rho +2B_1^{\nabla \rho} \rho \right)$ & $B_1^{\tau} \rho$ & $B_1^{\tau} \rho + B_{1}^{j} \rho$ \\
(1,0) & $2 B_0^s \rho + 2B_0^{\tau s} \tau + {\bf q}^2 \left(\frac{1}{2} B_0^{J} \rho +2 B_0^{\nabla s} \rho \right)$ & $B_0^{T} \rho$ & $B_0^{T} \rho + B_0^{J} \rho$ \\
(1,1) & $2 B_1^s \rho + 2B_{10}^{\tau s} \tau + {\bf q}^2 \left( \frac{1}{2} B_1^J \rho + 2B_1^{\nabla s} \rho \right)$ & $B_1^T \rho$ & $B_1^T \rho +B_1^J \rho$ \\
\hline
\hline
\end{tabular}
\end{table}

\begin{table}[H]
\caption{\label{tab:Skyrme_int:3bodyEDF:coeff}
Coefficients entering in coefficients $W_{i=1,2,3}^{(\alpha)}$ with respect to interaction parameters.} 
\begin{tabular}{r c c c c c c | r c c c c c c}
\hline \hline \noalign{\smallskip}
&$u_0$&$u_1$&$u_1y_1$&$u_2$&$u_2y_{21}$&$u_2y_{22}$
& &$u_0$&$u_1$&$u_1y_1$&$u_2$&$u_2y_{21}$&$u_2y_{22}$\\
\noalign{\smallskip}  \hline \noalign{\smallskip}
$B^\rho_0$&$+\frac{3}{16}$&&&&& 
&$B^\rho_1$&$-\frac{3}{16}$&&&&&\\[0.3mm]
$B^\tau_0$&&$+\frac{3}{32}$&&$+\frac{15}{64}$&$+\frac{3}{16}$&$+\frac{3}{32}$
&$B^\tau_1$&&$-\frac{1}{16}$&$-\frac{1}{32}$&$+\frac{1}{32}$&$+\frac{1}{16}$&$-\frac{1}{16}$\\[0.3mm]
$B^{\nabla\rho}_0$&&$+\frac{15}{128}$&&$-\frac{15}{256}$&$-\frac{3}{64}$&$-\frac{3}{128}$
&$B^{\nabla\rho}_1$&&$-\frac{5}{128}$&$-\frac{1}{32}$&$-\frac{7}{256}$&$-\frac{1}{32}$&$-\frac{5}{128}$\\[0.3mm]
$B^J_0$&&$+\frac{1}{32}$&$-\frac{1}{16}$&$-\frac{7}{64}$&$-\frac{1}{8}$&$+\frac{1}{32}$
&$B^J_1$&&$+\frac{1}{32}$&&$-\frac{7}{64}$&$-\frac{1}{16}$&$-\frac{1}{32}$\\[0.3mm]
$B^s_0$&$-\frac{3}{16}$&&&&&
&$B^s_1$&$-\frac{3}{16}$&&&&&\\[0.3mm]
$B^T_0$&&$-\frac{1}{16}$&$+\frac{1}{32}$&$+\frac{1}{32}$&$+\frac{1}{16}$&$+\frac{1}{8}$
&$B^T_1$&&$-\frac{1}{16}$&&$+\frac{1}{32}$&&\\[0.3mm]
$B^{\tau s}_0$&&$-\frac{1}{32}$&$-\frac{1}{32}$&$-\frac{5}{64}$&$-\frac{1}{16}$&$+\frac{5}{32}$
&$B^{\tau s}_{10}$&&$-\frac{1}{32}$&&$-\frac{5}{64}$&$-\frac{1}{16}$&$-\frac{1}{32}$\\[0.3mm]
$B^{\nabla s}_0$&&$-\frac{5}{128}$&$+\frac{1}{32}$&$-\frac{7}{256}$&$-\frac{1}{32}$&$+\frac{1}{128}$
&$B^{\nabla s}_1$&&$-\frac{5}{128}$&&$-\frac{7}{256}$&$-\frac{1}{64}$&$-\frac{1}{128}$\\[0.3mm]
$B^j_0$&&$-\frac{3}{32}$&&$-\frac{15}{64}$&$-\frac{3}{16}$&$-\frac{3}{32}$
&$B^j_1$&&$+\frac{1}{32}$&$+\frac{1}{16}$&$-\frac{7}{64}$&$-\frac{1}{8}$&$-\frac{5}{32}$\\[1.5mm]
\hline \hline
\end{tabular}
\end{table}

\begin{figure}[H]
\begin{center}
  \includegraphics[width=0.43\textwidth,angle=-90]{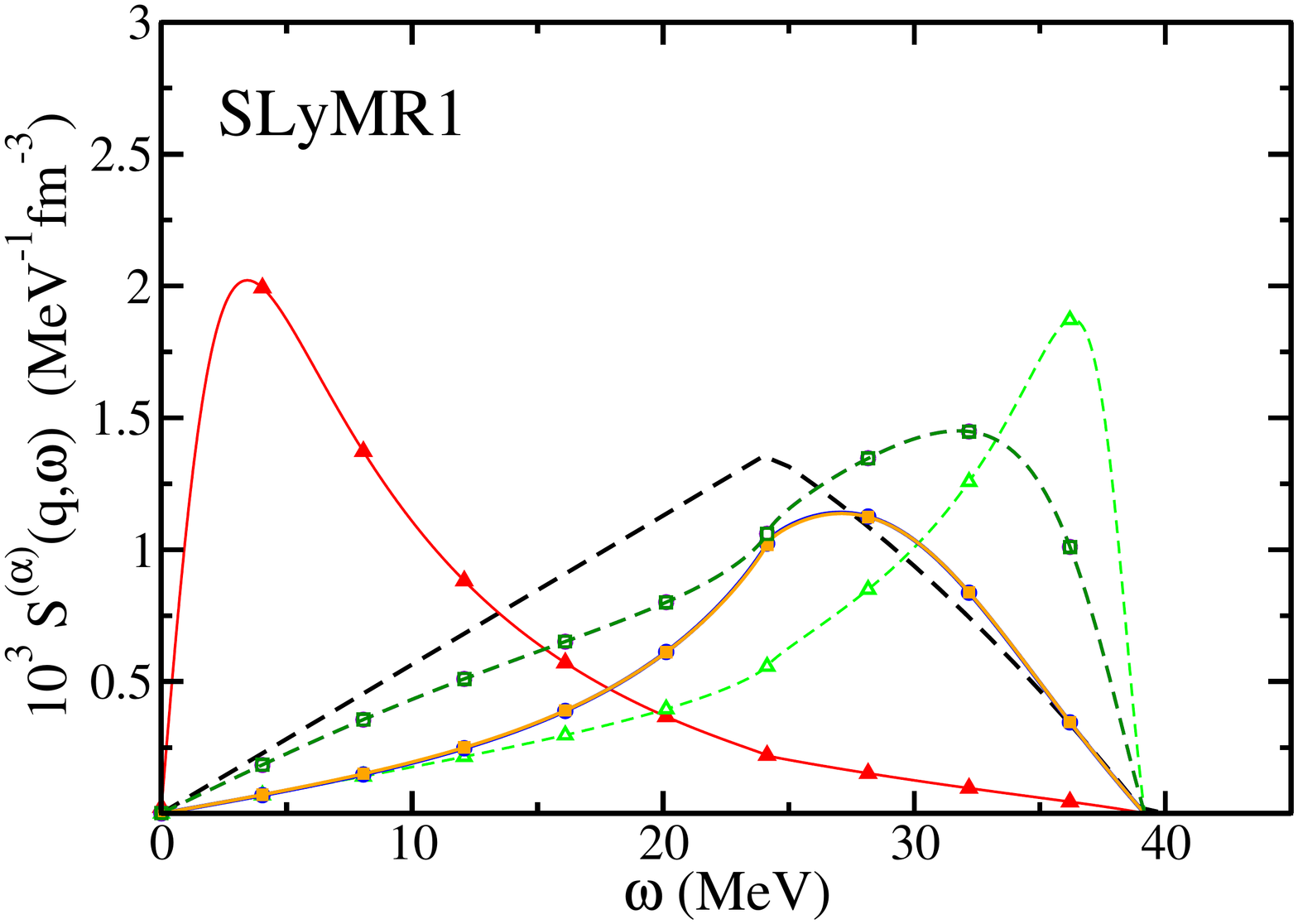}
\hspace{-1.89cm}
  \includegraphics[width=0.43\textwidth,angle=-90]{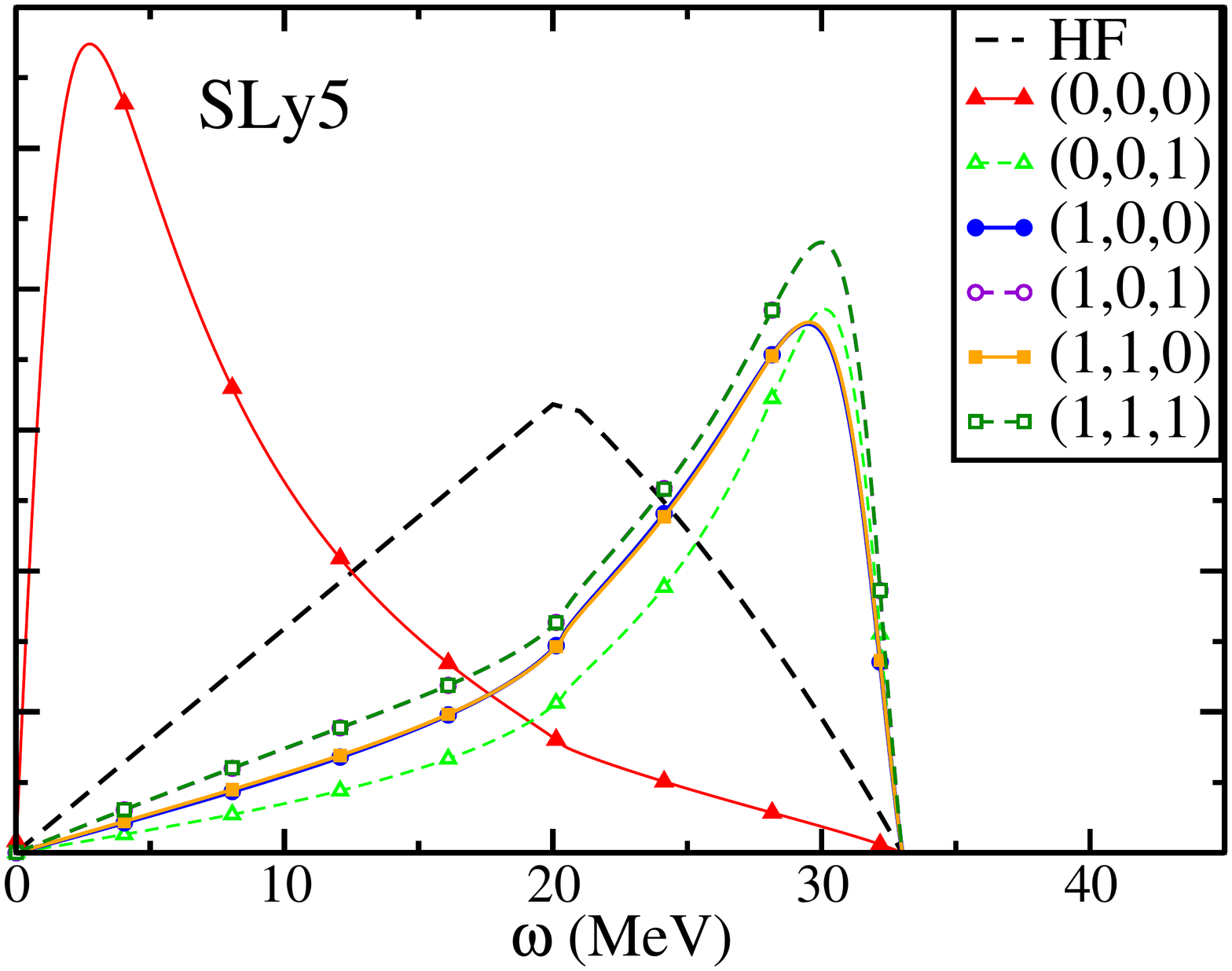}
  \caption{(Color online) Strength functions  for pseudo-potentials SLyMR1 (left panel) and SLy5 (right panel) at $q=0.5$~fm$^{-1}$ and density values $\rho=$ 0.08 fm$^{-3}$}
\label{Resp:3b4b}
\end{center}
\end{figure}

\begin{figure}[H]
\begin{center}
    \includegraphics[width=0.43\textwidth,angle=-90]{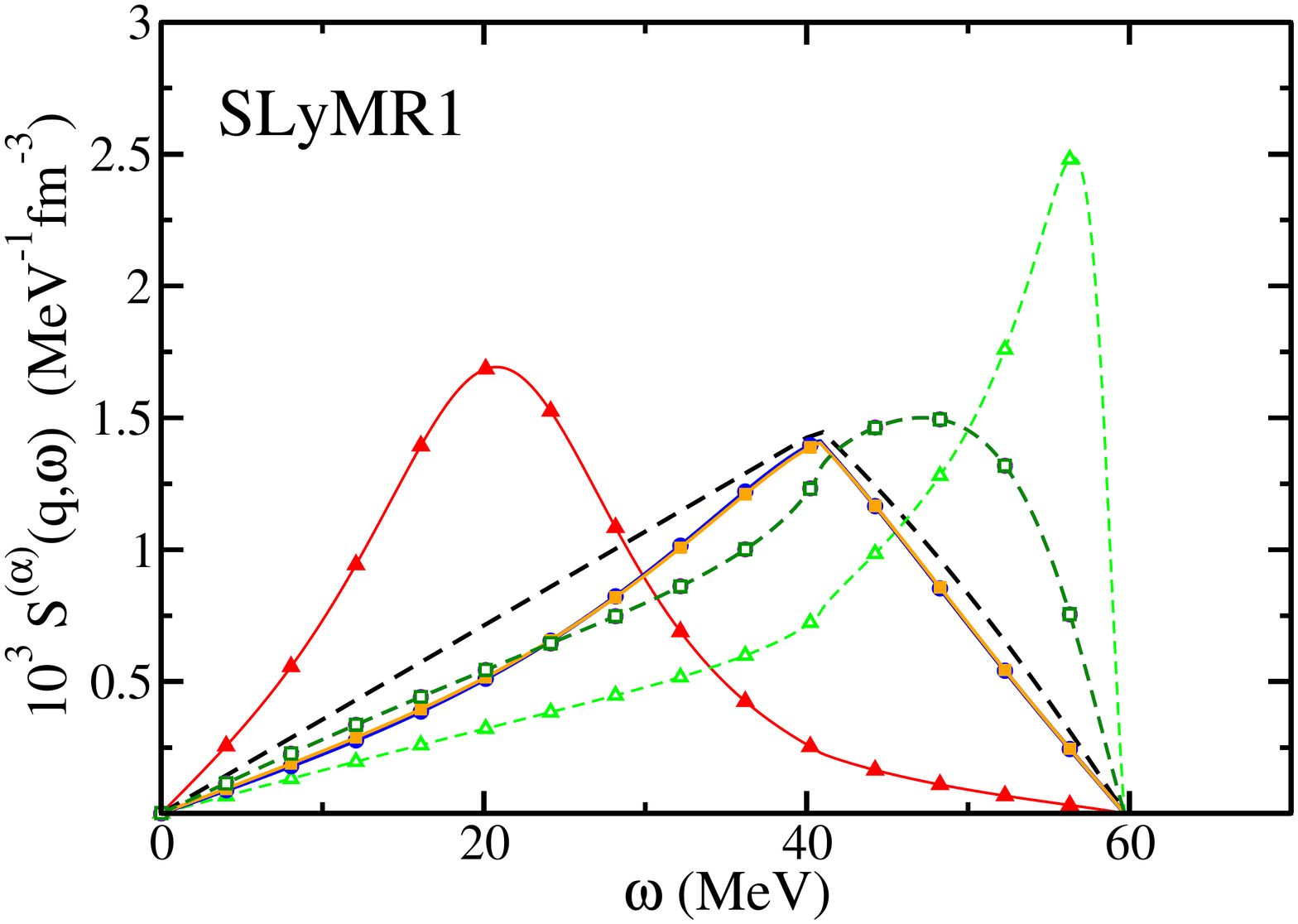}
\hspace{-1.89cm}
  \includegraphics[width=0.43\textwidth,angle=-90]{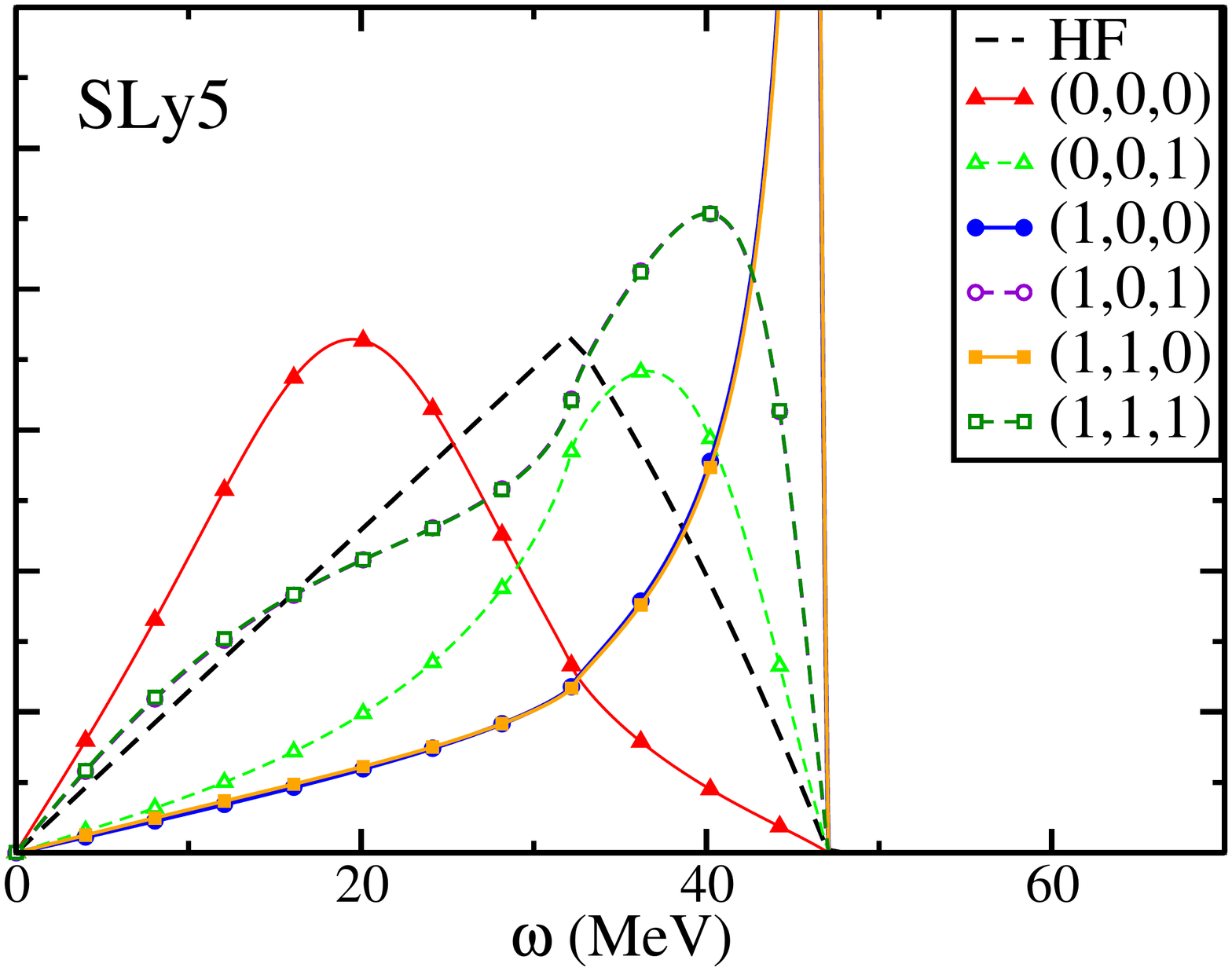}
  \vspace{0.5cm}
  \caption{(Color online) Same as Fig.\ref{Resp:3b4b}, but for $\rho=$ 0.16 fm$^{-3}$.}
\label{Resp:3balone}
\end{center}
\end{figure}

On top of the attractive three-body contribution, a contact four-body term has been also considered \cite{sad13,sad12}. The adopted form is $v_{0}\hat{\delta}_{r}$,  $\hat{\delta}_{r}$  being a shorthand notation for all the possible combinations of the form $\delta_{r_{1}r_{2}}\delta_{r_{2}r_{3}}
\delta_{r_{3}r_{4}}$ (see Refs.\cite{sad13,sad12} for more details). The only modification occurs in $W_{1}^{(\alpha)}$. In practice one has to add the following quantities
\begin{eqnarray}
\frac{1}{4}W_{1,4b}^{(0,0)}&=&  \frac{3}{8} v_{0} \rho_{0}^{2} \,, \nonumber\\
\frac{1}{4}W_{1,4b}^{(0,1)}&=& -\frac{3}{16}v_{0} v_{0} \rho_{0}^{2} \,,\nonumber\\
\frac{1}{4}W_{1,4b}^{(1,0)}&=& -\frac{3}{16}v_{0} v_{0} \rho_{0}^{2}\,, \nonumber\\
\frac{1}{4}W_{1,4b}^{(1,1)}&=& -\frac{3}{16}v_{0} v_{0} \rho_{0}^{2} \,.
\end{eqnarray}
We present here some results based on the effective interaction SLyMR1 \cite{Karim}, which includes three- and four-body terms. All parameters have been recently fitted  following the standard protocole used for the SLy family. 
In Fig. \ref{Resp:3b4b} is displayed the SNM strength function at  half saturation density  for $q=0.5k_{F}$ for the SLyMR1 (left panel) and for the SLy5 functional (right panel). The first point is that due to the absence of the tensor term and to the low value of transferred momentum, the spin orbit term is not important enough to split the projection of the total spin. The two $M$ projections are thus essentially on top of each other.
In Fig.~\ref{Resp:3balone}, we show the response function at saturation density and $q=0.5k_{F}$.

\begin{figure}[H]
\begin{center}
  \includegraphics[width=0.43\textwidth,angle=-90]{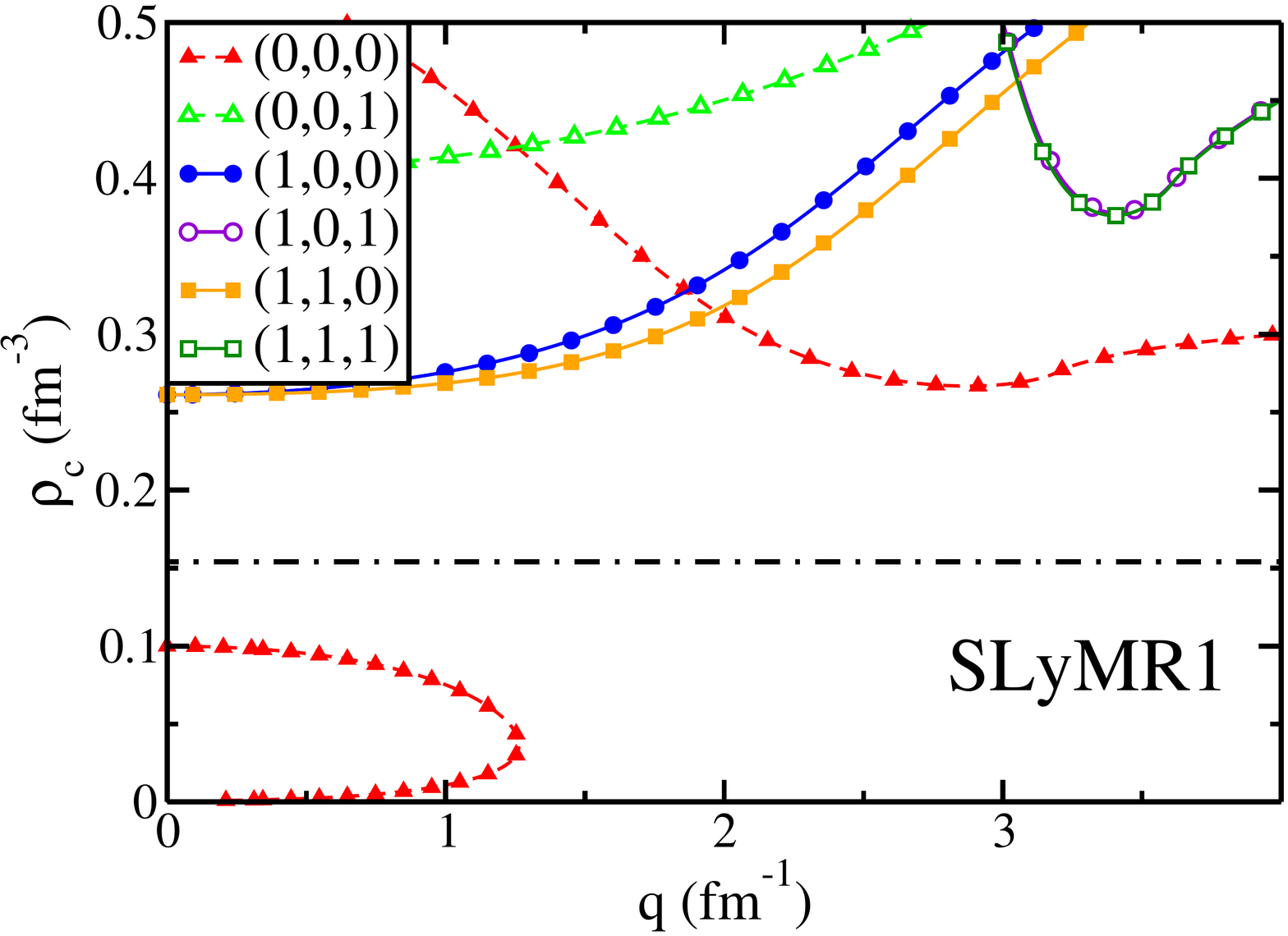}
\hspace{-1.9cm}
    \includegraphics[width=0.43\textwidth,angle=-90]{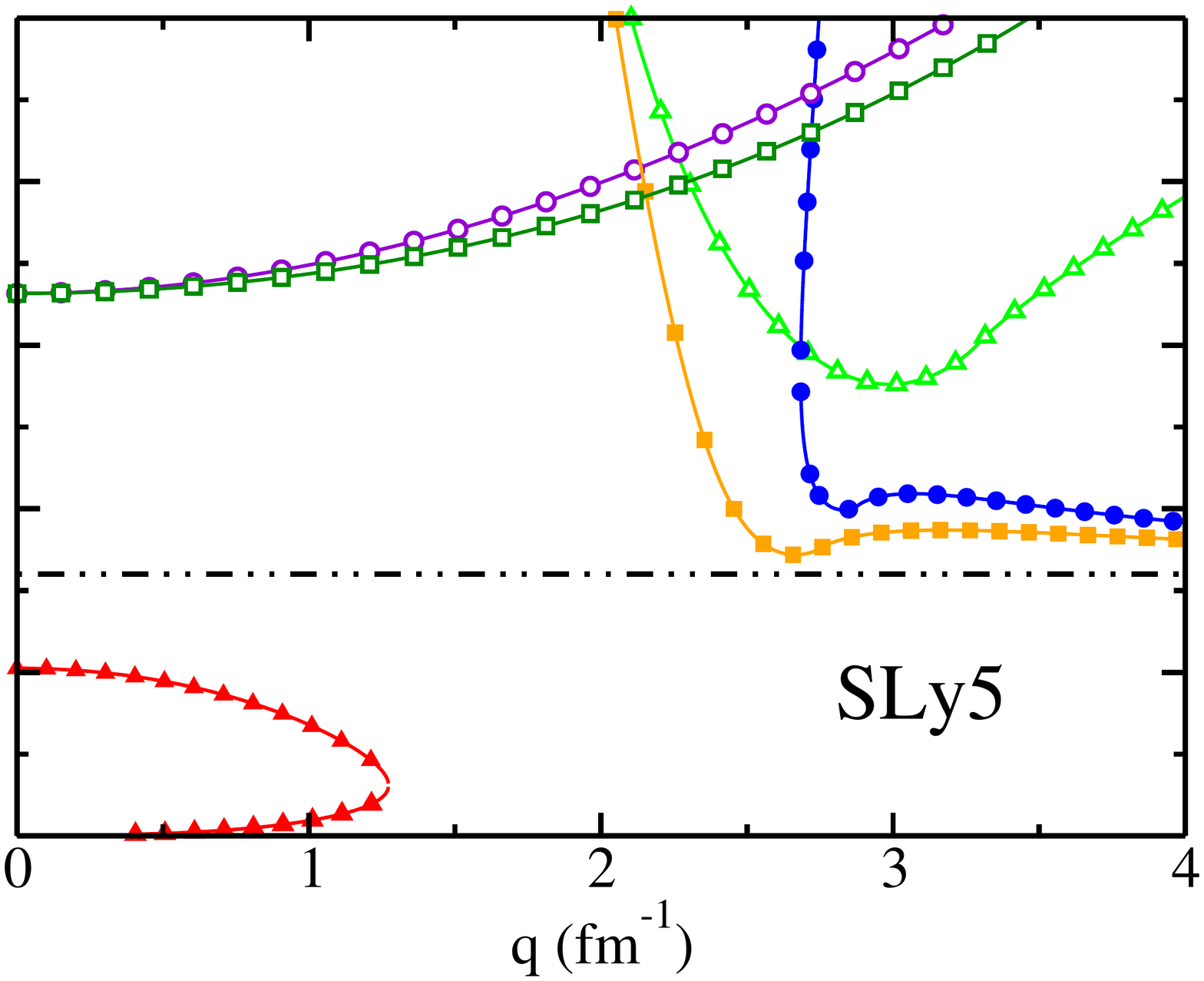}
  \caption{(Color online) Critical densities for the two pseudo-potentials discussed above.}
\label{Crit:3b4b}
\end{center}
\end{figure}

In Fig.\ref{Crit:3b4b}, we show the position of the instabilities in the $(\rho_{c},q)$ plane. It is worth noticing that the SLyMR1 functional has been constructed using the additional stability criterion defined in Ref.~\cite{hel13}.
The SLy5 functional presents a low density instability at $\rho_{c}\approx0.17$~fm$^{-3}$ and $q\approx2.6$~fm$^{-1}$ in the $(S=1,M=1,I=0)$ channel, which also manifests in finite-nuclei calculations~\cite{fra12}. On the contrary, all the instabilities for the SLyMR1 functional, apart from the spinodal one, have been pushed at higher values of $\rho_{c}$, by adopting the numerical criterion defined in Ref.~\cite{hel13} directly into the optimization procedure as explained in Ref.~\cite{pas13c}.


\section{Conclusions}\label{sec:conclusions}  

We have presented and discussed in some detail a method to obtain the response functions in infinite nuclear matter in the framework of the random phase approximation, based on a general Skyrme energy density functional. Its particular momentum dependence allows one to get relatively simple algebraic expressions, which permit in turn to derive odd-order energy weighted sum rules in a compact form. Adopting six representative Skyrme interactions, we have presented results for the response functions, paying particular attention on collective states and strength distribution, at different values of the density, transferred momentum, temperature and asymmetry of the system. Because of the specific form of the Skyrme pseudo-potentials, one should mention at this stage that our results are in principle limited to relatively small values of the transferred momentum (typically $1.5 - 2 k_F$). A possible extension of this range of validity may arise from N2LO or N3LO which simulate finite-range effects. In this case, our formalism can be applied, but it will be technically involved, although possible, to obtain analytical results.

We have shown that the tensor interaction has a very strong impact on the response functions for both the scalar $S=0$ (because of the spin-orbit coupling) and vector ($S=1$) channels. For $S=1$, we can further discriminate the longitudinal and transverse response functions, but only general trends emerged from our calculations. This is due to the fact that only recent Skyrme parameterizations with tensor terms have been derived from a merit function~\cite{bro06,les07}. Since their fitting protocols are mainly based on ground state properties of finite nuclei, it will be thus important in the future to find more adequate observables to constrain the tensor terms directly into the fitting procedure. 

We have also discussed in detail the appearance of instabilities in an infinite medium for arbitrary values of transferred momentum. We have provided tools for their detection and investigated their evolution as a function of the asymmetry. We have found that there is no general rule and  the smallest values for critical densities can appear for SNM, PNM or any asymmetry. Among these instabilities there is the spinodal, which is related to the physical characteristics of nuclear matter in the scalar-isoscalar channel. For this particular instability, our formalism provides results comparable to those obtained within a thermodynamical approach. In the other channels, we have shown that the (non physical) instabilities arise from some particular terms of the residual interaction which are proportional to the transferred momentum squared and thus can not be detected in the so-called {\it long-wavelength} limit. In that case,  we have used the inverse-weighted sum rule $M_{-1}$ to detect their presence in infinite matter. 

An infinite homogeneous medium has been considered along this work. This may represent of course quite a severe approximation when dealing with finite nuclei. However, it has been shown recently that the detection of instabilities in infinite matter is also relevant for nuclei:  even if one has to be cautious since there are only numerical evidences of this statement, one can say that an instability below $1.2-1.3$ saturation density will reveal itself in nuclei~\cite{hel13}.
The tensor part, which is an essential component of the nucleon-nucleon interaction, made this development even more crucial since it induces more pronounced instabilities : spin-isospin response functions exhibit quite systematically some poles at zero transferred energy for densities close or less than the saturation density in the spin channels. Combining the results obtained in the present report with those of Ref.~\cite{hel13}, it is thus possible to derive a simple numerical tool to avoid instabilities by construction if included into the optimization procedure used to derive the coupling constants of the functional. A first test has been presented in ~\cite{pas13c}: it has been shown that starting from a spin-unstable functional (SLy5) it was possible to derive a new functional (SLy5*) free from instabilities and with similar spectroscopic properties.
It is also worth mentioning that in Ref.~\cite{nav96}, the authors have investigated the extended Thomas-Fermi expansion for the nuclear response function in the infinite system. Although such a method is not very competitive for the calculations of the RPA response function in finite nuclei, it could be very useful to calculate the response function for nuclei in the inner crust of a neutron star~\cite{cham08}. Compared to fully microscopic calculations~\cite{kha05,bar10}, the main advantage of a semiclassical method is the reduced computational time. The formalism presented in this report could then be directly applied for the calculation of the vibrational spectrum of neutron rich nuclei within the inner crust.

Quantities of astrophysical interest, such as the neutrino mean free path (examples have been given in this report) or the specific heat (some recent results are now available in the literature~\cite{kel13,mar14}), can be directly computed from the response functions. For this kind of applications, the responses are usually calculated making use of a simplified residual interaction written in terms of $\ell=0, 1$ Landau parameters. In this report, we have presented results which includes higher order Landau parameters, obtained from effective finite-range potentials, as Gogny and Nakada, or from realistic NN potentials based on different microscopic methods. We have shown that well-converged response functions are obtained considering up to $\ell=2$ central and $\ell=1$ tensor Landau parameters.    

Following the recent findings on the limits of standard Skyrme functionals~\cite{kor13}, we have also investigated the possible extensions of our method to include new terms in the functional. In particular we have paid special attention to the introduction of a real central three-body term. Several recent examples have spotted the important role of these terms not only in the description of infinite matter properties~\cite{coe70,heb10}, but also in calculations of finite nuclei~\cite{hol12,ham13}. This has been, among others, the motivation to explore the possibility of adopting a real three-body term in Skyrme functional and thus removing the corresponding density dependent one. We have shown that our method is flexible enough to include these new terms in a natural way. Therefore, it could be easily included into the fitting procedure used to determine the coefficients of these extra terms, being a useful tool to avoid unphysical instabilities.

The RPA approximation is used throughout this report for the calculation of response functions. One may then wonder to which extent this approximation gives a good picture of the correlations in the system. Actually, it has been outlined numerous times with strong evidences that RPA is a well-suited tool to take into account correlations in nuclear physics. Actually, the potential problem is rather the correlations themselves than the tool to incorporate them. In that respect, it has been shown in neutrino physics that \emph{2p-2h} and in general \emph{np-nh} excitations represent not only sizable corrections to the nuclear response function calculated with \emph{1p-1h} excitations only, but are actually essential for understanding the experimental data~\cite{mar09,mar10,mar11}. These higher order excitations would surely affect the shape (the distribution of strength) of the response functions and consequently the position of the instabilities. However, the technical difficulty for incorporating them is such that it goes beyond the scope of this report. Moreover, as already mentioned, there is a clear indication that instabilities determined in infinite matter with 1p-1h excitations only reveal themselves in finite nuclei so that the method used here is trustworthy for the objectives of this report. Among other possible extensions, but also beyond the scope of this report are those including pairing effects ($i.e.$ QRPA formalism) which are known to play an important role under certain circumstances, for example in the crust of neutron stars~\cite{cham08}.

\section*{Acknowledgments}
We acknowledge  K. Bennaceur, M. Bender, N. Chamel, G. Col\`o, T. Duguet, M. Ericson, Nguyen Van Giai, S. Goriely, J. Margueron, M. Martini, J. Meyer, A. Polls and  J.M. Richard for interesting discussions and useful comments. 
This work was supported by Mineco (Spain), grant FIS2011-28617-C02-02. 

\appendix

\section{Local densities}
\label{loc-den}
A particle is characterized by the symbol $\ma \equiv({\mathbf x_a}, \sigma_a, q_a)$, indicating its position $\mathbf{x_a}$, 
spin $\sigma_a$ and isospin $q_a$ variables. We use the following symbols
\bea
 \si_{\ma} &\equiv& \int d^3 {\mathbf x_a} \sum_{\sigma_a} \sum_{q_a}\,,  \\
 \si_{\ma,\map} &\equiv& \int d^3 {\mathbf x_a} d^3 {\mathbf x_{a'}}
 \sum_{\sigma_a, \sigma_{a'}} \sum_{q_a, q_{a'}} \,, \\ 
 \delta( \ma,\map) &\equiv& \delta({\mathbf x}_{a}-\mathbf{x}_{a'}) 
 \delta(\sigma_a, \sigma_{a'}) \delta(q_a, q_{a'}) \,, \\
  \si_{\ma,\map}^{ \mathbf{ (r)}} &\equiv& \int d^3 {\mathbf x_a} d^3 {\mathbf x_{a'}}
 \sum_{\sigma_a, \sigma_{a'}} \sum_{q_a, q_{a'}} \delta( \mathbf{r} - \mathbf{x}_a) \,.
 \eea
The HF density matrix is written as
\begin{equation}
\hat{\rho}(\map, \ma)=\sum_{k \in F}\phi^{*}_{k}(\map) \phi_{k}(\ma) \, ,
\end{equation}
where  $\phi_{k}$ refers to the HF single particle wave function, and the sum in taken over the HF occupied levels ($F$ means Fermi sea). 
We also introduce ${\text{\boldmath$\nabla_a$}}$ and ${\text{\boldmath$\nabla_{a'}$}}$ for derivation with respect to 
$\mathbf{x}_a$ and $\mathbf{x}_{a'}$ respectively, and ${\text{\boldmath$\sigma$}}_{a' a}$ and 
${\text{\boldmath$\tau$}}_{a' a}$ for the Pauli matrices acting on spin and isospin space, respectively.
We use the notation  $\vec{\rho}$ of \cite{per04} to indicate that we take the full $\tau$ operator in isospin space and not just the 3rd component $\hat{\text{\boldmath{$\tau$}}}^z$.

The different local densities are classified according to their behavior under time reversal and isospin space and 
are written as~\cite{roh10}.

\subsection*{Time-even, isoscalar (TE-IS)}
\be
\rho_{0}(\mathbf{r}) = \si_{\ma,\map}^{ \mathbf{ (r)}} 
 \delta(\map, \ma) \ \hat{\rho}(\map, \ma) \,,
 \ee
 \be
\tau_{0}(\mathbf{r}) = \si_{\ma,\map}^{ \mathbf{ (r)}} 
\delta(\map, \ma) \ 
 \text{\boldmath$\nabla$}_{a'} \cdot  \text{\boldmath$\nabla$}_a \ \hat{\rho}(\map, \ma) \,,
 \ee
 \be
 \mathbb{J}_{0}(\mathbf{r}) = \si_{\ma,\map}^{ \mathbf{ (r)}} 
 \delta(\mathbf{x}_{a'} - \mathbf{x}_a) \delta(q_{a'}, q_a)  
 \frac{1}{2i}\left(\text{\boldmath$\nabla$}_{a'} -\text{\boldmath$\nabla$}_a \right) \otimes  \text{\boldmath$\sigma$}_{a' a} \ \hat{\rho}(\map,\ma) \,,
 \ee
 \be
J_{0}^{(0)}(\mathbf{r}) = \si_{\ma,\map}^{ \mathbf{ (r)}} 
 \delta(\mathbf{x}_{a'} - \mathbf{x}_a) \delta(q_{a'}, q_a)  
 \frac{1}{2i}\left(\text{\boldmath$\nabla$}_{a'} -\text{\boldmath$\nabla$}_a \right)  \text{\boldmath$\sigma$}_{a' a} 
  \ \hat{\rho}(\map,\ma) \,,
 \ee
 \be
\mathbf{J}_{0}(\mathbf{r}) = \si_{\ma,\map}^{ \mathbf{ (r)}}   
\delta(\map, \ma)
 \frac{1}{2i}\left(\text{\boldmath$\nabla$}_{a'} -\text{\boldmath$\nabla$}_a \right) \times  \text{\boldmath$\sigma$}_{a' a} 
 \ \hat{\rho}(\map,\ma) \,.
\ee

\subsection*{Time-even, isovector (TE-IV)}
\be
\vec{\rho}(\mathbf{r}) = \si_{\ma,\map}^{ \mathbf{ (r)}} 
\delta(\mathbf{x}_{a'}-\mathbf{x}_{a}) \delta(\sigma_{a'}, \sigma_a) \ {\text{\boldmath$\tau$}}_{a' a}
\ \hat{\rho}(\map, \ma) \,,
\ee
\be
\vec{\tau}(\mathbf{r}) = \si_{\ma,\map}^{ \mathbf{ (r)}}  
\delta(\mathbf{x}_{a'}-\mathbf{x}_{a}) \delta(\sigma_{a'}, \sigma_a) \ {\text{\boldmath$\tau$}}_{a' a}
 \text{\boldmath$\nabla$}_{a'} \cdot  \text{\boldmath$\nabla$}_a 
 \ \hat{\rho}(\map, \ma) \,,
 \ee
 \be
\vec{\mathbb{J}}(\mathbf{r}) = \si_{\ma,\map}^{ \mathbf{ (r)}}  
\delta(\mathbf{x}_{a'}-\mathbf{x}_{a}) {\text{\boldmath$\tau$}}_{a' a} 
 \frac{1}{2i}\left(\text{\boldmath$\nabla$}_{a'} -\text{\boldmath$\nabla$}_a \right) \otimes  
 \text{\boldmath$\sigma$}_{a' a} \ \hat{\rho}(\map,\ma) \,,
 \ee
 \be
\vec{\mathbf{J}}(\mathbf{r}) = \si_{\ma,\map}^{ \mathbf{ (r)}}  
\delta(\mathbf{x}_{a'}-\mathbf{x}_{a}) \delta(\sigma_a \sigma_{a'}) \ {\text{\boldmath$\tau$}}_{a' a}
 \frac{1}{2i}\left(\text{\boldmath$\nabla$}_{a'} -\text{\boldmath$\nabla$}_a \right) \times 
  \text{\boldmath$\sigma$}_{a' a} \ \hat{\rho}(\map,\ma) \,.
\ee

\subsection*{Time-odd, isoscalar (TO-IS)}
\begin{equation}
\mathbf{s}_{0}(\mathbf{r})= \si_{\ma,\map}^{ \mathbf{ (r)}} 
\delta(\mathbf{x}_{a'}-\mathbf{x}_a) \delta(q_{a'}, q_a) \ 
\text{\boldmath$\sigma$}_{a' a}  \ \hat{\rho}(\map, \ma)\,,
\end{equation}
\begin{equation}
\mathbf{T}_{0}(\mathbf{r})= \si_{\ma,\map}^{ \mathbf{ (r)}} 
\delta(\mathbf{x}_{a'}-\mathbf{x}_a) \delta(q_{a'},q_a) \ \text{\boldmath$\sigma$}_{a' a} \ 
\text{\boldmath$\nabla$}_{a'} \cdot \text{\boldmath$\nabla$}_a \hat{\rho}(\map, \ma)\,,
\end{equation}
\begin{equation}
\mathbf{j}_{0}(\mathbf{r})= \si_{\ma,\map}^{ \mathbf{ (r)}} 
\delta(\mathbf{a'},\mathbf{a}) \ \frac{1}{2i}\left(\text{\boldmath$\nabla$}'-\text{\boldmath$\nabla$}\right)
\ \hat{\rho}(\map, \ma)\,,
\end{equation}
\begin{equation}
\mathbf{F}_{0}(\mathbf{r})= \si_{\ma,\map}^{ \mathbf{ (r)}} 
\delta(\mathbf{x}_{a'}-\mathbf{x}_a) \delta(q_{a'}, q_a) \ 
\frac{1}{2}\left( \text{\boldmath$\nabla$}_a \otimes \text{\boldmath$\nabla$}_{a'} 
+ \text{\boldmath$\nabla$}_{a'} \otimes \text{\boldmath$\nabla$}_a \right) \cdot 
\text{\boldmath$\sigma$}_{a' a} \ \hat{\rho}(\map, \ma)\,.
\end{equation}

\subsection*{Time-odd, isovector (TO-IV)}
\begin{equation}
\vec{\mathbf{s}}(\mathbf{r})= \si_{\ma,\map}^{ \mathbf{ (r)}} 
\delta(\mathbf{x}_{a'}-\mathbf{x}_a) \ \text{\boldmath$\sigma$}_{a' a} 
\ {\text{\boldmath$\tau$}}_{a' a} \ \hat{\rho}(\map, \ma)\,,
\end{equation}
\begin{equation}
\vec{\mathbf{T}}(\mathbf{r})= \si_{\ma,\map}^{ \mathbf{ (r)}} 
\delta(\mathbf{x}_{a'}-\mathbf{x}_a) \ \text{\boldmath$\sigma$}_{a' a} \ {\text{\boldmath$\tau$}}_{a' a} \
\text{\boldmath$\nabla$}_{a'} \cdot \text{\boldmath$\nabla$}_a \ 
\hat{\rho}(\map, \ma)\,,
\end{equation}
\begin{equation}
\vec{\mathbf{j}}(\mathbf{r})= \si_{\ma,\map}^{ \mathbf{ (r)}} 
\delta(\mathbf{x}_{a'}-\mathbf{x}_a) \delta(\sigma_{a'} , \sigma_a) \
{\text{\boldmath$\tau$}}_{a' a} \ 
\frac{1}{2i}\left(\text{\boldmath$\nabla$}_{a'}-\text{\boldmath$\nabla$}_a \right)
\hat{\rho}(\map, \ma)\,.
\end{equation}

For completeness, we give some auxiliary quantities
\begin{equation}
\Delta \rho_{0}(\mathbf{r})= \si_{\ma,\map}^{ \mathbf{ (r)}} 
\delta(\map,\ma) 
\left(\text{\boldmath$\nabla$}^{2}_{a'} + 2 \text{\boldmath$\nabla$}_{a'} \cdot \text{\boldmath$\nabla$}_a 
+ \text{\boldmath$\nabla$}^{2}_a \right) \, \hat{\rho}(\map,\ma)\,,
\end{equation}
\bea
\text{\boldmath$\nabla$} \cdot \mathbf{J}_{0}(\mathbf{r}) &=& \si_{\ma,\map}^{ \mathbf{ (r)}} 
\delta(\mathbf{x}_{a'}-\mathbf{x}_a) \delta(q_{a'},q_a) \nnn
&& \frac{1}{2i} \left[ \text{\boldmath$\nabla$}_{a'} \cdot \left(\text{\boldmath$\nabla$}_a \times 
\text{\boldmath$\sigma$}_{a' a} \right) 
- \text{\boldmath$\nabla$}_a \cdot \left(\text{\boldmath$\nabla$}_{a'} \times 
\text{\boldmath$\sigma$}_{a' a} \right)\right] \, \hat{\rho}(\map,\ma)\,,
\eea
\begin{equation}
\text{\boldmath$\nabla$} \cdot \mathbf{s}_{0}(\mathbf{r})= \si_{\ma,\map}^{ \mathbf{ (r)}} 
\delta(\mathbf{x}_{a'}-\mathbf{x}_a) \delta(q_{a'},q_a)
\left[\left(\text{\boldmath$\nabla$}_{a'} + \text{\boldmath$\nabla$}_a \right) \cdot \text{\boldmath$\sigma$}_{a' a}\right] 
\, \hat{\rho}(\map,\ma)\,,
\end{equation}
\begin{equation}
\text{\boldmath$\nabla$} \times \mathbf{j}_{0}(\mathbf{r})= \si_{\ma,\map}^{ \mathbf{ (r)}} 
\delta(\map,\ma)  \left( -i \text{\boldmath$\nabla$}_{a'} \times \text{\boldmath$\nabla$}_a \right) 
\, \hat{\rho}(\map,\ma)\,,
\end{equation}
\begin{equation}
\text{\boldmath$\Delta$} \mathbf{s}_{0}(\mathbf{r})= \si_{\ma,\map}^{ \mathbf{ (r)}} 
\delta(\mathbf{x}_{a'}-\mathbf{x}_a) \delta(q_{a'},q_a) 
\left(\text{\boldmath$\nabla$}^{2}_{a'} + 2 \text{\boldmath$\nabla$}_{a'} \cdot \text{\boldmath$\nabla$}_a 
+\text{\boldmath$\nabla$}^{2}_a \right) \text{\boldmath$\sigma$}_{a' a}
\, \hat{\rho}(\map,\ma)\,.
\end{equation}
The corresponding isovector quantities can be obtained in a similar way.

\section{Coupling constants and Skyrme parameters}
\label{coupling-SK}

We suppose that the second order functional is derived from an effective Skyrme interaction
\begin{align}\label{Skyrme:force}
V^{Sk}
&= t_{0}(1+x_{0}\hat{P}_{\sigma})\delta(\mathbf{r})+ \tfrac{1}{6}\,t_{3}(1+x_{3}\hat{P}_{\sigma})
\rho_{0}^{\gamma}(\mathbf{R})\delta(\mathbf{r})\nonumber\\
&+ \tfrac{1}{2}t_{1}(1+x_{1}\hat{P}_{\sigma})
\left[ {\mathbf{k}\smash{'}}^2\delta(\mathbf{r})
+\delta(\mathbf{r})\mathbf{k}^{2}\right]+ t_{2}(1+x_{2}\hat{P}_{\sigma}) \mathbf{k}'\cdot
 \delta(\mathbf{r})\mathbf{k}\nonumber\\
&+ \mathrm{i}\,W_{0}(\hat{\boldsymbol\sigma}_{1}
+\hat{\boldsymbol\sigma}_{2})
\cdot \left[ \mathbf{k}'\times \delta(\mathbf{r})\mathbf{k}\right]\nonumber\\
&+\frac{1}{2}t_{e} \left\{ 
\left[3(\hat{\boldsymbol\sigma}_{1}\cdot \mathbf{k}')(\hat{\boldsymbol\sigma}_{2}\cdot \mathbf{k}')-(\hat{\boldsymbol\sigma}_{1}\cdot \hat{\boldsymbol\sigma}_{2} ) \mathbf{k}^{'2}\right]\delta(\mathbf{r})
+\delta(\mathbf{r}) \left[ 3(\hat{\boldsymbol\sigma}_{1}\cdot \mathbf{k})(\hat{\boldsymbol\sigma}_{2}\cdot \mathbf{k})-(\hat{\boldsymbol\sigma}_{1}\cdot \hat{\boldsymbol\sigma}_{2} ) \mathbf{k}^{2}\right]\right\} \nonumber\\
&+\frac{1}{2}t_{o}\left\{
\left[3(\hat{\boldsymbol\sigma}_{1}\cdot \mathbf{k}') \delta(\mathbf{r})  (\hat{\boldsymbol\sigma}_{2}\cdot \mathbf{k})-(\hat{\boldsymbol\sigma}_{1}\cdot \hat{\boldsymbol\sigma}_{2} ) \mathbf{k}^{'}\delta(\mathbf{r})\mathbf{k}\right]
+ \left[ 3(\hat{\boldsymbol\sigma}_{1}\cdot \mathbf{k}) \delta(\mathbf{r})  (\hat{\boldsymbol\sigma}_{2}\cdot \mathbf{k}')-(\hat{\boldsymbol\sigma}_{1}\cdot \hat{\boldsymbol\sigma}_{2} ) \mathbf{k}\delta(\mathbf{r}) \mathbf{k}'\right] \right\}\,,
\end{align}
where we used $\mathbf{r}\equiv \mathbf{r}_{1}-\mathbf{r}_{2}$ and
$\mathbf{R}=\frac{1}{2}\left(\mathbf{r}_{1}+\mathbf{r}_{2}\right)$ for
the relative and center of mass coordinates, respectively.
We also defined $\hat{P}_{\sigma}$ the spin exchange operator, while
$\mathbf{k}=-\frac{\mathrm i}{2}(\nabla_{1}-\nabla_{2})$ is the relative momentum
acting on the right and $\mathbf{k}'$ its complex conjugate acting on the left.
Finally $\rho_{0}(\mathbf{R})$ is the scalar-isoscalar density of the system.

\noindent We can thus express the coupling constants in terms of Skyrme parameters

\begin{eqnarray*}
C^{\rho}_{0}&=&\frac{3}{8}t_{0}+\frac{3}{48}t_{3}\rho_{0}^{\gamma}(\mathbf{r})\,, \\
C^{\rho}_{1}&=&-\frac{1}{4}t_{0}\left( \frac{1}{2}+x_{0}\right)-\frac{1}{24}t_{3}\left( \frac{1}{2}+x_{3}\right)\rho_{0}^{\gamma}(\mathbf{r})\,,\\
C^{\Delta \rho}_{0}&=&-\frac{9}{64}t_{1}+\frac{1}{16}t_{2}\left( \frac{5}{4}+x_{2}\right)\,,\\
C^{\Delta \rho}_{1}&=&\frac{3}{32}t_{1}\left(\frac{1}{2}+x_{1} \right)+\frac{1}{32}t_{2}\left(\frac{1}{2}+x_{2} \right)\,, \\
C^{\tau}_{0}&=&\frac{3}{16}t_{1}+\frac{1}{4}t_{2}\left(\frac{5}{4}+x_{2} \right)\,,\\
C^{\tau}_{1}&=&-\frac{1}{8}t_{1}\left( \frac{1}{2}+x_{1}\right)+\frac{1}{8}t_{2}\left( \frac{1}{2}+x_{2}\right)\,,
\end{eqnarray*}
\begin{eqnarray*}
C^{s}_{0}&=&-\frac{1}{4}t_{0}\left( \frac{1}{2}-x_{0}\right) -\frac{1}{24}t_{3}\left( \frac{1}{2}-x_{3}\right) \rho_{0}^{\gamma}(\mathbf{r})\,, \\
C^{s}_{1}&=&-\frac{1}{8}t_{0}-\frac{1}{48}t_{3}\rho_{0}^{\gamma}(\mathbf{r})\,,\\
C^{\Delta s}_{0}&=&\frac{3}{32}t_{1}\left( \frac{1}{2}-x_{1}\right)+\frac{1}{32}t_{2}\left( \frac{1}{2}+x_{2}\right)+\frac{3}{32}(t_{e}-t_{o})\,,\\
C^{\Delta s}_{1}&=&\frac{3}{64}t_{1}+\frac{1}{64}t_{2}-\frac{1}{32}(3t_{e}+t_{o})\,,\\
C^{T}_{0}&=&-\frac{1}{8}t_{1}\left( \frac{1}{2}-x_{1}\right)+\frac{1}{8}t_{2}\left( \frac{1}{2}+x_{2}\right)-\frac{1}{8}(t_{e}+3t_{o})\,,\\
C^{T}_{1}&=&-\frac{1}{16}t_{1}+\frac{1}{16}t_{2}+\frac{1}{8}(t_{e}-t_{o}) \,,
\end{eqnarray*}
\begin{eqnarray*}
C^{F}_{0}&=&\frac{3}{8}(t_{e}+3t_{o})\,,\\
C^{F}_{1}&=&-\frac{3}{8}(t_{e}-t_{o})\,,\\
C^{\nabla s}_{0}&=&\frac{9}{32}(t_{e}-t_{o})\,,\\
C^{\nabla s}_{1}&=&-\frac{3}{32}(3t_{e}+t_{o})\,,\\
C^{\nabla J}_{0}&=&-\frac{3}{4}W_{0}\,,\\
C^{\nabla J}_{1}&=&-\frac{1}{4}W_{0}\,.
\end{eqnarray*}

\noindent It is also useful to introduce a slightly different notation when we deal with density dependent coupling constants

\begin{eqnarray}\label{notation:cc}
C^{\rho}_{t}[\rho]&=&C^{\rho,0}_{t}+C^{\rho,\gamma}_{t} \rho^{\gamma}\,,\\
C^{s}_{t}[\rho]&=&C^{s,0}_{t}+C^{s,\gamma}_{t} \rho^{\gamma}\,,
\end{eqnarray}

\noindent with $t=0,1$.
Here we give the coupling constant for the extra terms of the Brussels-Montreal~\cite{gor10,cha10c} interactions given in Eq.\ref{extra:bsk:func}.

\begin{eqnarray}
\label{app:bru1}
C^{\Delta \rho,\beta_{1}}_{0}\rho^{\beta_{1}}&=& -\frac{3t_{1b}}{32}\rho^{\beta_{1}}\,,\\
C^{\Delta \rho,\beta_{1}}_{1}\rho^{\beta_{1}}&=&\frac{t_{1b}}{16}  \left(\frac{1}{2}+x_{1b} \right)\rho^{\beta_{1}}\,,\\
C^{\tau,\beta_{1}}_{0}\rho^{\beta_{1}}+C^{\tau,\beta_{2}}_{0}\rho^{\beta_{2}}&=& 3\frac{t_{1b}}{16} \rho^{\beta_{1}}+\frac{t_{2b}}{4}\rho^{\beta_{2}}\left[ \frac{5}{4}+x_{2b} \right]\,,\\
C^{\tau,\beta_{1}}_{1}\rho^{\beta_{1}}+C^{\tau,\beta_{2}}_{1}\rho^{\beta_{2}}&=&- \frac{t_{1b}}{8}  \left(\frac{1}{2}+x_{1b} \right)\rho^{\beta_{1}}+\frac{t_{2b}}{8} \rho^{\beta_{2}} \left[\frac{1}{2}+x_{2b} \right]\,,\\
C^{\nabla \rho,\beta_{1}}_{0}\rho^{\beta_{1}}+C^{\nabla \rho,\beta_{2}}_{0}\rho^{\beta_{2}}&=&3\frac{t_{1b}}{64}  \rho^{\beta_{1}}-\frac{t_{2b}}{16} \rho^{\beta_{2}}\left[ \frac{5}{4}+x_{2b} \right]\,,\\
\label{app:bru2}
C^{\nabla \rho,\beta_{1}}_{1}\rho^{\beta_{1}}+C^{\nabla \rho,\beta_{2}}_{1}\rho^{\beta_{2}}&=&- \frac{t_{1b}}{32} \left(\frac{1}{2}+x_{1b} \right)\rho^{\beta_{1}}-\frac{t_{2b}}{32} \rho^{\beta_{2}} \left[\frac{1}{2}+x_{2b} \right]\,.
\end{eqnarray}

Here we give the explicit relation between coupling constants and Skyrme terms in the case of extended density dependence as given in ref.~\cite{mar09c}

\begin{eqnarray}
\label{app:spi1}
C^{s\rho}_{0}&=& \frac{1}{16}t_{3}^{s}\,,\\
C^{s\rho}_{1}&=&-\frac{1}{48}t_{3}^{st}(2x_{3}^{st}+1)\,,\\
C^{ss}_{0}&=&\frac{1}{48}t_{3}^{s}(2x_{3}^{s}-1)\,,\\
C^{ss}_{1}&=&-\frac{1}{48}t_{3}^{st}\,,\\
C^{s\rho}_{10}&=&\frac{1}{16}t_{3}^{st}\,,\\
C^{s\rho}_{01}&=&-\frac{1}{48}t_{3}^{s}\,,\\
\label{app:spi2}
C^{ss}_{10}+C^{ss}_{01}&=&-\frac{1}{24}t_{3}^{st}x_{3}^{st}\,.
\end{eqnarray}

\section{$\beta$ functions}
\label{app:beta}
\setcounter{table}{0}
\setcounter{figure}{0}

As already discussed in the text, the non-interacting ground state can describe via a $ph$ propagator of the form

\begin{equation}\label{app:beta:GHF}
G^{(\tau\tau')}_{HF}(\mathbf{k},\mathbf{q},\omega,T)=\frac{n_{\tau}(\mathbf{k})-n_{\tau'}(\mathbf{k}+\mathbf{q})}{\omega-\left[ \varepsilon_{\tau'}(\mathbf{k}+\mathbf{q})- \varepsilon_{\tau}(\mathbf{k}) \right]+i\eta} \, ,
\end{equation}
where $\tau (\tau')$ are used to identify the hole (particle). In our paper, they specify the isospin: they can be equal ($i.e.$ $\tau=\tau'=n(p)$) as in the case of SNM, PNM or ANM with no charge-exchange processes or different ($i.e.$ $\tau=n(p),\tau'=p(n)$) when charge-exchange processes are involved. They can also represent spin index ($i.e.$ $\tau,\tau'=\pm$) if one considers polarized neutron matter by instance. Formula derived in this appendix are therefore general and can be useful for totally different systems. In the above equation, $n_{\tau}(\mathbf{k})$ is the usual Fermi distribution
\begin{equation}\label{app:beta:FDdistribution}
n_{\tau}(\mathbf{k})=\left[ 1+e^{(\varepsilon_{\tau}(k)-\mu_{\tau})/T}\right]^{-1} \,,
\end{equation}
which reduces to the step function $\theta(\mathbf{k}^{\tau}_{F}-\mathbf{k})$ at zero temperature.
All the notations are standard: $T$ stands for the temperature and $\mathbf{k}^{\tau}_{F}$ ($\mu_\tau$) for the Fermi momentum (chemical potential) of species $\tau$. Moreover, the single particle energy $\varepsilon_{\tau}(\mathbf{k})$ is given by 
\begin{equation}
\varepsilon_{\tau}(\mathbf{k})=\frac{\mathbf{k}^{2}}{2m_{\tau}}+U_{\tau} \, ,
\end{equation}
with $m_{\tau}$ the effective mass and $U_{\tau}$ the central potential obtained with the Skyrme functional.

As explained in the main text, the resolution of Bethe-Salpeter equations provides, as intermediate steps, expectation values of the type
\begin{equation}
\langle |\mathbf{k}|^a Y_{10}(\hat{k})^b |Y_{11}(\hat{k})|^{2c}G^{(\tau,\tau')}_{HF}(\mathbf{k},q,\omega,T) \rangle \equiv \int \frac{d^{3}\mathbf{k}}{(2\pi)^{3}} |\mathbf{k}|^a Y_{10}(\hat{k})^b |Y_{11}(\hat{k})|^{2c}G^{(\tau,\tau')}_{HF}(\mathbf{k}_{},q,\omega,T) \, .
\end{equation}
Such integrals can actually be re-expressed in terms of the $\beta^{(\tau,\tau')}_{i}$ functions, defined later on. The cases encountered in our calculations are written as
\begin{equation}
\label{abc}
\langle |\mathbf{k}|^a Y_{10}(\hat{k})^b |Y_{11}(\hat{k})|^{2c}G^{(\tau,\tau')}_{HF}(\mathbf{k},q,\omega,T) \rangle \equiv q^a \left(\frac{3}{4\pi}\right)^{b/2}\left(\frac{3}{8\pi}\right)^{c}\sum_{i=0}^8 x_i \, \beta^{(\tau,\tau')}_{i} \, ,
\end{equation}
with the coefficients $x_i$ listed in Table \ref{txi}. 
\begin{table}[H]
\begin{center}
\begin{tabular}{|c|c|c|c|c|c|c|c|c|c|} \hline
  (a,b,c)         &  $x_0$ & $x_1$ & $x_2$ & $x_3$ & $x_4$ & $x_5$ & $x_6$ & $x_7$ & $x_8$ \\ \hline
  (0,0,0)         &  1 & 0 & 0 & 0 & 0 & 0 & 0 & 0 & 0 \\ \hline
  (1,1,0)         &  0 & 1 & 0 & 0 & 0 & 0 & 0 & 0 & 0 \\ \hline
  (2,0,0)         &  0 & 0 & 1 & 0 & 0 & 0 & 0 & 0 & 0 \\ \hline
  (2,2,0)         &  0 & 0 & 0 & 1 & 0 & 0 & 0 & 0 & 0 \\ \hline
  (3,1,0)         &  0 & 0 & 0 & 0 & 1 & 0 & 0 & 0 & 0 \\ \hline
  (4,0,0)         &  0 & 0 & 0 & 0 & 0 & 1 & 0 & 0 & 0 \\ \hline
  (3,3,0)         &  0 & 0 & 0 & 0 & 0 & 0 & 1 & 0 & 0 \\ \hline
  (4,4,0)         &  0 & 0 & 0 & 0 & 0 & 0 & 0 & 1 & 0 \\ \hline
  (4,2,0)         &  0 & 0 & 0 & 0 & 0 & 0 & 0 & 0 & 1 \\ \hline
  (2,0,1)         &  0 & 0 & 1 & -1 & 0 & 0 & 0 & 0 & 0 \\ \hline
  (3,1,1)         &  0 & 0 & 0 &  1 & 0 & -1 & 0 & 0 & 0 \\ \hline
  (4,0,1)         &  0 & 0 & 0 &  0 & 0 & 1 & 0 & 0 & -1 \\ \hline
  (4,2,1)         &  0 & 0 & 0 &  0 & 0 & 0 & 0 & -1 & 1 \\ \hline
  (4,0,2)         &  0 & 0 & 0 &  0 & 0 & 1 & 0 & 1 & -2 \\ \hline
\end{tabular}
\caption{Coefficients $a, b, c$ for HF averages entering Eq. (\ref{abc}).}
\label{txi}
\end{center}
\end{table}

The $\beta^{(\tau,\tau')}_{i}$ functions are the following integrals 
\be
\label{app:beta:betafunct}
\beta_{i}^{(\tau,\tau')}(q,\omega,T) =\int \frac{d^{3} \mathbf{k} }{(2\pi)^{3}}G_{HF}^{(\tau,\tau')}(\mathbf{k},\mathbf{q},\omega,T)F_{i}(\mathbf{k},\mathbf{q})\,,
\ee
where
\be
F_{i=0,8}(\mathbf{k},\mathbf{q}) \equiv 1,\frac{\mathbf{k}\cdot \mathbf{q}}{q^{2}},\frac{k^{2}}{q^{2}},\left[\frac{\mathbf{k}\cdot \mathbf{q}}{q^{2}}\right]^{2},\frac{(\mathbf{k}\cdot \mathbf{q})k^{2}}{q^{4}},\frac{k^{4}}{q^{4}},\left[\frac{\mathbf{k}\cdot \mathbf{q}}{q^{2}}\right]^{3},\left[\frac{\mathbf{k}\cdot \mathbf{q}}{q^2}\right]^{4},\frac{(\mathbf{k}\cdot \mathbf{q})^{2}k^{2}}{q^{6}}.\nonumber
\ee
It is thus interesting to provide some analytic formula, when possible, for these integrals. A thorough examination
actually shows that Eq.~\ref{app:beta:betafunct} cannot be solved completely analytically when we work in the general case of asymmetric matter (or polarized neutron matter) at finite temperature. However, as already shown \cite{gar92, dav09,pas12}, in the limit case of zero temperature and for the two extreme cases of symmetric nuclear matter and pure neutron matter, analytic expressions can be obtained. This will be discussed at the end of this appendix. In the following we show how to deal with the general case.

Since in Eq.~\ref{app:beta:GHF}, the limit ${\eta\rightarrow 0^{+}}$ is as usual implicit, the imaginary part can be immediately written as 
\begin{equation}
{\rm Im} \left[\beta_{i}^{(\tau,\tau')}(q,\omega,T)\right] = -\pi \int_{0}^{\infty} \frac{k^{2} dk}{(2\pi)^{2}} 
\int_{-1}^{1}dx \; F_{i}(\mathbf{k},\mathbf{q}) \; \left[ n_{\tau}(\mathbf{k})-n_{\tau'}(\mathbf{k}+\mathbf{q})\right]\delta\left(\omega- \varepsilon_{\tau'}(\mathbf{q}+\mathbf{k})+ \varepsilon_{\tau}(\mathbf{k})\right) \,,
\label{app:partie-im}
\end{equation}
where $x = {\hat q} \cdot {\hat k}$.

By inspecting the argument of the Dirac distribution, we note an angular dependence. It induces a constraint on the possible range for the momentum so that the integral is not zero. The following limits come out naturally

\begin{eqnarray}\label{app:beta:e12n}
\varepsilon^{\tau}_{1,2}&=&\frac{q^{2}m_{\tau}}{2(\delta m)^{2}}\left[ 1\pm \sqrt{1-\frac{2k_{-}^{\tau'}\delta m}{qm_{\tau}}}\right]^{2}\,,\\
\varepsilon^{\tau'}_{1,2}&=&\frac{q^{2}m_{\tau'}}{2(\delta m)^{2}}\left[ 1\pm \sqrt{1-\frac{2(k_{-}^{\tau}+q)\delta m}{qm_{\tau}}}\right]^{2}\,,
\end{eqnarray}
where  $\delta m=m_{\tau'}-m_{\tau}$, $k_{-}^{\tau}=\frac{m_{\tau} \tilde{\omega}}{q}-\frac{q}{2}$ and $\tilde{\omega}=\omega-\left(U_{\tau'}-U_{\tau}\right)$. 
The above limits are defined only if
\begin{eqnarray}
\label{effmcon1}
1-\frac{2k_{-}^{\tau'}\delta m}{qm_{\tau}} > 0\,,\\
\label{effmcon2}
1-\frac{2(k_{-}^{\tau}+q)\delta m}{qm_{\tau'}} > 0.
\end{eqnarray}
When these inequalities are not satisfied, imaginary parts given in Eq. \ref{app:partie-im} are zero.
With these notations, it is straightforward to put the result under the following form

\begin{equation}
{\rm Im} \left[\beta_{i}^{(\tau,\tau')}(q,\omega,T)\right] =  -\frac{m_{\tau}^{*}m_{\tau'}^{*}}{4\pi q}\left[\int_{min(\varepsilon^{\tau}_{1},\varepsilon^{\tau}_{2})}^{max(\varepsilon^{\tau}_{1},\varepsilon^{\tau}_{2})}  n_{\tau}(\varepsilon) \; F_i^{\tau} \; d \varepsilon 
-\int_{min(\varepsilon^{\tau'}_{1},\varepsilon^{\tau'}_{2})}^{max(\varepsilon^{\tau'}_{1},\varepsilon^{\tau'}_{2})}  n_{\tau'}(\varepsilon) \; F_i^{\tau'} \;d \varepsilon\right]\,,
\label{app:partie-im}
\end{equation}
where functions $F_i^{\tau}$, $F_i^{\tau'}$ are summarized in Table \ref{tableFF}.

\begin{figure}[H]
\begin{center}
  \includegraphics[clip,scale=.4,angle=-90]{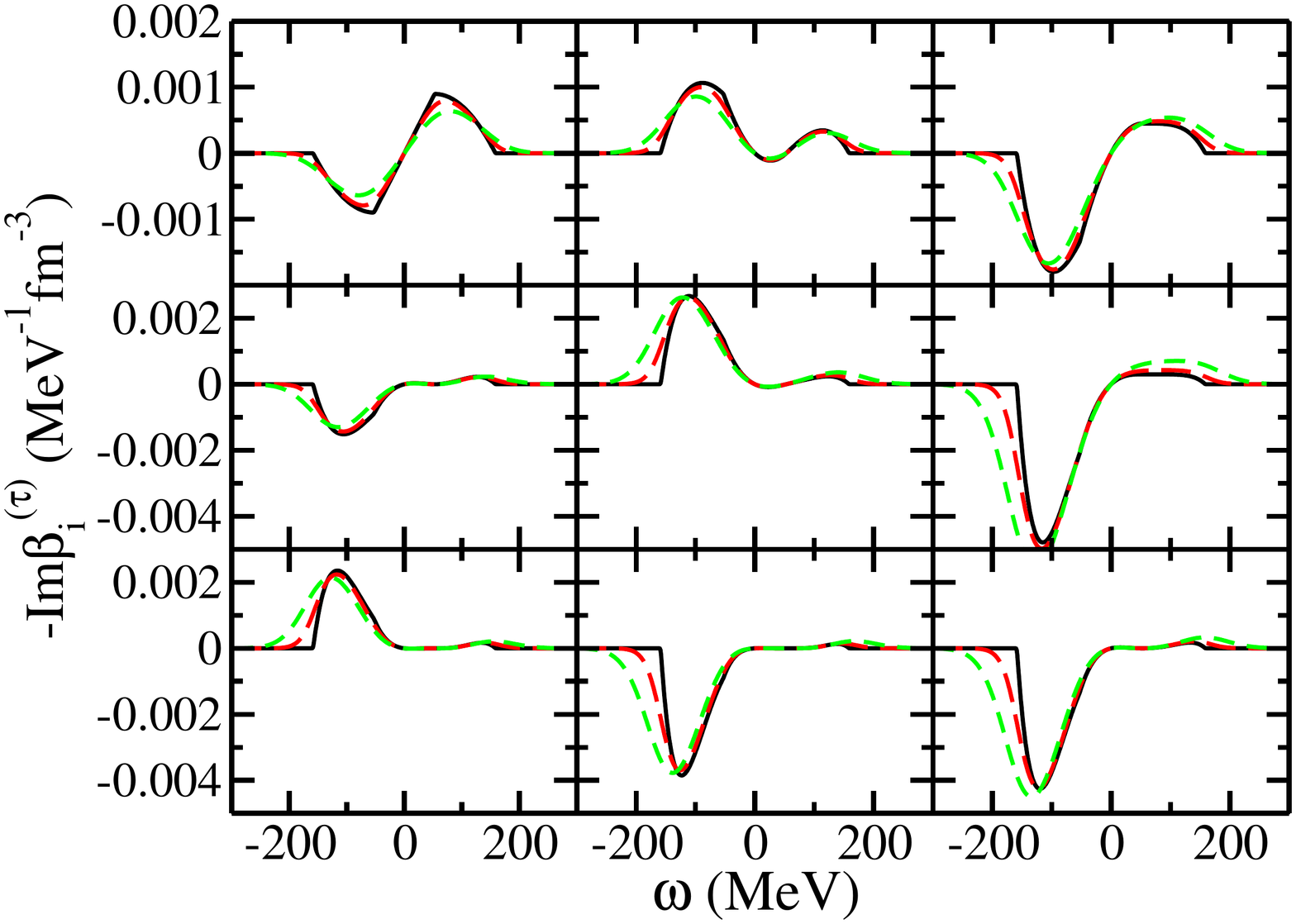}  
      \includegraphics[clip,scale=.4,angle=-90]{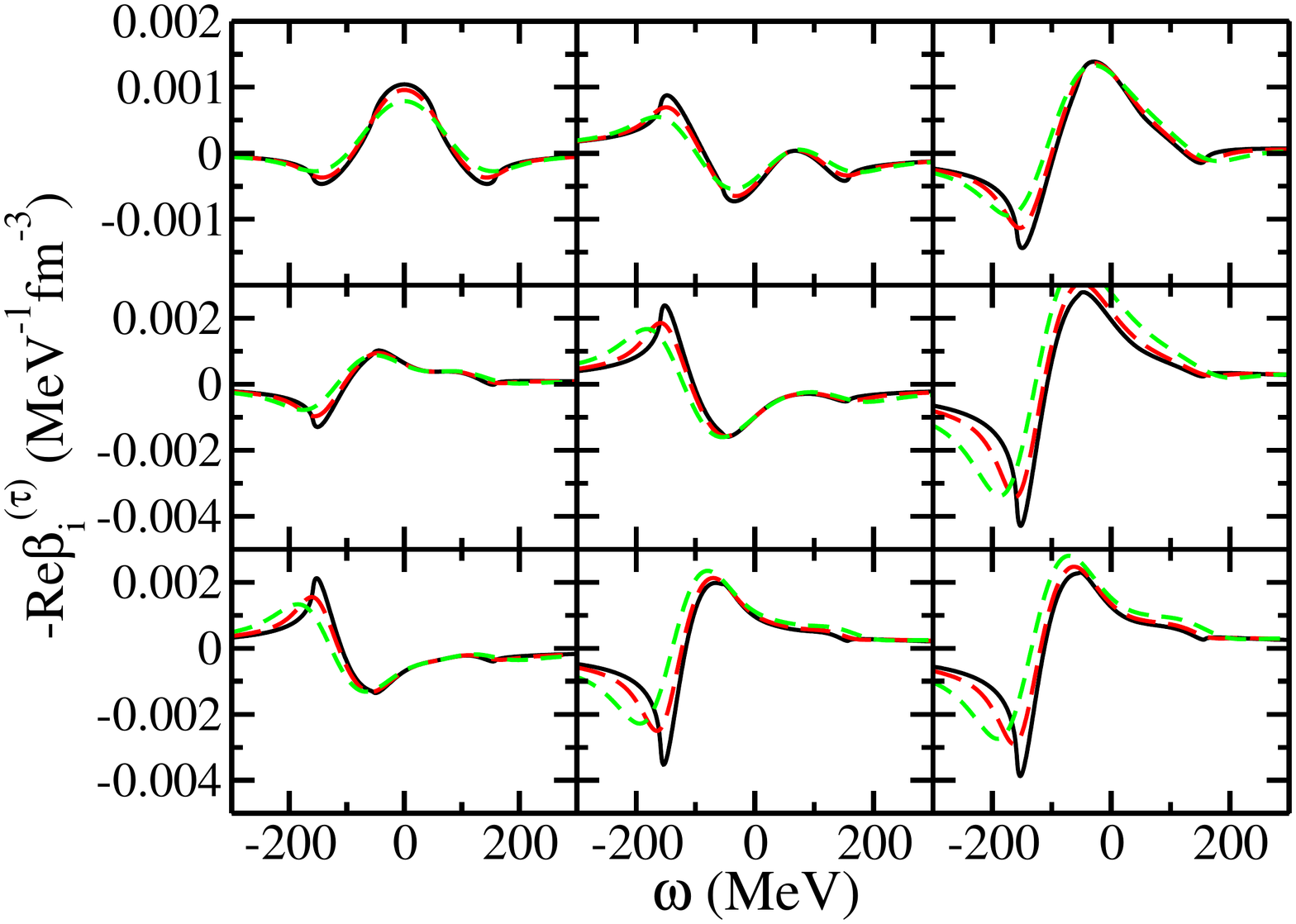}  
\caption{(Color online) Imaginary and real parts of functions $\beta_{i=0,..,8}^{(\tau)}$ (from left to right and top to bottom) for SLy5 interaction, calculated at $\rho=0.16$ fm$^{-3}$ and $q=k_F$, for temperatures
$T=0, 0.25$ and $0.5 \varepsilon_F$ (full, long-dashed and dashed lines respectively).}
\label{app:imbeta-T}
\end{center}
\end{figure}

\begin{figure}[H]
\begin{center}
  \includegraphics[clip,scale=.5,angle=-90]{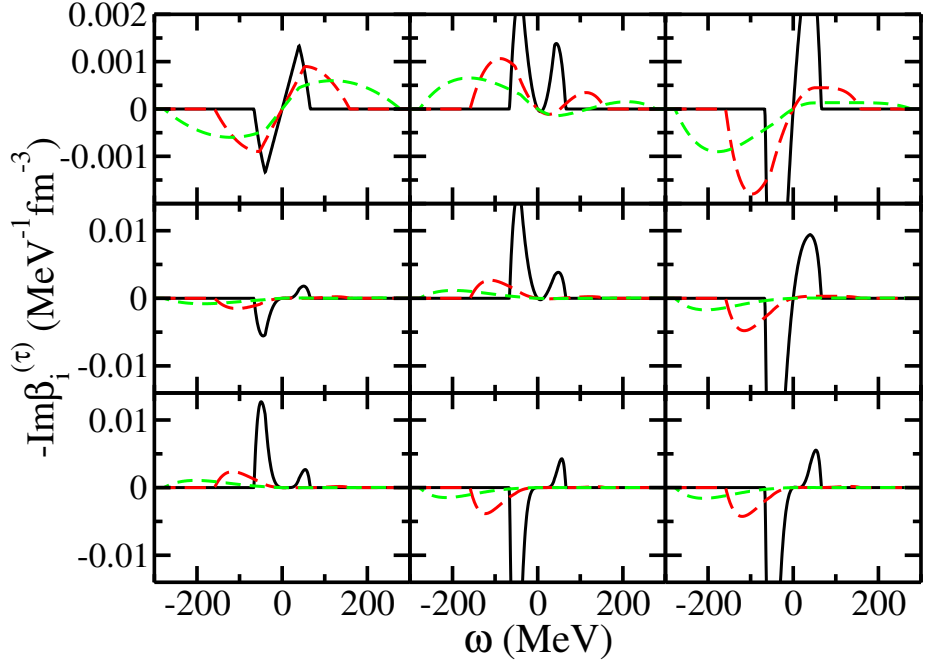}  
      \includegraphics[clip,scale=.5,angle=-90]{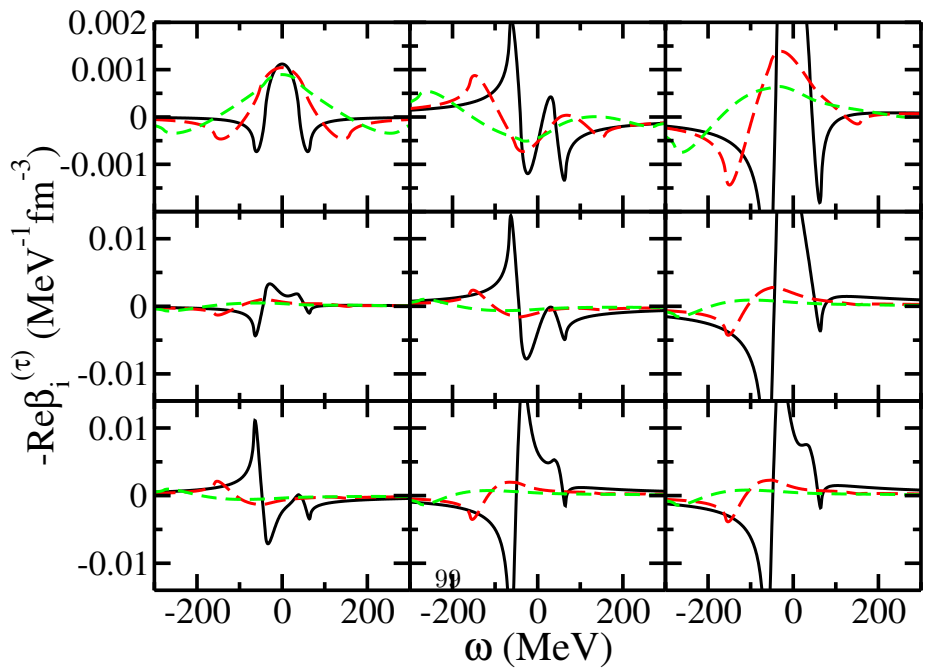}  
\caption{(Color online) Imaginary and real parts of functions $\beta_{i=0,..,8}^{(\tau)}$ (from left to right and top to bottom) for SLy5 interaction, calculated at $\rho=0.16$ fm$^{-3}$ and $T=0$, for transferred momenta
$q=0.5, 1$ and $1.5 k_F$ (full, long-dashed and dashed lines respectively).}      
\label{app:imbeta-q}
\end{center}
\end{figure}

\begin{figure}[H]
\begin{center}
  \includegraphics[clip,scale=.5,angle=-90]{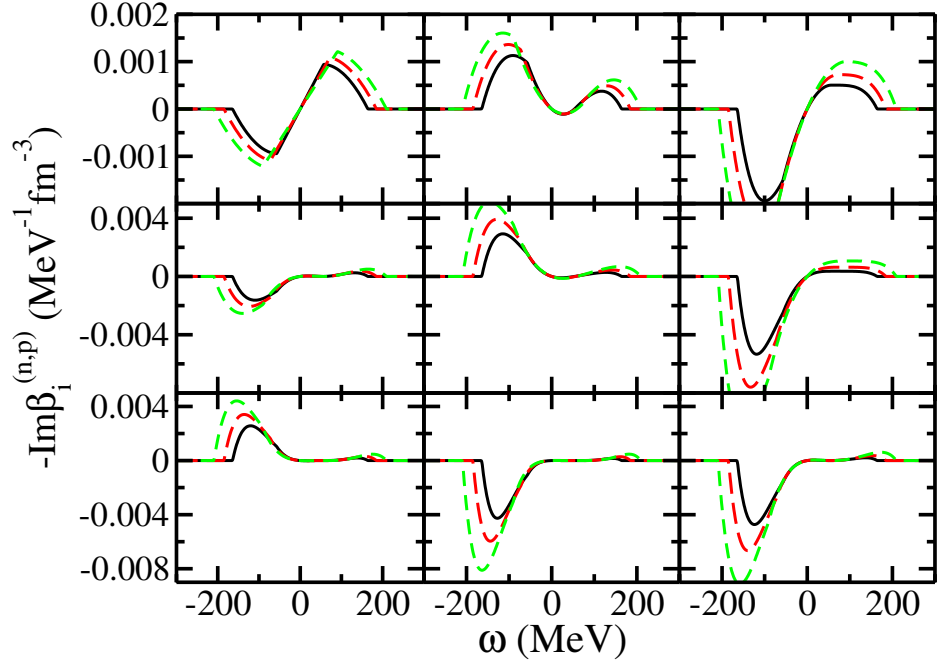}  
     \includegraphics[clip,scale=.5,angle=-90]{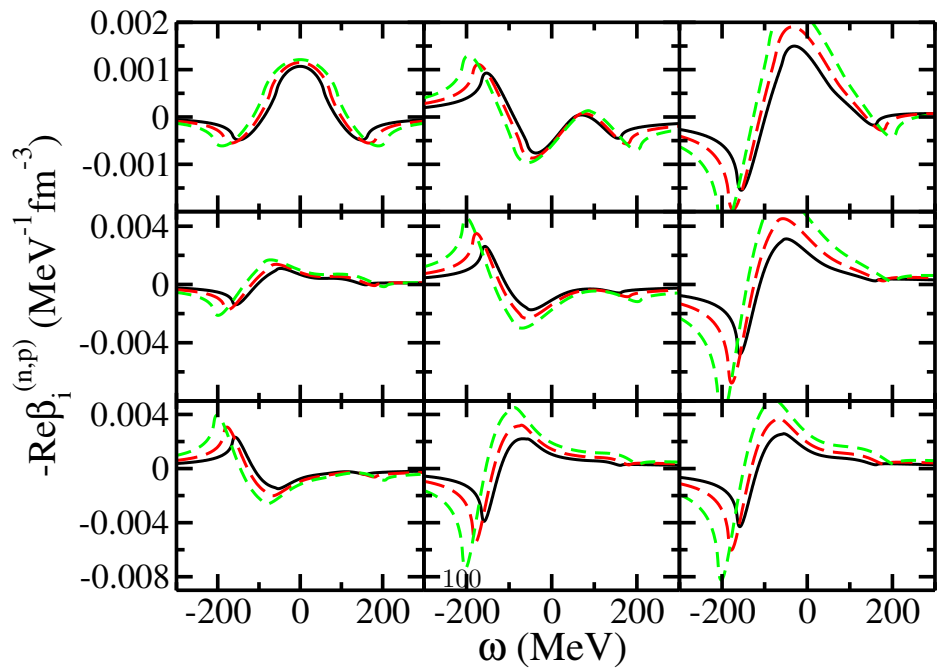}  
\caption{(Color online) Imaginary and real parts of functions $\beta_{i=0,..,8}^{(n,p)}$ (from left to right and top to bottom) for SLy5 interaction, calculated at $\rho=0.16$ fm$^{-3}$, $T=0$, $q=k_F$ for asymmetries 
$Y=0.1, 0.5$ and $1$ (full, long-dashed and dashed lines respectively) .}
\label{app:imbeta-y}
\end{center}
\end{figure}

\begin{table}[H]
\begin{center}
\begin{tabular}{c|c|c|c|c|c|c|c|c|c|} \cline{2-10}
           &  0 & 1 & 2 & 3 & 4 & 5 & 6 & 7 & 8  \\ \hline
    \multicolumn{1}{|c|}{}       &   &  &  &  &  &  &  &  &   \\
 \multicolumn{1}{|c|}{$F_i^{\tau}$}  & 1 & $K^{\tau'}$ & $\dfrac{2m_{\tau}}{q^2}\varepsilon$ & $\left[K^{\tau'} \right]^{2}$ & $\dfrac{2m_{\tau}}{q^{2}}\varepsilon K^{\tau'}$  & $\left[\dfrac{2m_{\tau}}{q^2}\varepsilon\right]^2$  & $\left[K^{\tau'} \right]^{3}$ & $\left[K^{\tau'} \right]^{4}$  & $\dfrac{2m_{\tau}}{q^{2}}\varepsilon\left[K^{\tau'} \right]^2$ 
 \\ 
     \multicolumn{1}{|c|}{}       &   &  &  &  &  &  &  &  &   \\ \hline
     \multicolumn{1}{|c|}{}       &   &  &  &  &  &  &  &  &   \\ 
\multicolumn{1}{|c|}{$F_i^{\tau'}$}  & 1 & $K^{\tau}$ & $\dfrac{2m_{\tau}}{q^2}(\varepsilon-\tilde{\omega})$  & $\left[K^{\tau} \right]^{2}$ & $\dfrac{2m_{\tau}}{q^{2}} (\varepsilon-\tilde{\omega})K^{\tau}$ & $\left[\dfrac{2m_{\tau}}{q^2}(\varepsilon-\tilde{\omega})\right]^2$& $\left[K^{\tau} \right]^{3}$ & $\left[K^{\tau}\right]^{4}$ & $\dfrac{2m_{\tau}}{q^{2}} (\varepsilon-\tilde{\omega})\left[K^{\tau}\right]^2$ \\ 
     \multicolumn{1}{|c|}{}       &   &  &  &  &  &  &  &  &   \\ \hline
\end{tabular}
\caption{Functions entering the imaginary part of $\beta_{i=0,..,8}^{(\tau,\tau')}$, with 
$K^\tau\equiv \dfrac{k_{-}^{\tau}q+ \delta m \; \varepsilon }{q^{2}}$. 
}
\label{tableFF}
\end{center}
\end{table}
Real parts can be then obtained though the dispersion relation 
\begin{equation}
{\rm Re}\left[\beta_{i}^{(\tau,\tau')}(q,\omega,T)\right]=-\frac{1}{\pi} \int_{-\infty}^{+\infty} d\omega' \frac{{\rm Im}\left[\beta_{i}^{(\tau,\tau')}(q,\omega',T)\right]}{\omega-\omega'}.
\end{equation}
Some typical curves obtained when $(\tau,\tau')=(p,n)$ are displayed on figures \ref{app:imbeta-T}--\ref{app:imbeta-y}  and will be discussed at the end of this appendix.

A very useful limit is $\tau=\tau'$. In this case, $\delta m =0$ (thus $m_\tau = m_{\tau'} \equiv m^*$) and $\tilde \omega = \omega$ so that
\begin{eqnarray}
\lim_{\delta m \rightarrow 0} \varepsilon_{\tau}^{1}&=&+\infty\,,\\
\lim_{\delta m \rightarrow 0} \varepsilon_{\tau}^{2}&=&\frac{1}{2m^{*}}\left[ \frac{m^{*}\omega}{q}-\frac{q}{2}\right]^{2}\equiv \varepsilon_{-}^{\tau}\,,\\
 \lim_{\delta m \rightarrow 0} \varepsilon_{\tau'}^{1}&=&+\infty\,,\\
\lim_{\delta m \rightarrow 0} \varepsilon_{\tau'}^{2} &=&\frac{1}{2m_{*}}\left[\frac{m^{*}\omega}{q}+\frac{q}{2} \right]^{2}\equiv \varepsilon_{+}^{\tau}\,.
\end{eqnarray}
The imaginary parts are highly simplified in that case. With the notation 
$\beta_{i}^{(\tau,\tau)} \equiv \beta_{i}^{(\tau)}$, we give the detailed expressions 
\begin{eqnarray}
{\rm Im}(\beta_{0}^{(\tau)}(q,\omega,T))&=&\frac{(m^{*})^{2}}{4\pi q}T\ln \left( \frac{1+\exp^{-\frac{\varepsilon_{+}^{\tau}-\mu_{\tau}}{T}}}{1+\exp^{-\frac{\varepsilon_{-}^{\tau}-\mu_{\tau}}{T}}} \right)\,,\\
{\rm Im}(\beta_{1}^{(\tau)}(q,\omega,T))&=&\left( \frac{k_{-}^{\tau}}{q}\right) {\rm Im}(\beta_{0}^{(\tau)}(q,\omega,T))\,,\\
{\rm Im}(\beta_{2}^{(\tau)}(q,\omega,T))&=&\frac{(m^{*})^{3}}{2\pi q^3}T \left\{  \varepsilon_{-}^{\tau}  \ln \left(\frac{1+\exp^{-\frac{\varepsilon_{+}^{\tau}-\mu_{\tau}}{T}}  }{1+\exp^{-\frac{\varepsilon_{-}^{\tau}-\mu_{\tau}}{T}}}\right)-\int_{\varepsilon_{-}^{\tau}}^{\varepsilon_{+}^{\tau}}d \varepsilon \ln\left( 1+\exp^{-\frac{\varepsilon-\mu_{\tau}}{T}}\right)  \right\}\,,\\
{\rm Im}(\beta_{3}^{(\tau)}(q,\omega,T))&=&\left[\frac{k_{-}^{\tau}}{q} \right]^{2} {\rm Im}(\beta_{0}^{(\tau)}(q,\omega,T))\,,\\
{\rm Im}(\beta_{4}^{(\tau)}(q,\omega,T))&=&\frac{k_{-}^{\tau}}{q}{\rm Im}(\beta_{2}^{(\tau)}(q,\omega,T))\,,\\
{\rm Im}(\beta_{5}^{(\tau)}(q,\omega,T))&=&-\frac{(m^{*})^4}{\pi q^{5}} T \left\{  -(\varepsilon_{-}^{\tau })^2 \ln \left( \frac{1+\exp^{-\frac{\varepsilon_{+}^{\tau}-\mu_{\tau}}{T}}}{1+\exp^{-\frac{\varepsilon_{-}^{\tau}-\mu_{\tau}}{T}}} \right)\right. \nonumber\\
& &\left. \; \; \; \; \; \; \; \; \; \; \; \; \; \; \; \; \; \; +2\int_{\varepsilon_{+}^{\tau}}^{\infty}d\varepsilon  \; (\omega+\varepsilon) \ln (1+\exp^{-\frac{\varepsilon-\mu_{\tau}}{T}}) \right\}\,,\\
{\rm Im}(\beta^{(\tau)}_{6}(q,\omega,T))&=&\left(\frac{k_{-}^{\tau}}{q}\right)^{3} {\rm Im}(\beta^{(\tau)}_{0}(q,\omega,T))\,,\\
{\rm Im}(\beta^{(\tau)}_{7}(q,\omega,T))&=&\left(\frac{k_{-}^{\tau}}{q}\right)^{4} {\rm Im}(\beta^{(\tau)}_{0}(q,\omega,T))\,,\\
{\rm Im} (\beta_{8}^{(\tau)}(q,\omega,T))&=&\left(\frac{k_{-}^{\tau}}{q}\right)^{2}{\rm Im} (\beta_{2}^{(\tau)}(q,\omega,T))\,.
\end{eqnarray}

Finally, one has often to deal with energy-weighted sum rules. As already mentioned in the main text, some of these sum rules can be determined with an explicit calculation of a double commutator averaged on the ground state. In that context, expressions for $\beta_{i}^{(\tau,\tau')}$ for zero temperature are mandatory. In this particular limit, we can perform analytic calculations for both real and imaginary parts. In the general case where $\tau \neq \tau'$, imaginary parts can be cast into the following form 

\begin{eqnarray}
{\rm Im} \, \beta^{(\tau,\tau')}_i
&=& - \frac{1}{8\pi q B_{\tau}}   G_i (k) \bigg|^{{\rm Inf} \big[k_F^{\tau'},
\frac{q B_{\tau} + \sqrt{\Delta}}{|A|}\big]}_ { \left| \frac{q B_{\tau} - \sqrt{\Delta}}{A} \right|}\; \; \theta \left( {\rm Inf}\left(k_F^{\tau'},
\frac{q B_{\tau} + \sqrt{\Delta}}{|A|}\right) -  \left| \frac{q B_{\tau} - \sqrt{\Delta}}{A} \right| \; \right)\nnn
& & +  \frac{1}{8\pi q B_{\tau'}}   \tilde{G}_i(k) \bigg|_{ \left| \frac{q B_{\tau'} - \sqrt{\Delta}}{A} \right|}^{{\rm Inf} \big[k_F^{\tau},
\frac{q B_{\tau'} + \sqrt{\Delta}}{|A|}\big]} \; \; \theta \left( {\rm Inf}\left(k_F^{\tau},
\frac{q B_{\tau'} + \sqrt{\Delta}}{|A|}\right) -  \left| \frac{q B_{\tau'} - \sqrt{\Delta}}{A} \right| \; \right) \,,\nn
\end{eqnarray}
with $B_{\tau} = 1/(2 m_\tau)$, $A = B_{\tau'} - B_{\tau}$, $C_{\tau} = \tilde{\omega} - B_{\tau} q^2$, 
$C_{\tau'} = \tilde{\omega} + B_{\tau'} q^2$ and $\Delta_\tau = (B_\tau)^2 q^2 - A C_{\tau}$.
Notice that: $\Delta_{\tau'} = \Delta_\tau \equiv \Delta = B_{\tau} B_{\tau'} \left(q^2 - 2(m_{\tau}-m_{\tau'}) \tilde{\omega} \right)$. Functions $G_i$ and $\tilde{G}_i$ are listed in Table~\ref{app:tableFG}.

\begin{table}[H]
\begin{center}
\begin{tabular}{c|c|c|} \cline{2-3}
     \multicolumn{1}{c|}{}       &   &   \\ 
 \multicolumn{1}{c|}{}     & $G_i$  & $\tilde{G}_i$ \\ \hline
     \multicolumn{1}{|c|}{}       &   &   \\ 
 \multicolumn{1}{|c|}{0} & $\frac{k^2}{2}$ & $\frac{k^2}{2}$ \\ 
       \multicolumn{1}{|c|}{}      &   &   \\ \hline
     \multicolumn{1}{|c|}{}       &   &   \\ 
\multicolumn{1}{|c|}{1} & $\frac{A k^4+2 k^2 C_{\tau}}{8 q^2 B_{\tau}}$ & $\frac{k^2 \left(A k^2-4 q^2 B_{\tau'}+2 C_{\tau'}\right)}{8 q^2 B_{\tau'}}$ \\       \multicolumn{1}{|c|}{}      &   &   \\ \hline
     \multicolumn{1}{|c|}{}       &   &   \\ 
\multicolumn{1}{|c|}{2} & $\frac{k^4}{4 q^2}$ & $\frac{k^2 \left(-A k^2+B_{\tau'} \left(k^2+2 q^2\right)-2 C_{\tau'}\right)}{4 q^2 B_{\tau'}}$ \\       \multicolumn{1}{|c|}{}      &   &   \\ \hline
     \multicolumn{1}{|c|}{}       &   &   \\ 
\multicolumn{1}{|c|}{3} & $\frac{A^2 k^6+3 A k^4 C_{\tau}+3 k^2 C_{\tau}^2}{24 q^4 B_{\tau}^2}$ & $\frac{A^2 k^6+3 A k^4 \left(C_{\tau'}-2 q^2 B_{\tau'}\right)+3 k^2 \left(C_{\tau'}-2 q^2 B_{\tau'}\right){}^2}{24 q^4 B_{\tau'}^2}$ 
\\       \multicolumn{1}{|c|}{}      &   &   \\ \hline
     \multicolumn{1}{|c|}{}       &   &   \\ 
\multicolumn{1}{|c|}{4} & $\frac{2 A k^6+3 k^4 C_{\tau}}{24 q^4 B_{\tau}}$ & $-\frac{k^2 \left(2 A^2 k^4-B_{\tau'} \left(A k^2 \left(2 k^2+9 q^2\right)+3 C_{\tau'} \left(k^2+6 q^2\right)\right)+6 C_{\tau'} \left(A k^2+C_{\tau'}\right)+6 q^2 B_{\tau'}^2 \left(k^2+2 q^2\right)\right)}{24 q^4 B_{\tau'}^2}$ \\       \multicolumn{1}{|c|}{}      &   &   \\ \hline
     \multicolumn{1}{|c|}{}       &   &   \\ 
\multicolumn{1}{|c|}{5} & $\frac{k^6}{6 q^4}$ & $\frac{k^2 \left(3 k^2 \left(B_{\tau'}-A\right) \left(q^2 B_{\tau'}-C_{\tau'}\right)+k^4 \left(A-B_{\tau'}\right){}^2+3 \left(C_{\tau'}-q^2 B_{\tau'}\right){}^2\right)}{6 q^4 B_{\tau'}^2}$ \\       \multicolumn{1}{|c|}{}      &   &   \\ \hline
     \multicolumn{1}{|c|}{}       &   &   \\ 
\multicolumn{1}{|c|}{6} & $\frac{\left(A k^2+C_{\tau}\right){}^4-C_{\tau}^4}{64 A q^6 B_{\tau}^3}$ & $\frac{\left(A k^2-2 q^2 B_{\tau'}+C_{\tau'}\right){}^4-\left(C_{\tau'}-2 q^2 B_{\tau'}\right){}^4}{64 A q^6 B_{\tau'}^3}$ \\       \multicolumn{1}{|c|}{}      &   &   \\ \hline
     \multicolumn{1}{|c|}{}       &   &   \\ 
\multicolumn{1}{|c|}{7} & $\frac{\left(A k^2+C_{\tau}\right){}^5-C_{\tau}^5}{160 A q^8 B_{\tau}^4}$ & $\frac{k^2 \left(A^4 k^8-5 \left(2 q^2 B_{\tau'}-C_{\tau'}\right) \left(A k^2-2 q^2 B_{\tau'}+C_{\tau'}\right) \left(A^2 k^4-\left(2 q^2 B_{\tau'}-C_{\tau'}\right) \left(A k^2-2 q^2 B_{\tau'}+C_{\tau'}\right)\right)\right)}{160 q^8 B_{\tau'}^4}$ \\       \multicolumn{1}{|c|}{}      &   &   \\ \hline
     \multicolumn{1}{|c|}{}       &   &   \\ 
\multicolumn{1}{|c|}{8} & $\frac{3 A^2 k^8+8 A k^6 C_{\tau}+6 k^4 C_{\tau}^2}{96 q^6 B_{\tau}^2}$ & $-\frac{3 A^2 k^8 \left(A-B_{\tau'}\right) +4 A k^6 \left(C_{\tau'} \left(3 A-2 B_{\tau'}\right)+q^2 B_{\tau'} \left(4 B_{\tau'}-5 A\right)\right)}{96 q^6 B_{\tau'}^3}$ \\
 \multicolumn{1}{|c|}{}  & &$+\frac{6 k^4 \left(2 q^2 B_{\tau'}-C_{\tau'}\right) \left(C_{\tau'} \left(B_{\tau'}-3 A\right)-2 q^2 B_{\tau'} \left(B_{\tau'}-2 A\right)\right)-12 k^2 \left(q^2 B_{\tau'}-C_{\tau'}\right) \left(C_{\tau'}-2 q^2 B_{\tau'}\right){}^2}{96 q^6 B_{\tau'}^3}$ \\       \multicolumn{1}{|c|}{}      &   &   \\ \hline
\end{tabular}
\caption{Functions entering the imaginary part of $\beta_{i=0,..,8}^{(\tau,\tau')}$ at zero temperature.}
\label{app:tableFG}
\end{center}
\end{table}

When there is no asymmetry, we have checked that usual expressions (given by instance in \cite{gar92,dav09}) are recovered. By instance
\bea
{\rm Im} \, \beta^{(\tau,\tau')}_0 \bigg|_{\tau=\tau'} 
&=& - \frac{m_\tau (k_F^\tau)^2}{8 \pi q} \bigg\{ 
\theta[1-(\nu_\tau-k_\tau)^2]  \left[ 1 - (\nu_\tau-k_\tau)^2 \right] - 
\theta[1-(\nu_\tau+k_\tau)^2] \left[ 1 - (\nu_\tau+k_\tau)^2 \right] \bigg\}\,,
\nn
\eea
with the adimensional variables $k_\tau=\frac{q}{2k^\tau_{F}}$ and $\nu_\tau=\frac{m_\tau\omega}{qk^\tau _{F}}$.

For completeness we also give here the general structure of real parts
\bea
{\rm Re} \, \beta^{(\tau,\tau')}_i
&=& \alpha_i^{(1)} + \alpha_i^{(2)} + \beta_i^{(1)} \Theta(-\Delta)\left[ \arctan \left( \frac{ A k_F^{\tau'} - B_{\tau} q}{\sqrt{- \Delta}} \right)
 + \arctan \left( \frac{ A k_F^{\tau'} + B_{\tau} q}{\sqrt{-\Delta}} \right)
  \right]\nnn
& &  + \beta_i^{(2)} \Theta(-\Delta)\left[ \arctan \left( \frac{ A k_F^{\tau} - B_{\tau'} q}{\sqrt{- \Delta}} \right)
 + \arctan \left( \frac{ A k_F^{\tau} + B_{\tau'} q}{\sqrt{-\Delta}} \right)
  \right]\nnn
& & + \gamma_i^{(1)} \Theta(\Delta)
  \log\left|\frac{(\sqrt{\Delta}-A k_F^{\tau'})^2-q^2 B_{\tau})^2}{(\sqrt{\Delta}+A k_F^{\tau'})^2-q^2 B_{\tau}^2))}\right| + \gamma_i^{(2)} \Theta(\Delta)
  \log\left|\frac{(\sqrt{\Delta}-A k_F^{\tau})^2-q^2 B_{\tau'})^2}{(\sqrt{\Delta}+A k_F^{\tau})^2-q^2 B_{\tau'}^2))}\right|\nnn
  & & + \delta_i^{(1)}  \log \left|\frac{A (k_F^{\tau'})^2 + 2 B_{\tau} k_F^{\tau'} q + C_{\tau}}{A k_F^2(n) - 2 B_{\tau} k_F^{\tau'} q + C_{\tau}}\right|  + \delta_i^{(2)}  \log \left|\frac{A (k_F^{\tau})^2 + 2 B_{\tau'} k_F^{\tau} q + C_{\tau'}}{A k_F^2(p) - 2 B_{\tau'} k_F^{\tau} q + C_{\tau'}}\right|.  \nnn
\eea

The $i=0$ case reads
\begin{table}[H]
\begin{center}
\begin{tabular}{cccc}
$\alpha_0^{(1)} = \frac{k_{F}^{\tau'}}{4 \pi ^2 A}$, &
$\beta_0^{(1)} =  -\frac{\sqrt{-\Delta }}{4 \pi ^2 A^2}$, &
$\gamma_0^{(1)} =  \frac{\sqrt{\Delta }}{8 \pi ^2 A^2}$, &
$\delta_0^{(1)} =  \frac{2 q^2 B_\tau^2-A \left(A (k^{\tau'}_{F})^2+C_\tau\right)}{16 \pi ^2 A^2 q B_\tau}$, \\
$\alpha_0^{(2)} = -\frac{k_{F}^{\tau}}{4 \pi ^2 A}$, &
$\beta_0^{(2)} =  -\frac{\sqrt{-\Delta }}{4 \pi ^2 A^2}$, &
$\gamma_0^{(2)} =  \frac{\sqrt{\Delta }}{8 \pi ^2 A^2}$, &
$\delta_0^{(2)} =   \frac{A \left(A (k^\tau_{F})^2+C_{\tau'}\right)-2 q^2 B_{\tau'}^2}{16 \pi ^2 A^2 q B_{\tau'}}$. \\
\end{tabular}
\end{center}
\end{table}

For $i\neq0$, expressions are rather cumbersome and are not written here. 

The limit $\tau=\tau'$ is simpler. In this case, one can advantage of the analytic properties of generalized Lindhardt functions (see \cite{gar92} for notations)

\begin{equation}
\Pi_{2i}^\tau(q,\omega)=\int\frac{d^4 p}{(2\pi)^4} G^{(\tau)}_{HF}(p,\mathbf{q},\omega)\frac{1}{2}\left[\left(\frac{\mathbf{p}^2}{(k^\tau_F)^2}\right)^i+\left(\frac{(\mathbf{p}+\mathbf{q})^2}{(k^\tau_F)^2}\right)^i\right]\,.
\label{app:definition-Pi}
\end{equation}

\noindent These functions are tabulated for $i=0,1,2$ in \cite{gar92}. It is then a simple task to express 
$\beta^{(\tau)}_i$ as \cite{dav09}

\vfill\eject

\begin{eqnarray}
\beta^{(\tau)}_{0}&=&\Pi^{\tau}_{0}\,,\\
2k_\tau \; \beta^{(\tau)}_{1}&=&(\nu_\tau-k_\tau)\Pi^{\tau}_{0}\,,\\
4k_\tau^{2} \; \beta^{(\tau)}_{2}&=&\Pi^{\tau}_{2}-2k_\tau\nu_\tau\Pi^{\tau}_{0}\,,\\
4k_\tau^{2} \; \beta^{(\tau)}_{3}&=&(\nu_\tau-k_\tau)^{2}\Pi^{\tau}_{0}-\frac{m_\tau k^\tau_{F}}{6\pi^{2}}\,,\\
8k_\tau^{3} \; \beta^{(\tau)}_{4}&=&2k_\tau\nu_\tau(k_\tau-\nu_\tau)\Pi^{\tau}_{0}+(\nu_\tau-k_\tau)\Pi^{\tau}_{2}+\frac{m_\tau k^\tau_{F}}{3\pi^{2}}k_\tau\,,\\
16k_\tau^{4} \; \beta^{(\tau)}_{5}&=&\Pi^{\tau}_{4}-4k_\tau\nu_\tau\Pi^{\tau}_{2}\,,\\
8k_\tau^{3} \; \beta^{(\tau)}_{6}&=&(\nu_\tau-k_\tau)^{3}\Pi^{\tau}_{0}+(3k_\tau-\nu_\tau)\frac{m_\tau k^\tau_{F}}{6\pi^{2}}\,,\\
16k_\tau^{4} \; \beta^{(\tau)}_{7}&=&(\nu_\tau-k_\tau)^{4}\Pi^{\tau}_{0}-\frac{m_\tau k^\tau_{F}}{2\pi^{2}}\left[ k_\tau^{2}+\frac{1}{5}+\frac{1}{3}(2k_\tau-\nu_\tau)^{2}\right]\,,\\
16k_\tau^{4} \; \beta^{(\tau)}_{8}&=&(\nu_\tau-k_\tau)^{2}\Pi^{\tau}_{2}-2k_\tau\nu_\tau(\nu_\tau-k_\tau)^{2}\Pi^{\tau}_{0}
-\frac{m_\tau k^\tau_{F}}{6\pi^{2}}\left[1+2k_\tau(3k_\tau-\nu_\tau) \right]\,.
\end{eqnarray}

Since $\beta_{i}^{(\tau,\tau')}$ are the basic ingredients of response functions, we now come to a rapid description and discussion on the influence on several parameters as temperature,  transfer momentum and asymmetry.
On figures \ref{app:imbeta-T} are displayed respectively the imaginary part and the real part of $\beta_{i=0,..,8}^{(\tau)}$ for different temperatures ($T=0, 0.25$ and $0.5 \varepsilon_F$) for $q=k_F$ and $\rho=0.16$ fm$^{-3}$. We clearly see that the deviation from the zero-temperature limit is small even for temperatures of order of the evaporation energy. The only effect is, as expected, a relative widening. This implies of course that the effect of temperature on response functions is expected to be moderate.

On the contrary, the effect of the transfer momentum is important as shown in \ref{app:imbeta-q}: both the shape and the amplitude are modified significantly. The detailed structure appear to be largely affected becoming smoother and smoother with increasing transfer momentum. This evolution has of course some consequences on response functions themselves and the general features described here will be also observed in different situations.

Finally, we illustrate the effect of asymmetry for typical values of the transferred momentum ($q=k_F$) and density ($\rho=0.16$ fm$^{-3}$) at zero temperature on figures \ref{app:imbeta-y}. Here again the effect is sizable, leading to a two-peak structure that evolves for increasing value of $Y$. An interesting point is that the evolution is continuous: all the curves go from SNM case to PNM case when we vary $Y$. This is of course reasonable, but at the same time, it explains partially why asymmetric matter can be understood, from the response functions or critical densities point of view, as an intermediate state between SNM and PNM.


\section{Landau parameters}
\label{app:landau}
\setcounter{table}{0}

In this Appendix are given the Landau parameters for the effective finite-range interactions of Nakada \cite{nak03}, and Gogny \cite{dec80}, currently used in mean-field calculations. One has to write the $ph$ interaction in momentum space, set the transferred momentum $q=0$, put the hole momenta ${\bf k}_{1,2}$ on the Fermi surface, project on Legendre polynomials $P_{\ell}({\bf {\hat k}}_1 \cdot {\bf {\hat k}}_2)$, and identify the Landau parameters by comparing to Eq. (\ref{landau-Vph}). We only give the final expressions. 

For completeness, we give in Table \ref{app:lan-sky} the Landau parameters deduced from the standard Skyrme interaction. The contribution of the density-dependent term  is the same for all these three interactions, apart from 
a factor of six in the definition of Nakada and Gogny interactions.  

\begin{table}[h]
\begin{center}
\begin{tabular}{c|ccc}
\hline
$(\alpha)$ & $f_0^{(\alpha)}+f_1^{(\alpha)}$ & $f_1^{(\alpha)}$ & $h_0^{(\alpha)}$ \\
\hline
(0,0) & $\frac{3}{4} t_0 + \frac{1}{16} (\gamma+1)(\gamma+2) t_3 \rho^{\gamma}$ 
& $- \frac{1}{8} k_F^2 \left( 3 t_1 + (5+4 x_2) t_2 \right)$ & --  \\
(1,0) & $ -\frac{1}{4} (1-2 x_0) t_0 - \frac{1}{24} (1- 2x_3) t_3 \rho^{\gamma}$ 
& $- \frac{1}{8} k_F^2 \left( - (1- 2x_1) t_1 + (1+2 x_2) t_2 \right)$ & 
$ \frac{1}{8} k_F^2 \left( t_e+3 t_o \right)$  \\
(0,1) & $-\frac{1}{4} (1+ 2 x_0) t_0 - \frac{1}{24} (1 + 2 x_3) t_3 \rho^{\gamma} $ 
& $- \frac{1}{8} k_F^2 \left( -(1+ 2 x_1) t_1 + (1+2 x_2) t_2 \right)$ & --  \\
(1,1) & $ -\frac{1}{4} t_0 - \frac{1}{24} t_3 \rho^{\gamma} $ 
& $- \frac{1}{8} k_F^2 \left( -t_1+t_2 \right) $ & $\frac{1}{8} k_F^2 \left( -t_e+t_o \right)$  \\
\hline
(0,n) & $\frac{1}{2} (1-x_0) t_0 + \frac{1}{24} (\gamma+1)(\gamma+2) (1-x_3) t_3 \rho^{\gamma}$ 
 & $- \frac{1}{4} k_F^2 \left( (1-x_1) t_1 + 3 (1+ x_2) t_2 \right)$ & -- \\
(1,n) & $\frac{1}{2} (-1+ x_0) t_0 + \frac{1}{12} (-1 + x_3) t_3 \rho^{\gamma} $ 
& $- \frac{1}{4} k_F^2 \left( -(1-x_1) t_1+ (+x_2) t_2 \right) $ & $\frac{1}{2} k_F^2 t_o$ \\
\hline
\end{tabular}
\end{center}
\caption{Landau parameters for the standard Skyrme interaction}
\label{app:lan-sky}
\end{table}

The two-body Nakada interaction contains central, density-dependent, tensor and spin-orbit parts. As the latter one does not contribute to Landau parameters, one has to consider only the following contributions 
\bea
V^{(C)}_{12} &=& \sum_n \left[ 
t_n^{(SE)} P_{SE} + t_n^{(TE)} P_{TE} + t_n^{(SO)} P_{SO} + t_n^{(TO)} P_{TO}
\right] f_n^{(C)}(r_{12})\,,  \\
V^{(TN)}_{12} &=& \sum_n \left[ 
t_n^{(TNE)} P_{TE} + t_n^{(TNO)} P_{TO}
\right] f_n^{(TN)}(r_{12}) \, r^2_{12} S_T(\hat{r}_{12}) \label{nak-tensor} \,, \\
V^{(DD)}_{12} &=& t^{(DD)} \left( 1 + x^{(DD)} P_{\sigma} \right) 
\rho^{\gamma} \delta(r_{12})  \,.
\eea
The interaction is written in terms of spin singlet/triplet and space even/odd projectors, and the tensor operator is 
\bea
S_T(\hat{r}_{12}) &=& 3 (\vec{\sigma}_1 \cdot \hat{r}_{12})
(\vec{\sigma}_2 \cdot \hat{r}_{12}) - (\vec{\sigma}_1 \cdot \vec{\sigma}_2 ) \,.
\eea
For the sake of simplicity, in the following we omit indices $n$ in the central and tensor contributions. A sum over $n$ is to be understood in the final expressions.   

The Landau parameters can be written as
\bea
f_{\ell}^{(\alpha)} &=& \delta(\ell,0) D^{(\alpha)} F_C(0) + E_C^{(\alpha)} J^C_{\ell} + \delta(\ell,0) R^{(\alpha)} 
\label{landau-f} \,, \\
h_{\ell}^{(\alpha)} &=& E_T^{(\alpha)} J^T_{\ell} \,.
\label{landau-h}
\eea
We have defined the following functions
\bea
F_C(q) &=& 4 \pi \int {\rm d}r \, r^2 j_0(qr) f^C(r) \,, \\
F_T(q) &=& 4 \pi \int {\rm d}r \, r^4 j_2(qr) f^{TN}(r)  \,,\\
J^C_{\ell} &=& \frac{2 \ell +1}{2} \int_{-1}^1 {\rm d}x P_{\ell}(x)
F_C\left( \sqrt{2 k_F^2 (1-x)}\right) \,, \\
J^T_{\ell} &=& \frac{2 \ell +1}{2} \int_{-1}^1 {\rm d}x P_{\ell}(x) \frac{1}{2(1-x)} 
F_T\left( \sqrt{2 k_F^2 (1-x)}\right)  \,.
\eea

The coefficients $D$ and $E$ comes from the direct and exchange contributions to the $ph$ interaction, respectively. Notice that the tensor parameters results only from the exchange contribution. These coefficients are given in Table \ref{app:nakada} in terms of the interaction parameters. The coefficient $R$ comes from the density-dependent contribution, and it is immediately deduced from Table \ref{app:lan-sky}, by replacing $t_3 \to 6 t_{DD}, x_3 \to x_{DD}$.  

\begin{table}[h]
\begin{center}
\begin{tabular}{c|ccc}
\hline
$(\alpha)$ & $D^{(\alpha)}  $ & $E_C^{(\alpha)} $ & $E_T^{(\alpha)} $ \\
\hline
(0,0) &  $  3 t^{(SE)} + 3 t^{(TE)} + t^{(SO)} + 9 t^{(TO)} $ & $ 3 t^{(SE)} + 3 t^{(TE)} 
-  t^{(SO)} - 9 t^{(TO)} $  & - \\
(1,0) & $- 3 t^{(SE)} +  t^{(TE)} -  t^{(SO)} + 3 t^{(TO)} $ & $  -3 t^{(SE)} +  t^{(TE)} 
+  t^{(SO)} - 3 t^{(TO)}  $ &   $-\frac{1}{4} t^{TNE} + \frac{3}{4} t^{TNO}$\\
(0,1) & $  t^{(SE)} - 3 t^{(TE)} - t^{(SO)} + 3 t^{(TO)} $ & $  t^{(SE)} - 3 t^{(TE)} 
+  t^{(SO)} - 3 t^{(TO)} $  & - \\ 
(1,1) & $ -  t^{(SE)} -  t^{(TE)} +  t^{(SO)} +  t^{(TO)} $ & $  - t^{(SE)} -  t^{(TE)} 
-  t^{(SO)} -  t^{(TO)} $ & $ \frac{1}{4} t^{TNE} + \frac{1}{4} t^{TNO}$ \\
\hline
(0,n) & $ \frac{1}{4} t^{(SE)} + \frac{3}{4} t^{(TO)}  $ & $ \frac{1}{4} t^{(SE)} - \frac{3}{4} t^{(TO)} $  & -   \\
(1,n) & $ - \frac{1}{4}  t^{(SE)} + \frac{1}{4}  t^{(TO)} $ & $ - \frac{1}{4}  t^{(SE)} - \frac{1}{4}  t^{(TO)}$  & $ t^{TNO}$ \\
\hline
\end{tabular}
\end{center}
\caption{Combinations $D^{(\alpha)}$ and $E_{C,T}^{(\alpha)}$ for Nakada interaction.}
\label{app:nakada}
\end{table}

Both radial parts are taken as Yukawa functions ${\rm e}^{- r \mu} / (r \mu)$, and then   
\bea
F_C(q) &=& \frac{4 \pi}{\mu} \frac{1}{\mu^2+q^2}\,, \nonumber \\
F_T(q) &=&  \frac{32 \pi}{\mu} \frac{q^2}{\left( \mu^2+q^2 \right)^3}\,. \nonumber
\eea
We give the first multipoles $J^C_{\ell}$ and $J^T_{\ell}$
\bea
J^C_0 &=& \frac{\pi}{\mu^3 z^2} \ln (1+4 z^2)\,, \\
J^C_1 &=& \frac{3 \pi}{2 \mu^3 z^4} \left\{ - 4 z^2 + (1+2z^2) \ln (1+4z^2) \right\} \,,\\
J^C_2 &=& \frac{5 \pi}{8 \mu^3 z^6} \left\{ -12 z^2 (1+2 z^2) +
(3 + 12 z^2 + 8 z^4) \ln (1+4 z^2) \right\}\,,\\
J^T_0 &=& \frac{64 \pi}{\mu^5} \frac{z^2}{(1+4 z^2)^2} \,,\\
J^T_1 &=& \frac{12 \pi}{ \mu^5 z^4} \left\{  4 z^2 \frac{1+6z^2+4z^4}{(1+4z^2)^2}- \ln (1+4z^2) \right\} \,,
\eea
with $z=k_F/\mu$.

We turn now to the standard Gogny interaction, whose central and density-dependent terms are written as
\bea
V_{12}^{(C)} &=& \sum_n \left\{ W_n + B_n P^{\sigma}_{12} - H_n P^{\tau}_{12} 
- M_n P^{\sigma \tau}_{12} \right\} f_n^C(r_{12}) \,, \\
V^{(DD)}_{12} &=& t_3 \left( 1 + x_3 P_{\sigma} \right)  \rho^{\gamma} \delta(r_{12}) \,.
\eea
We add a tensor contribution as in \cite{ang12}
\bea
V^{TN}_{12} = \left\{ V_{T1} + V_{T2} P_{\tau}
\right\} V_T(r_{12}) \,  S_T(\hat{r}_{12})  \,.
\label{gog-tensor}
\eea

The coefficients $D$ and $E$ are given in Table \ref{app:gogny}. The coefficient $R$  comes from the density-dependent contribution, and it is immediately deduced from Table \ref{app:lan-sky}, by simply multiplying $t_3$ by a factor of 6.  

\begin{table}[h]
\begin{center}
\begin{tabular}{c|ccc}
\hline
$(\alpha)$ & $D^{(\alpha)} $ & $E_C^{(\alpha)} $  & $E_T^{(\alpha)} $ \\
\hline
(0,0) & $ W+ \frac{1}{2} B- \frac{1}{2}  H- \frac{1}{4} M$ & $ - \frac{1}{4} W- \frac{1}{2}  B+ \frac{1}{2}  H+ M$  & - \\
(1,0) & $ \frac{1}{2} B- \frac{1}{4} M$ & $- \frac{1}{4} W+ \frac{1}{2} H$ & $\frac{1}{2} V_{T1} + V_{T2}$\\
(0,1) & $- \frac{1}{2} H - \frac{1}{4} M$ & $- \frac{1}{4} W- \frac{1}{2} B$ & - \\
(1,1) & $-  \frac{1}{4} M$ & $- \frac{1}{4} W$ & $\frac{1}{2} V_{T1}$  \\
\hline
(0,n) & $ W+ \frac{1}{2} B- H- \frac{1}{2} M$ & $-  \frac{1}{2} W - B + \frac{1}{2}  H + M$ & - \\
(1,n) &$ \frac{1}{2} B- \frac{1}{2} M$ & $- \frac{1}{2} W + \frac{1}{2} H$ &  $ V_{T1} + V_{T2}$ \\
\hline
\end{tabular}
\end{center}
\caption{Combinations $D^{(\alpha)}$ and $E_{C,T}^{(\alpha)}$ for Gogny interaction. Indices $n$ are omitted for simplicity.}
\label{app:gogny}
\end{table}

Both central and tensor radial parts are taken as Gaussian functions ${\rm e}^{- r^2/ \mu^2}$. The contribution of the central parts is straightforward, with the results   
\bea
F_C(q) &=& \pi \sqrt{\pi} \mu^3 {\rm e}^{-\frac{1}{4} q^2 \mu^2} \,,\nonumber \\
J^C_0 &=& \frac{\pi \sqrt{\pi} \mu^3}{z^2} \left[ 1 - {\rm e}^{-z^2} \right] \,, \\
J^C_1 &=& \frac{3 \pi \sqrt{\pi} \mu^3}{z^2} \left[ \left( 1 - \frac{2}{z^2} \right) + \left( 1 + \frac{2}{z^2} \right) {\rm e}^{-z^2} \right] \,, \\
J^C_2 &=& \frac{5 \pi \sqrt{\pi} \mu^3}{z^2} \left[ \left( 1 - \frac{6}{z^2} + \frac{12}{z^4} \right) - \left( 1 + \frac{6}{z^2} + \frac{12}{z^4} \right) {\rm e}^{-z^2} \right] \,.
\eea
The tensor part is a little bit more complicated as compared to that of Nakada. It is no longer possible to give compact analytical expressions, because due to the Gaussian radial functions the final expressions involve integrals of error functions, which are best to calculate numerically. We give now some hints for the calculation. First of all, notice that due to the $r^2$ factor missing in the definition (\ref{gog-tensor}) as compared to (\ref{nak-tensor}), the function $F_T(q)$ is properly defined as
\be
F_T(q) = 4 \pi \int {\rm d}r \, r^2 j_2(qr) f^{TN}(r) \,.
\ee
The tensor part of the $ph$ interaction is
\be
\left\{ E_T^{(\alpha)} \, \frac{k_F^2}{{\bf k}_{12}^2} \bf F_T(k_{12}) \right\}_{{\bf k}_i=k_F}
 \, \frac{{\bf k}^2_{12}}{k_F^2} S_T({\hat k}_{12})\,.
\ee
The expansion in Legendre polynomials of the factor in curved brackets provides us with the Landau tensor parameters. The function $J_{\ell}^T$ entering (\ref{landau-h}) is 
\be
J^T_{\ell} =  (2 \ell+1) \int_{-1}^1 {\rm d}x \, P_{\ell}(x) \frac{2 \pi}{1-x} \, \int_0^{\infty} {\rm d} r \, r^2 
j_2( r \sqrt{2 k_F^2(1-x)}) F^{TN}(r) \,.
\ee
For the tensor interaction introduced in \cite{ang11}, based on a regularized Argonne potential, the Landau parameters are obtained by numerically calculating this double integral. In the case of a Gaussian form for the function $f^{(TN)}$, the integral can be written as
\be
J^T_{\ell} =  (2 \ell+1) \pi \mu^3 \int_{-1}^1 {\rm d}x \, \frac{P_{\ell}(x)}{1-x} \, \int_0^{\infty} {\rm d} y \, y^2 
j_2( y \sqrt{2 z(1-x)}) {\rm e}^{-y^2} \,,
\ee
with $z= k_F \mu$.

\bibliographystyle{phcpc}
\bibliography{biblio}

\end{document}